\documentstyle[astrobib,times,psfig]{mn_daw}

\def\figegs{
\begin{figure*}
	\parbox{0.49\textwidth}{
		\psfig{figure=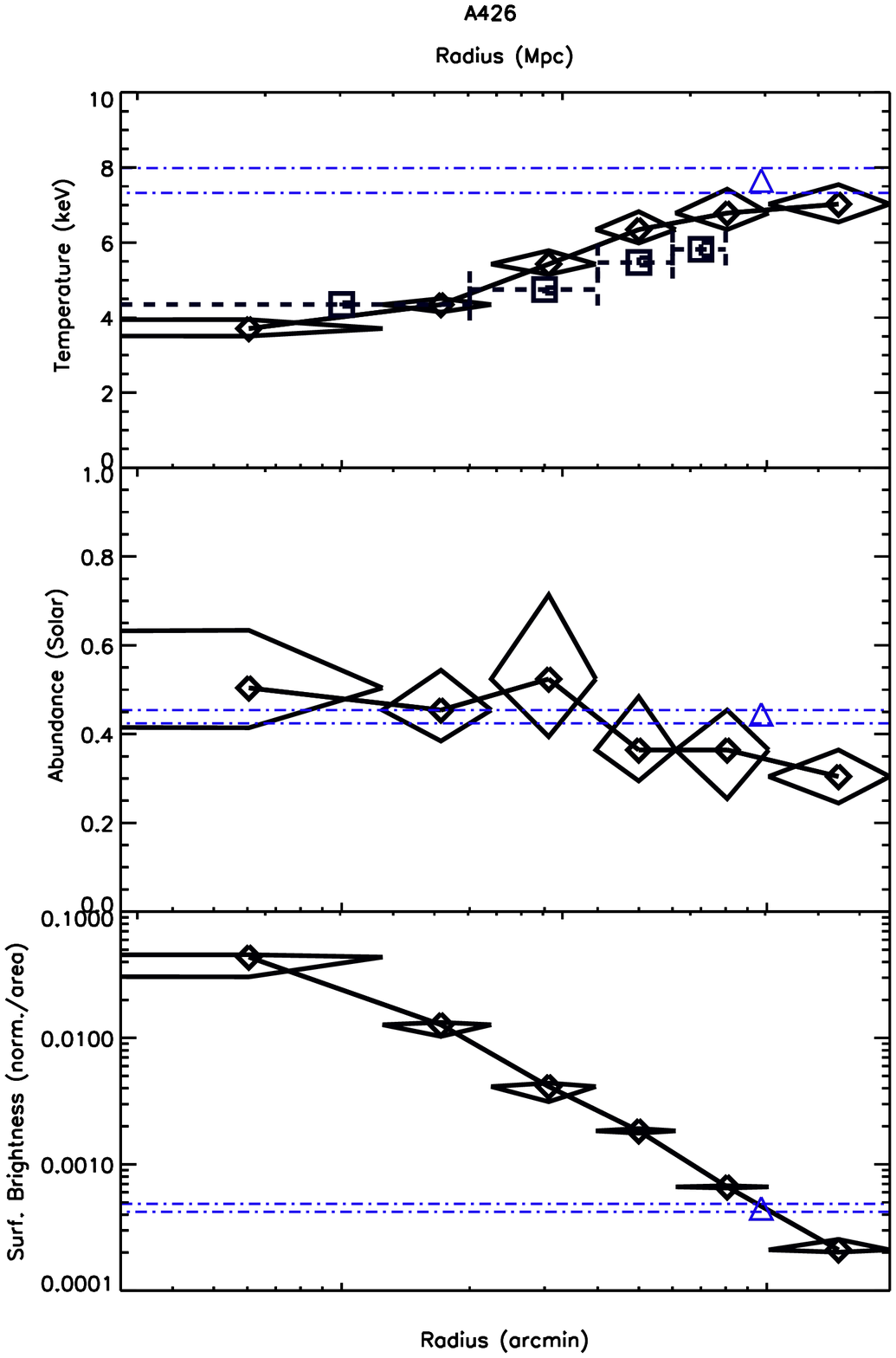,angle=0,width=0.49\textwidth,height=0.79\textheight}
		\figsmallfont \normalsize } \parbox{0.49\textwidth}{
		\psfig{figure=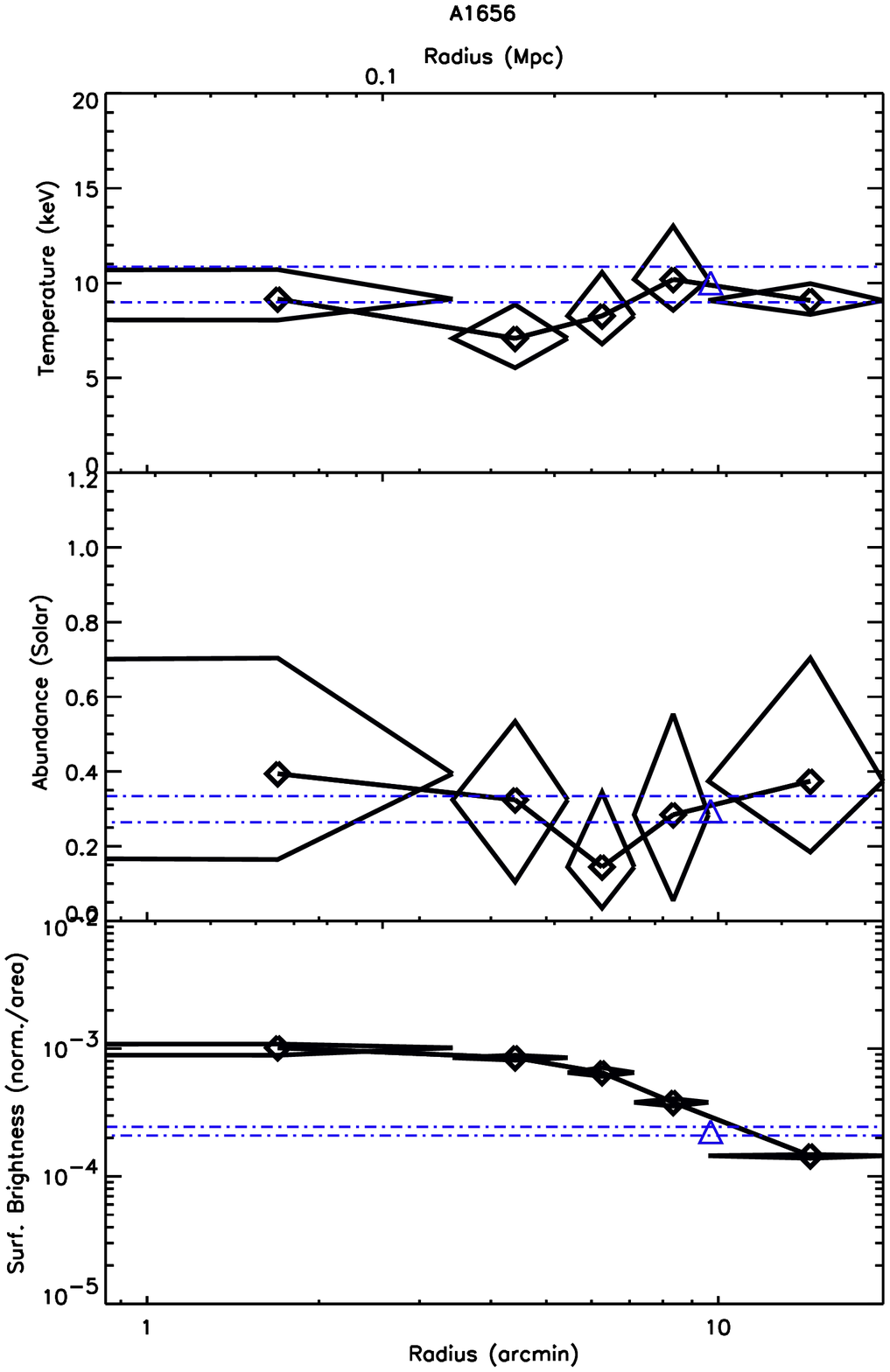,angle=0,width=0.49\textwidth,height=0.79\textheight}
		\figsmallfont \normalsize } \parbox{0.99\textwidth}{
		\hfill\parbox{0.95\textwidth}{
		\captionfont\caption{\label{figure:figegs} This figure
		presents example of the spectral analysis results on a
		cooling flow cluster, A426 (Perseus), and non
		cooling-flow A1656 (Coma Berenices). The heavy
		solid-lines are the average of the spectral fit
		results for a single-phase plasma applied to the GIS2
		and GIS3 deconvolution results. The single triangle
		data-points, with error-bars which span the whole
		radius range, are the the cooling-flow spectral fits
		to the non-deconvolved data. The additional
		temperature profile (square symbols) for A426 is from
		Beppo-SAX (see the end of
		Section~\ref{section:spectral} for the reference).}}}
		\normalsize
\end{figure*}
}

\def\figcmp{
\begin{figure}
	\parbox{0.49\textwidth}{
		\psfig{figure=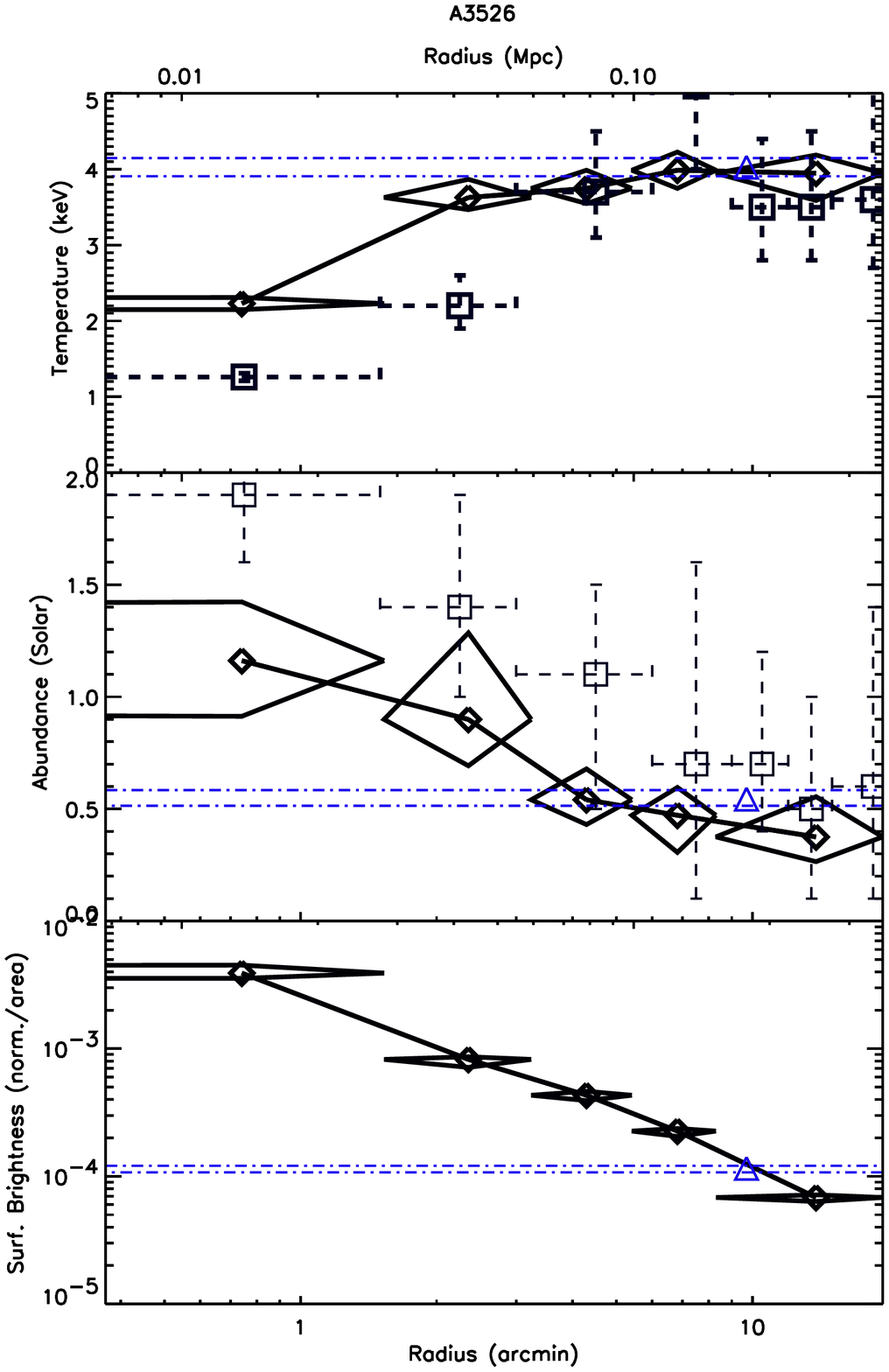,angle=0,width=0.49\textwidth,height=0.79\textheight}
		\parbox{0.45\textwidth}{
		\captionfont\caption{\label{figure:figcmp} This figure
		shows the detailed results for A3526 (Centaurus
		cluster).  Additional temperature and abundance
		profiles from \ROSAT\ are shown with square symbols
		(see the end of Section~\ref{section:spectral} for the
		reference).}}} \normalsize
\end{figure}
}

\def\figiso{
\begin{figure*}
	\parbox{0.49\textwidth}{
		\psfig{figure=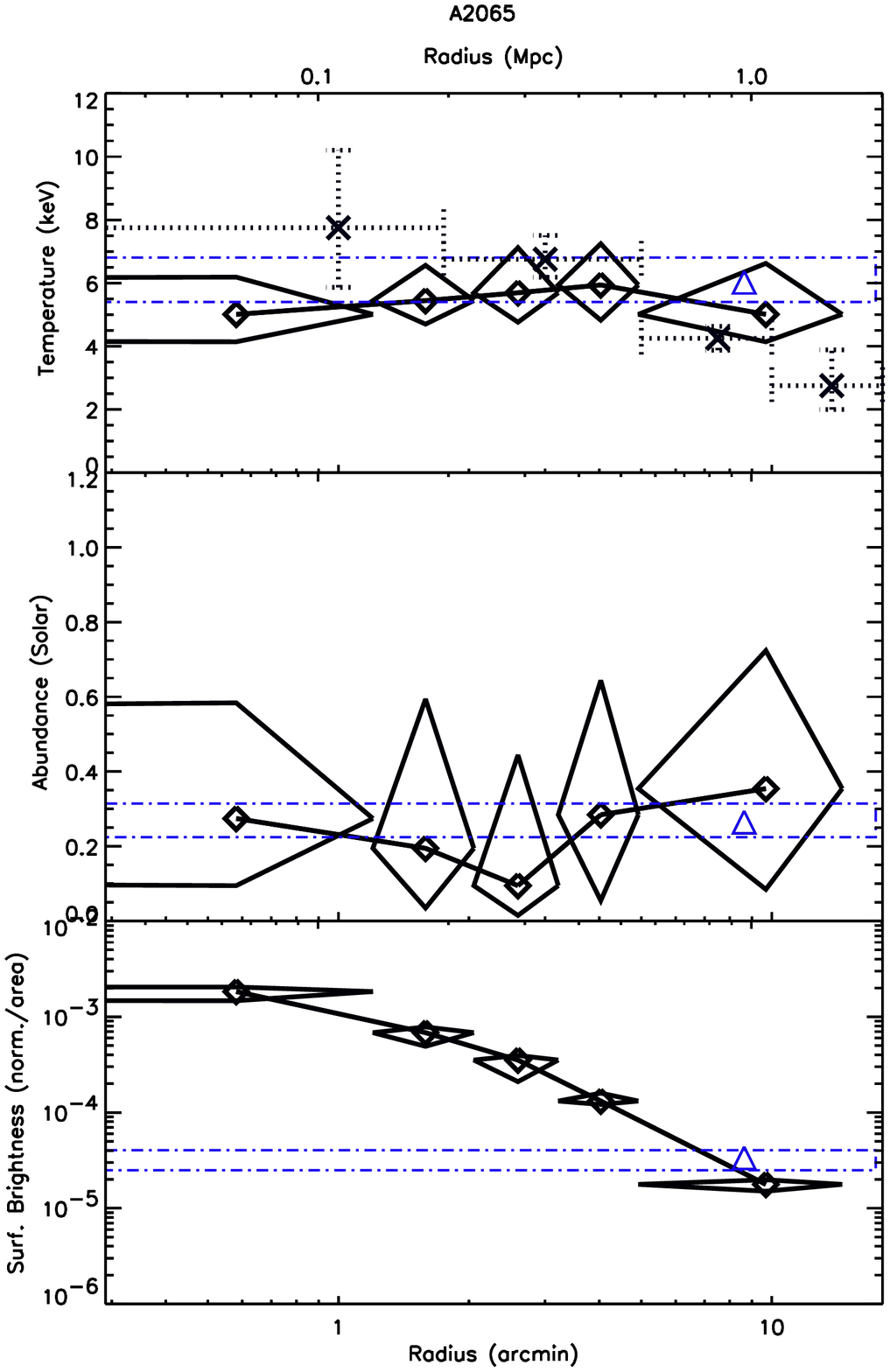,angle=0,width=0.49\textwidth,height=0.79\textheight}
		\figsmallfont \normalsize } \parbox{0.49\textwidth}{
		\psfig{figure=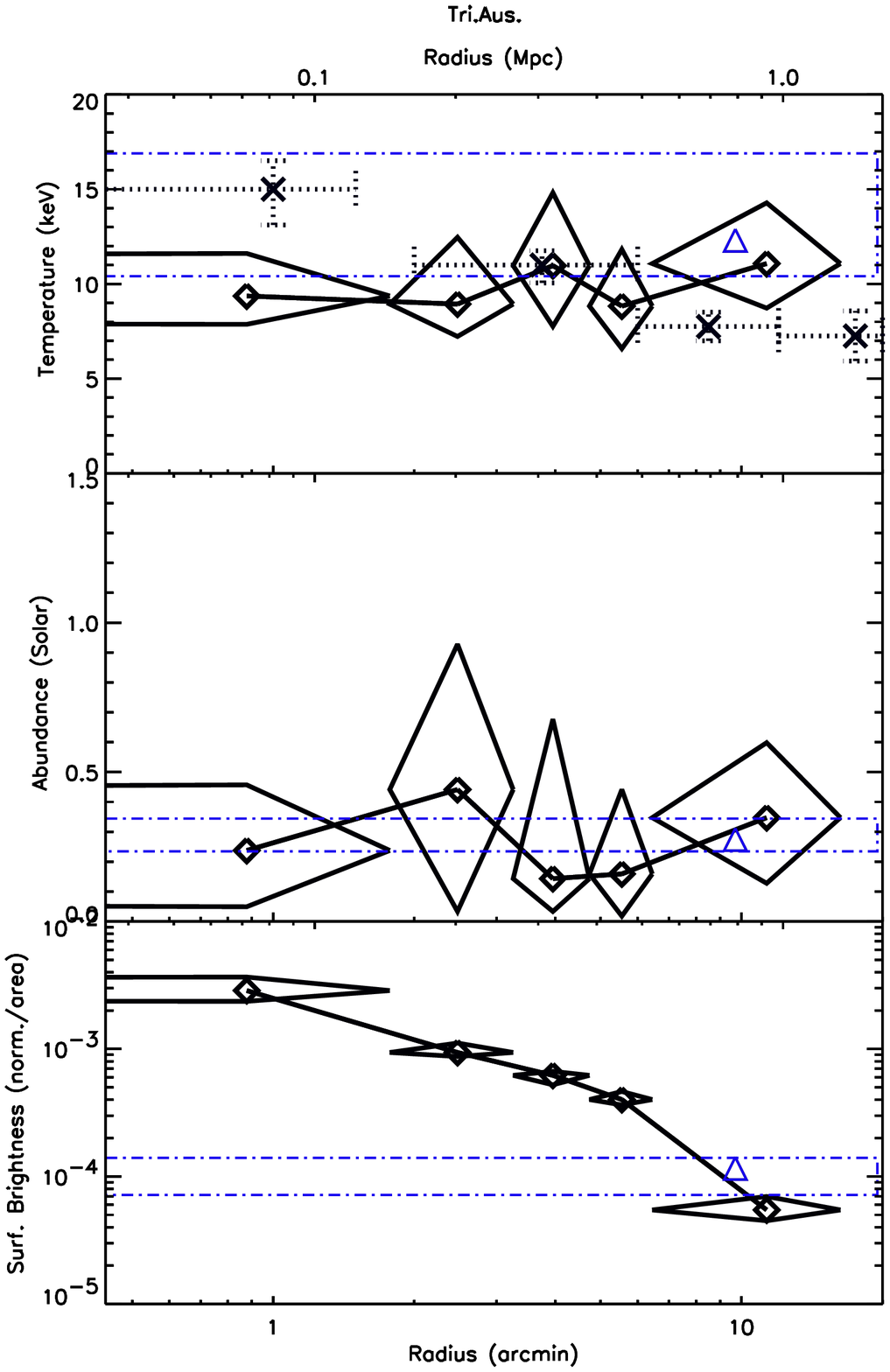,angle=0,width=0.49\textwidth,height=0.79\textheight}
		\figsmallfont \normalsize } \parbox{0.99\textwidth}{
		\hfill\parbox{0.95\textwidth}{
		\captionfont\caption{\label{figure:figiso} This figure
		shows examples of the discrepancy between our results
		and those presented by \MFSV\ for the A2065 and
		Triangulum Australis clusters. Our results indicate
		isothermal temperature profiles, contrary to the
		\MFSV\ results which show temperature declines (which
		are shown by the cross symbols with dotted
		error-bars). }}} \normalsize
\end{figure*}
}

\def\figegs{
\begin{figure*}
	\parbox{0.49\textwidth}{
		\psfig{figure=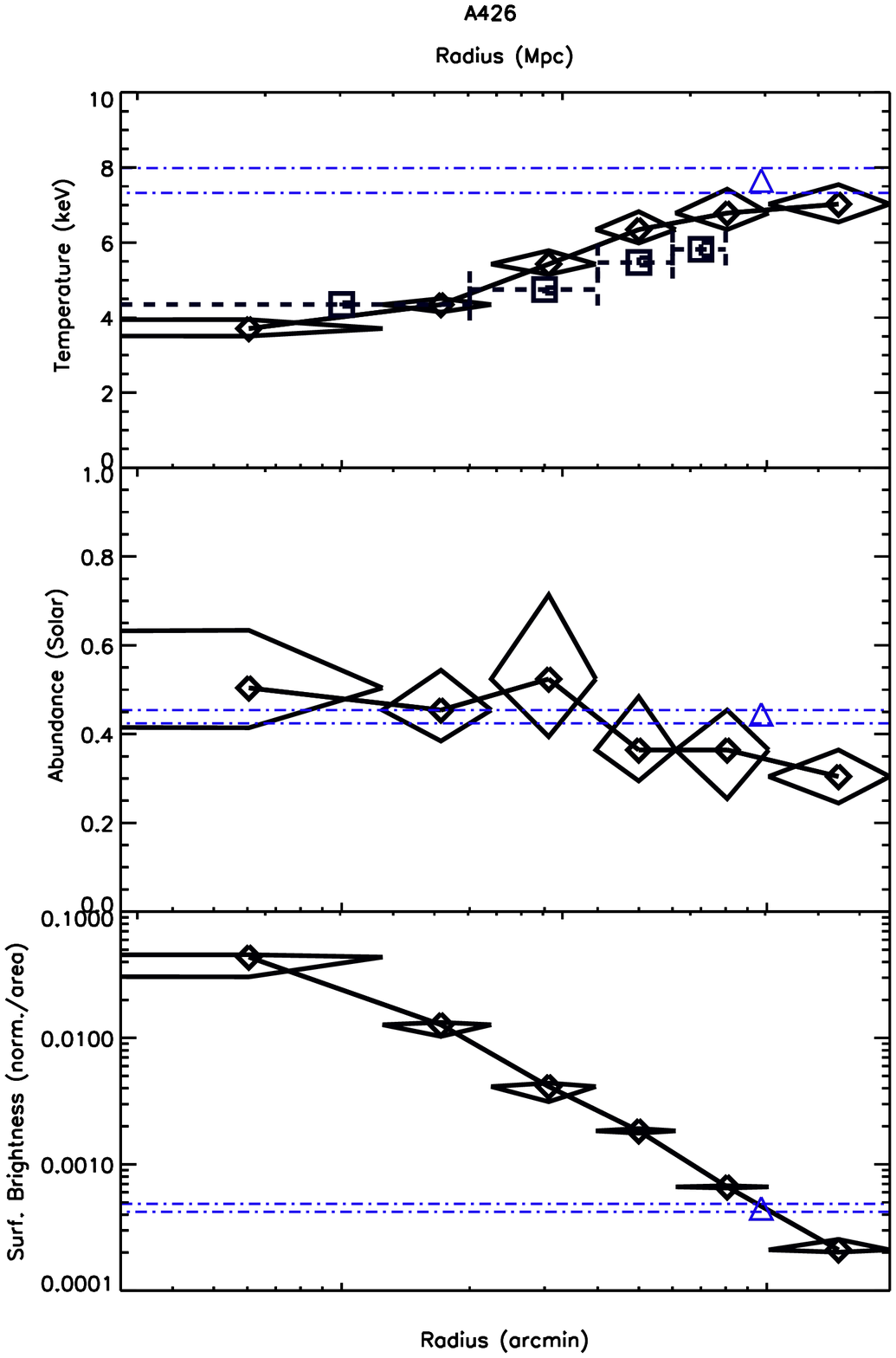,angle=0,width=0.49\textwidth,height=0.79\textheight}
		\figsmallfont \normalsize } \parbox{0.49\textwidth}{
		\psfig{figure=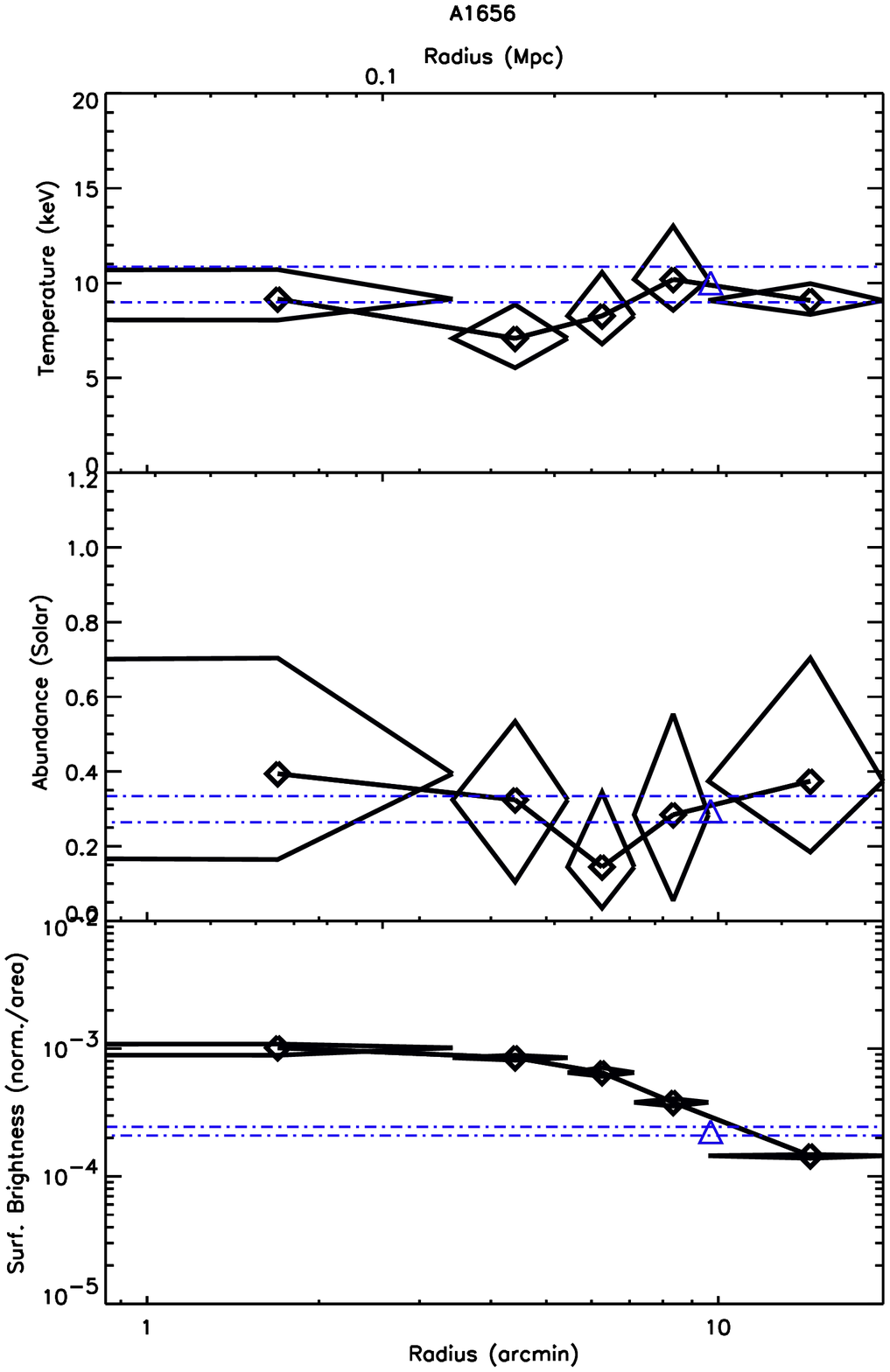,angle=0,width=0.49\textwidth,height=0.79\textheight}
		\figsmallfont \normalsize } \parbox{0.99\textwidth}{
		\hfill\parbox{0.95\textwidth}{
		\captionfont\caption{\label{figure:figegs} This figure
		presents example of the spectral analysis results on a
		cooling flow cluster, A426 (Perseus), and non
		cooling-flow A1656 (Coma Berenices). The heavy
		solid-lines are the average of the spectral fit
		results for a single-phase plasma applied to the GIS2
		and GIS3 deconvolution results. The single triangle
		data-points, with error-bars which span the whole
		radius range, are the the cooling-flow spectral fits
		to the non-deconvolved data. The additional
		temperature profile (square symbols) for A426 is from
		Beppo-SAX (see the end of
		Section~\ref{section:spectral} for the reference).}}}
		\normalsize
\end{figure*}
}

\def\figsax{
\begin{figure}
	\parbox{0.49\textwidth}{
		\psfig{figure=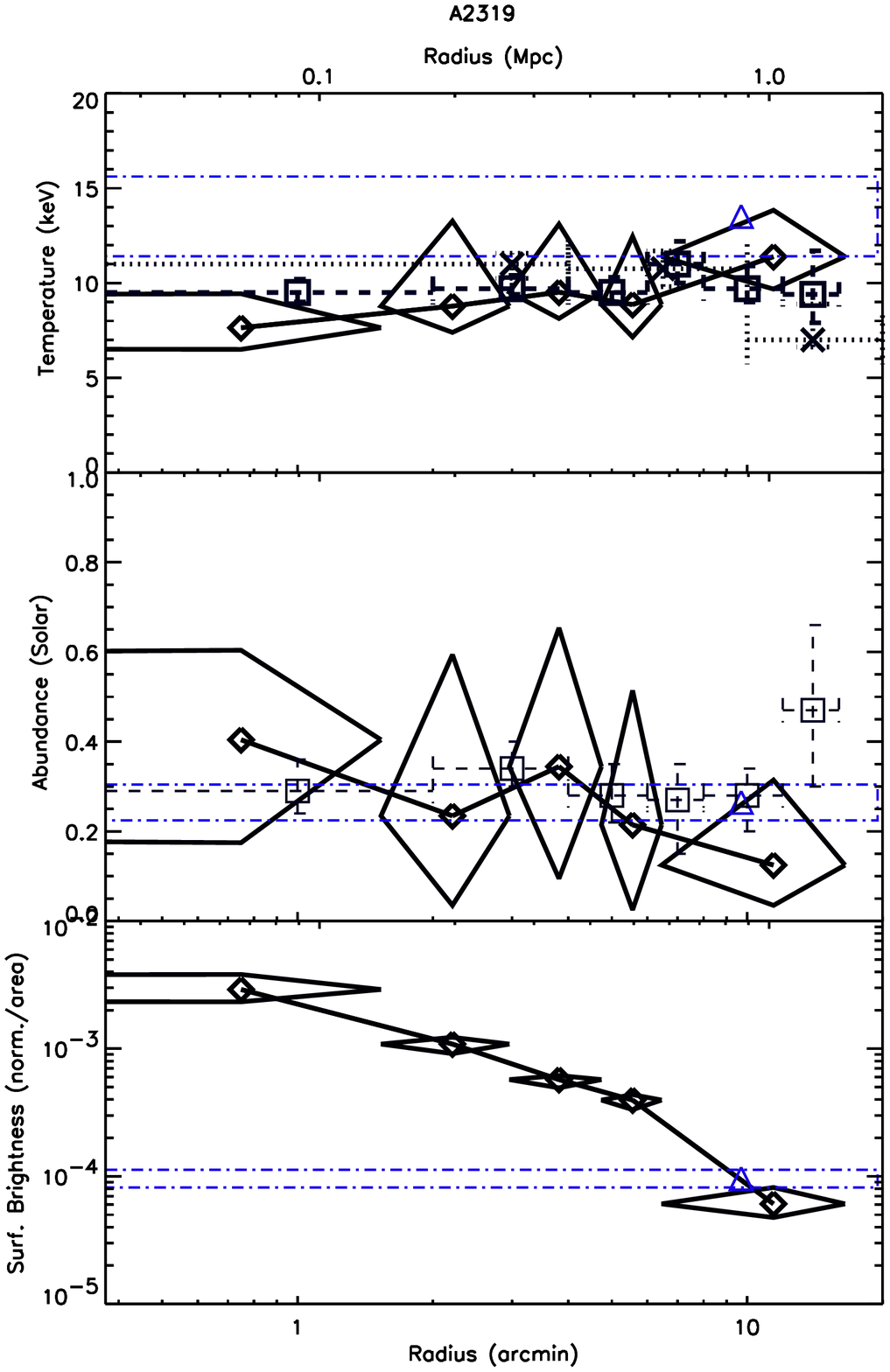,angle=0,width=0.49\textwidth,height=0.79\textheight}
		\parbox{0.45\textwidth}{
		\captionfont\caption{\label{figure:figsax} This figure
		compares the results from our \SID\ analysis (solid
		line, triangle symbols), those of Markevitch \etal\
		(dotted line, 3 cross symbols for temperature
		profile), and the Beppo-SAX results from Molendi
		\etal\ (dashed line, 6 square symbols for temperature
		and abundance profiles).  }}} \normalsize
\end{figure}
}

\def\figavg{
\begin{figure*}
	\parbox{0.59\textwidth}{
		\psfig{figure=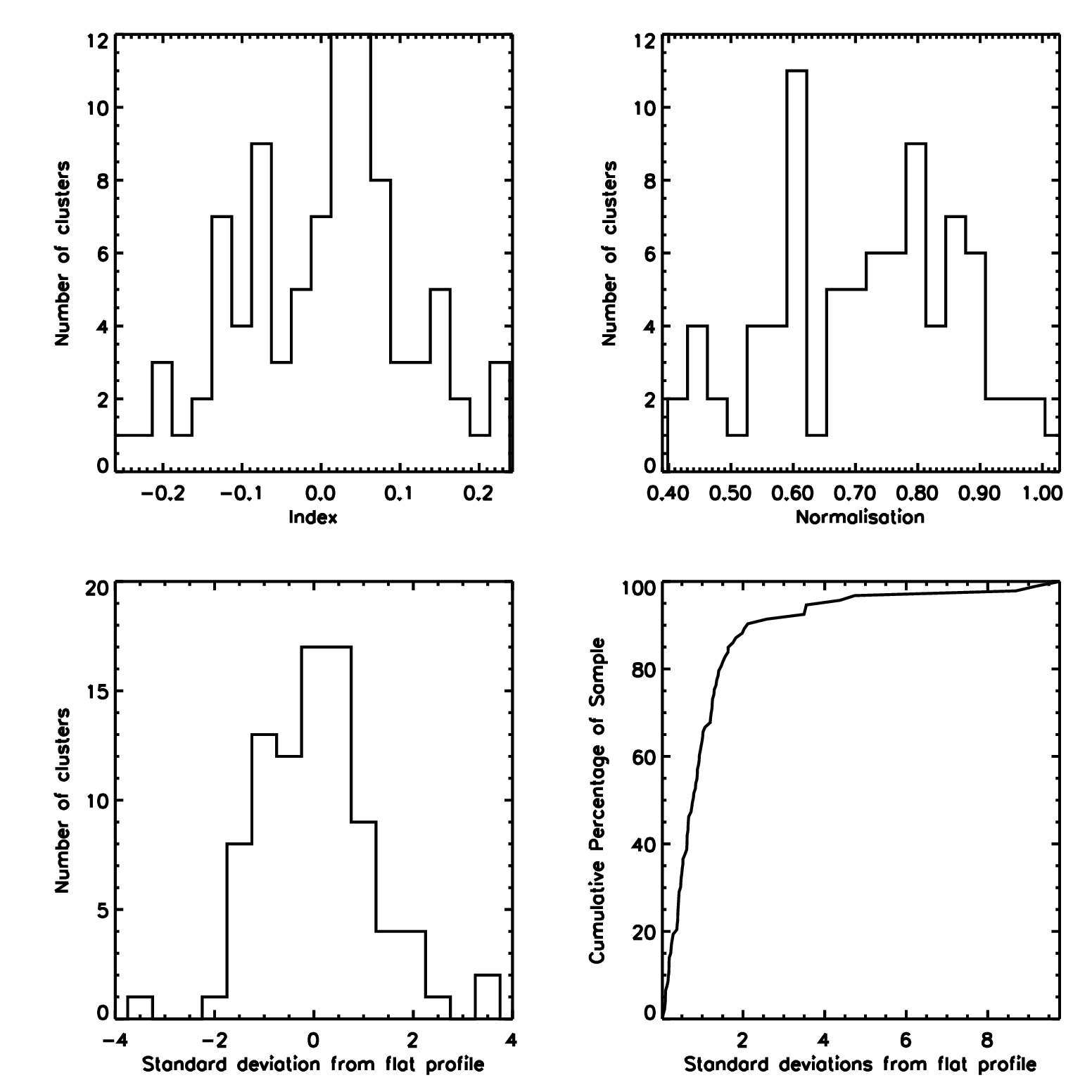,angle=0,width=0.59\textwidth,height=0.59\textwidth}
	\centerline{(a) Temperature Profile Fit Parameters} }
	\parbox{0.59\textwidth}{
		\psfig{figure=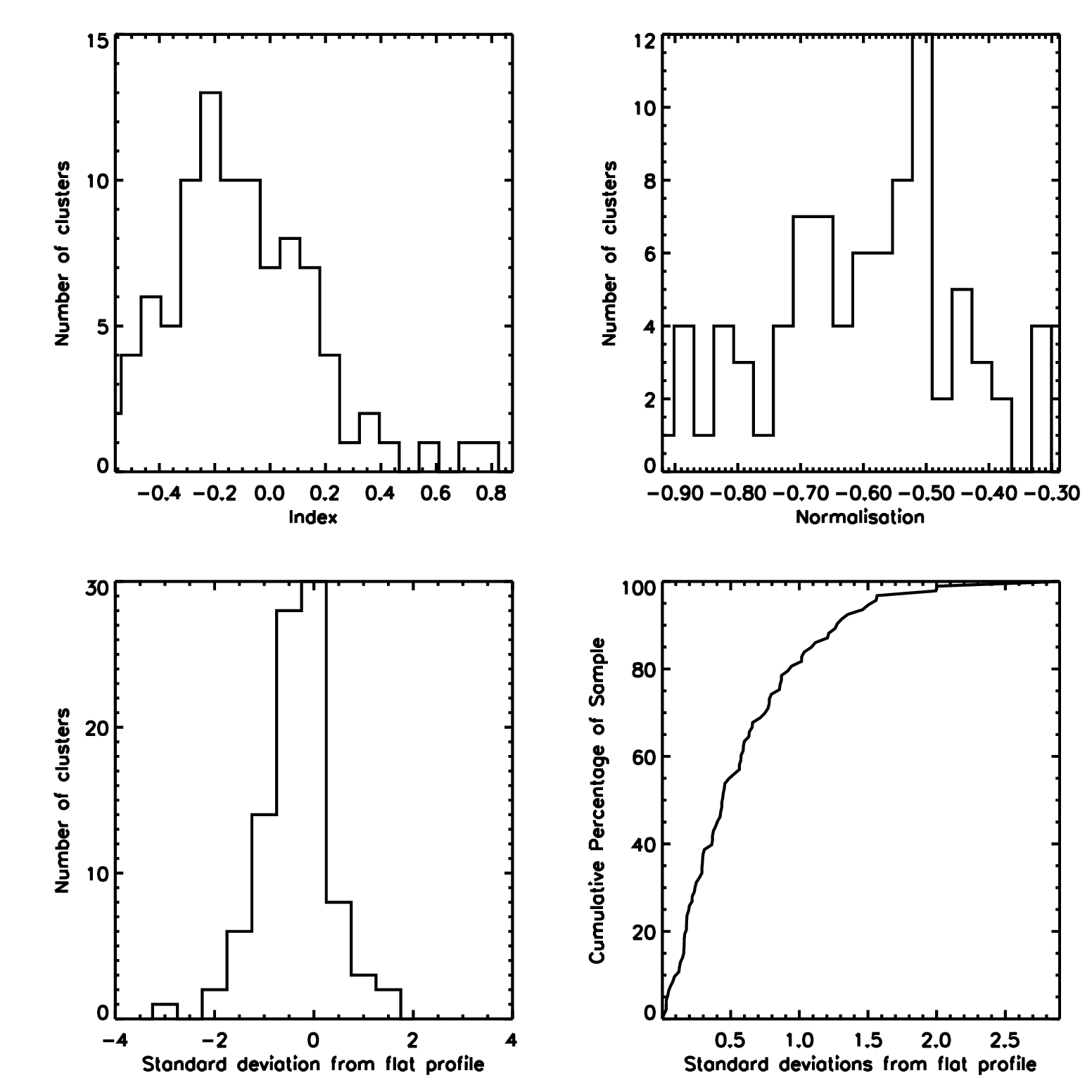,angle=0,width=0.59\textwidth,height=0.59\textwidth}
	\centerline{(b) Metallicity Profile Fit Parameters} }
	\parbox{0.99\textwidth}{ \hfill\parbox{0.99\textwidth}{
	\captionfont\caption{\label{figure:figavg} This figure
	summarises the slopes of the (a) temperature and (b) abundance
	profiles of all the clusters in our sample.  For each group of
	four plots we show the distribution of slopes and
	normalisation of the power-law model fitted to each cluster's
	profile; the number of standard-deviations of the fitted slope
	from a flat profile (\ie\ isothermality for the temperature
	profiles); and finally the cumulative distribution of these
	number of standard-deviations. From figure (a) we conclude that the
	sample is on average isothermal, and from figure (b) that there is
	a general abundance gradients in which
	metallicity decreases with radius, but that it is not
	statistically significant for the sample as a whole. }}}
\normalsize \end{figure*} }

\def\figwidth{0.32\textwidth}
\def\figheight{0.25\textheight}
\def\resfig{
\clearpage
\begin{figure*}
\parbox{\textwidth}{
    \psfig{figure=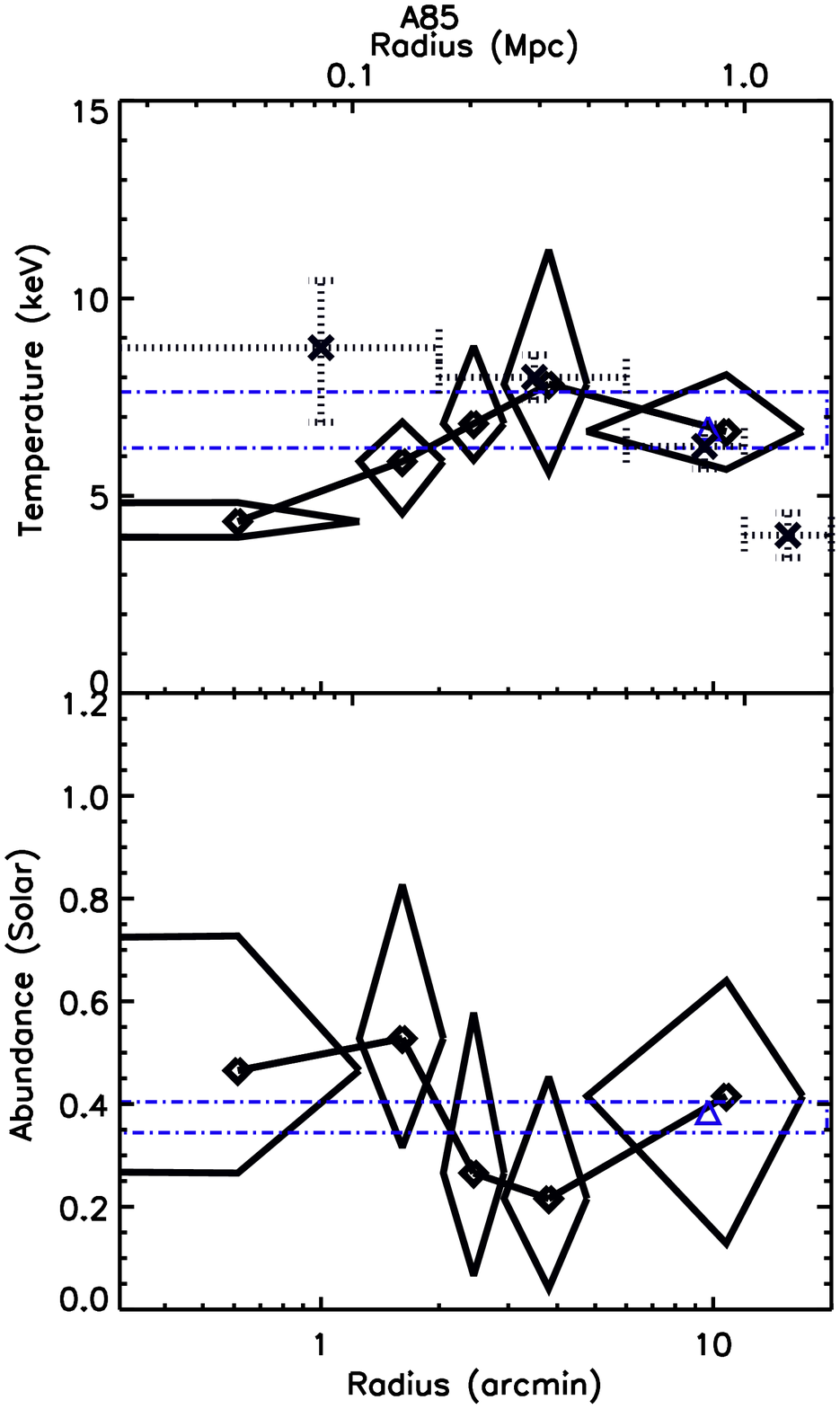,angle=0,width=\figwidth,height=\figheight}
    \psfig{figure=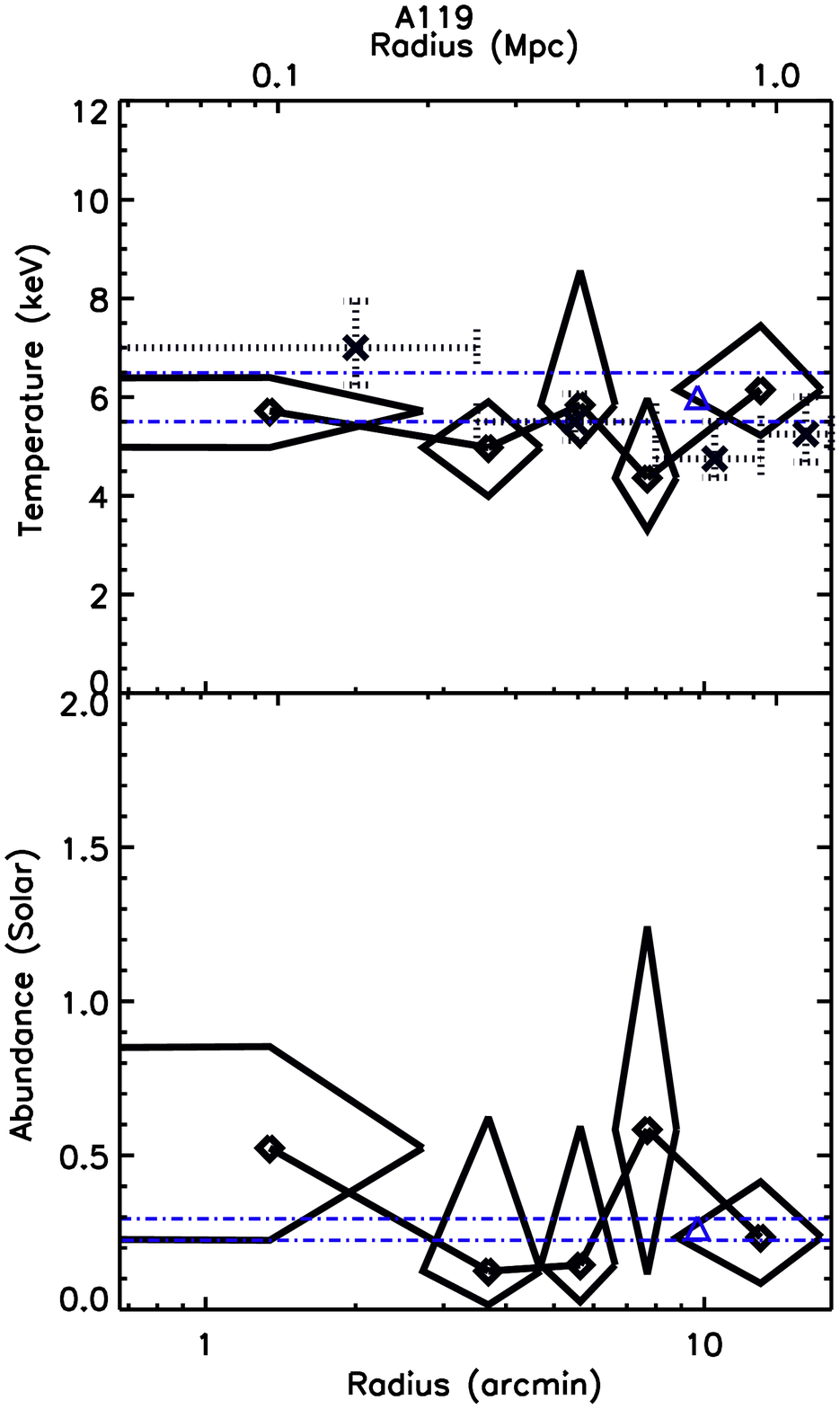,angle=0,width=\figwidth,height=\figheight}
    \psfig{figure=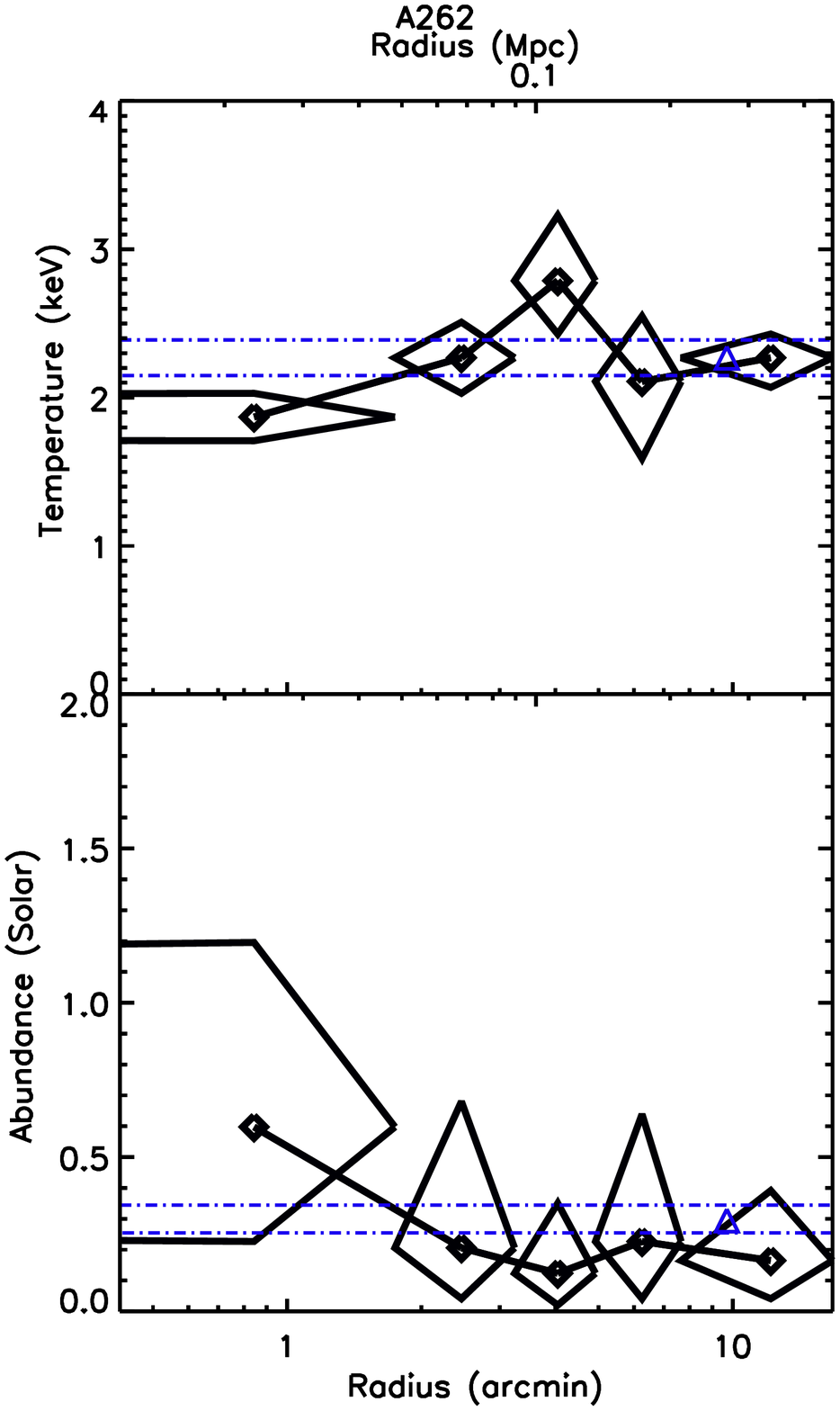,angle=0,width=\figwidth,height=\figheight}
    \psfig{figure=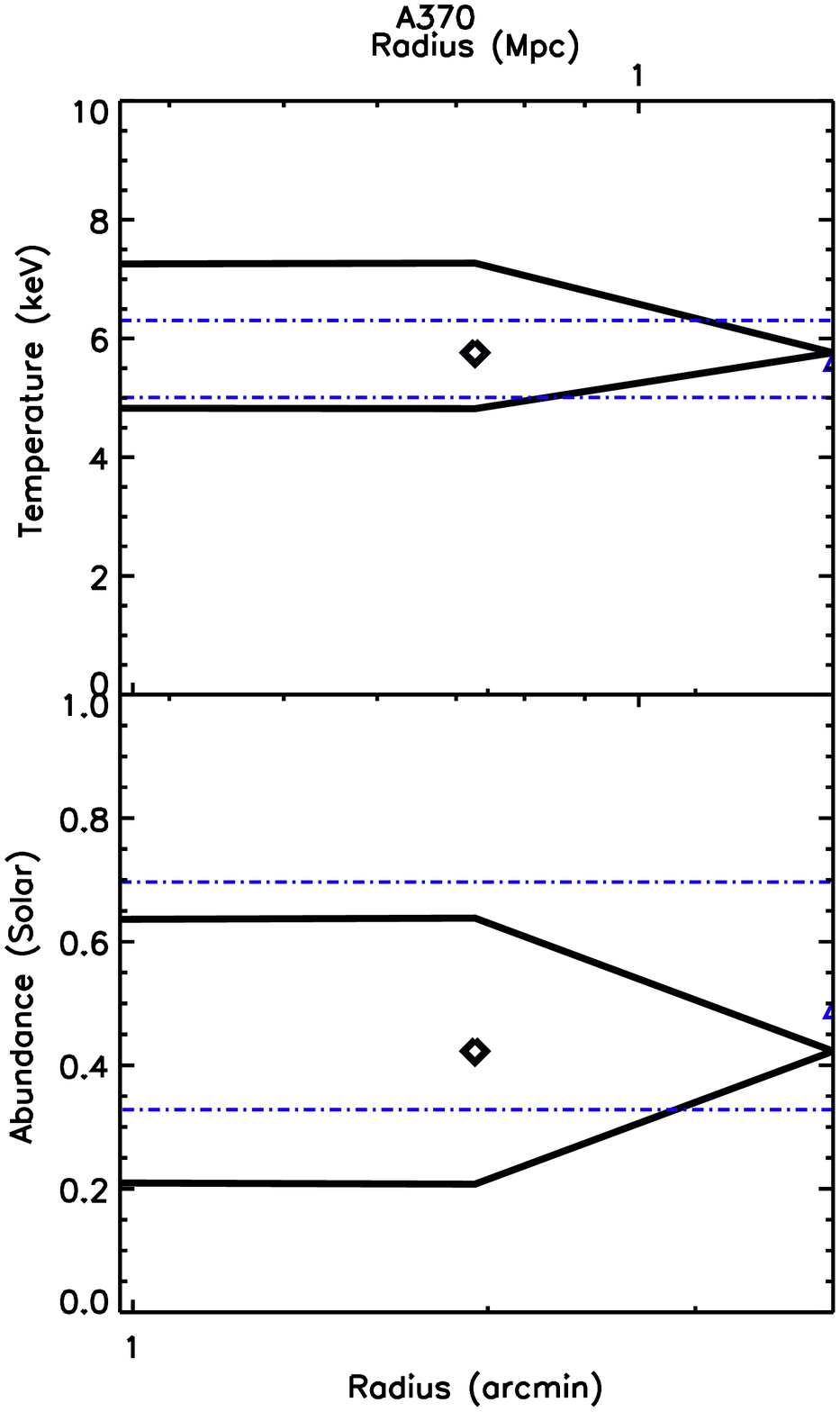,angle=0,width=\figwidth,height=\figheight}
  }
\parbox{\textwidth}{
    \psfig{figure=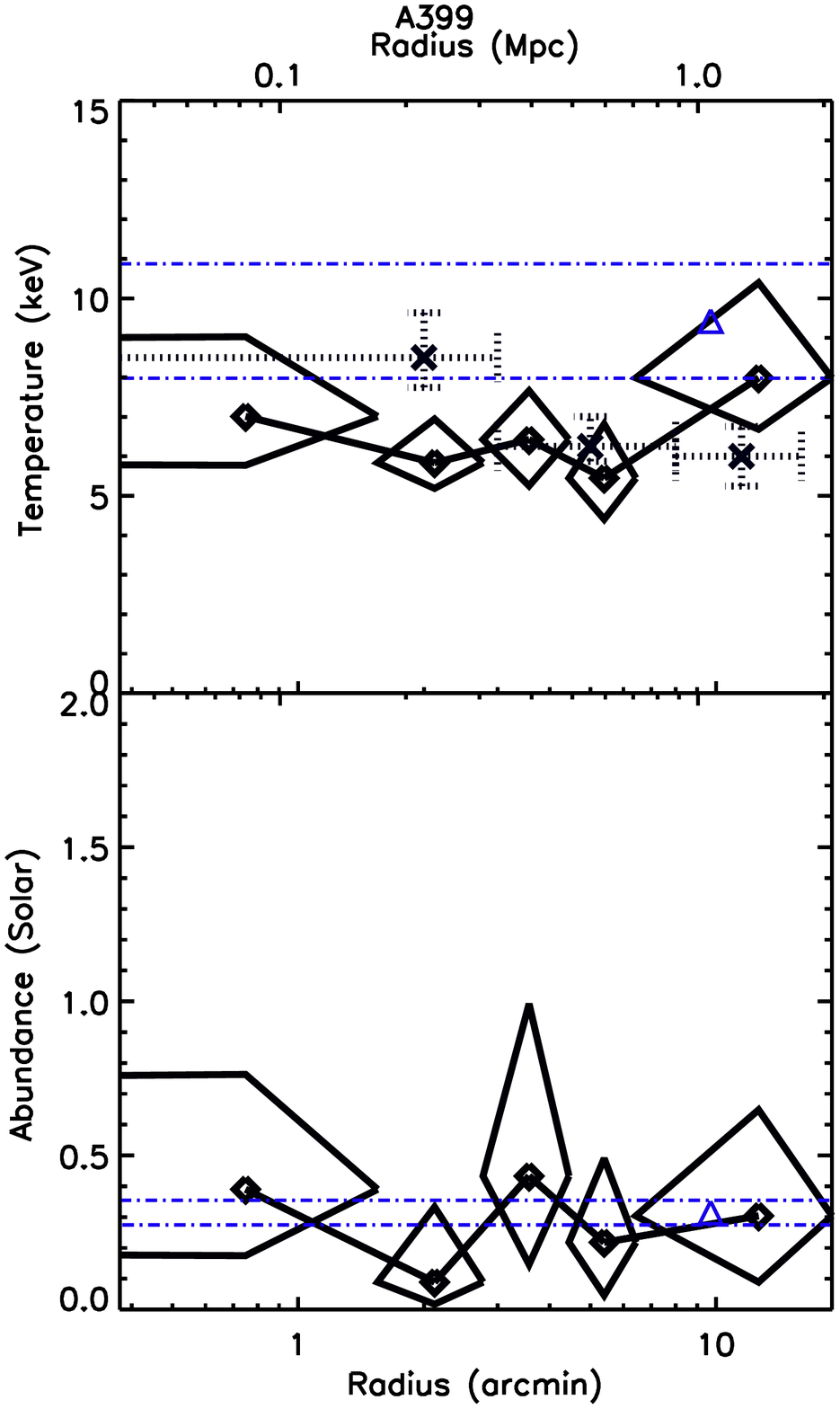,angle=0,width=\figwidth,height=\figheight}
    \psfig{figure=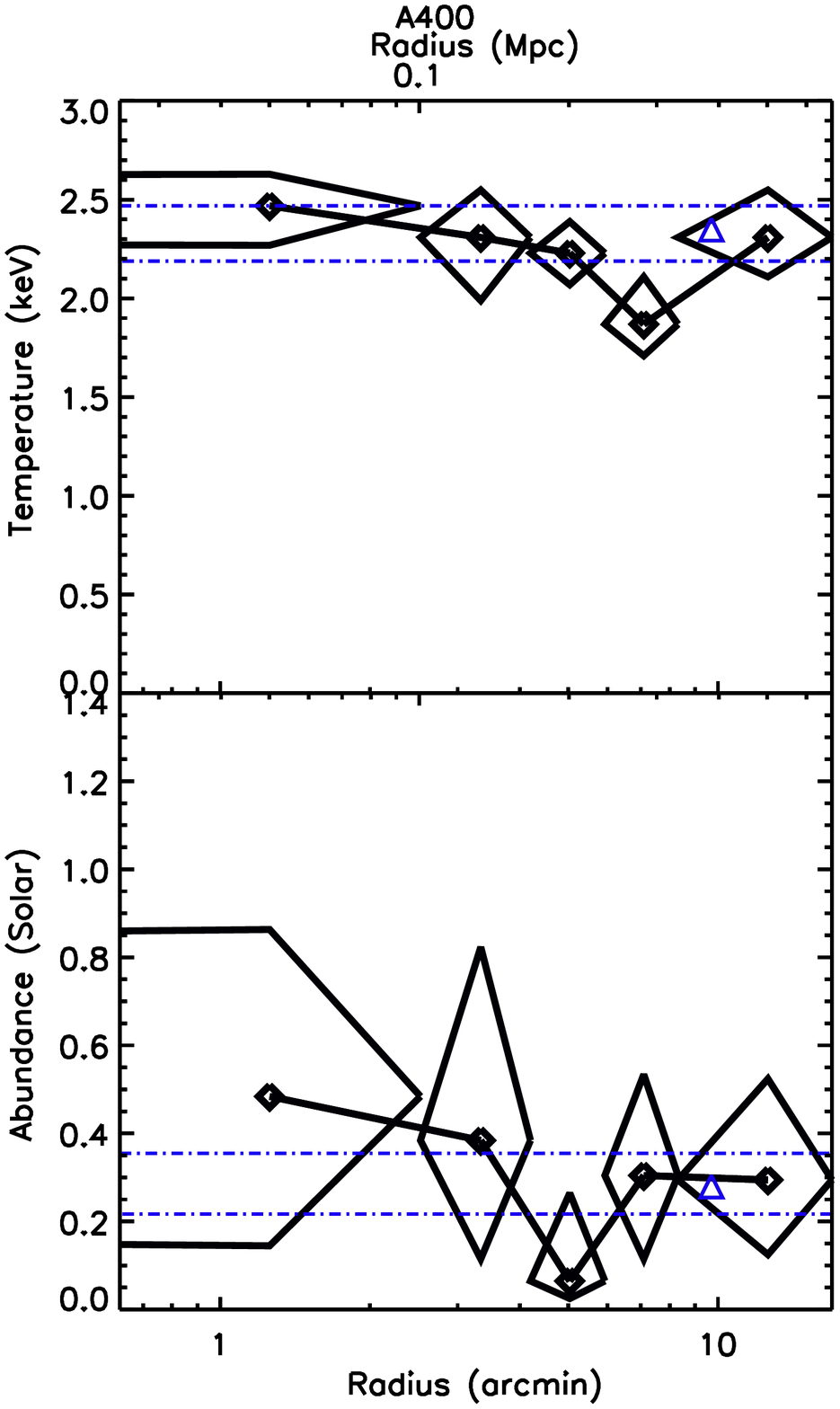,angle=0,width=\figwidth,height=\figheight}
    \psfig{figure=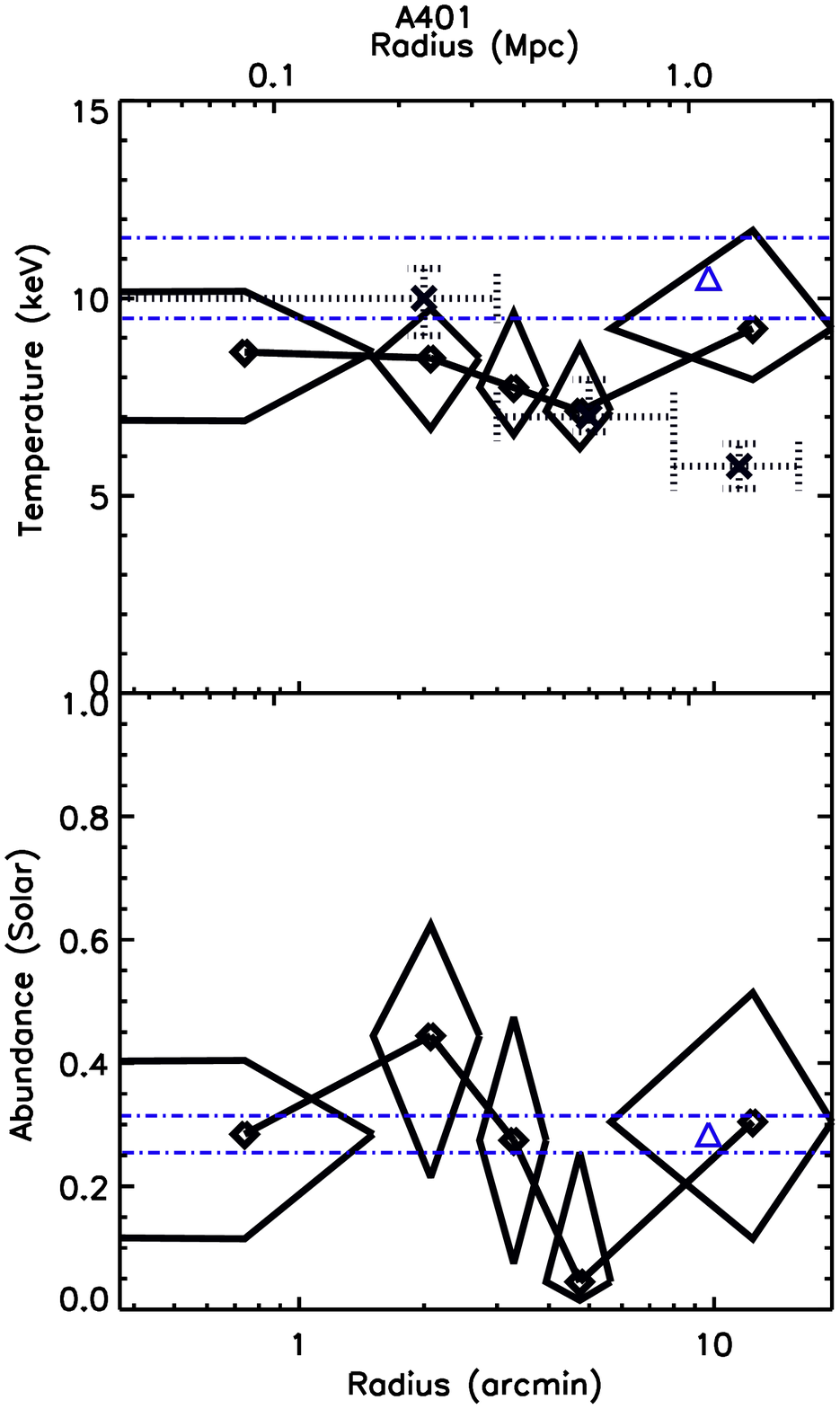,angle=0,width=\figwidth,height=\figheight}
    \psfig{figure=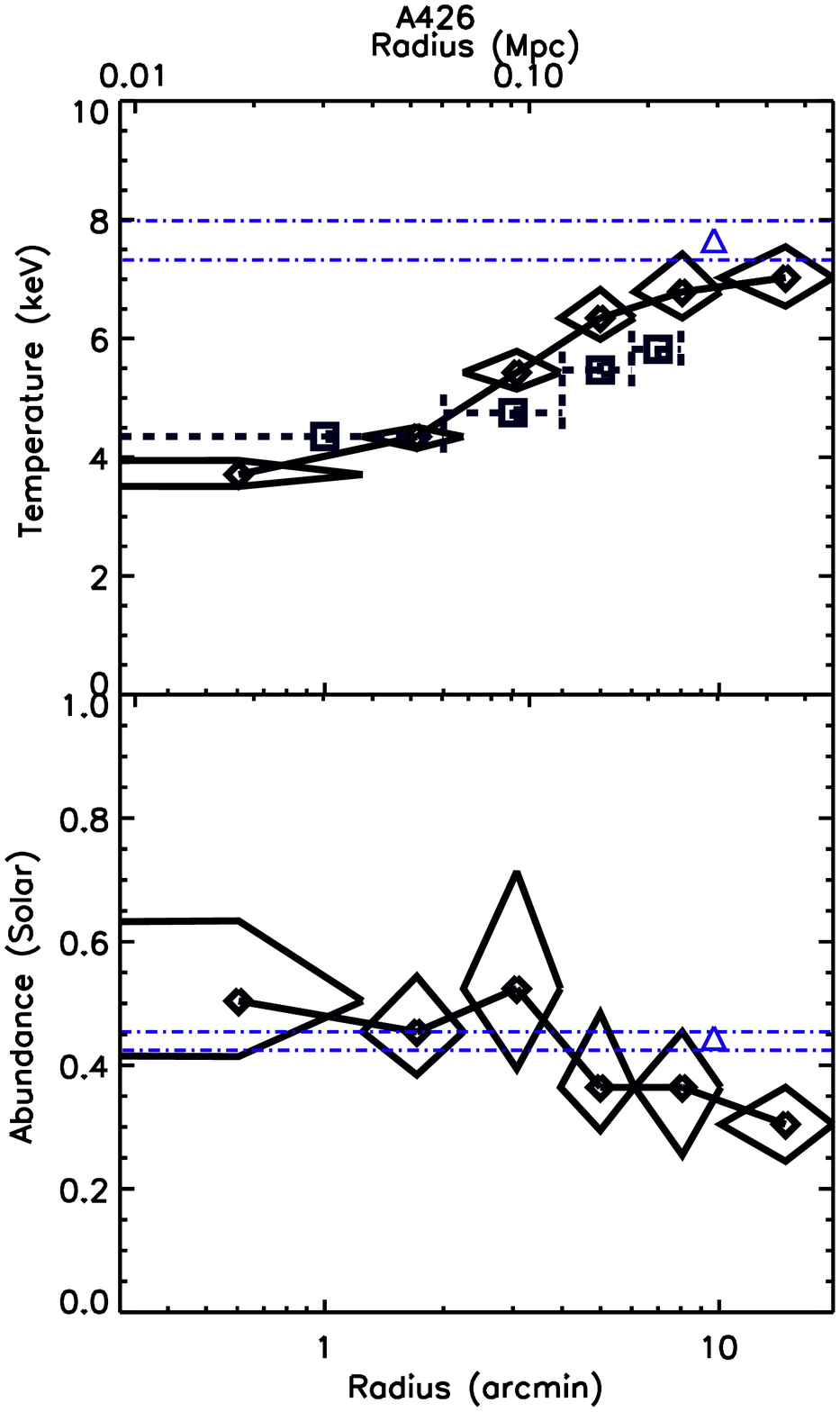,angle=0,width=\figwidth,height=\figheight}
  }
\parbox{\textwidth}{
    \psfig{figure=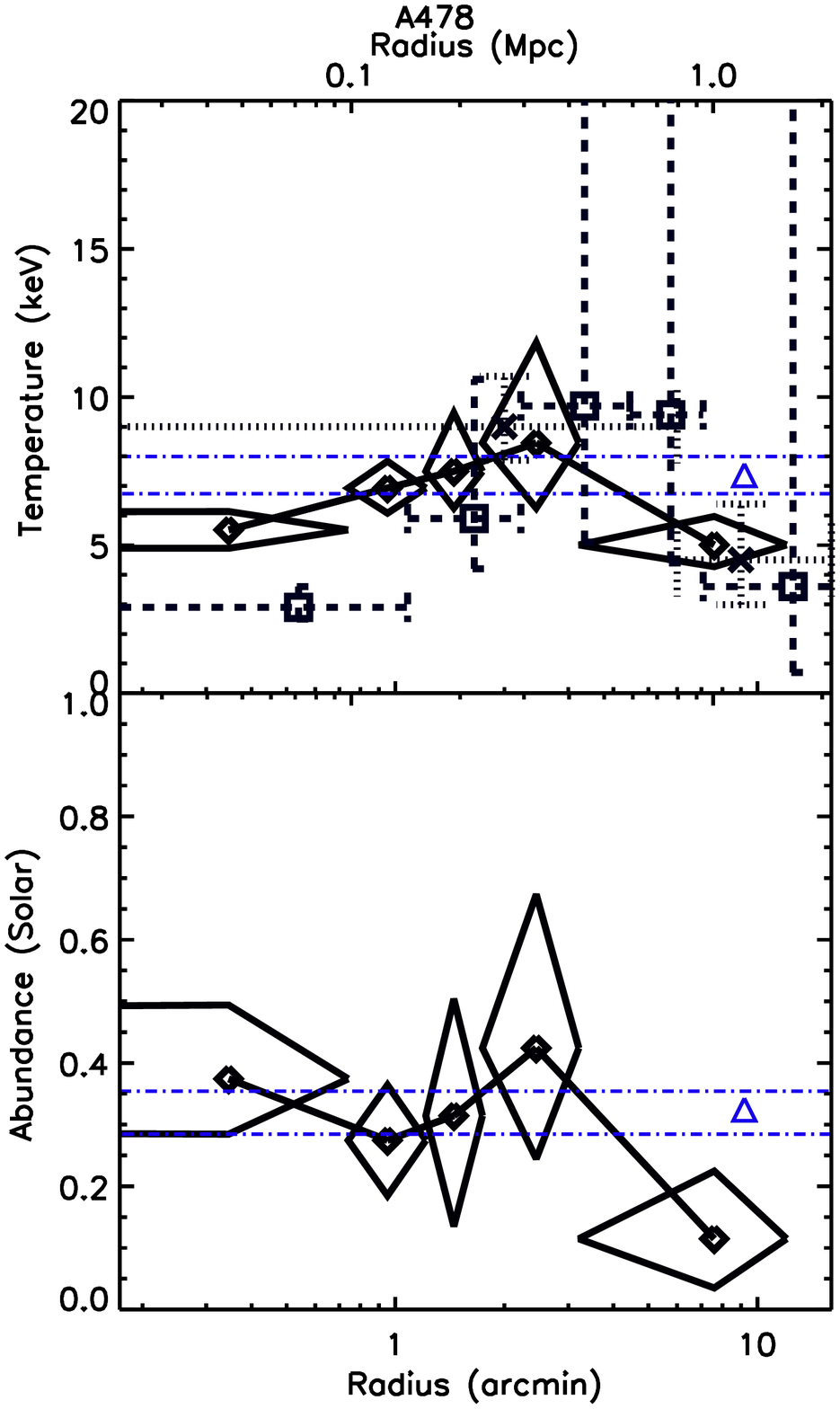,angle=0,width=\figwidth,height=\figheight}
    \psfig{figure=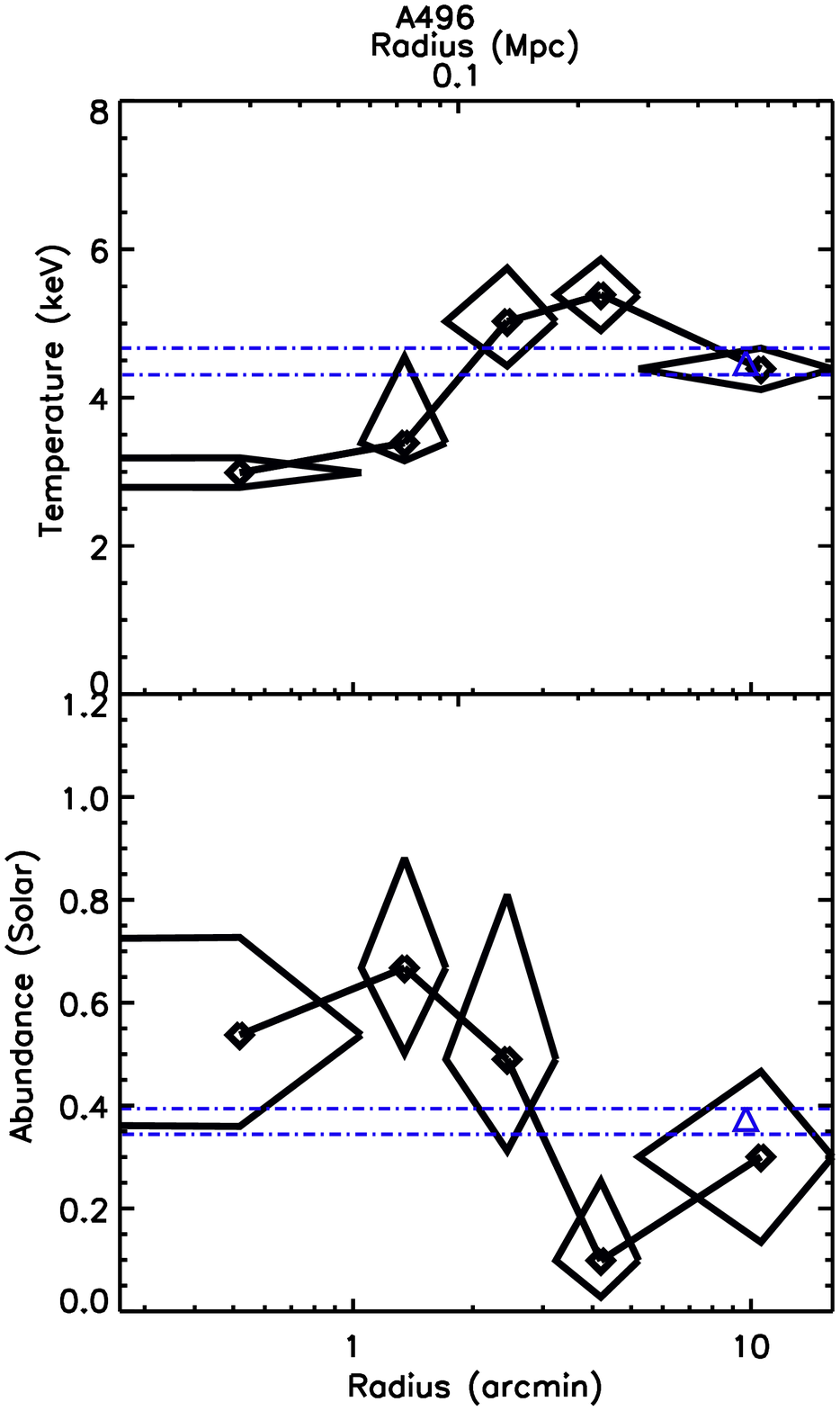,angle=0,width=\figwidth,height=\figheight}
    \psfig{figure=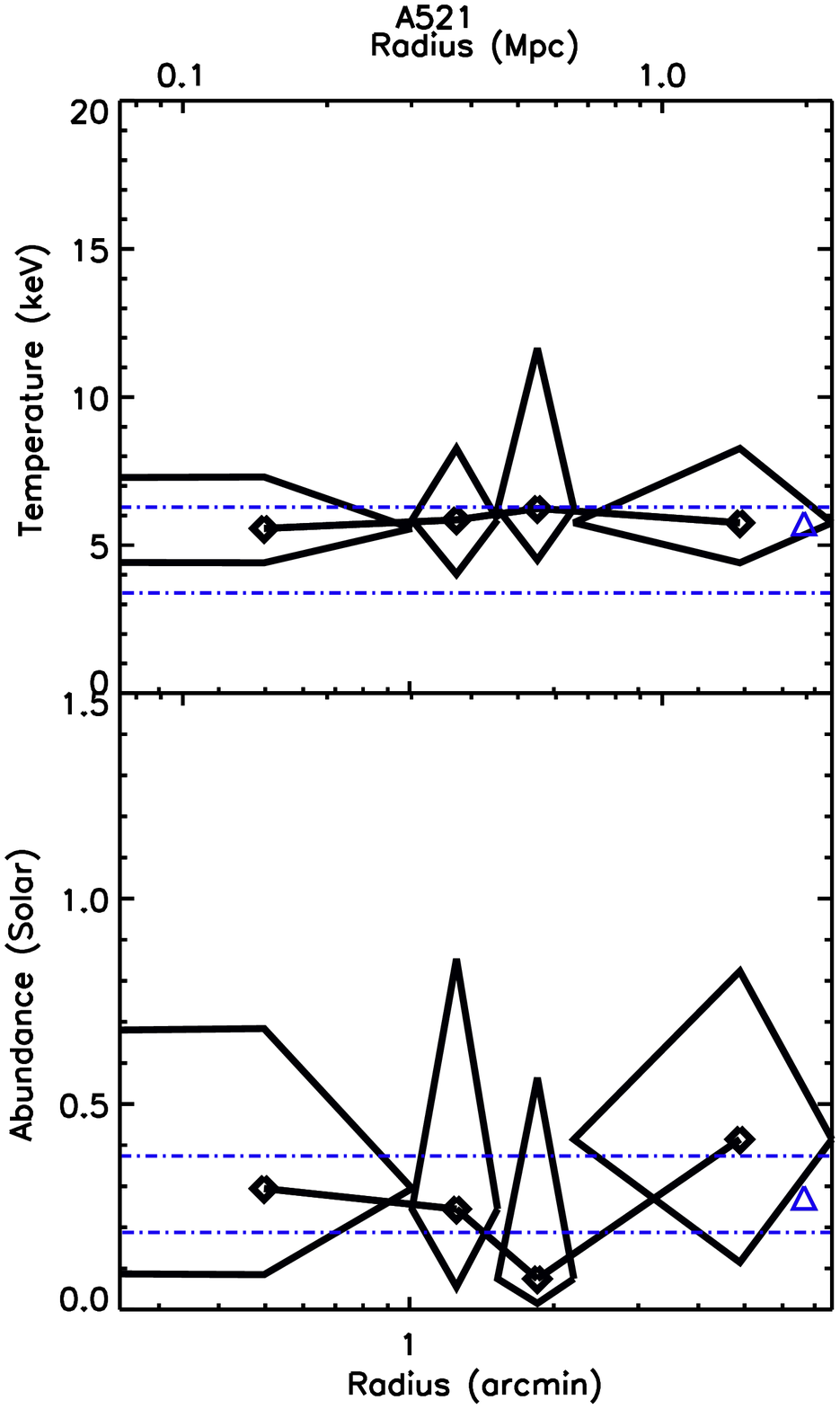,angle=0,width=\figwidth,height=\figheight}
    \psfig{figure=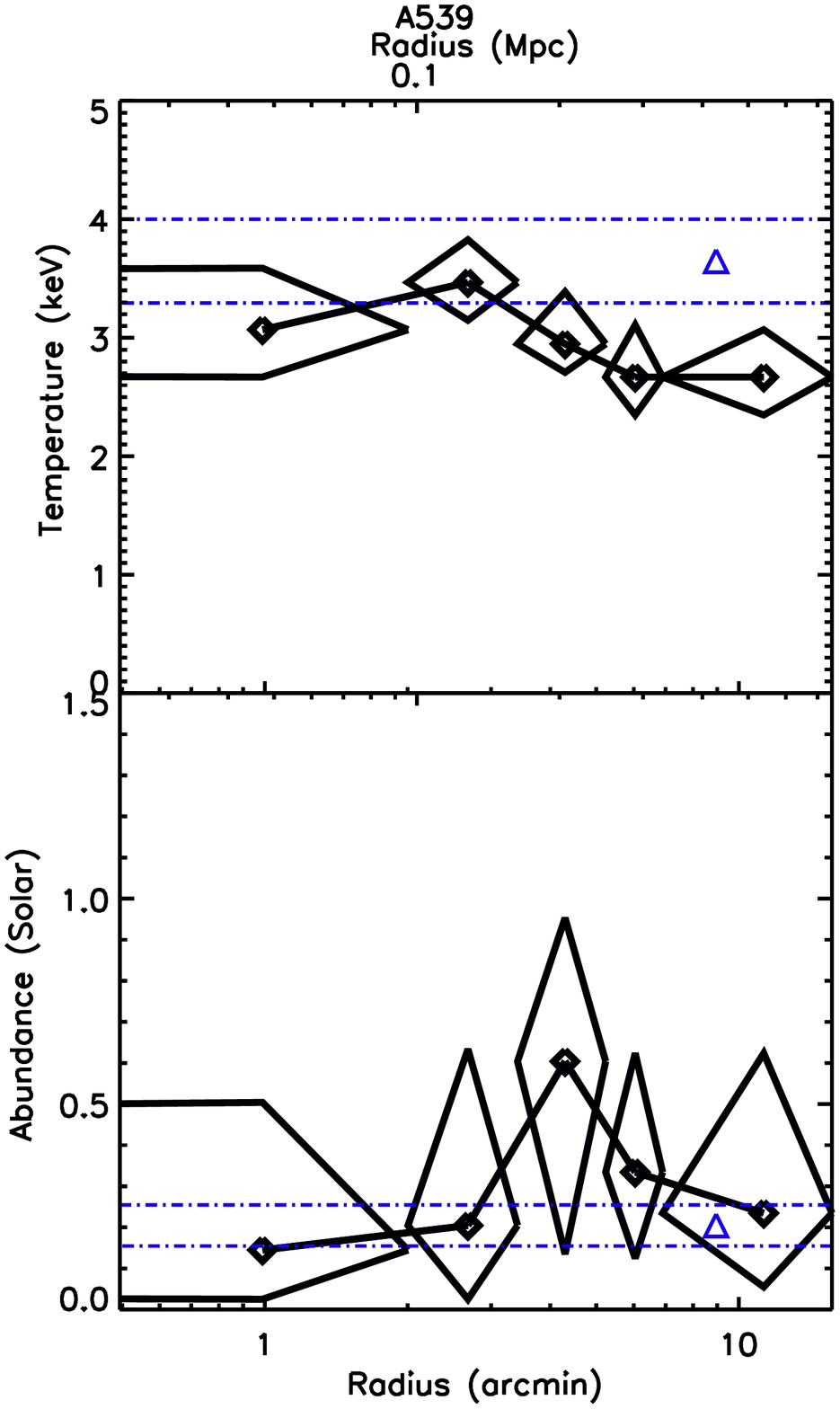,angle=0,width=\figwidth,height=\figheight}
  }
\end{figure*}
\clearpage
\begin{figure*}
\parbox{\textwidth}{
    \psfig{figure=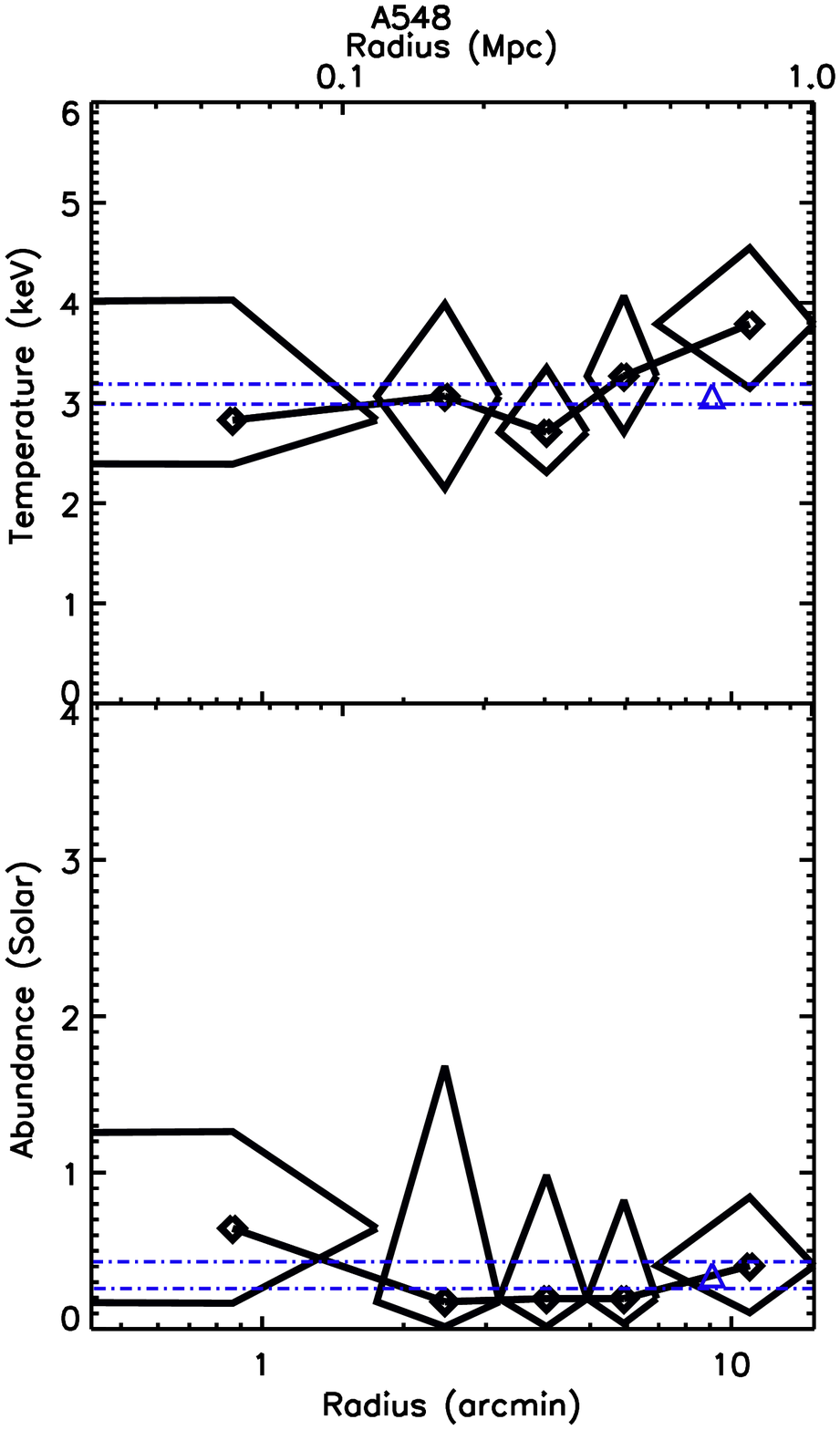,angle=0,width=\figwidth,height=\figheight}
    \psfig{figure=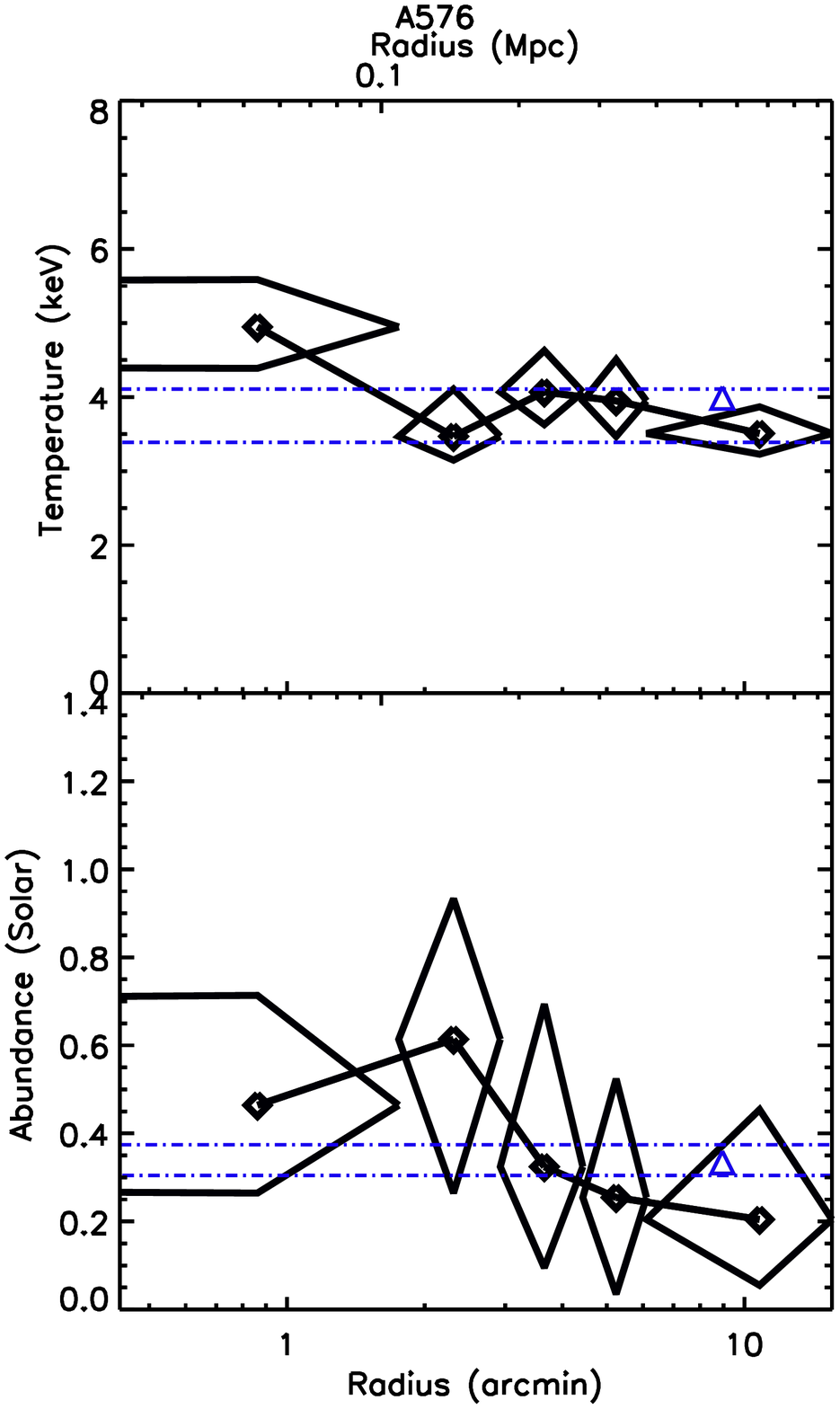,angle=0,width=\figwidth,height=\figheight}
    \psfig{figure=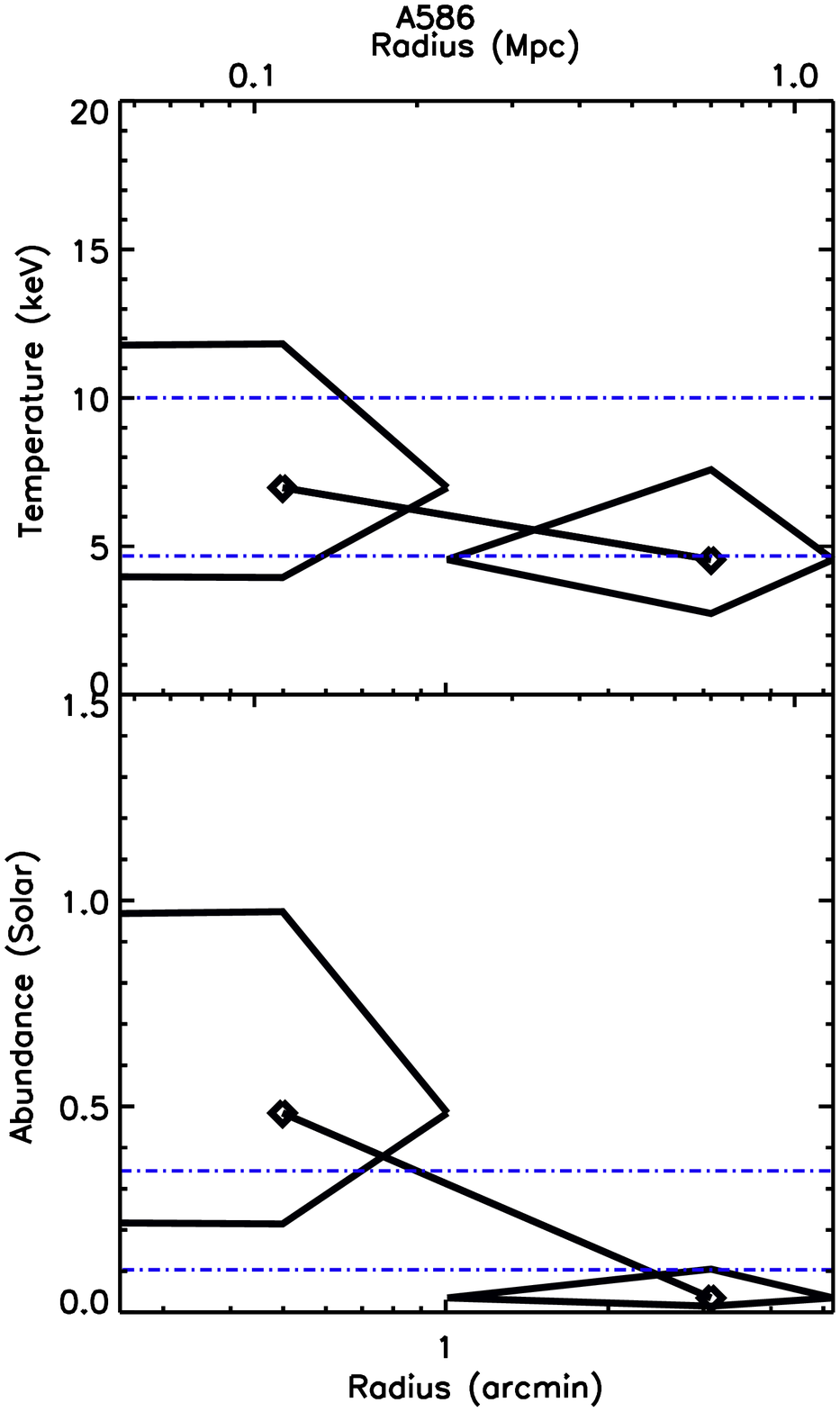,angle=0,width=\figwidth,height=\figheight}
    \psfig{figure=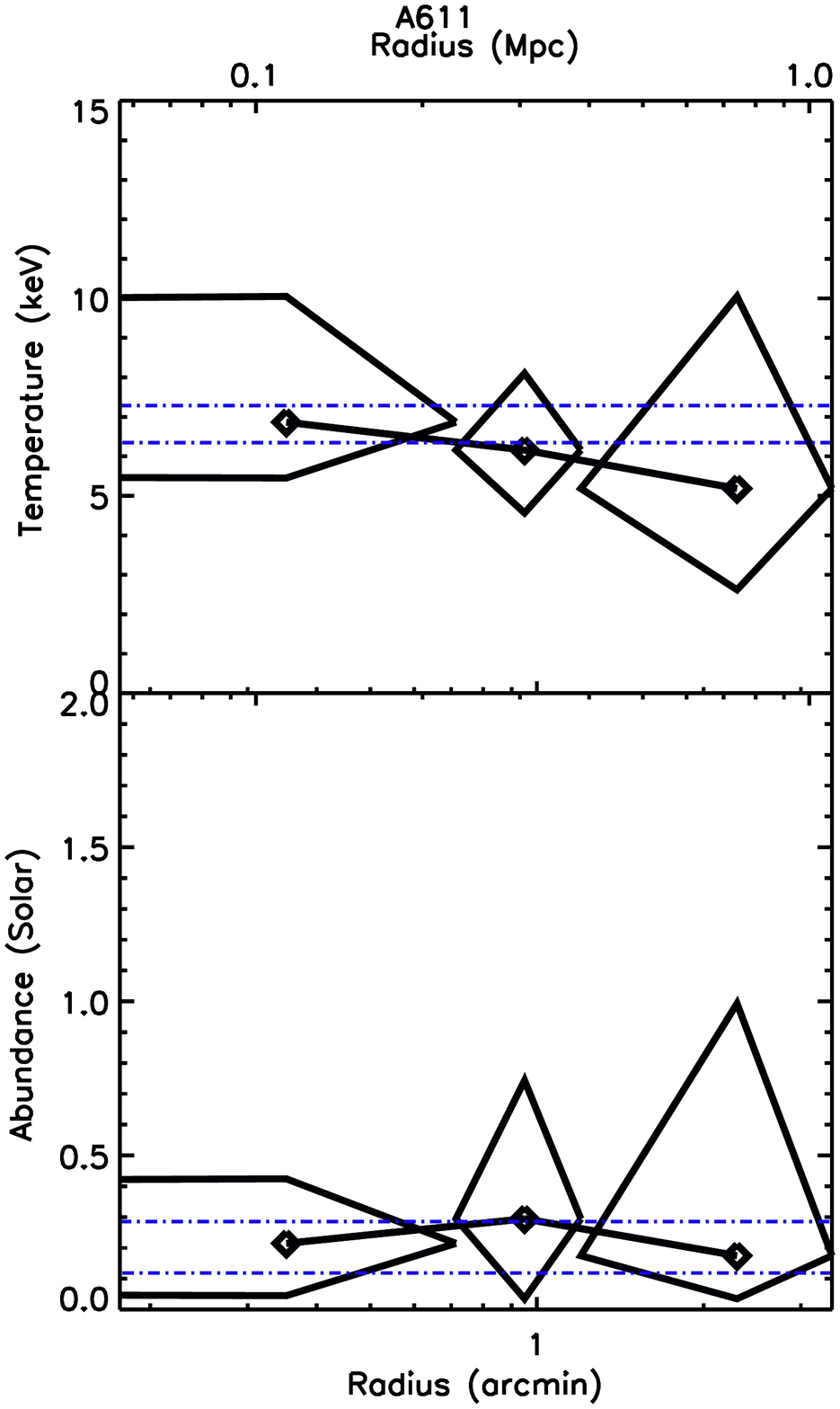,angle=0,width=\figwidth,height=\figheight}
  }
\parbox{\textwidth}{
    \psfig{figure=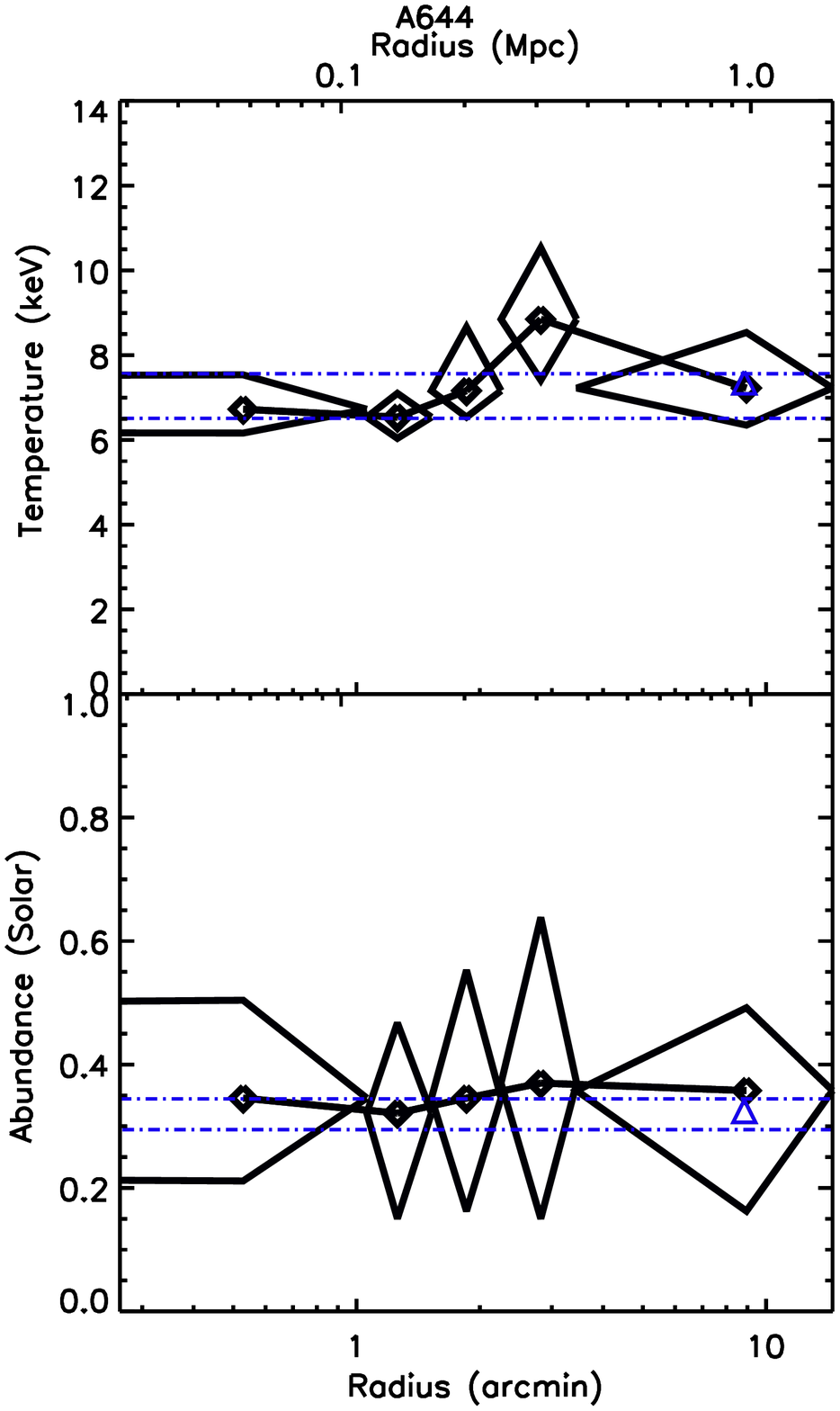,angle=0,width=\figwidth,height=\figheight}
    \psfig{figure=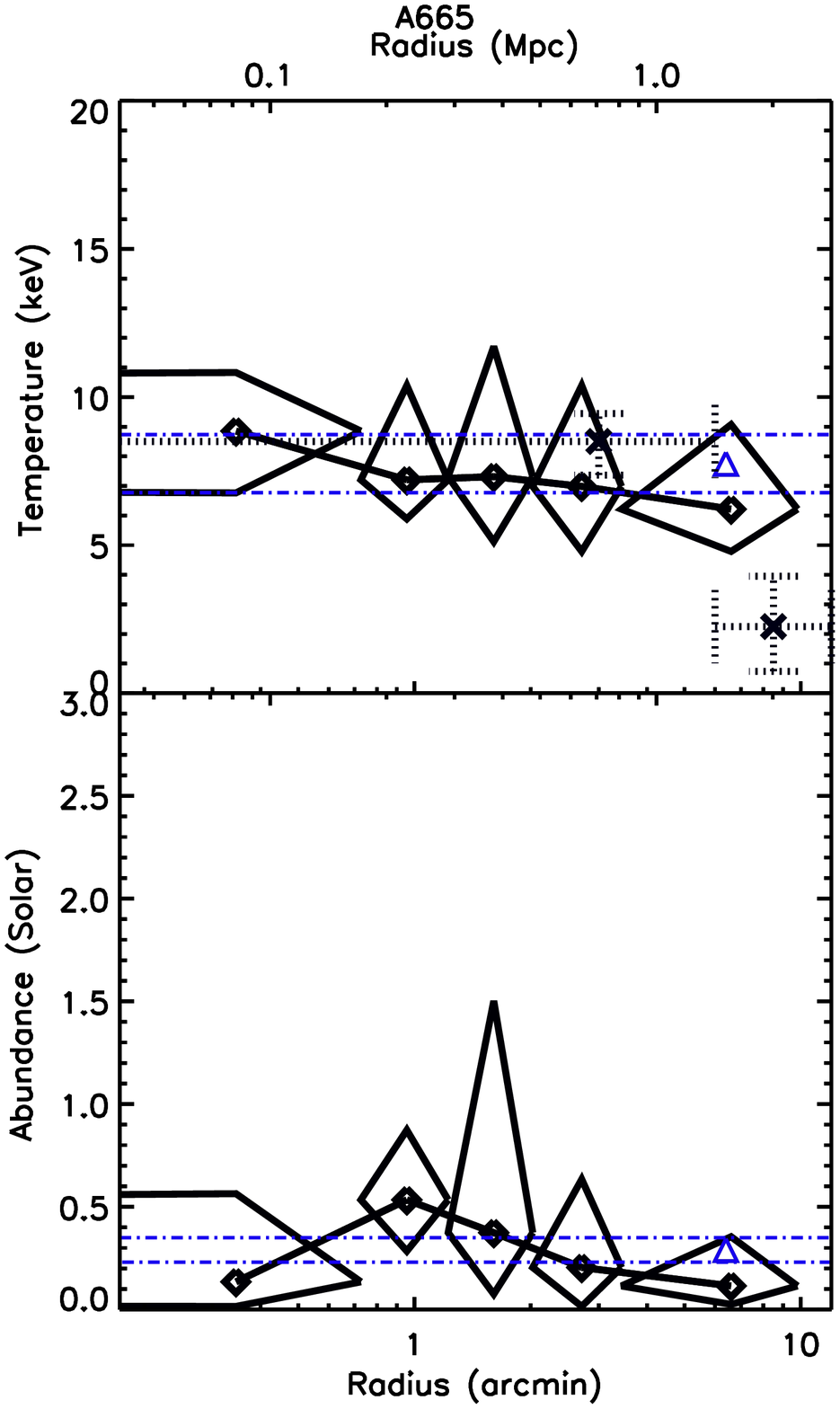,angle=0,width=\figwidth,height=\figheight}
    \psfig{figure=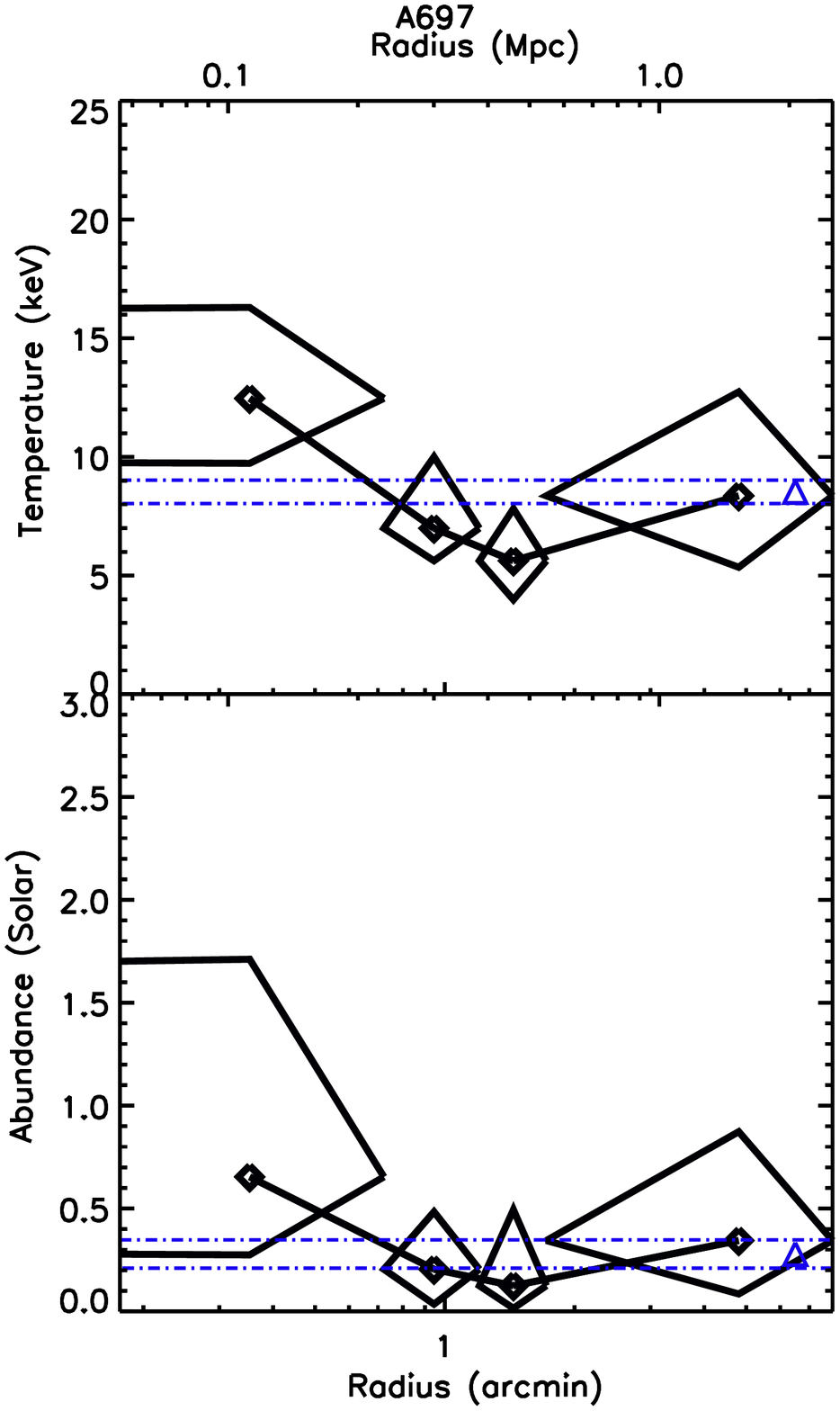,angle=0,width=\figwidth,height=\figheight}
    \psfig{figure=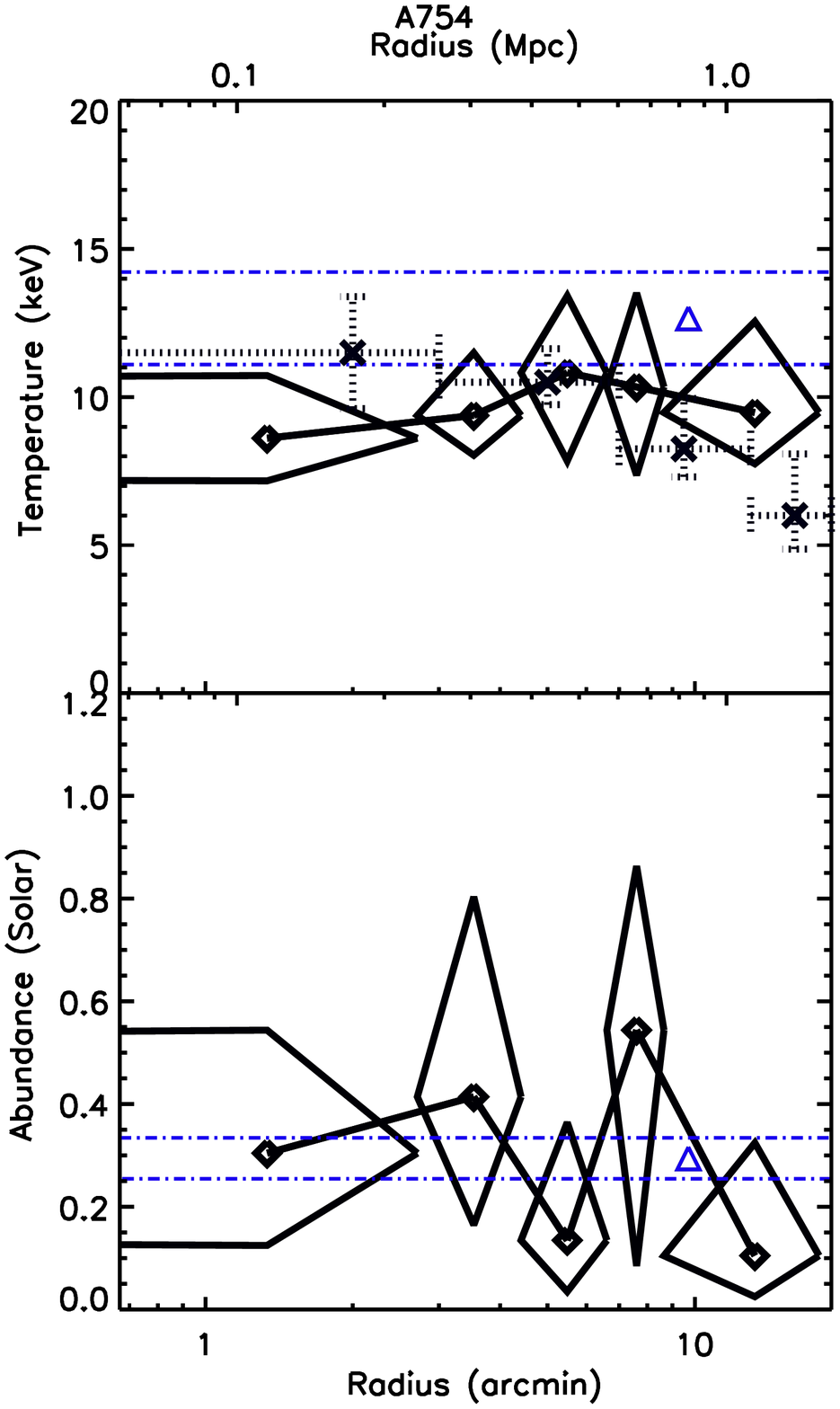,angle=0,width=\figwidth,height=\figheight}
  }
\parbox{\textwidth}{
    \psfig{figure=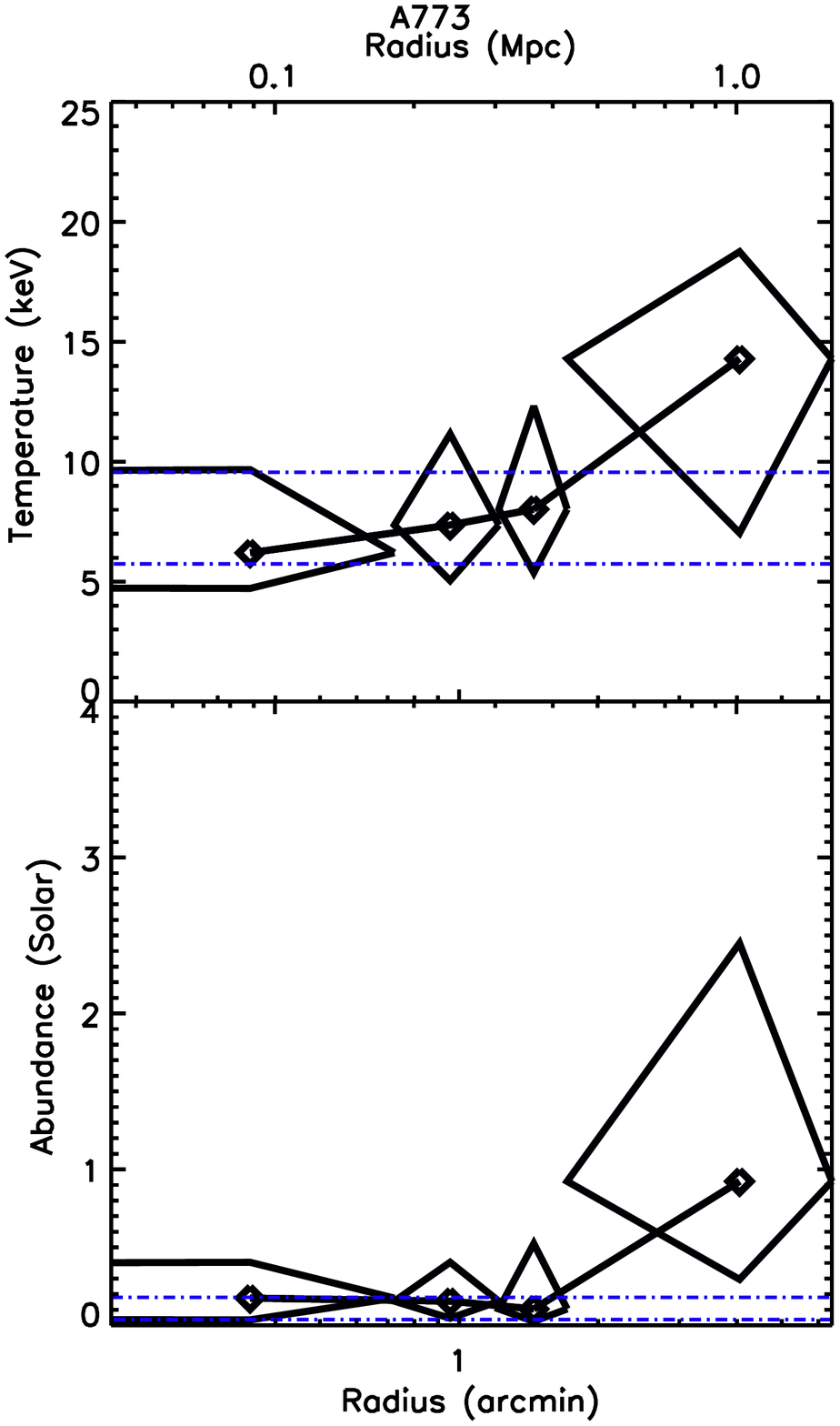,angle=0,width=\figwidth,height=\figheight}
    \psfig{figure=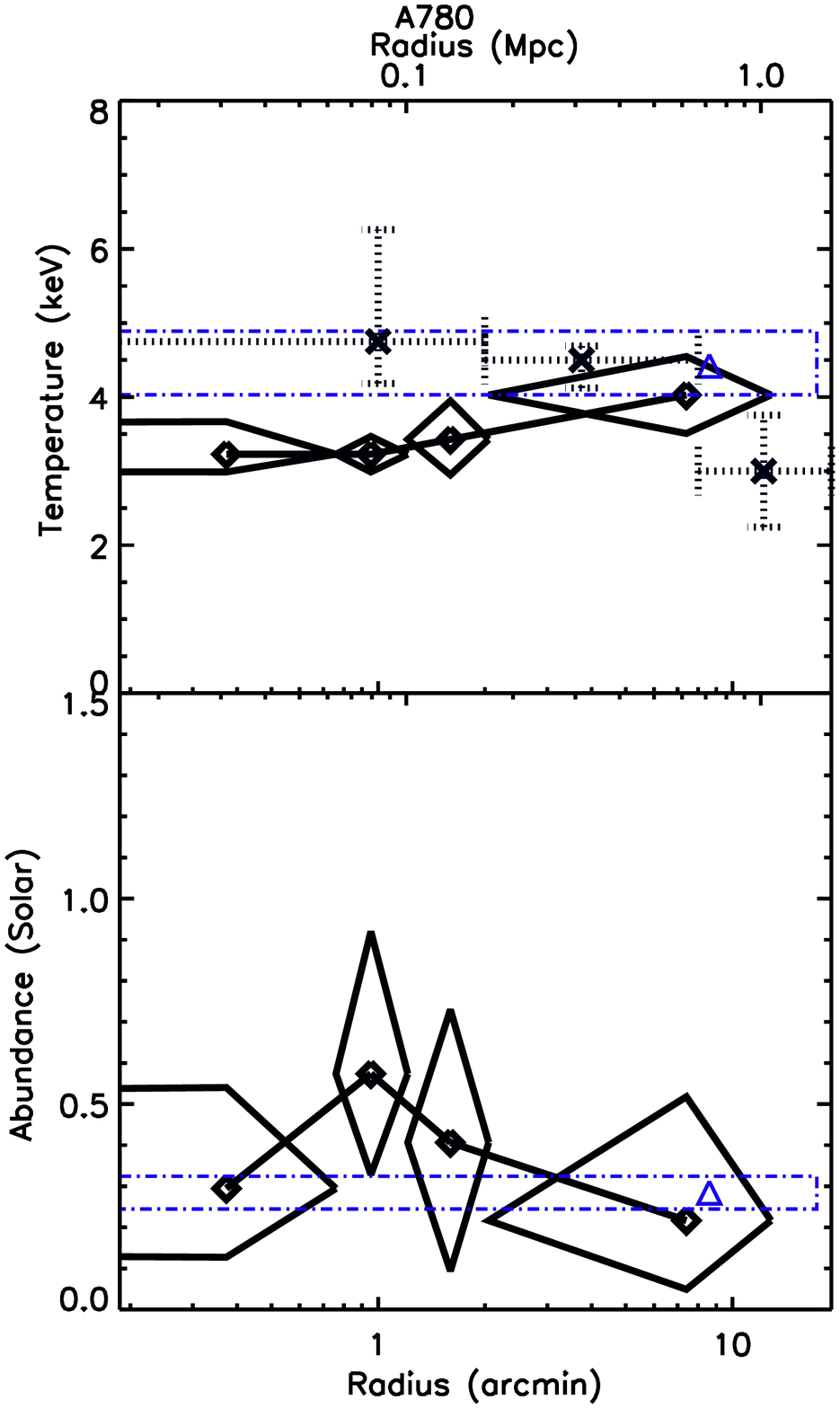,angle=0,width=\figwidth,height=\figheight}
    \psfig{figure=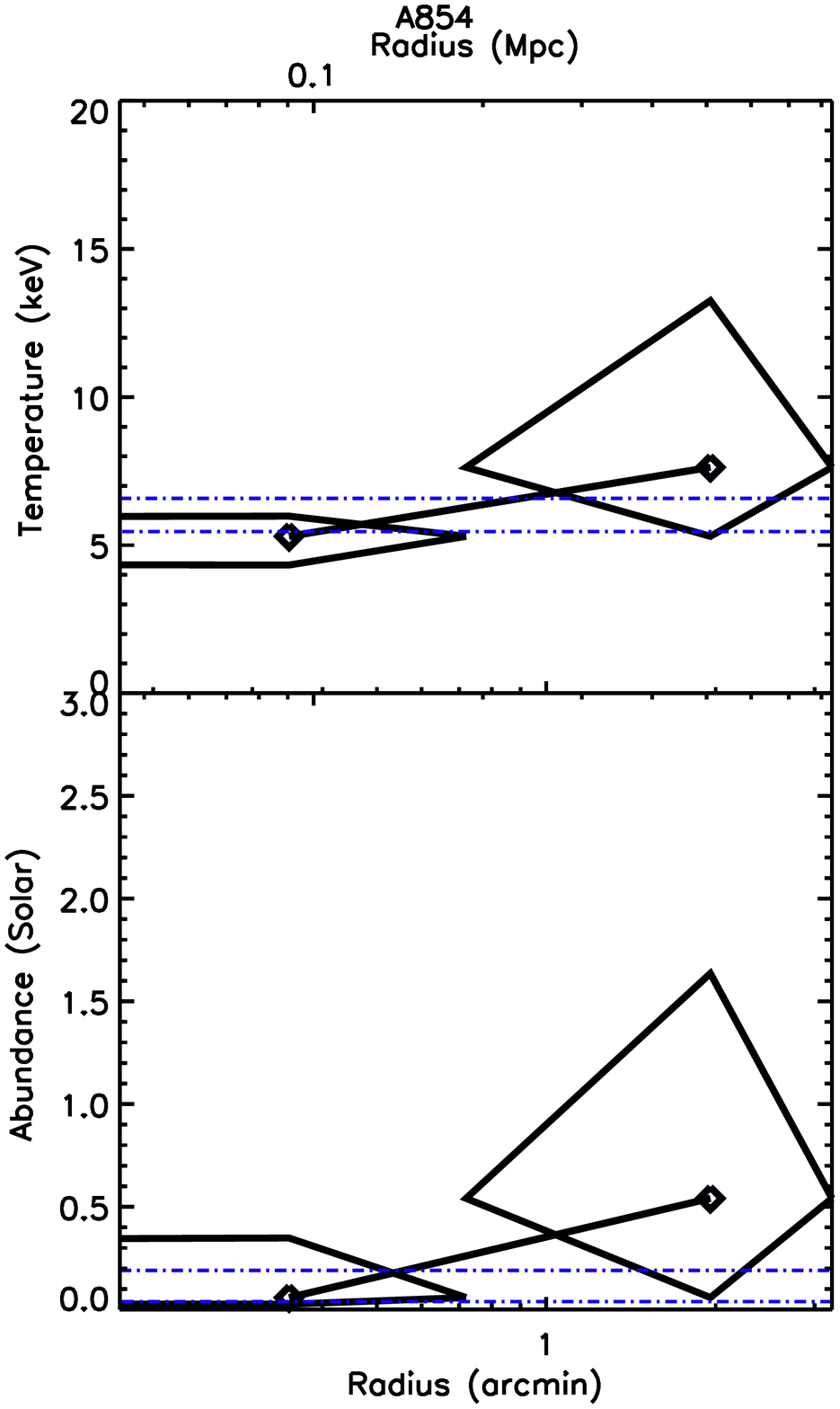,angle=0,width=\figwidth,height=\figheight}
    \psfig{figure=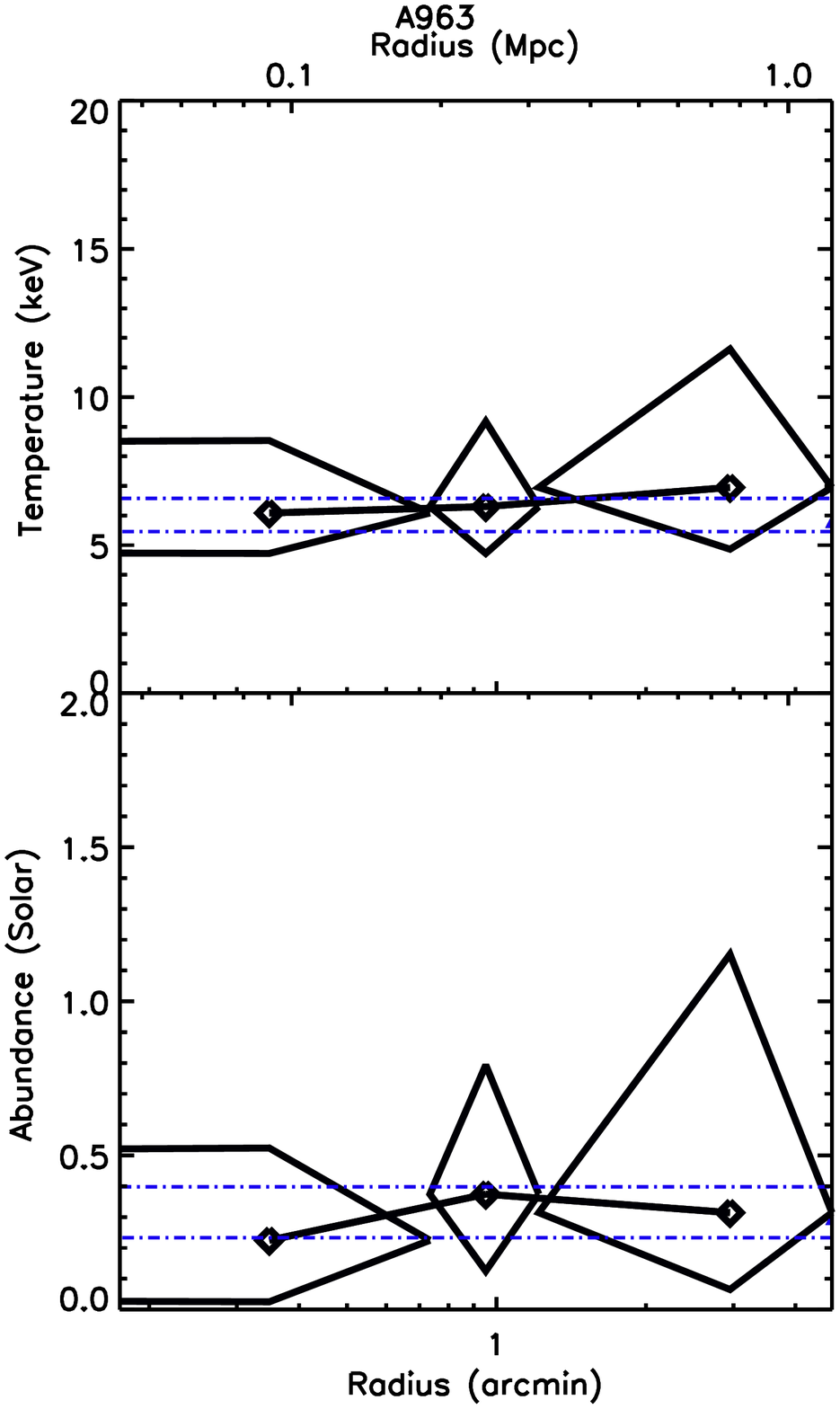,angle=0,width=\figwidth,height=\figheight}
  }
\end{figure*}
\clearpage
\begin{figure*}
\parbox{\textwidth}{
    \psfig{figure=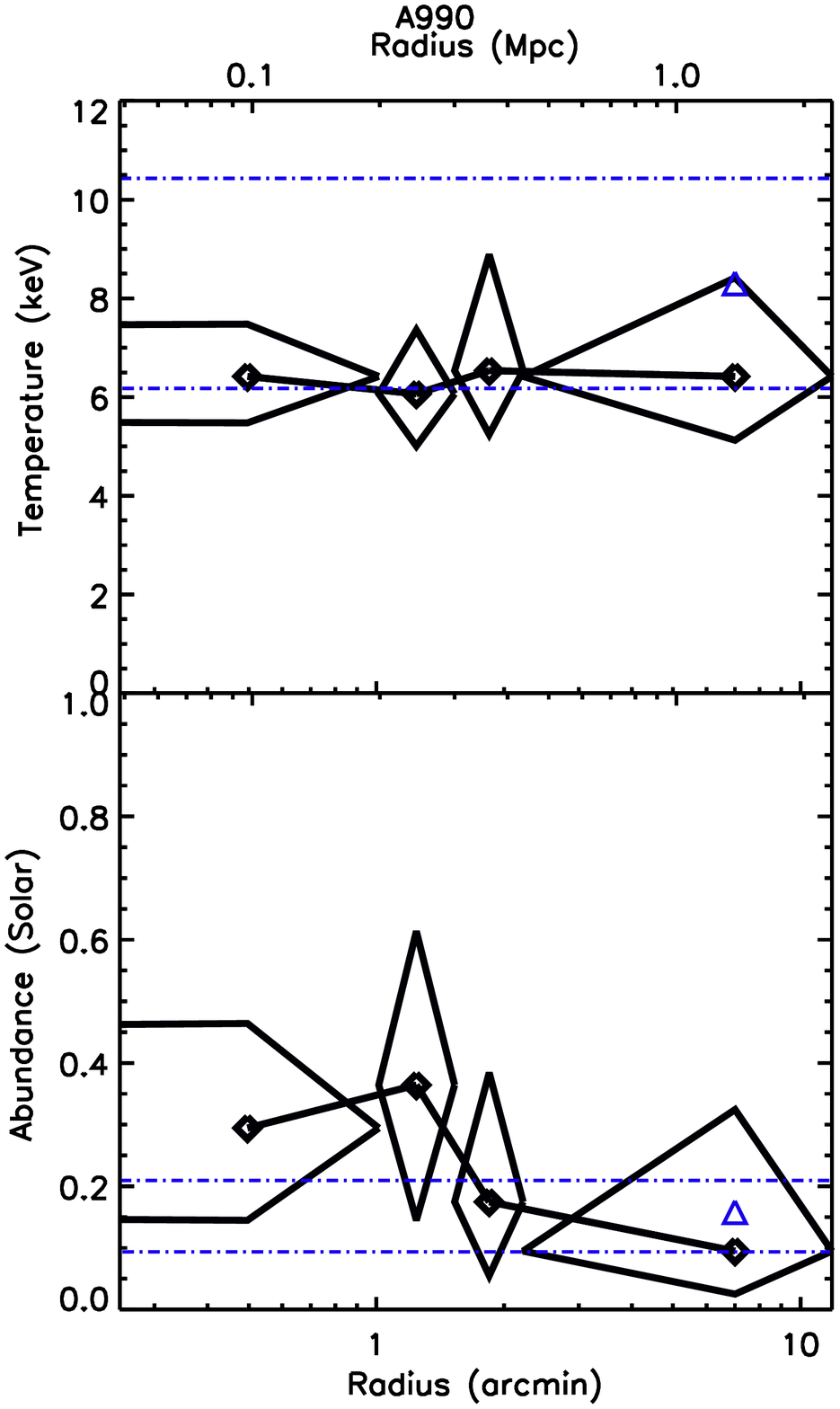,angle=0,width=\figwidth,height=\figheight}
    \psfig{figure=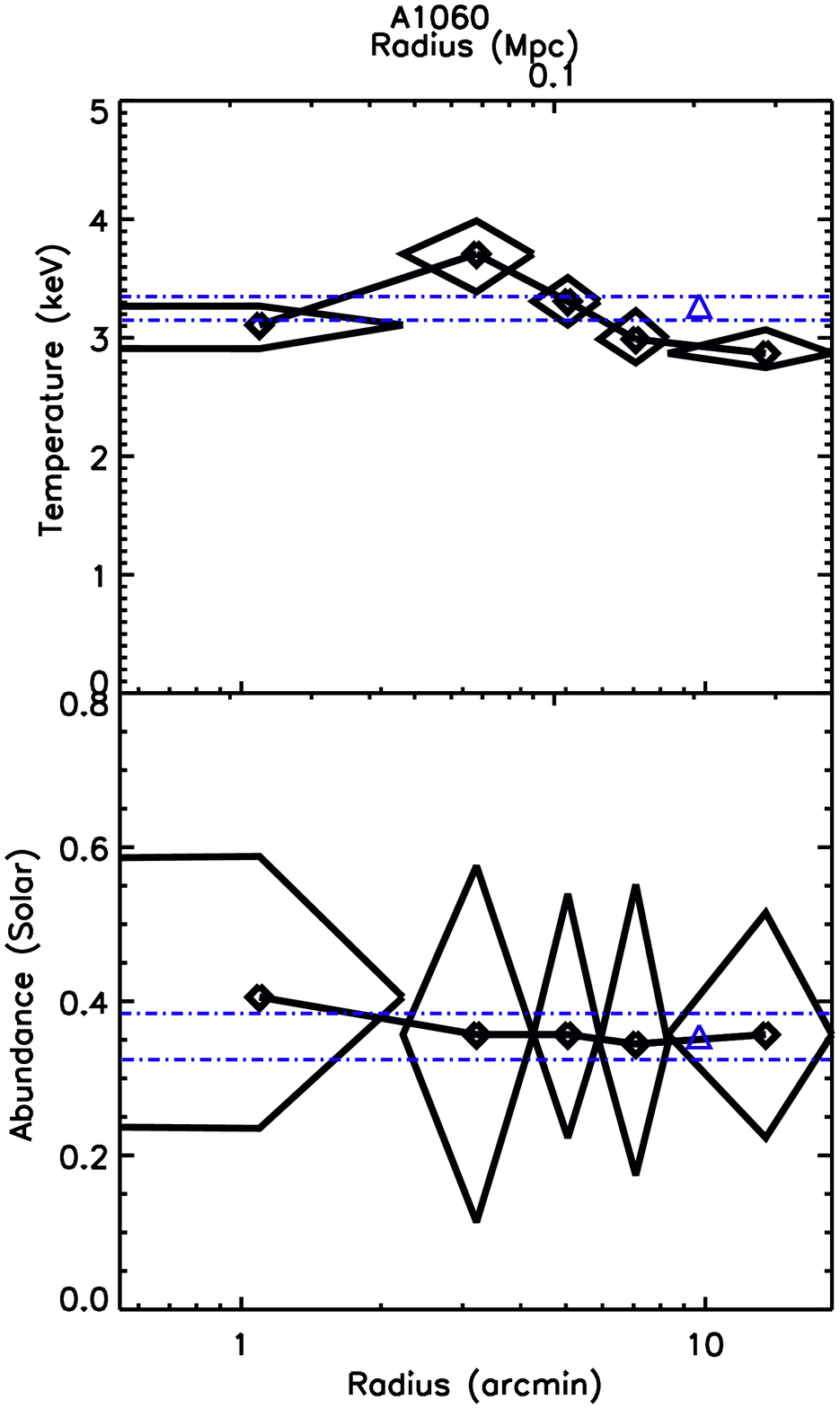,angle=0,width=\figwidth,height=\figheight}
    \psfig{figure=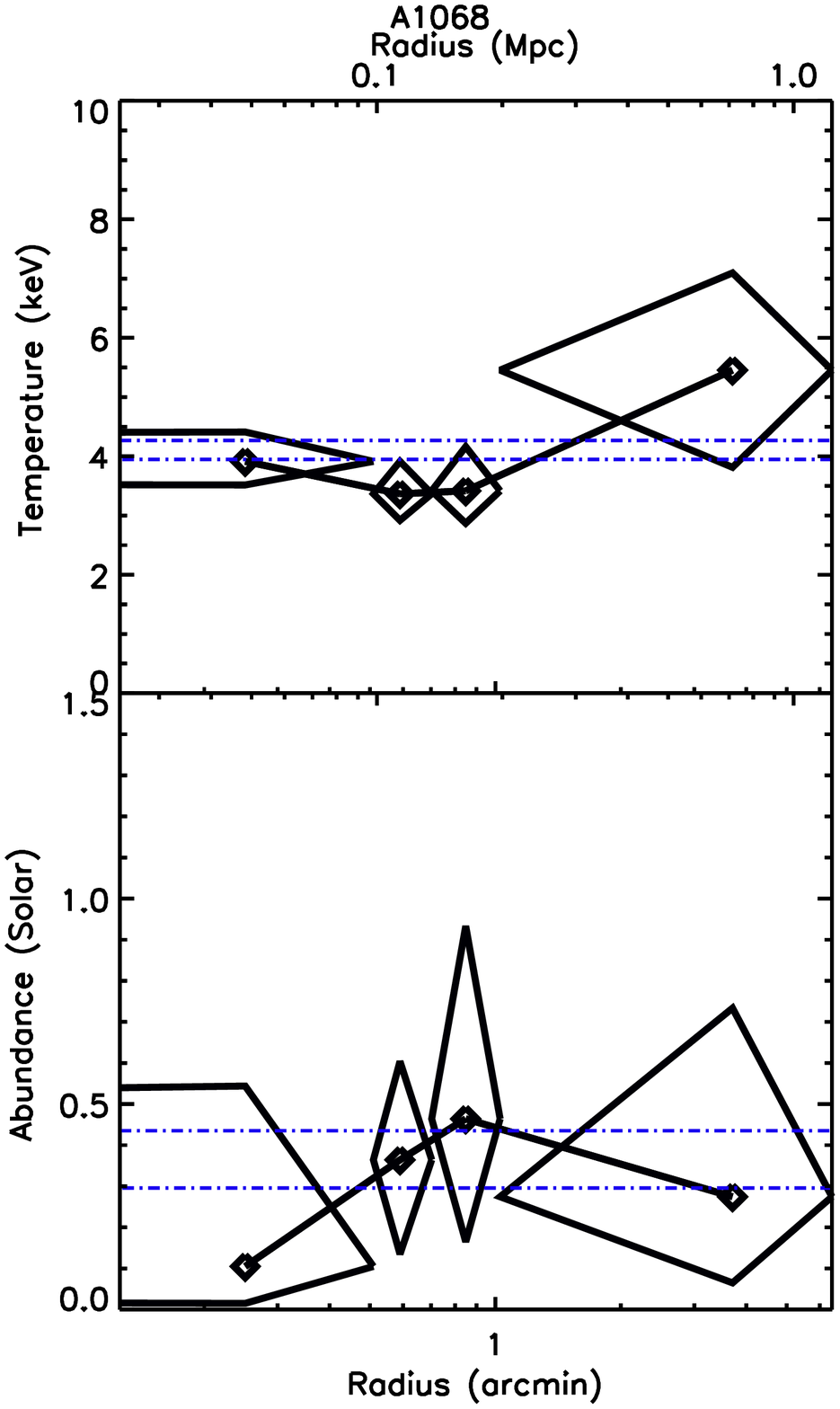,angle=0,width=\figwidth,height=\figheight}
    \psfig{figure=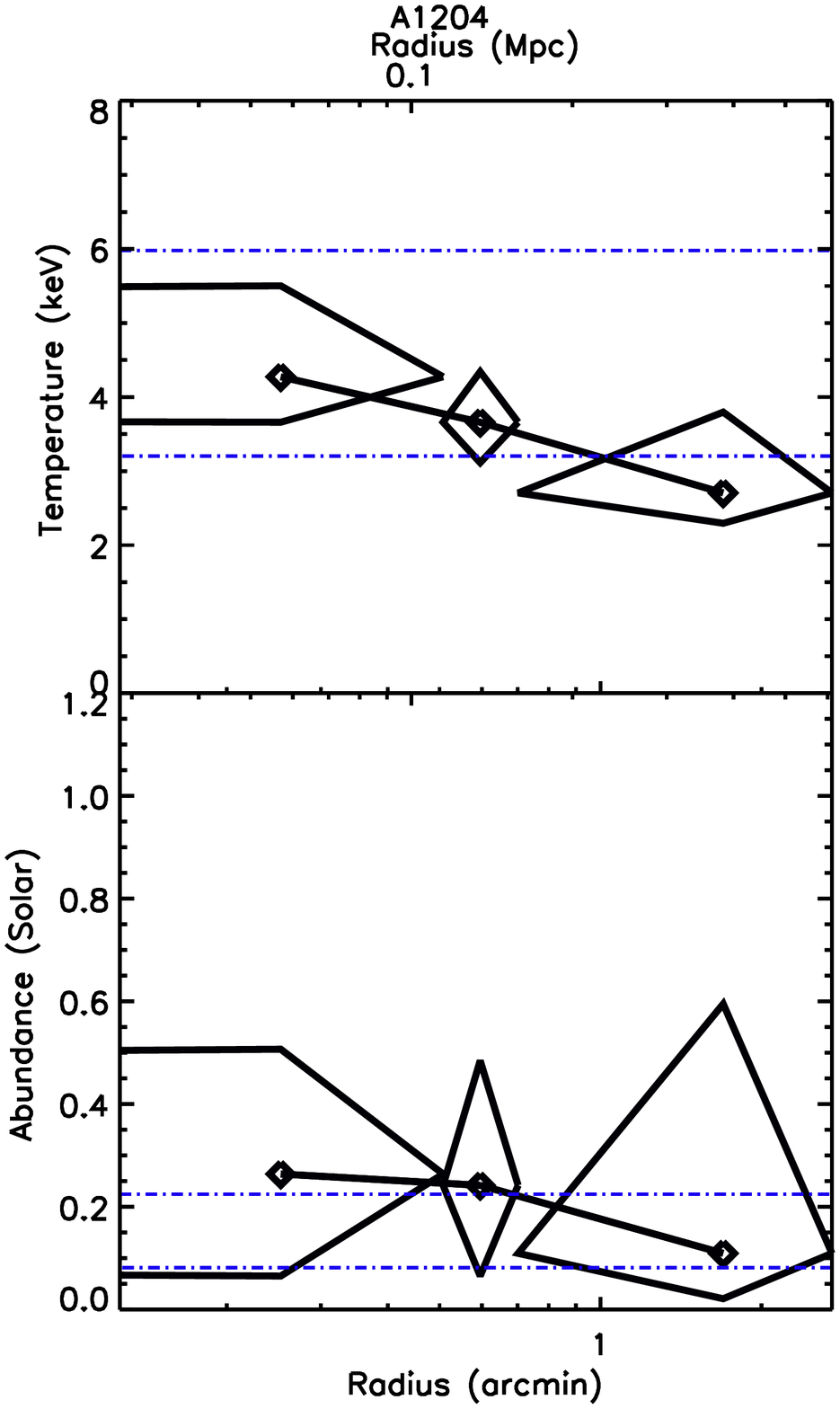,angle=0,width=\figwidth,height=\figheight}
  }
\parbox{\textwidth}{
    \psfig{figure=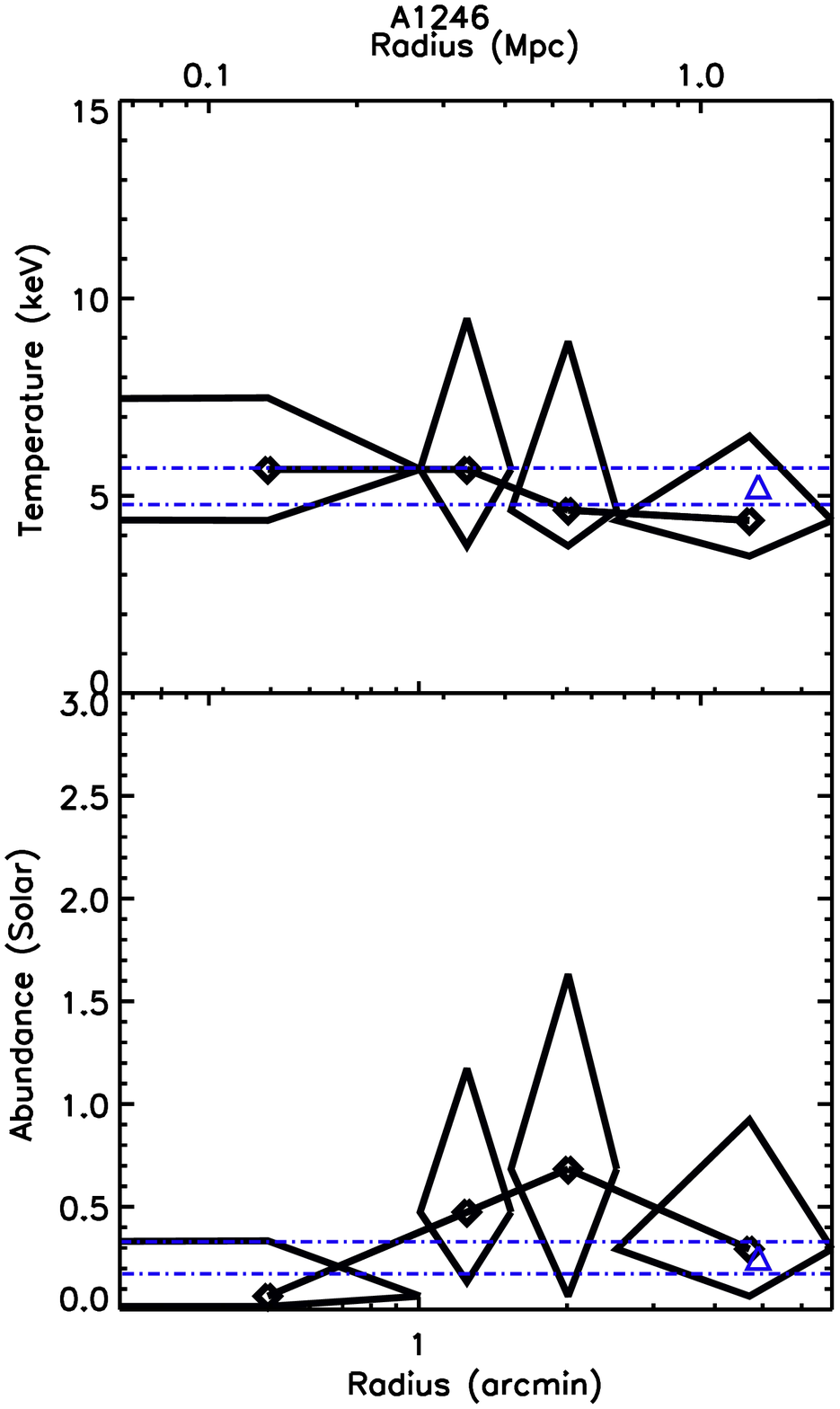,angle=0,width=\figwidth,height=\figheight}
    \psfig{figure=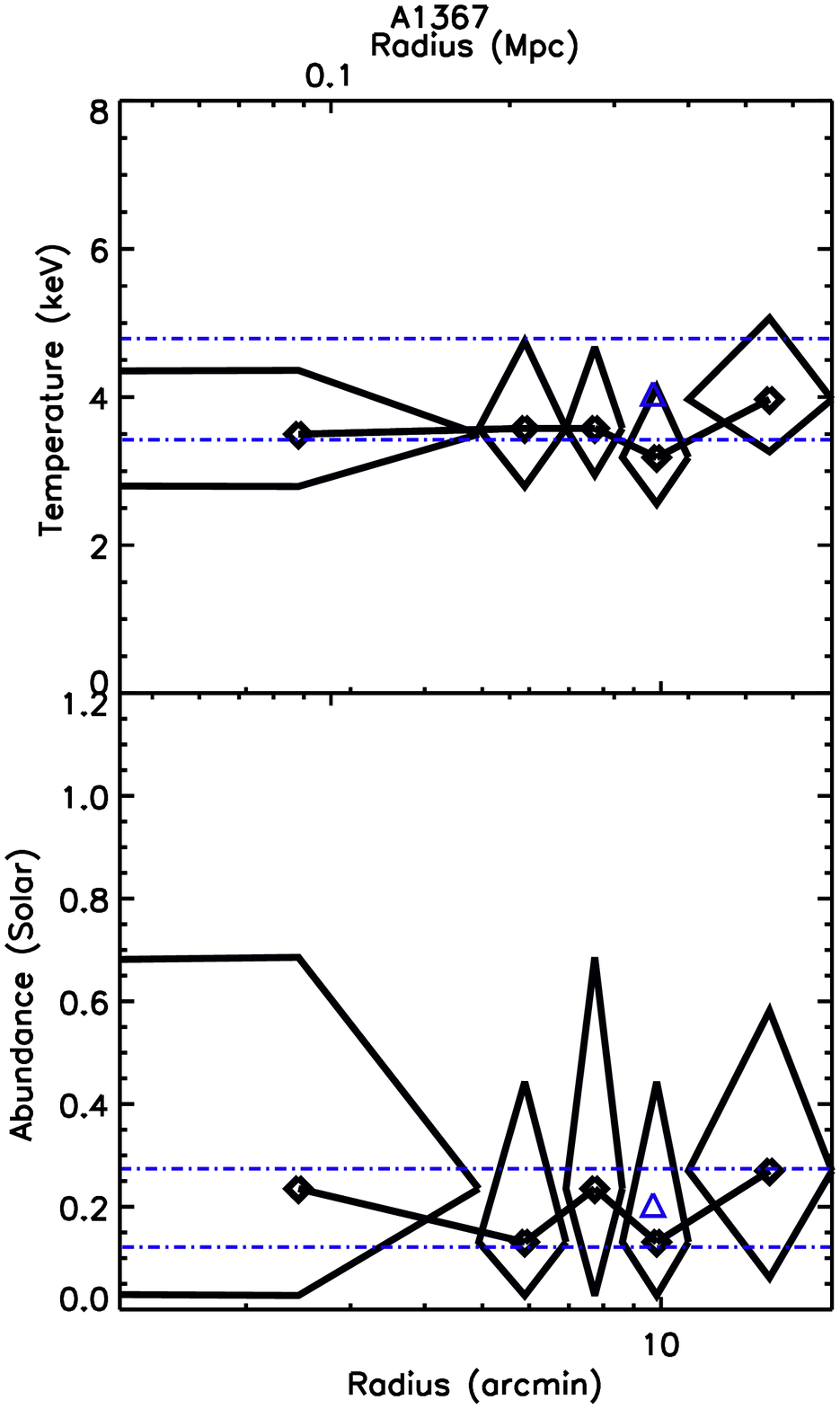,angle=0,width=\figwidth,height=\figheight}
    \psfig{figure=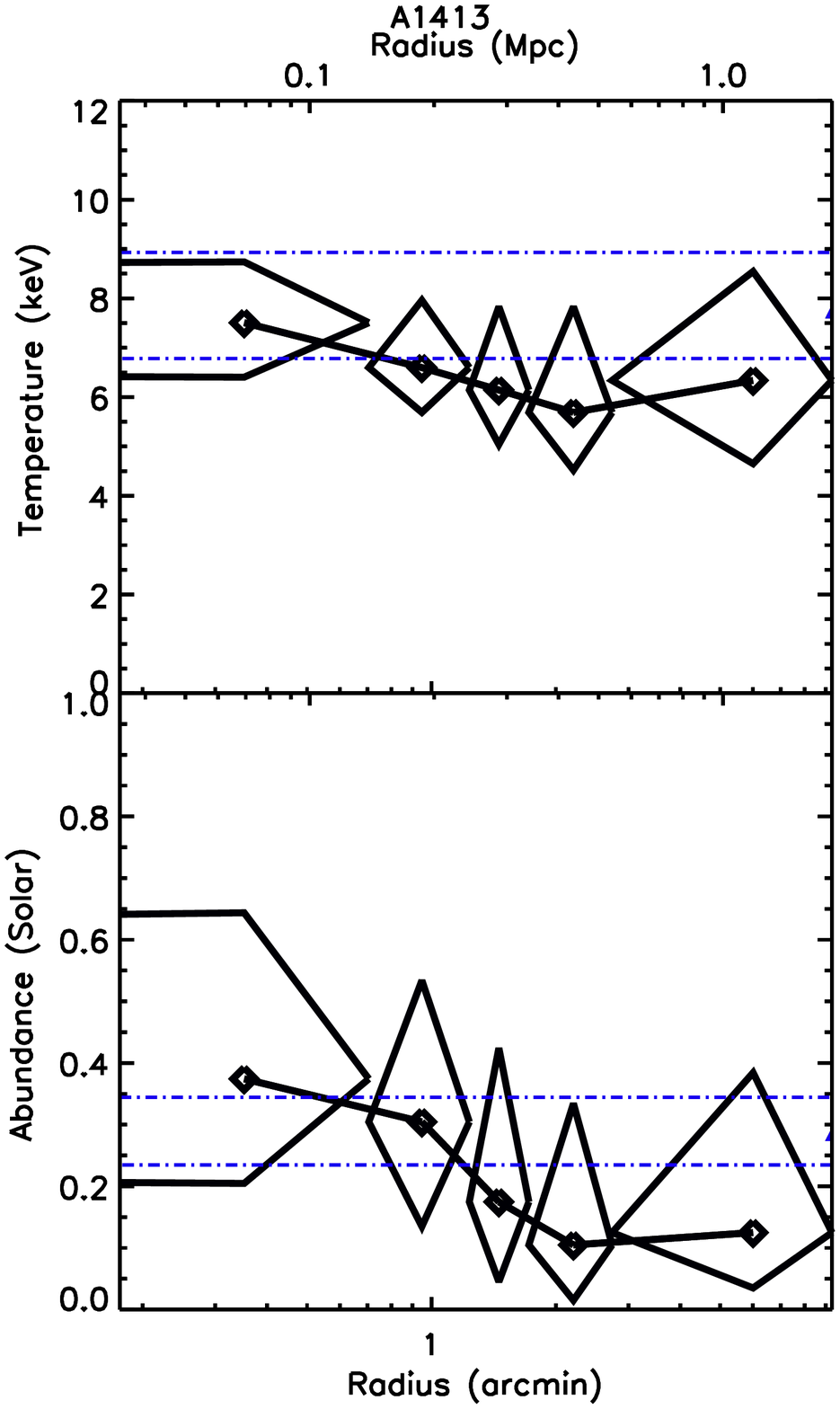,angle=0,width=\figwidth,height=\figheight}
    \psfig{figure=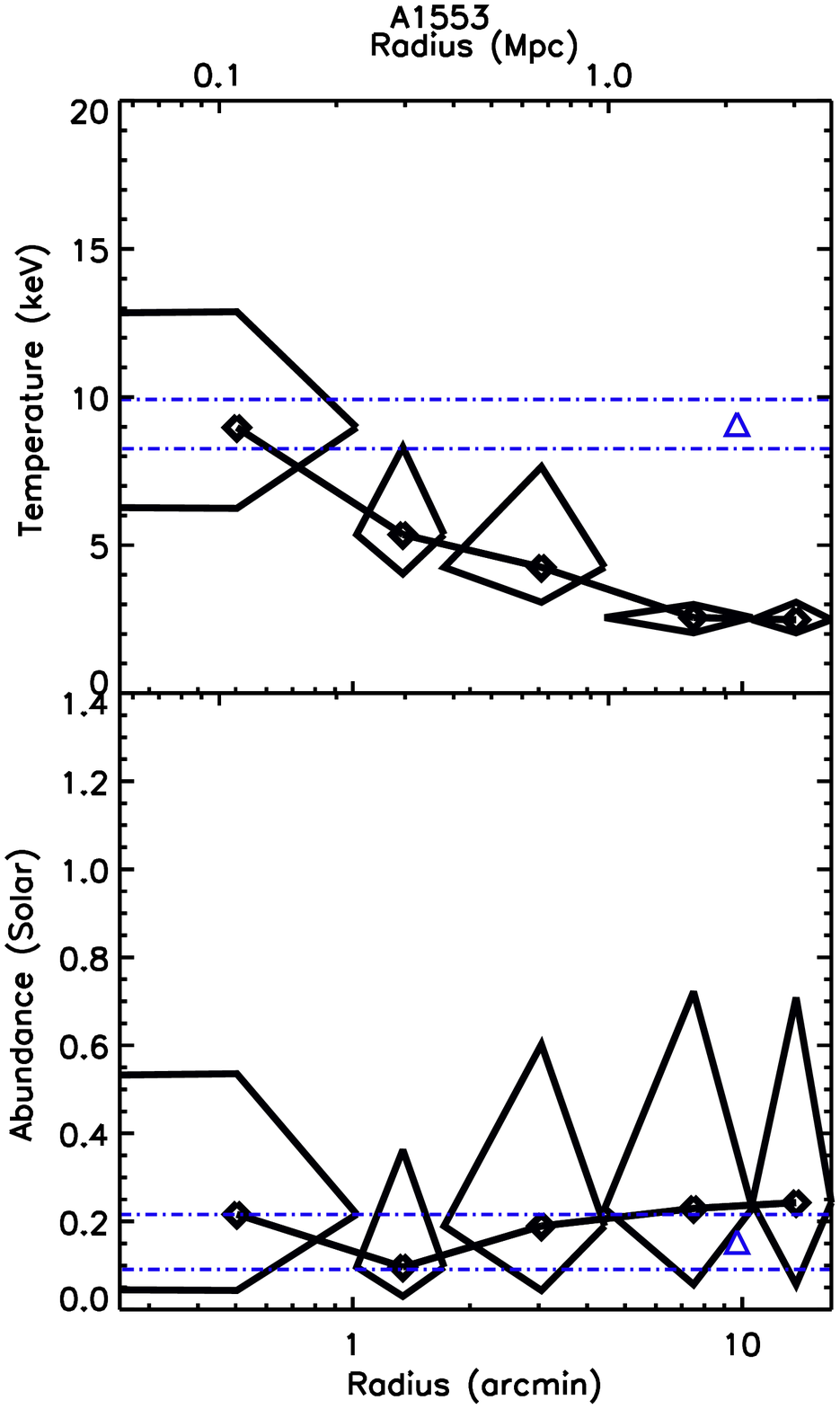,angle=0,width=\figwidth,height=\figheight}
  }
\parbox{\textwidth}{
    \psfig{figure=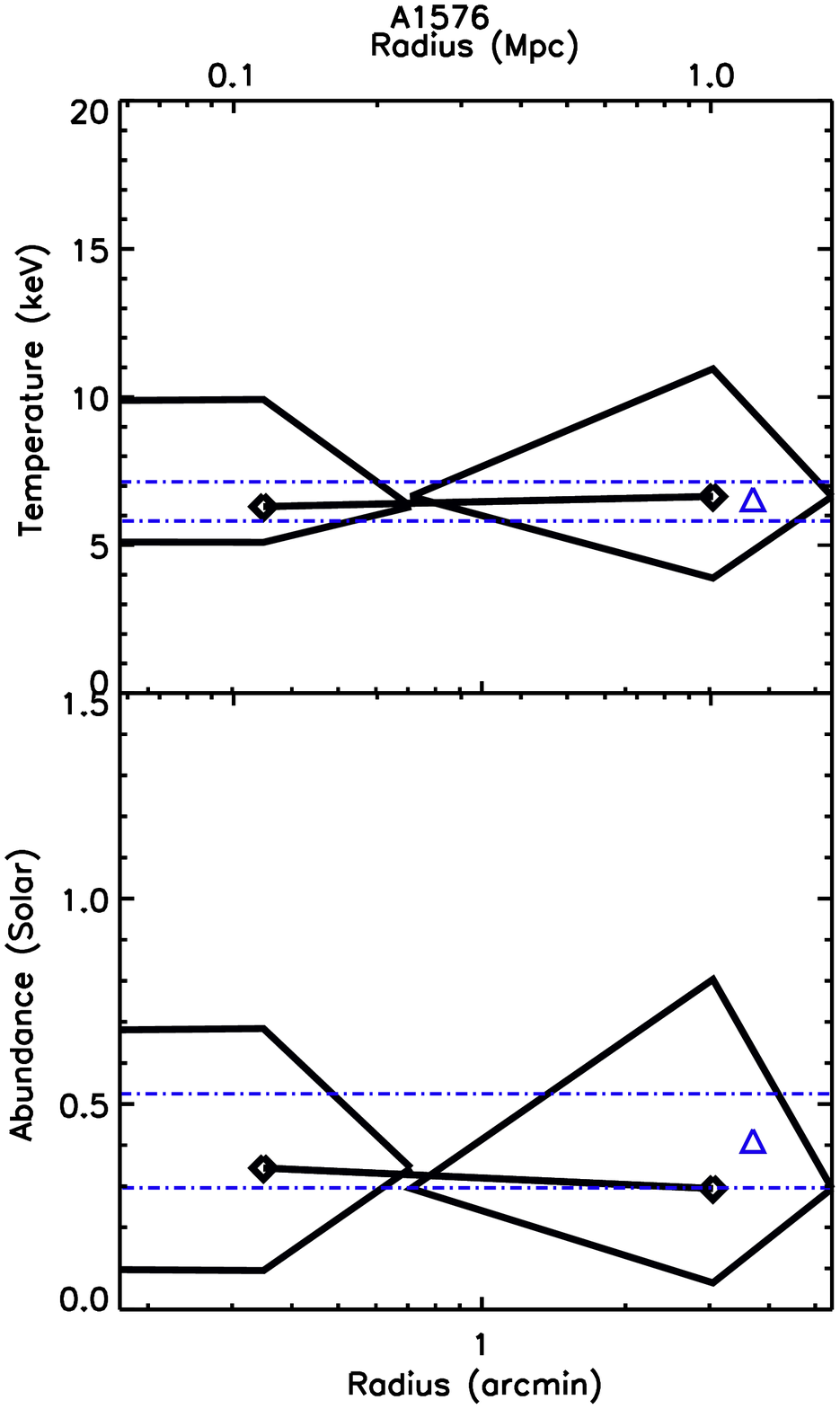,angle=0,width=\figwidth,height=\figheight}
    \psfig{figure=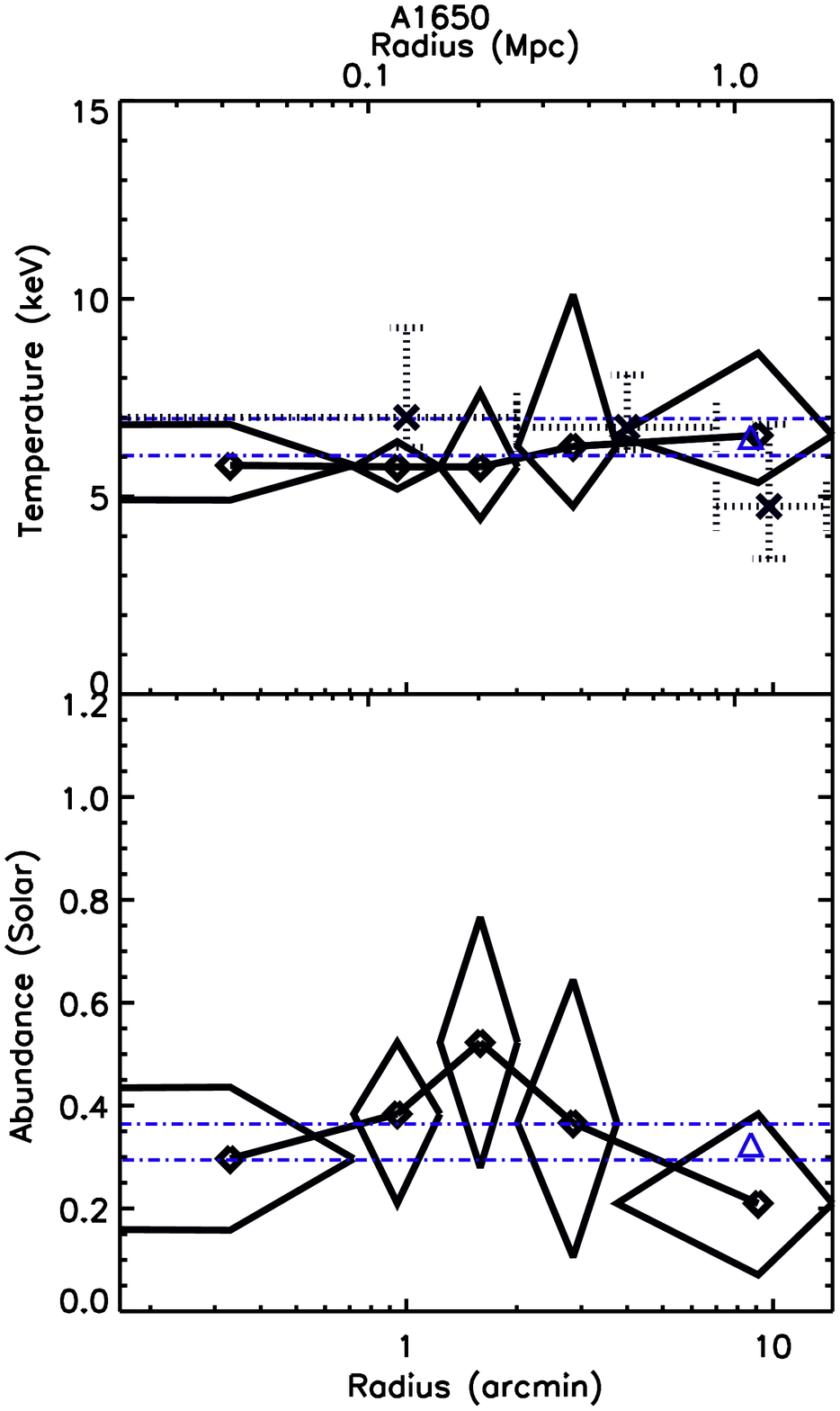,angle=0,width=\figwidth,height=\figheight}
    \psfig{figure=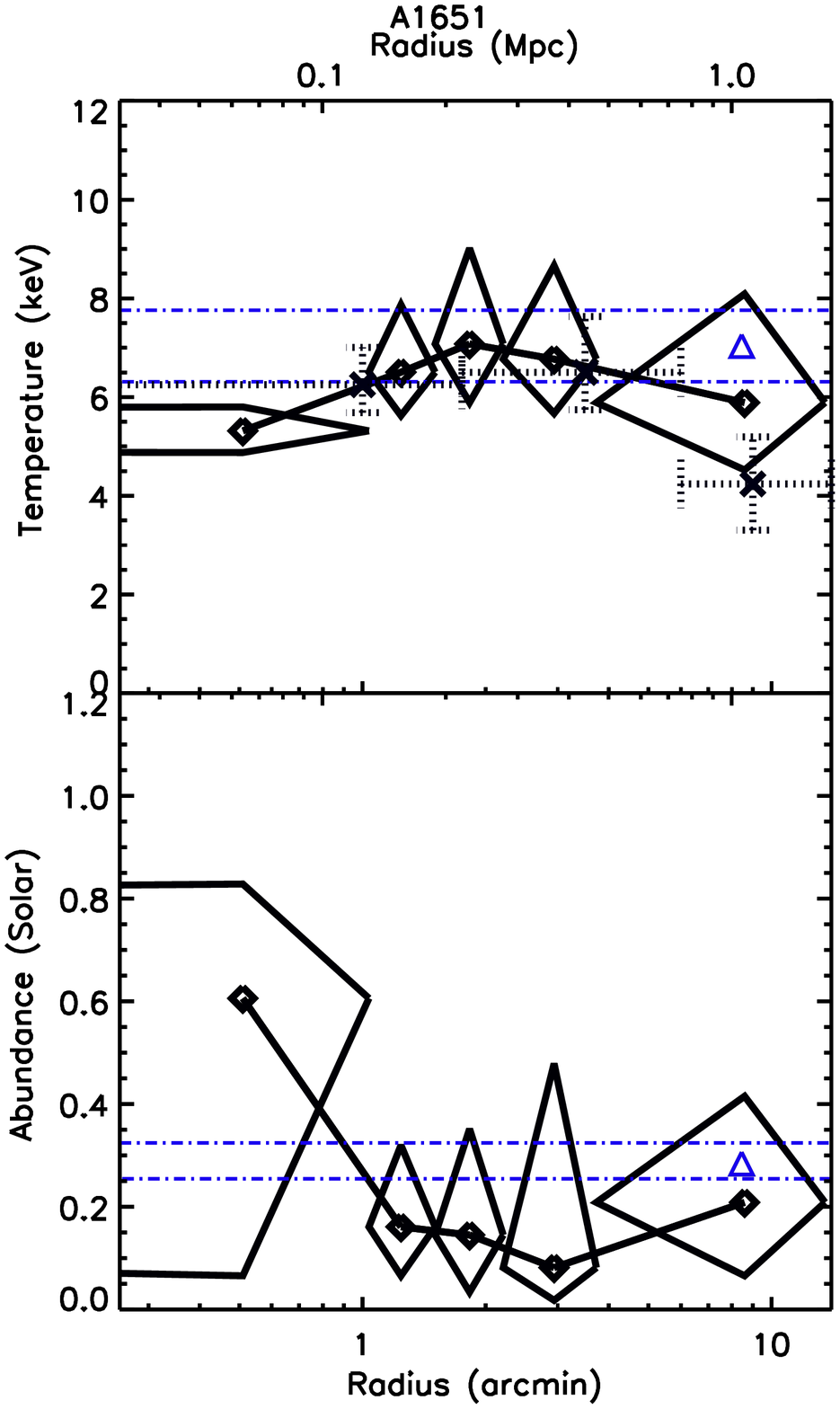,angle=0,width=\figwidth,height=\figheight}
    \psfig{figure=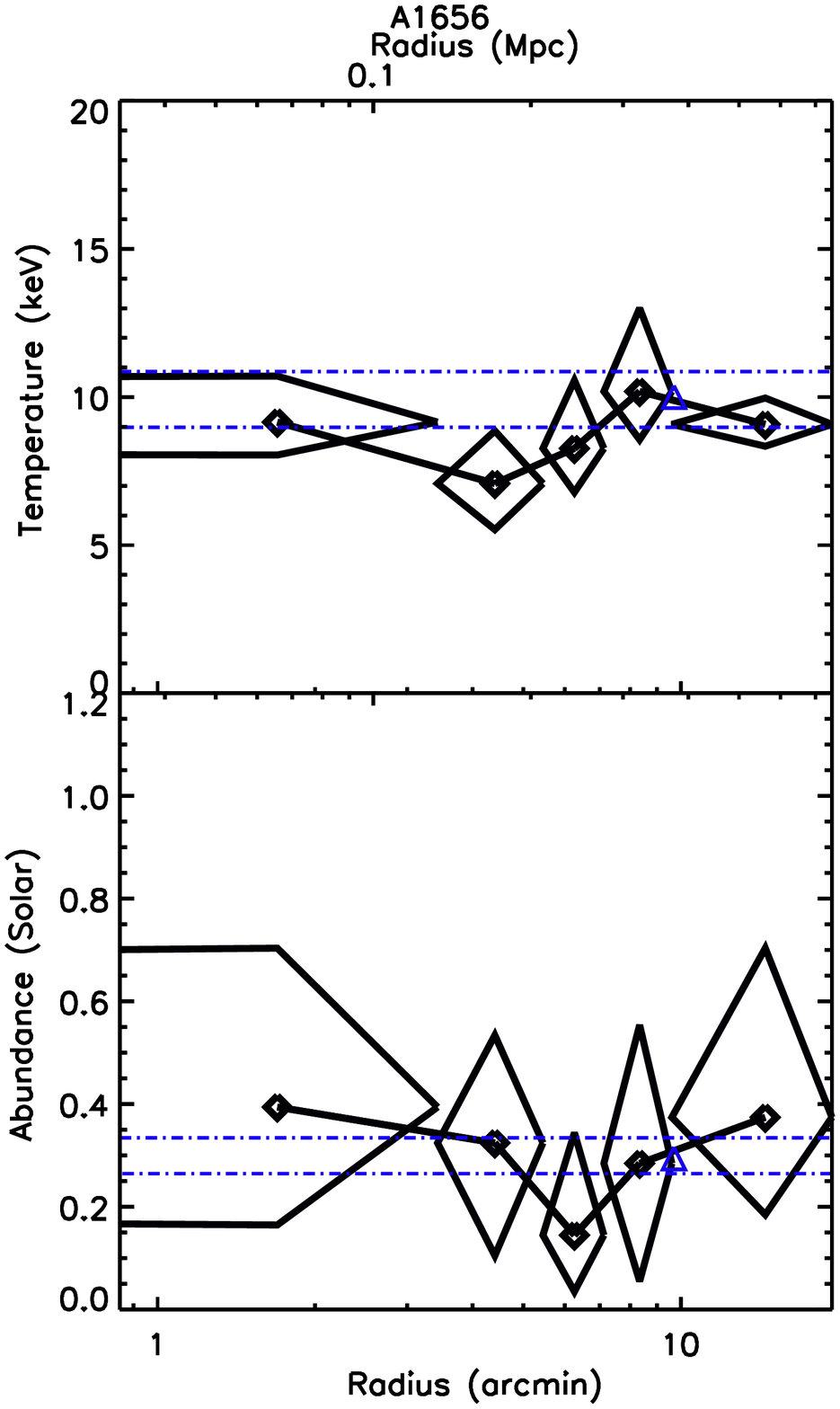,angle=0,width=\figwidth,height=\figheight}
  }
\end{figure*}
\clearpage
\begin{figure*}
\parbox{\textwidth}{
    \psfig{figure=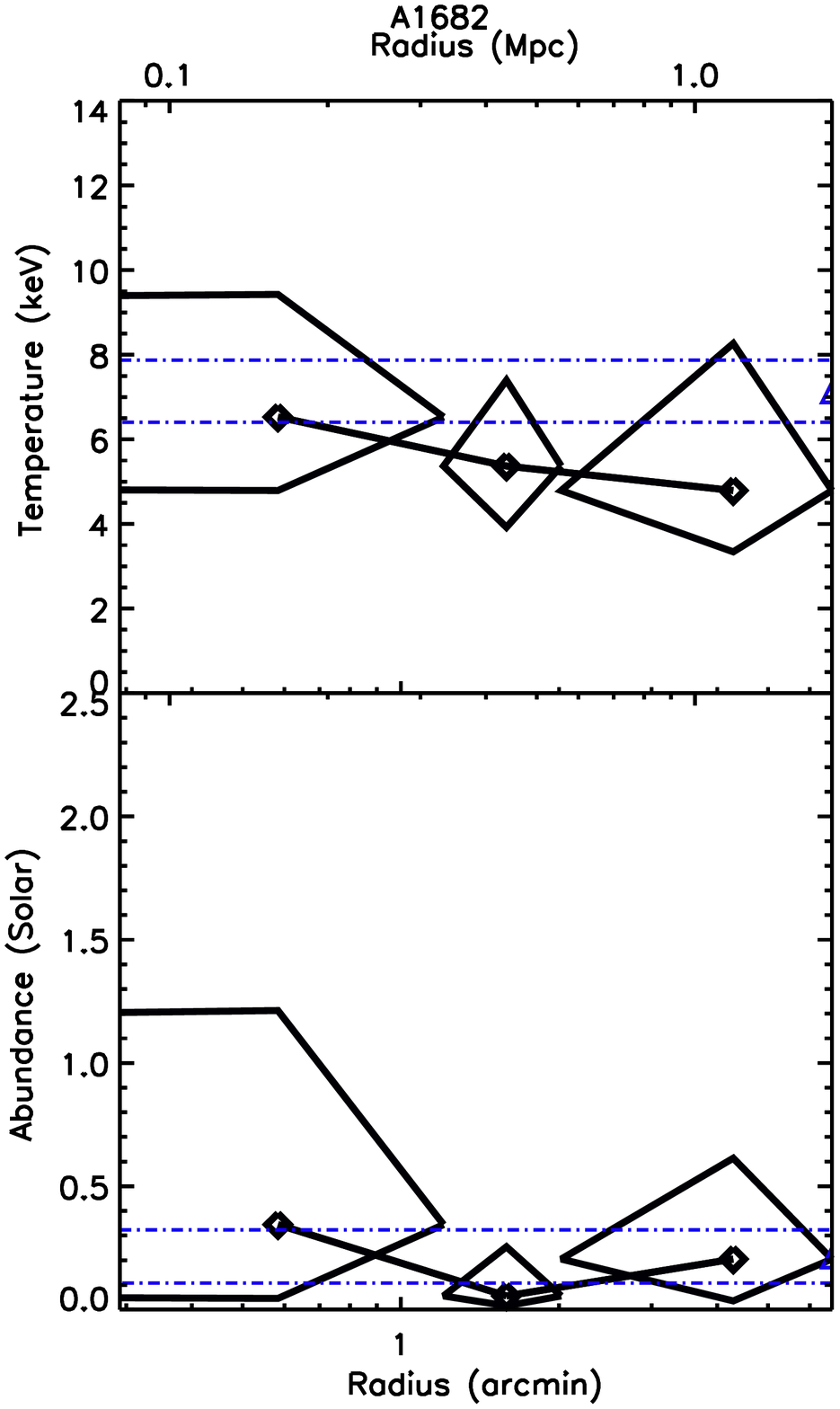,angle=0,width=\figwidth,height=\figheight}
    \psfig{figure=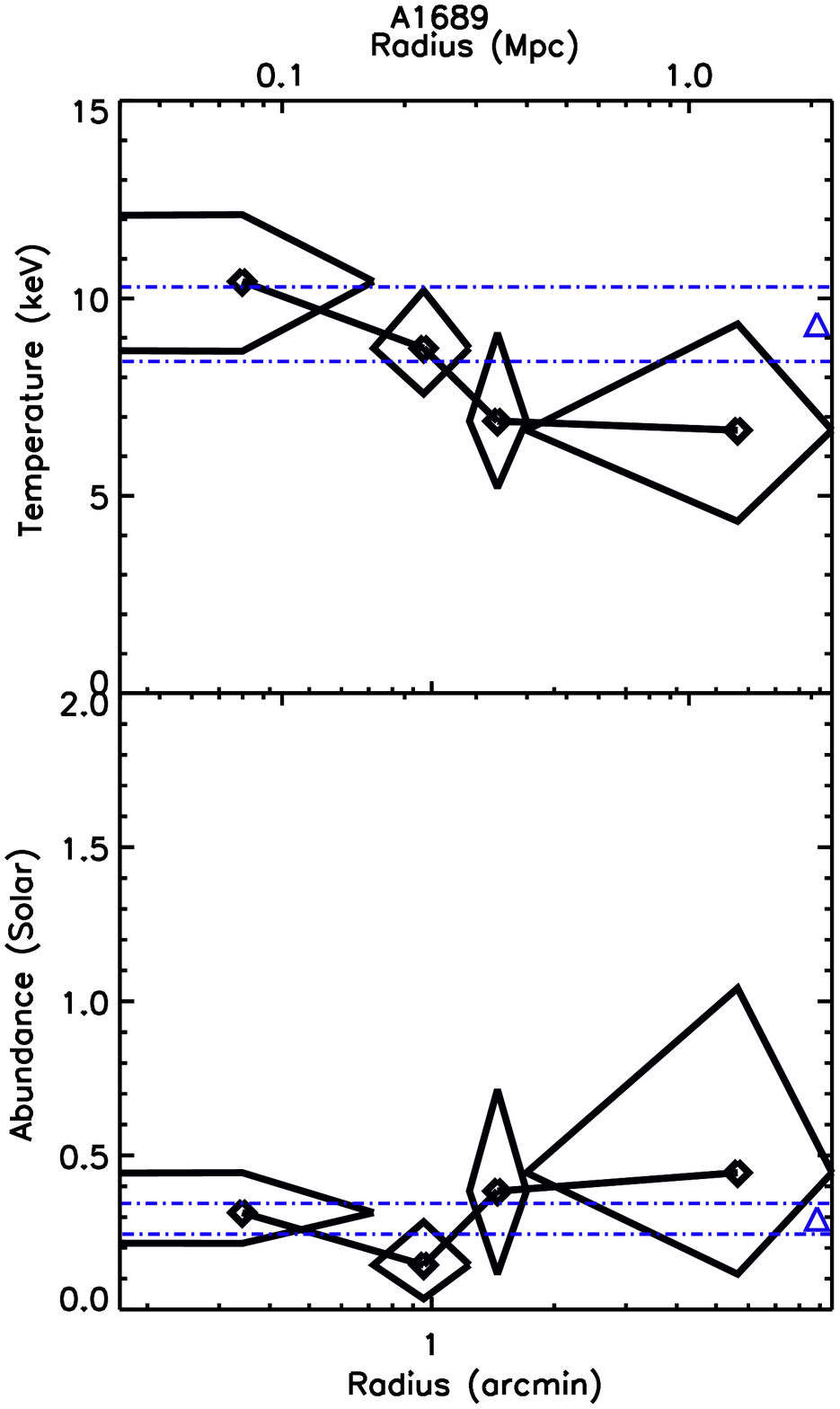,angle=0,width=\figwidth,height=\figheight}
    \psfig{figure=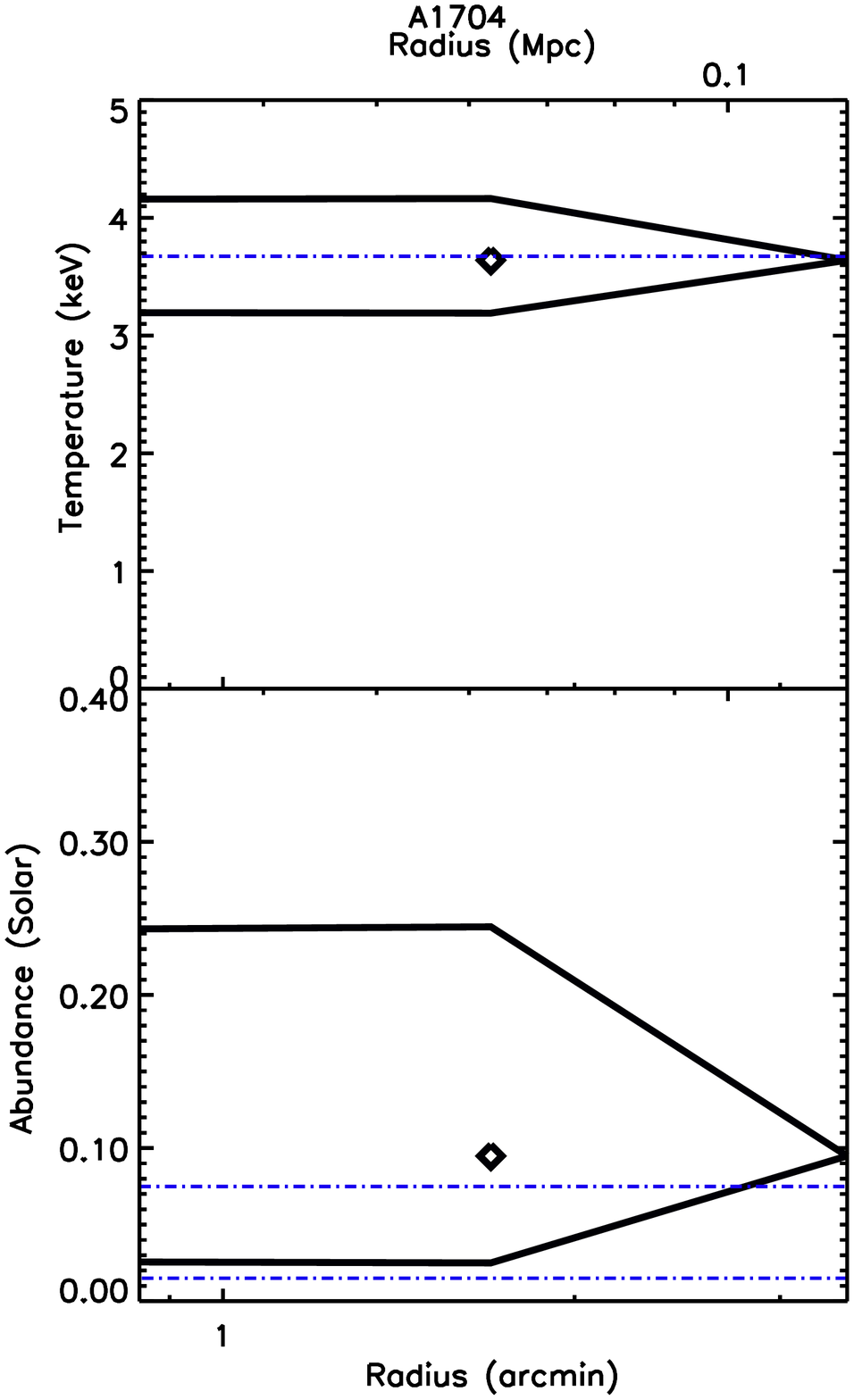,angle=0,width=\figwidth,height=\figheight}
    \psfig{figure=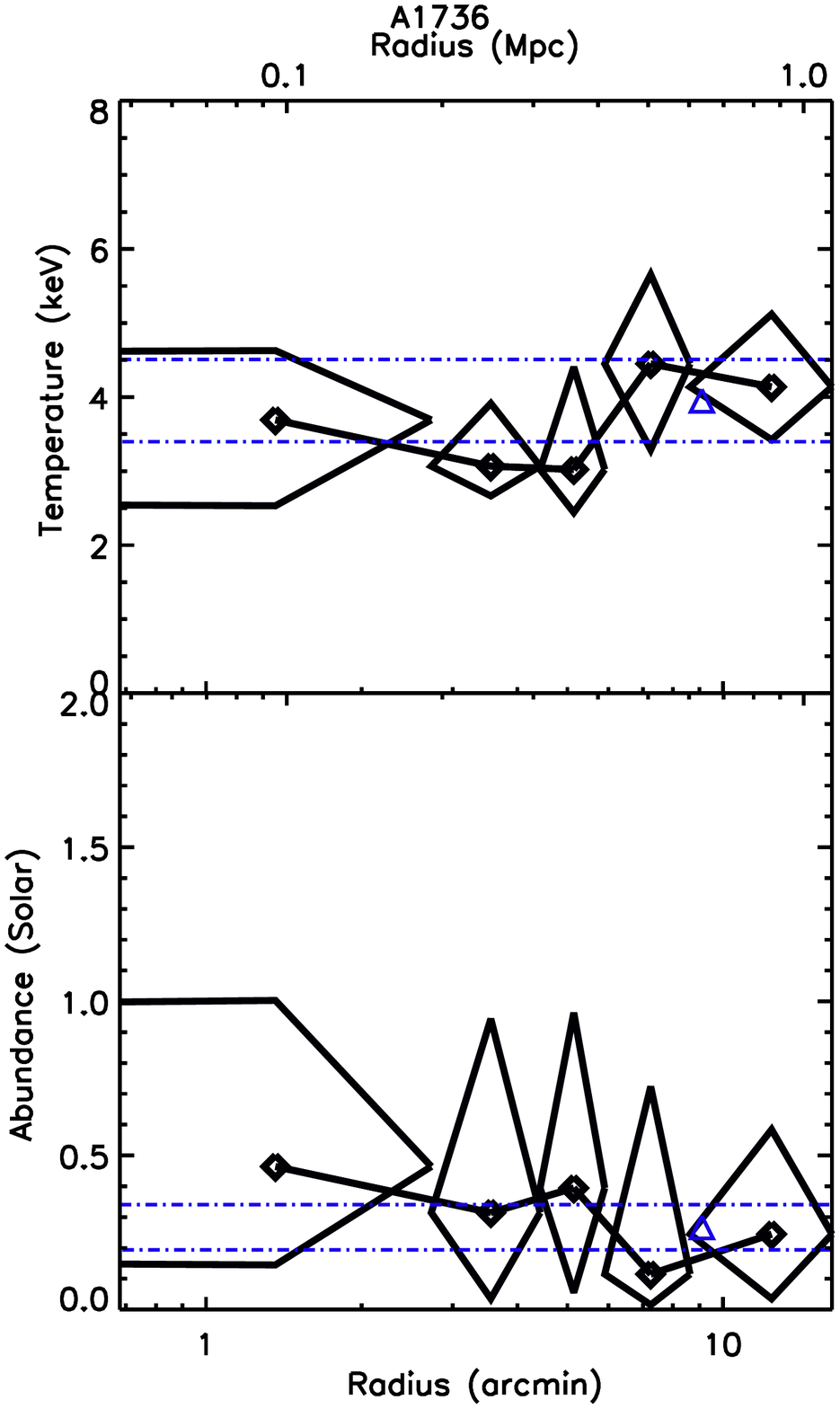,angle=0,width=\figwidth,height=\figheight}
  }
\parbox{\textwidth}{
    \psfig{figure=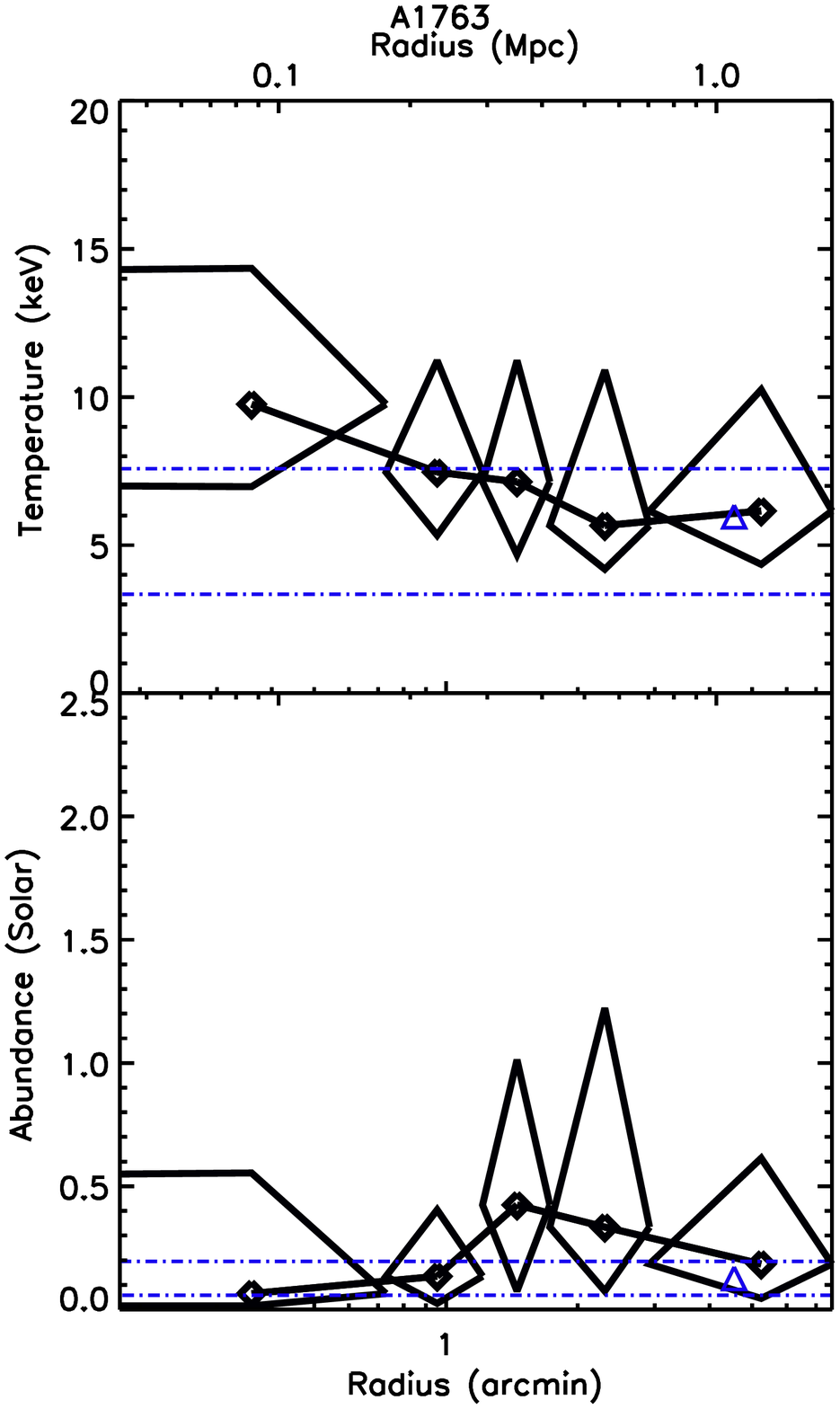,angle=0,width=\figwidth,height=\figheight}
    \psfig{figure=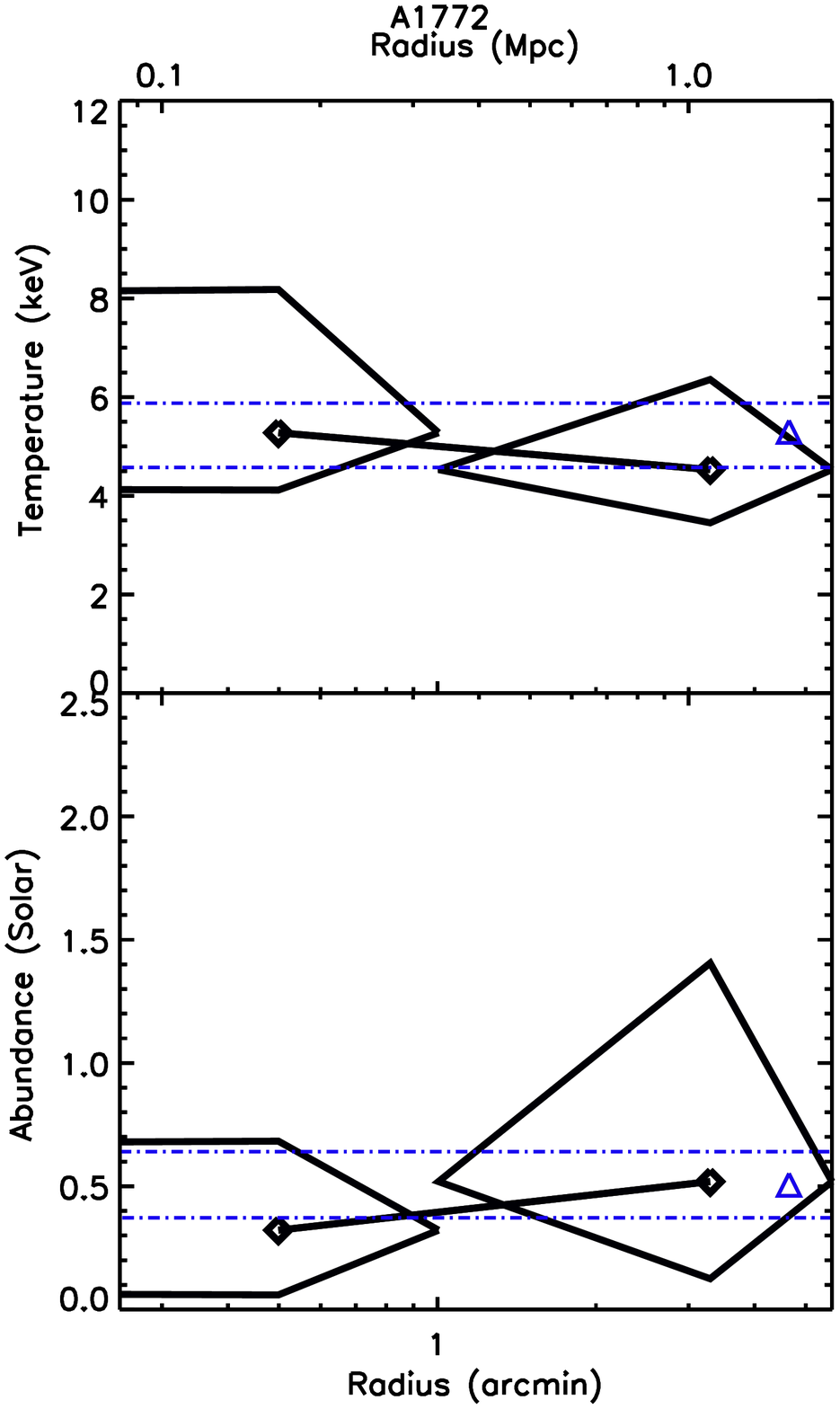,angle=0,width=\figwidth,height=\figheight}
    \psfig{figure=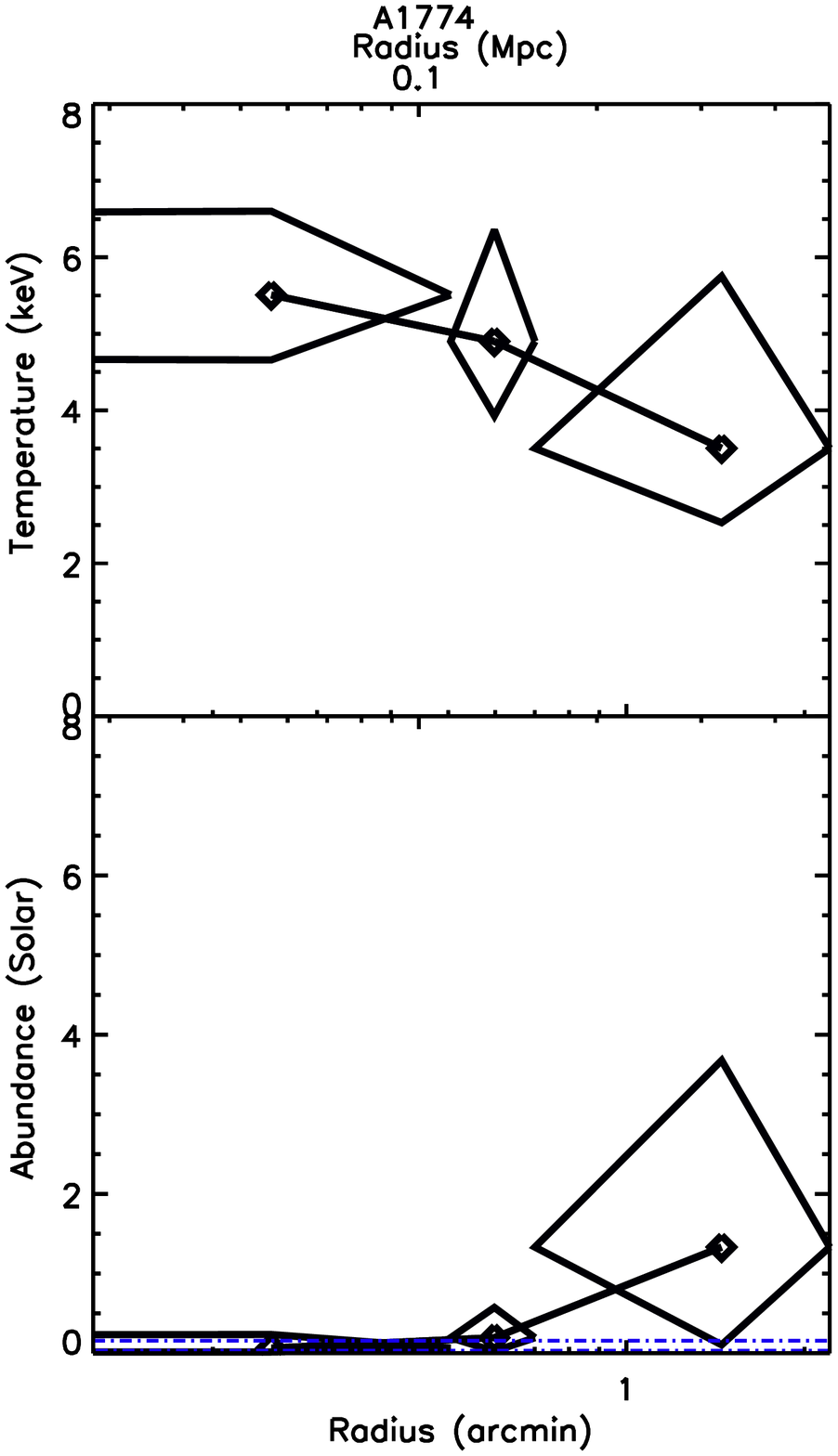,angle=0,width=\figwidth,height=\figheight}
    \psfig{figure=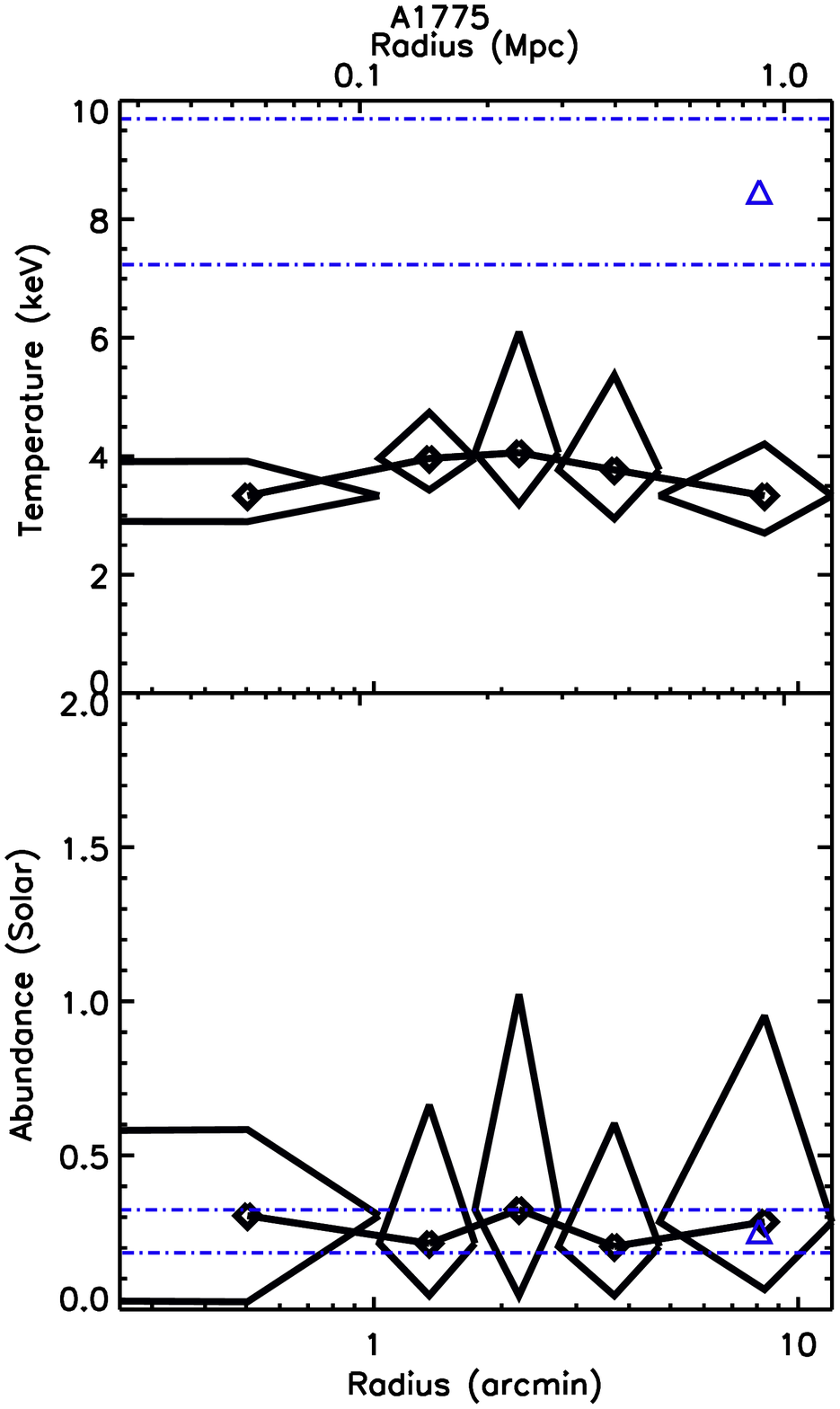,angle=0,width=\figwidth,height=\figheight}
  }
\parbox{\textwidth}{
    \psfig{figure=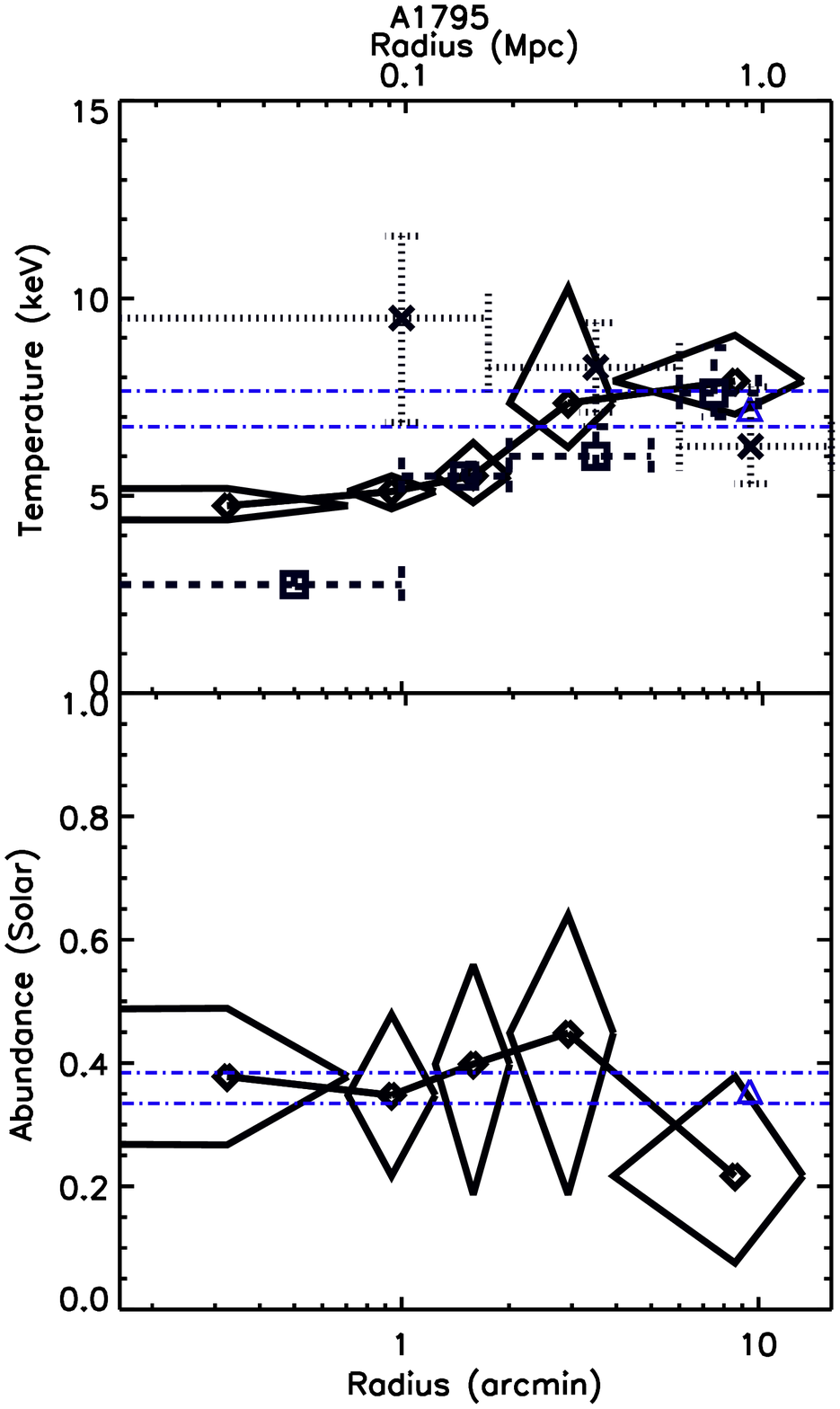,angle=0,width=\figwidth,height=\figheight}
    \psfig{figure=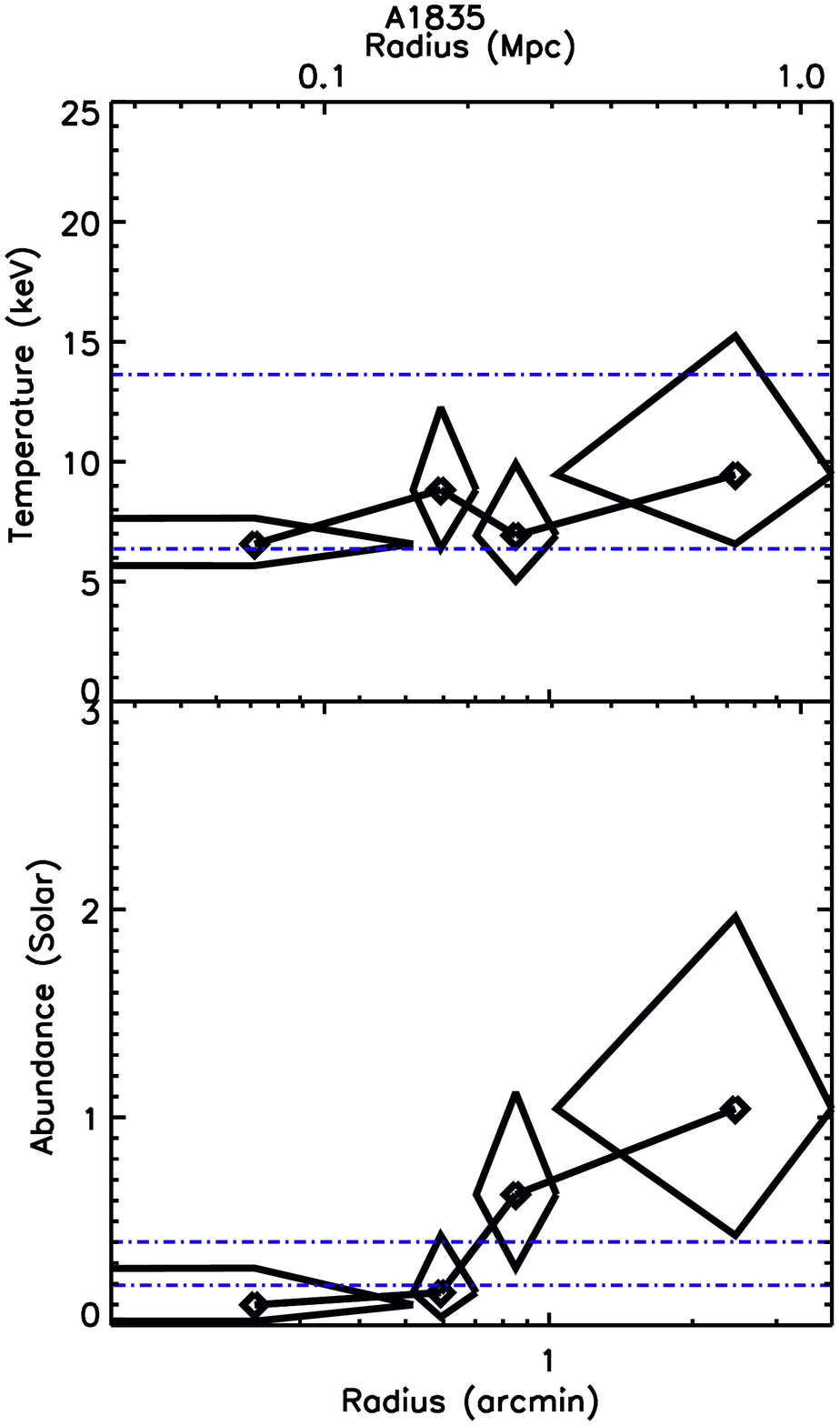,angle=0,width=\figwidth,height=\figheight}
    \psfig{figure=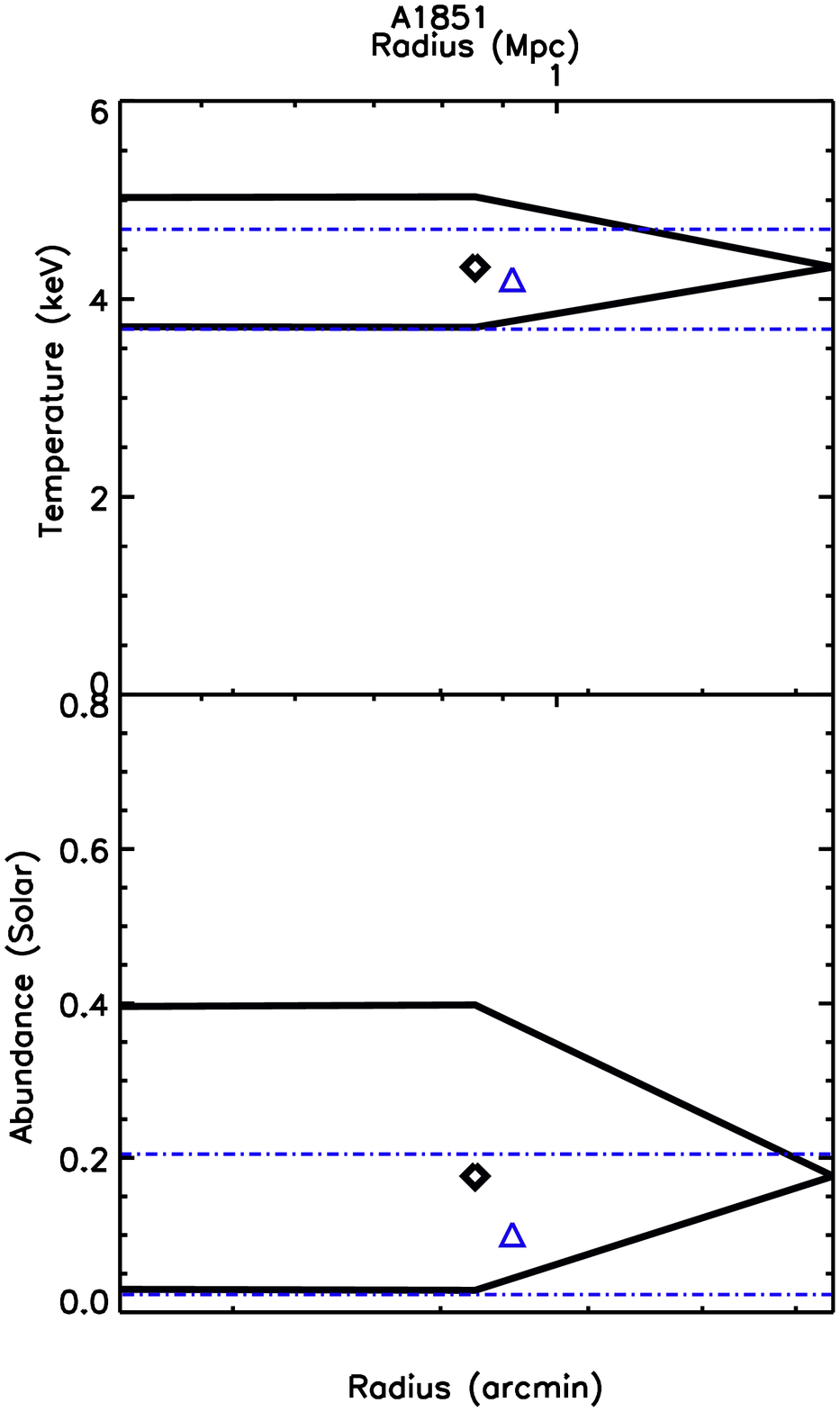,angle=0,width=\figwidth,height=\figheight}
    \psfig{figure=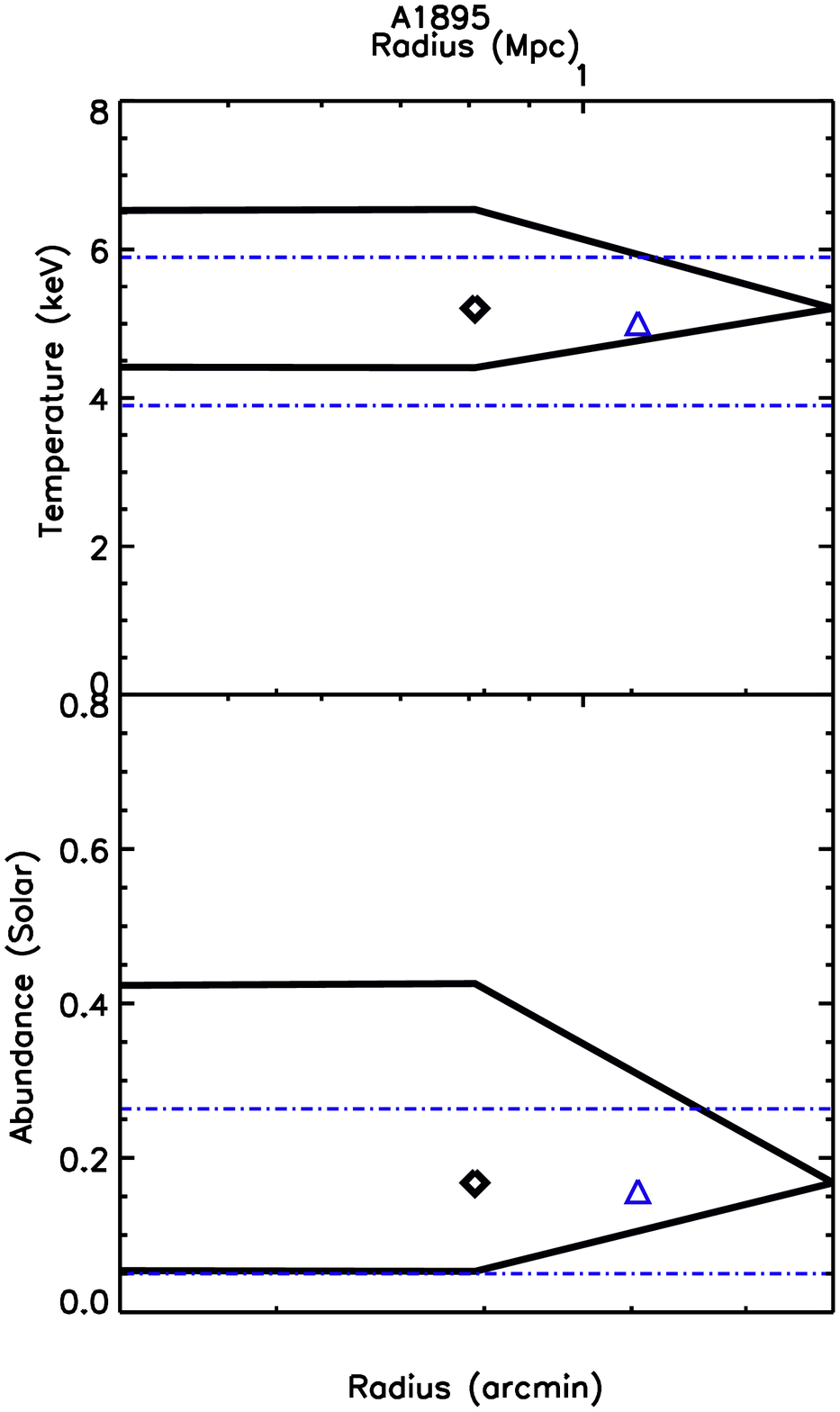,angle=0,width=\figwidth,height=\figheight}
  }
\end{figure*}
\clearpage
\begin{figure*}
\parbox{\textwidth}{
    \psfig{figure=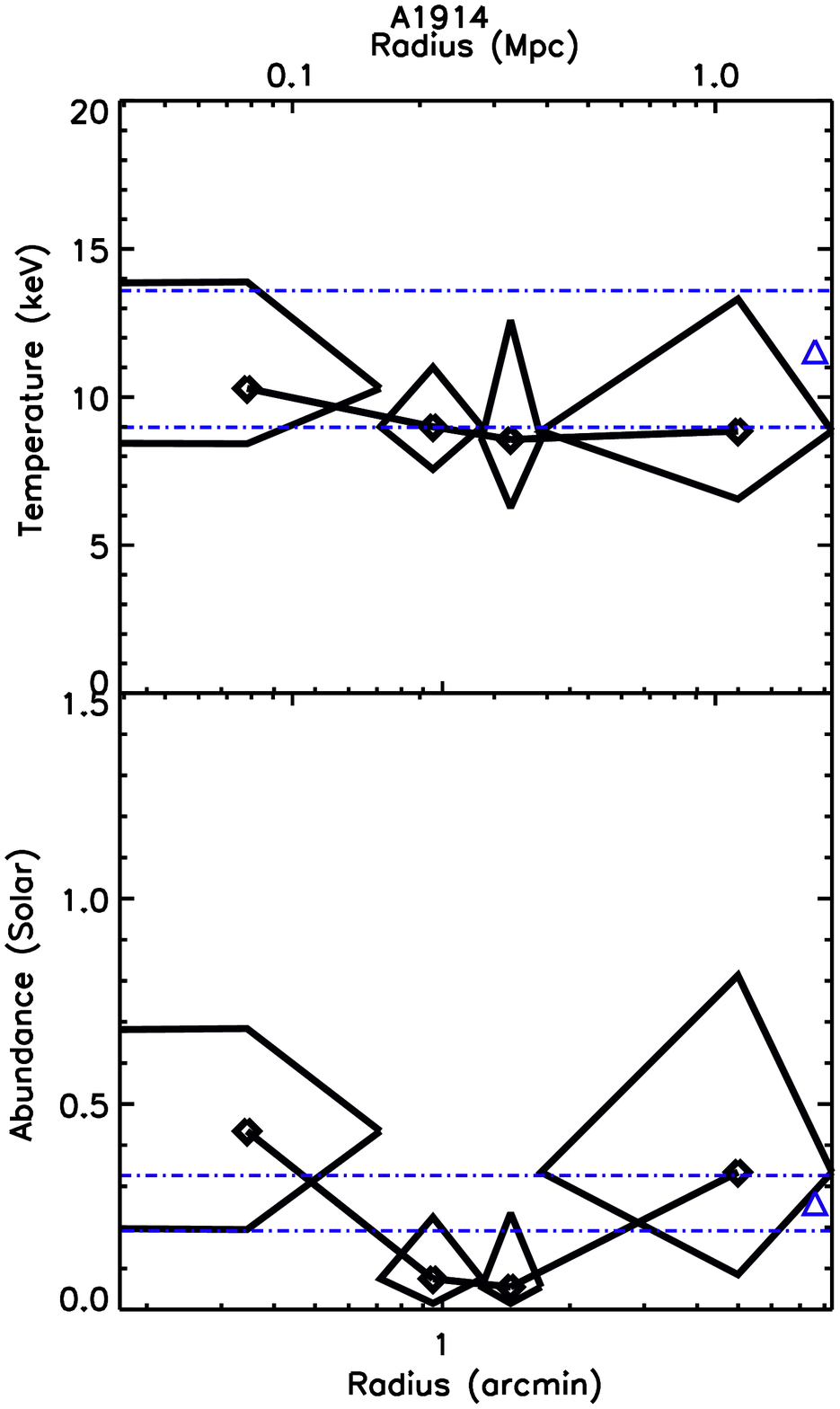,angle=0,width=\figwidth,height=\figheight}
    \psfig{figure=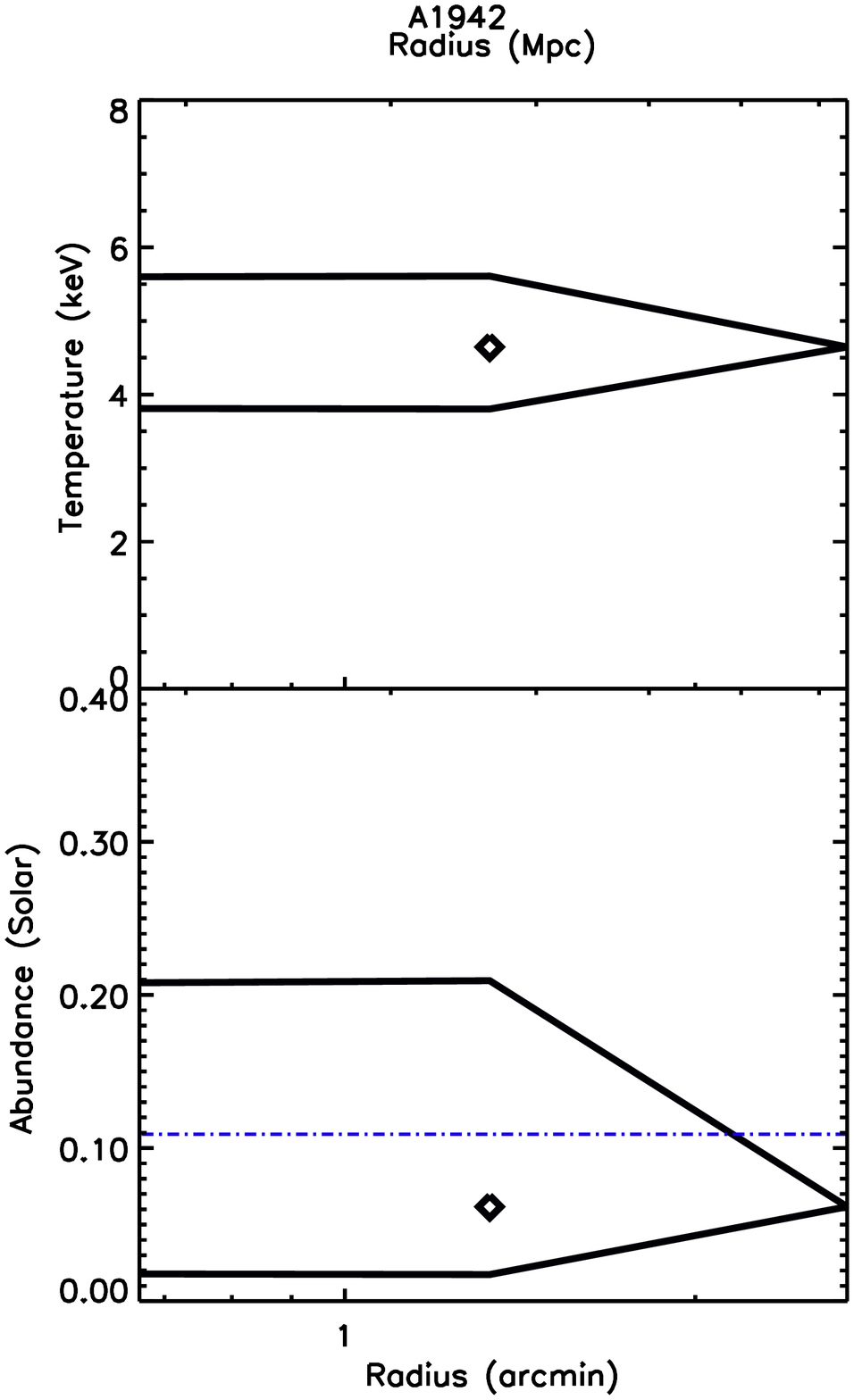,angle=0,width=\figwidth,height=\figheight}
    \psfig{figure=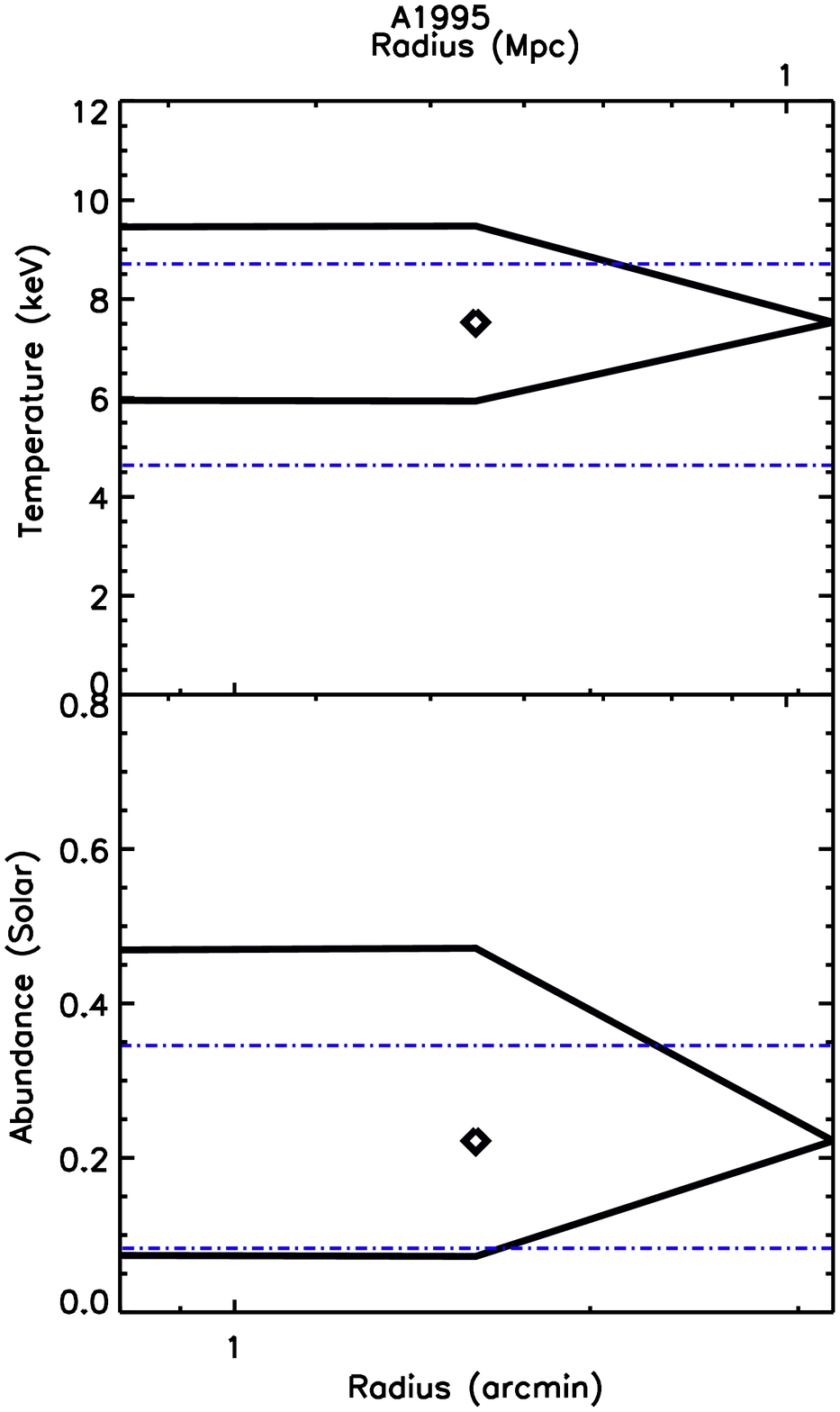,angle=0,width=\figwidth,height=\figheight}
    \psfig{figure=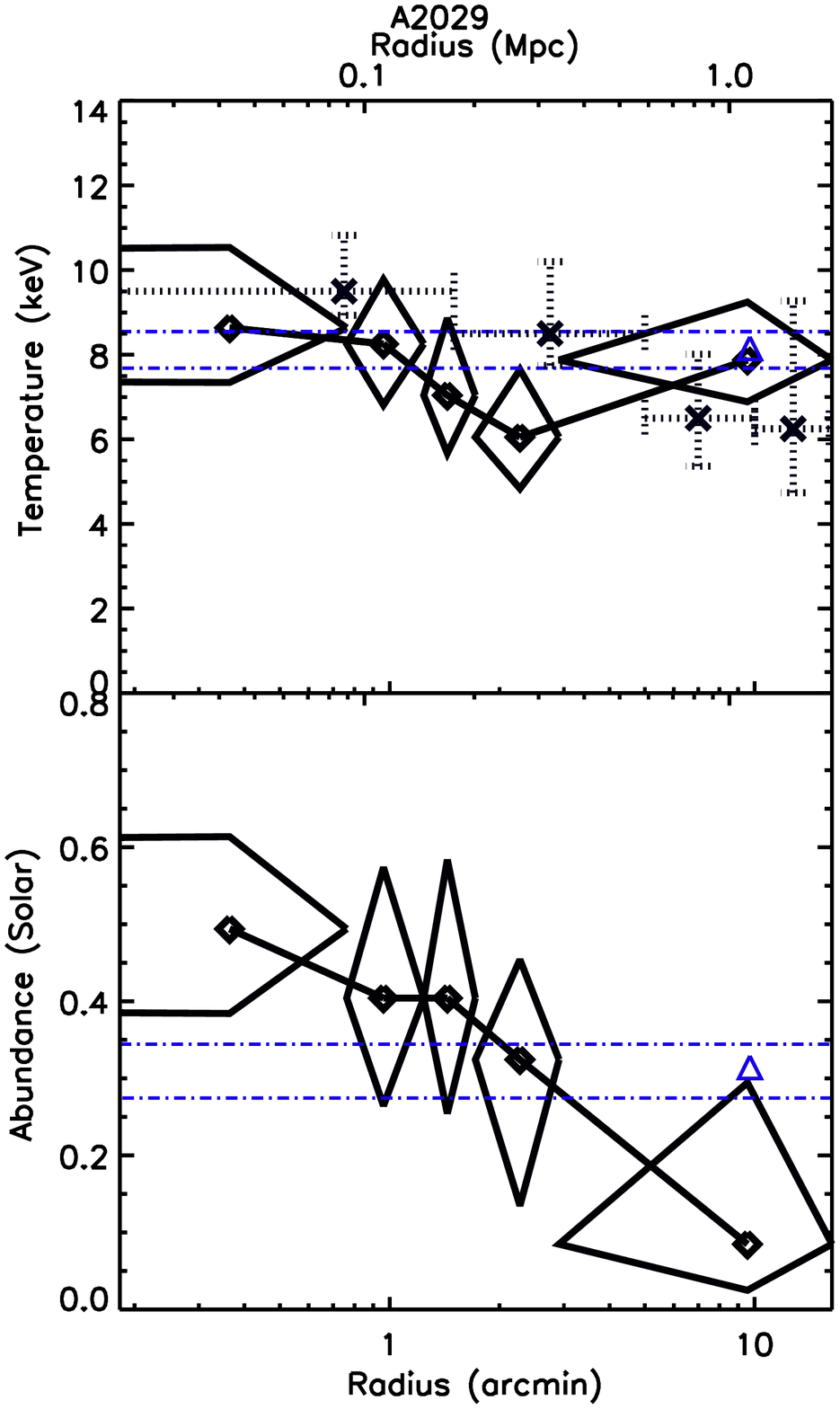,angle=0,width=\figwidth,height=\figheight}
  }
\parbox{\textwidth}{
    \psfig{figure=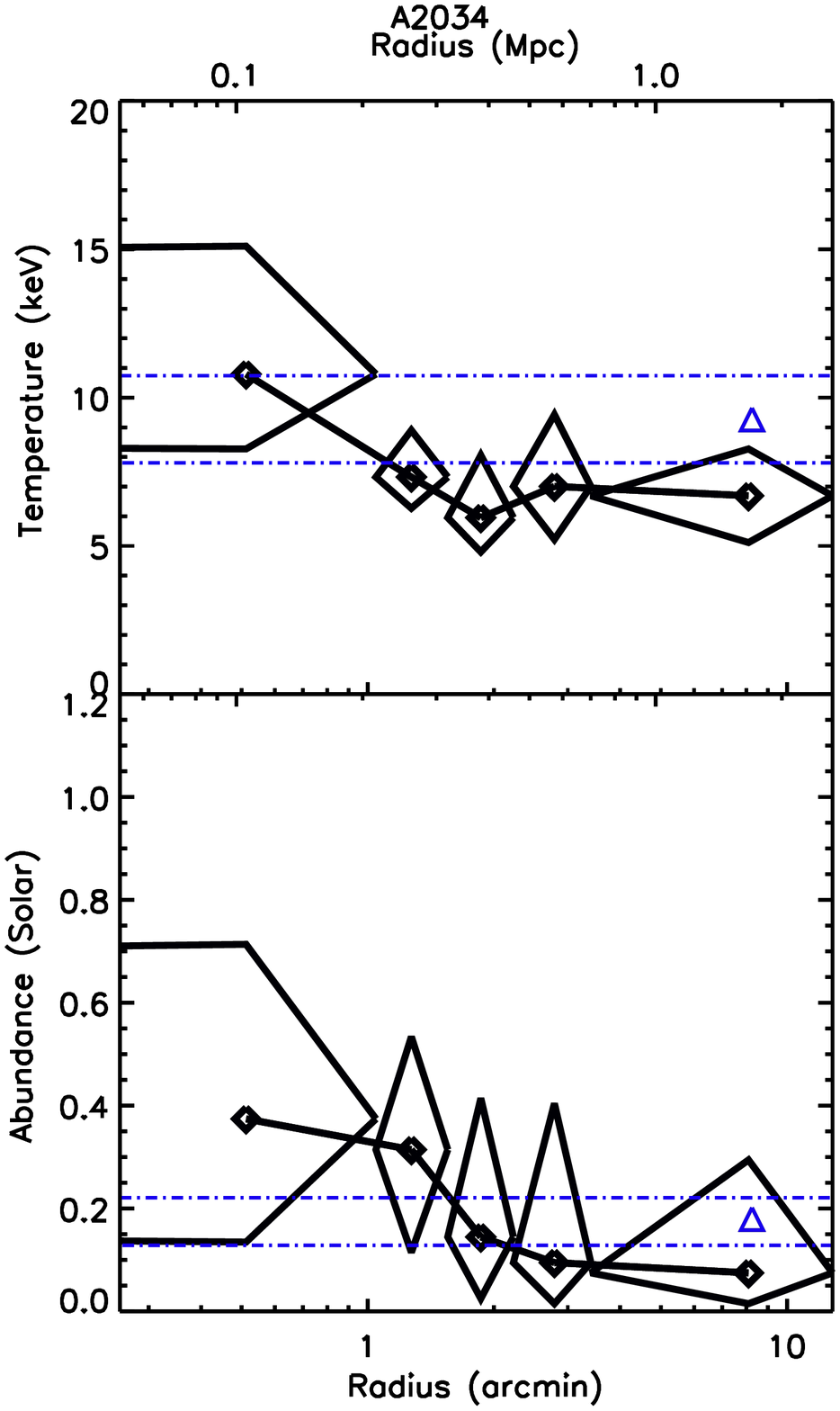,angle=0,width=\figwidth,height=\figheight}
    \psfig{figure=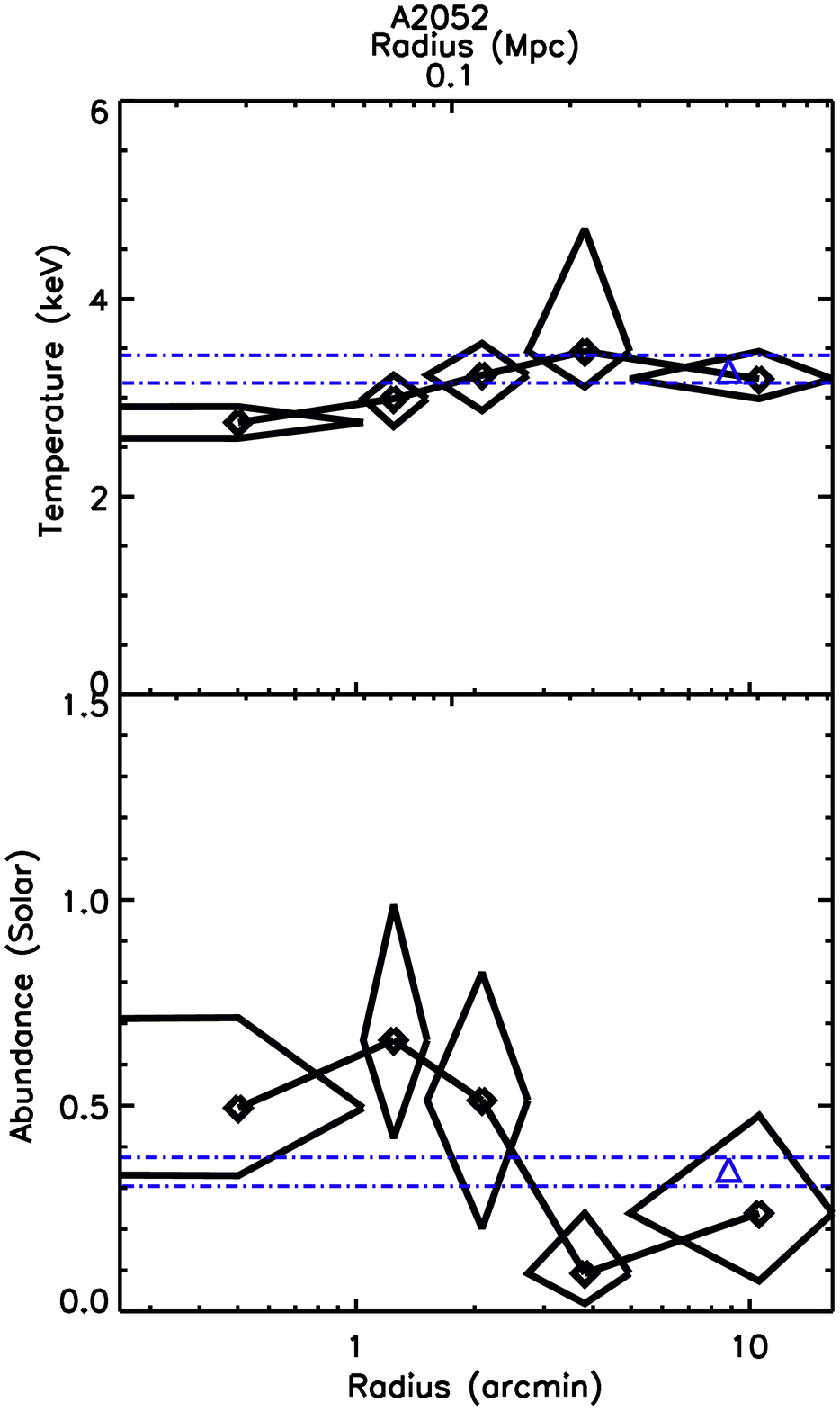,angle=0,width=\figwidth,height=\figheight}
    \psfig{figure=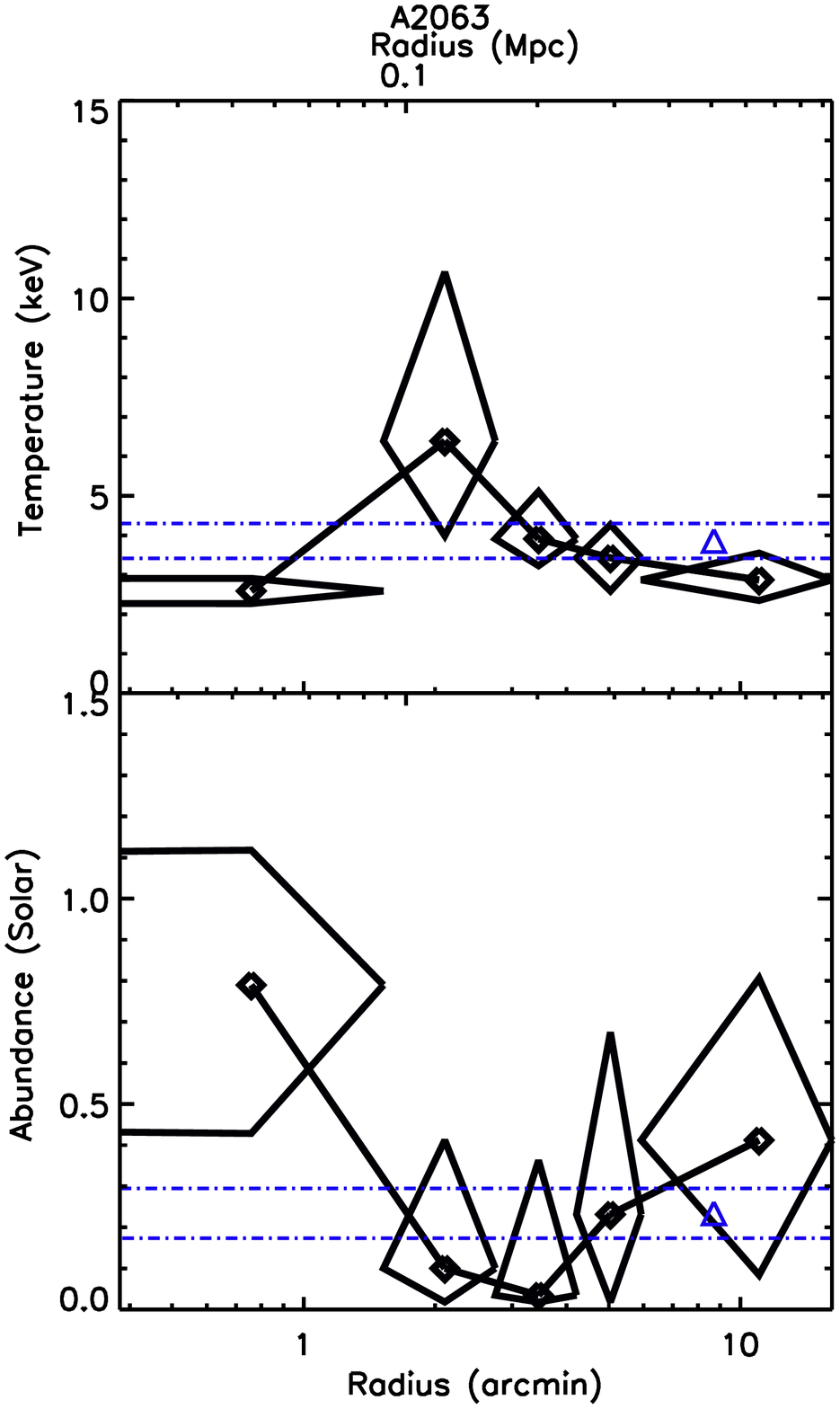,angle=0,width=\figwidth,height=\figheight}
    \psfig{figure=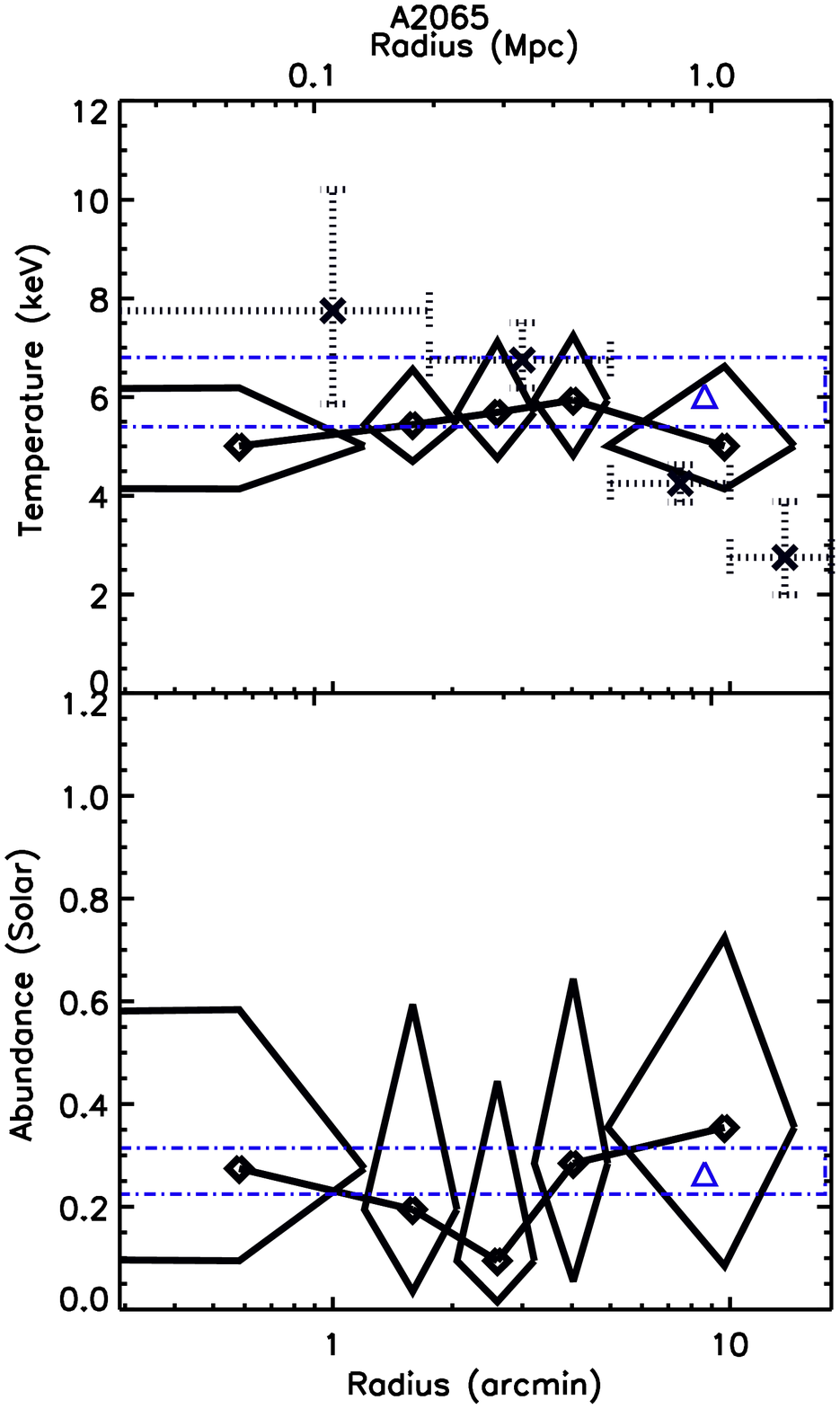,angle=0,width=\figwidth,height=\figheight}
  }
\parbox{\textwidth}{
    \psfig{figure=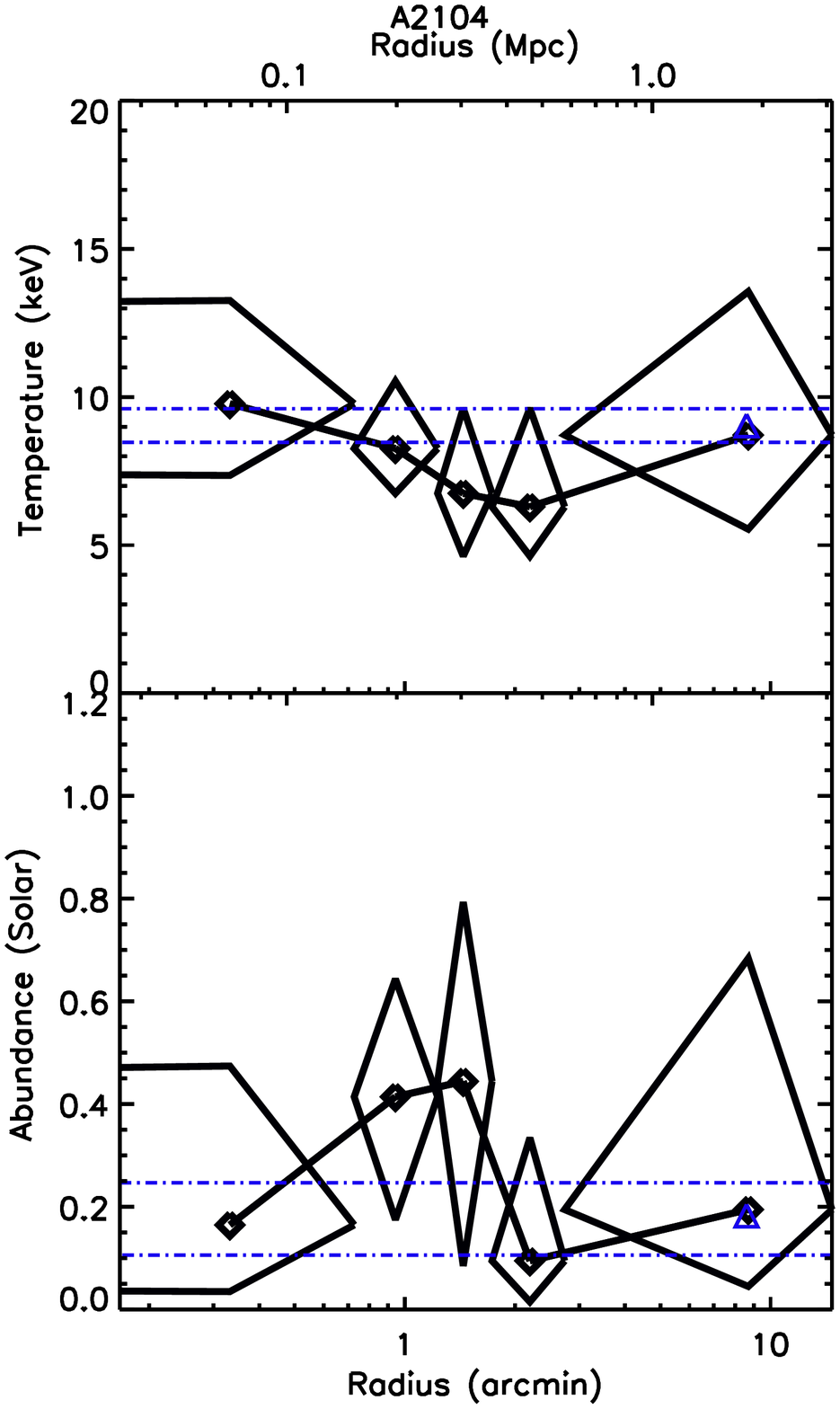,angle=0,width=\figwidth,height=\figheight}
    \psfig{figure=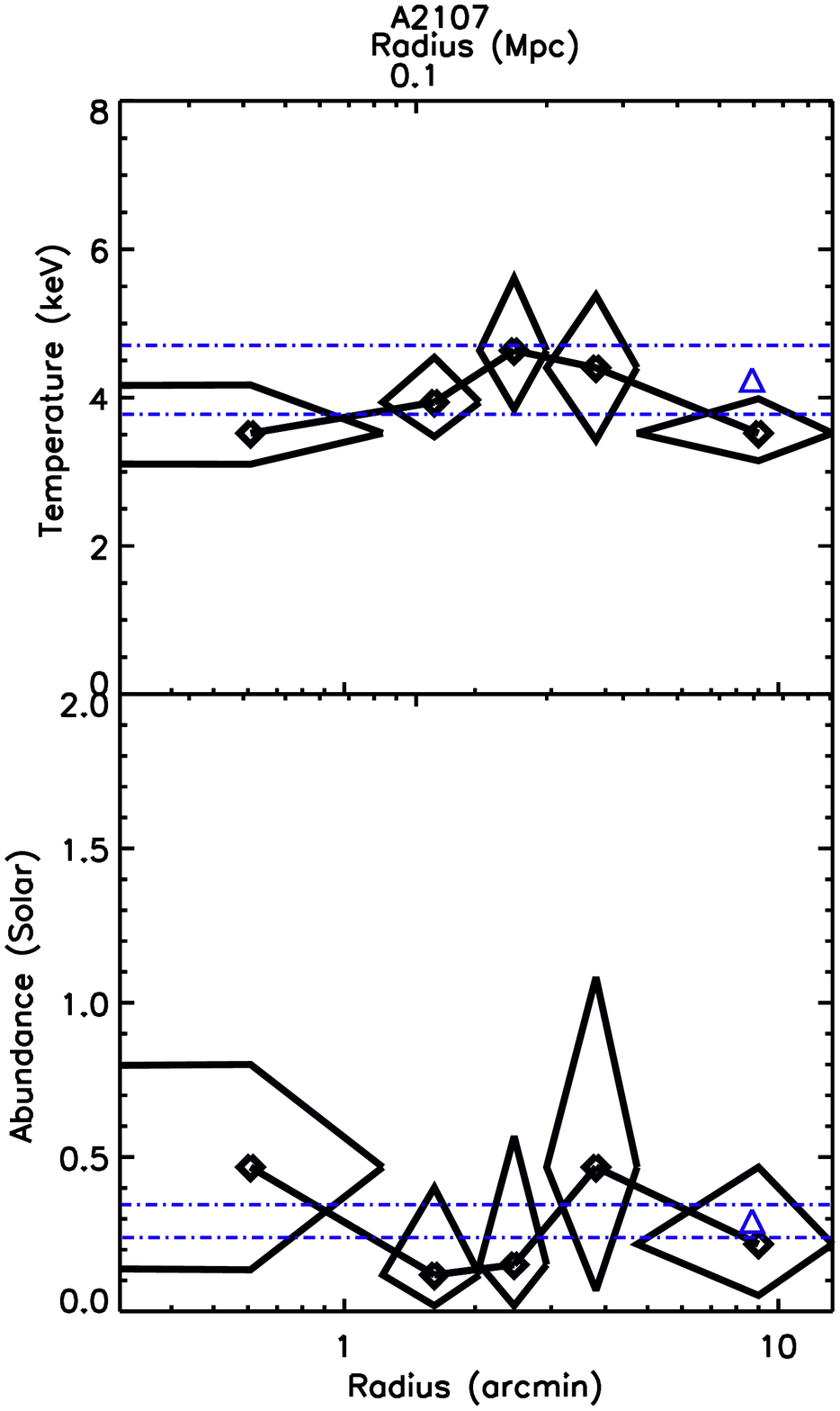,angle=0,width=\figwidth,height=\figheight}
    \psfig{figure=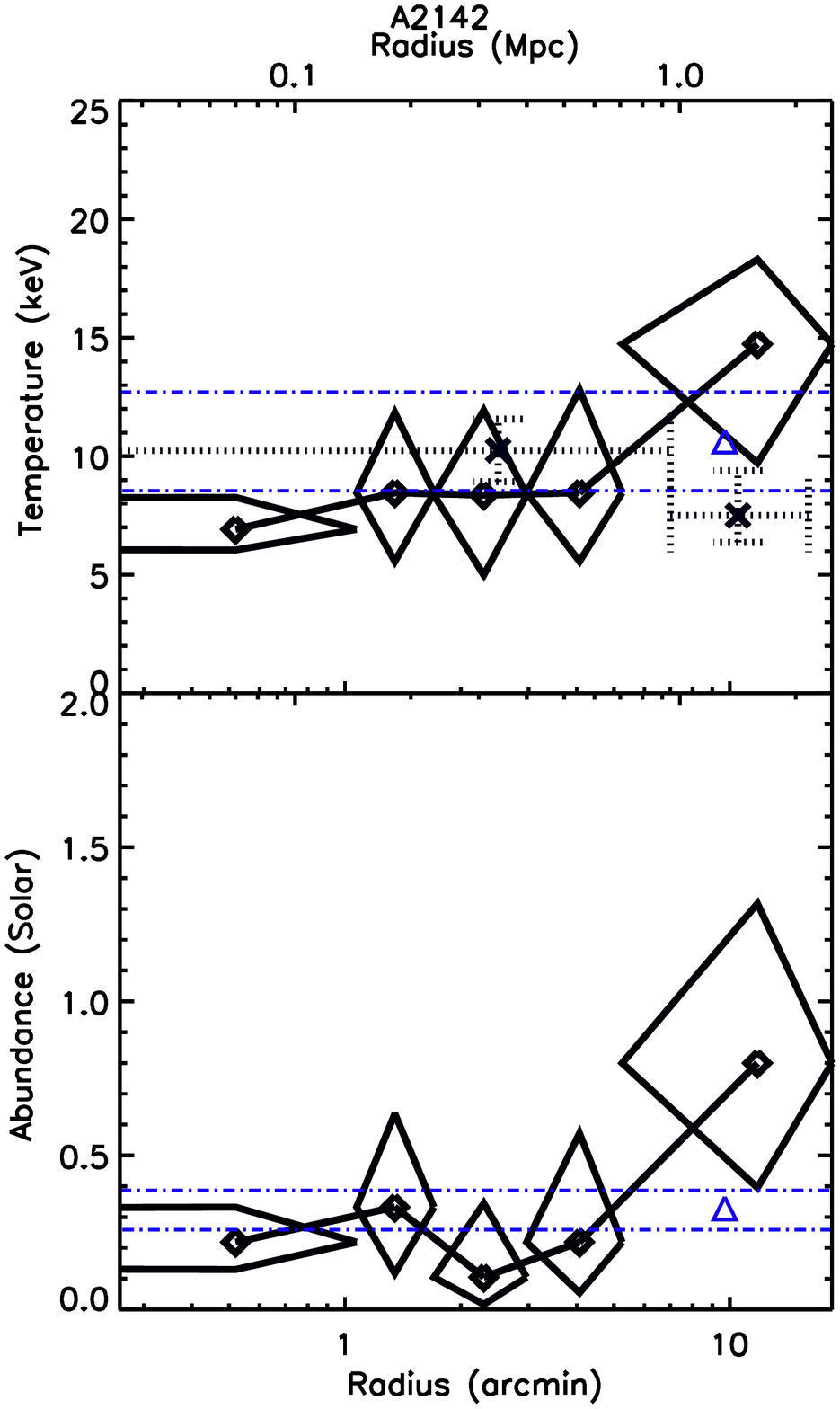,angle=0,width=\figwidth,height=\figheight}
    \psfig{figure=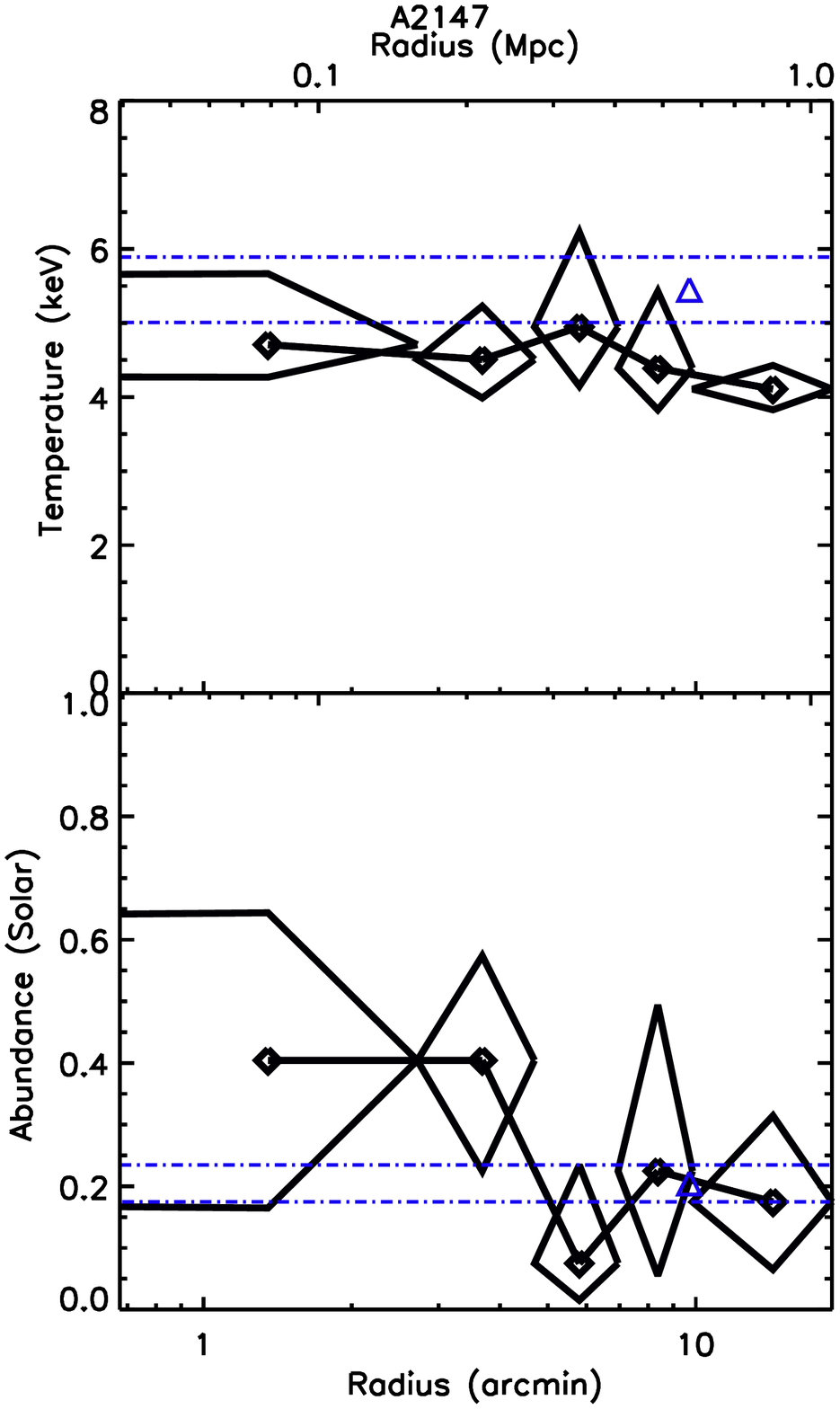,angle=0,width=\figwidth,height=\figheight}
  }
\end{figure*}
\clearpage
\begin{figure*}
\parbox{\textwidth}{
    \psfig{figure=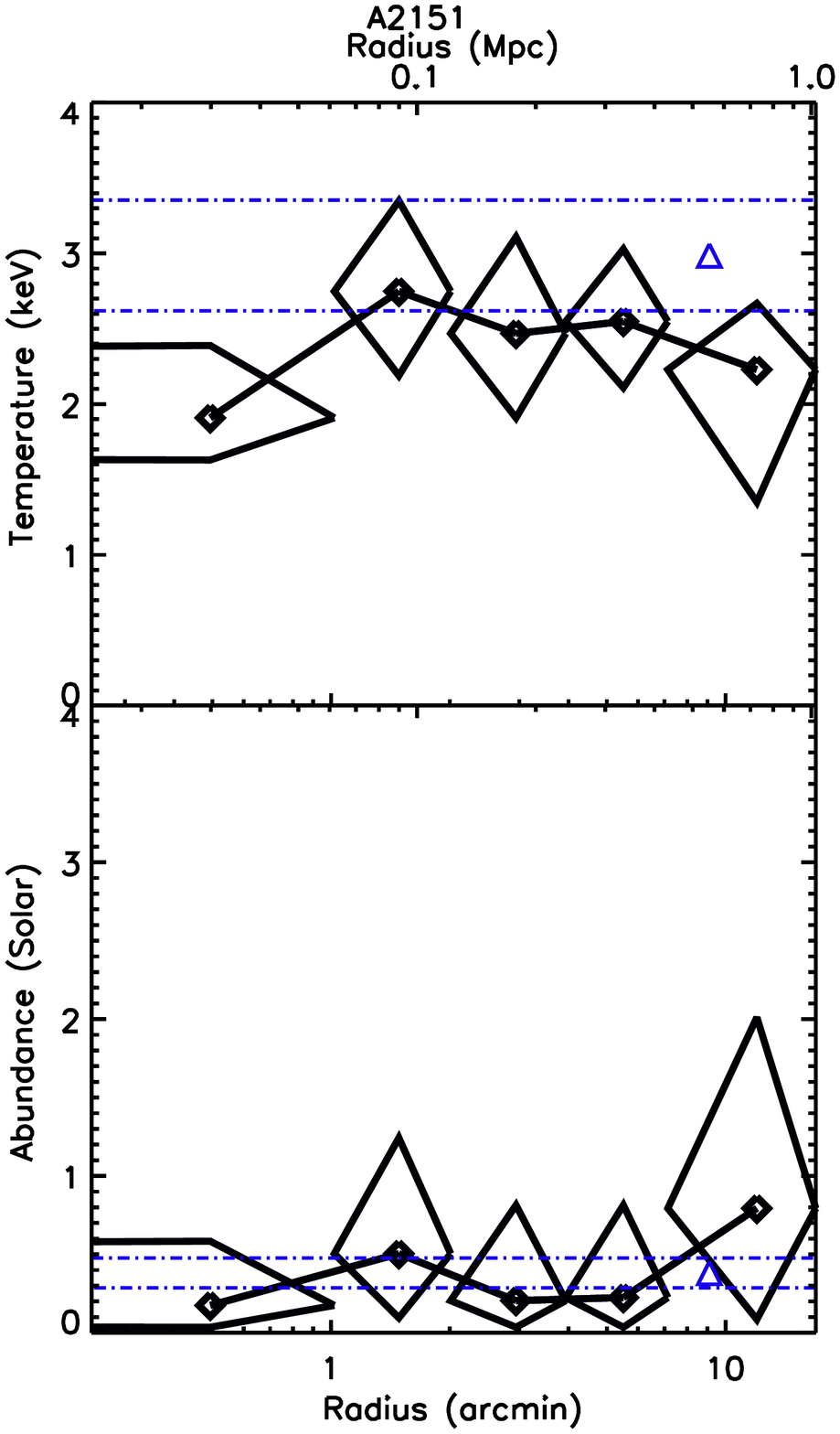,angle=0,width=\figwidth,height=\figheight}
    \psfig{figure=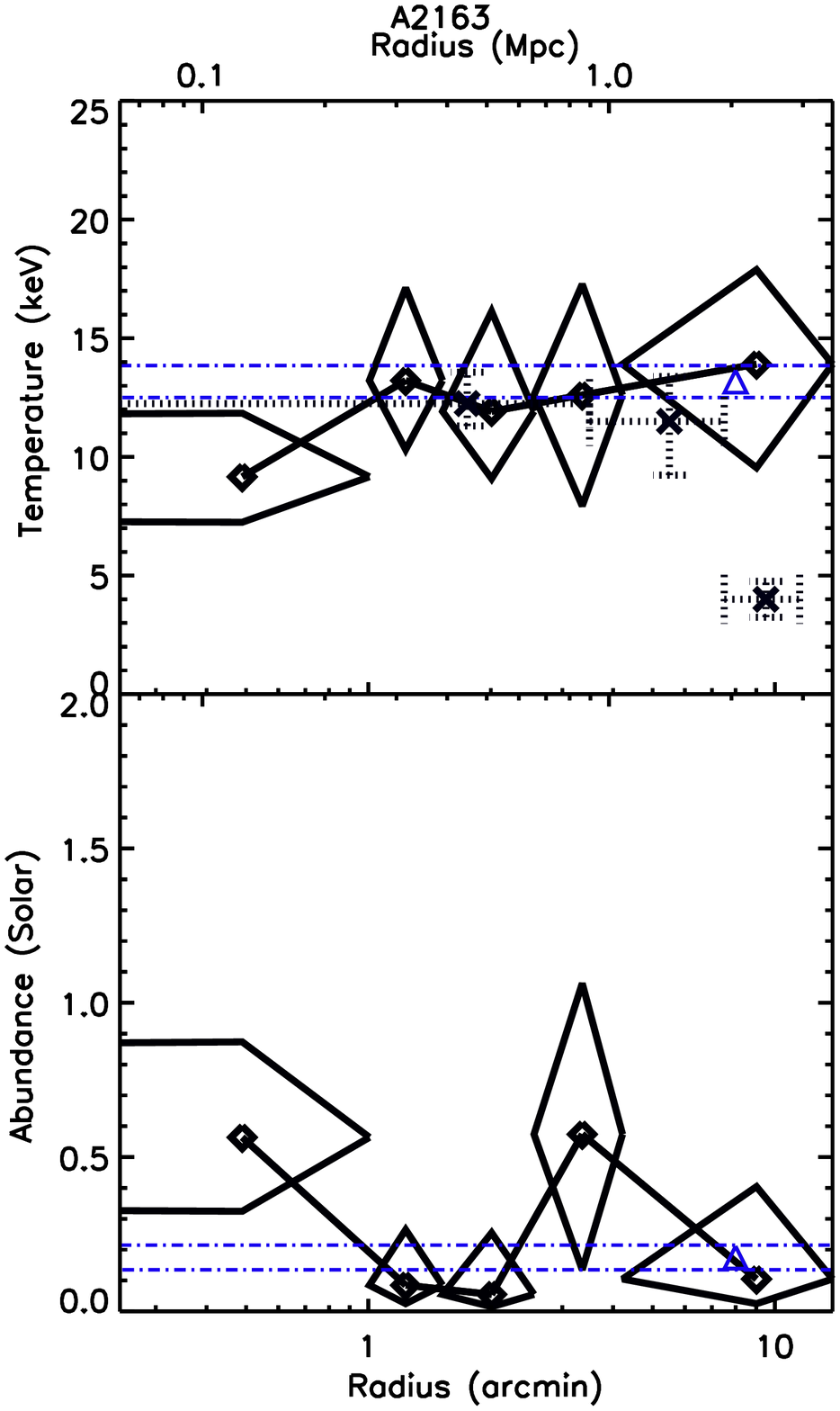,angle=0,width=\figwidth,height=\figheight}
    \psfig{figure=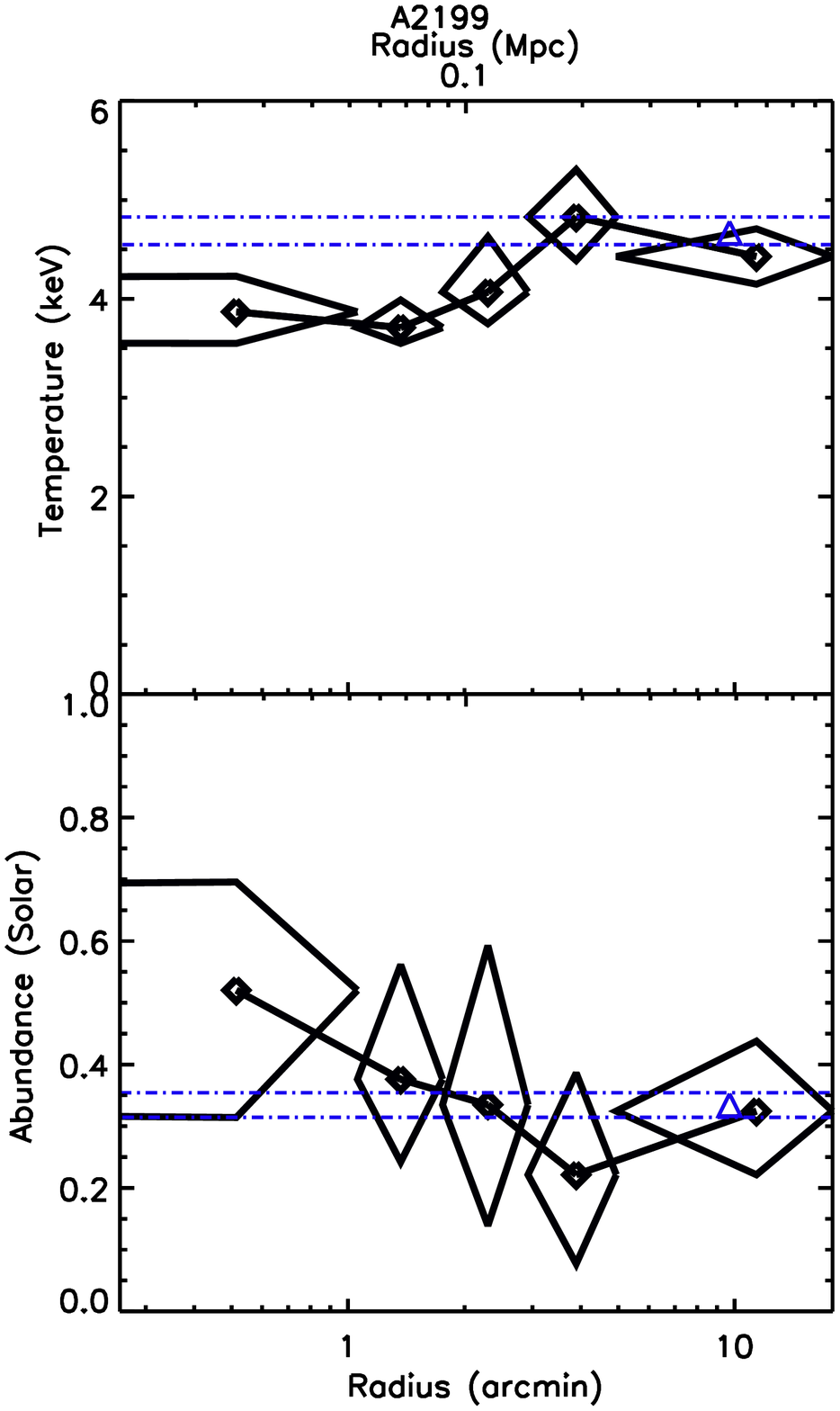,angle=0,width=\figwidth,height=\figheight}
    \psfig{figure=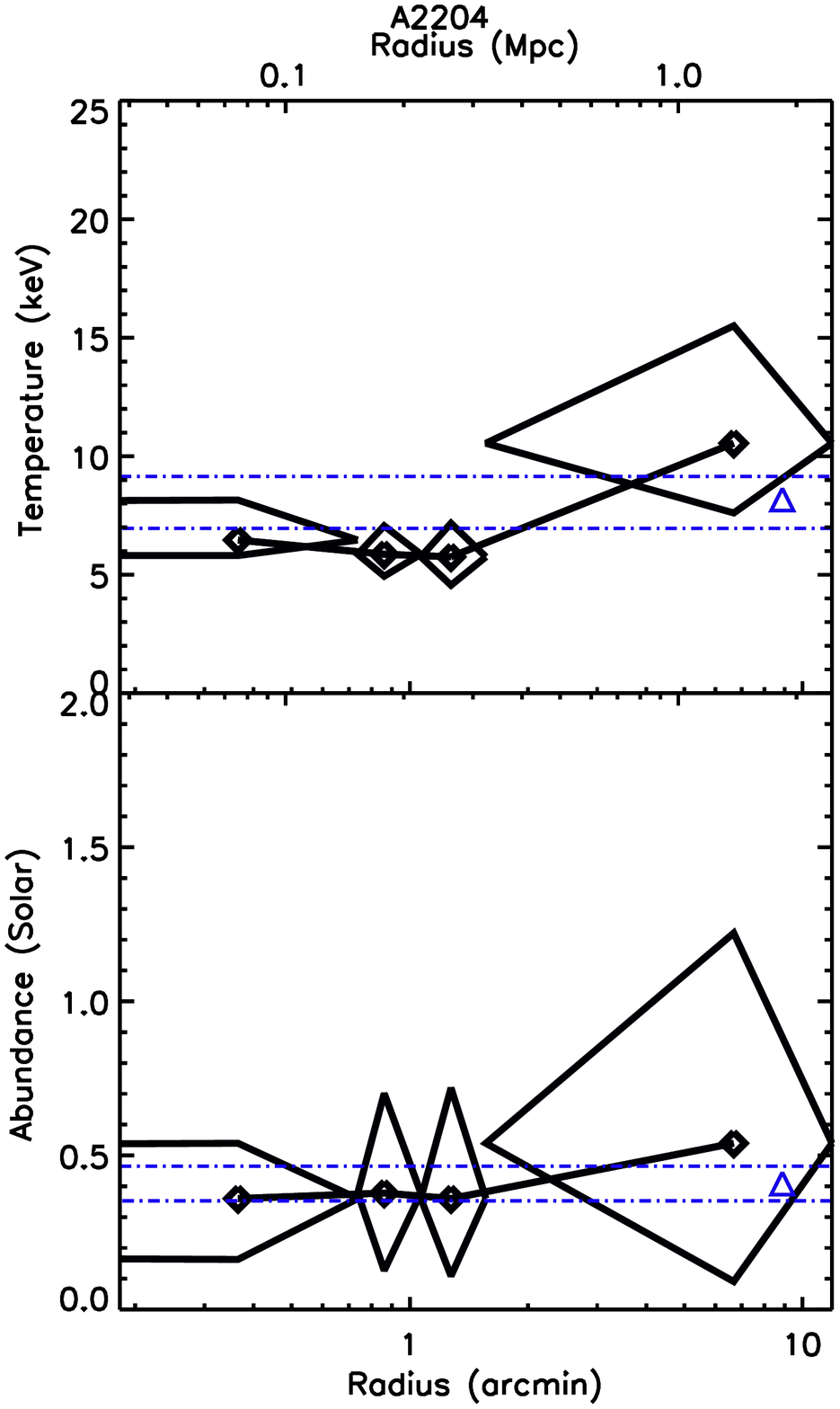,angle=0,width=\figwidth,height=\figheight}
  }
\parbox{\textwidth}{
    \psfig{figure=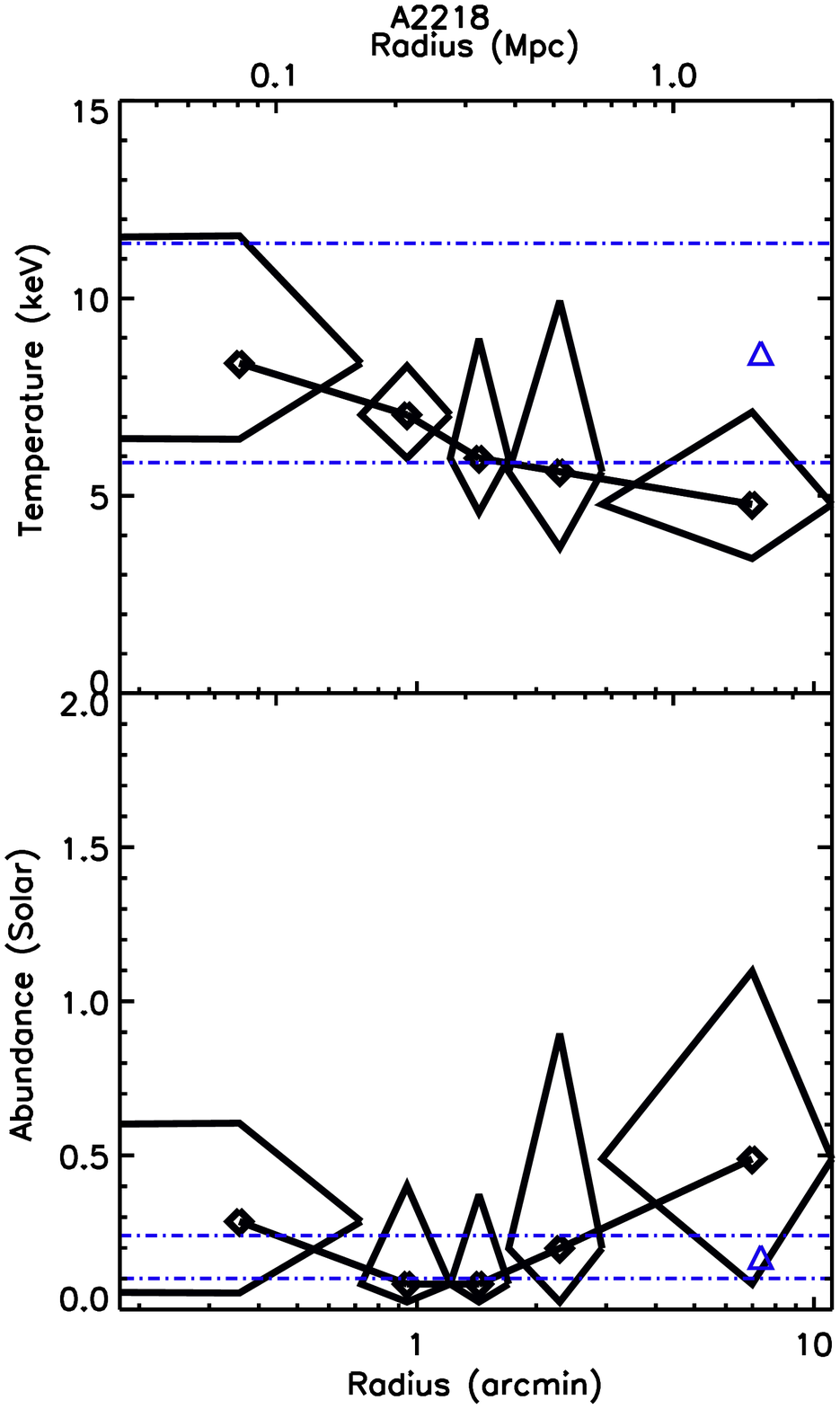,angle=0,width=\figwidth,height=\figheight}
    \psfig{figure=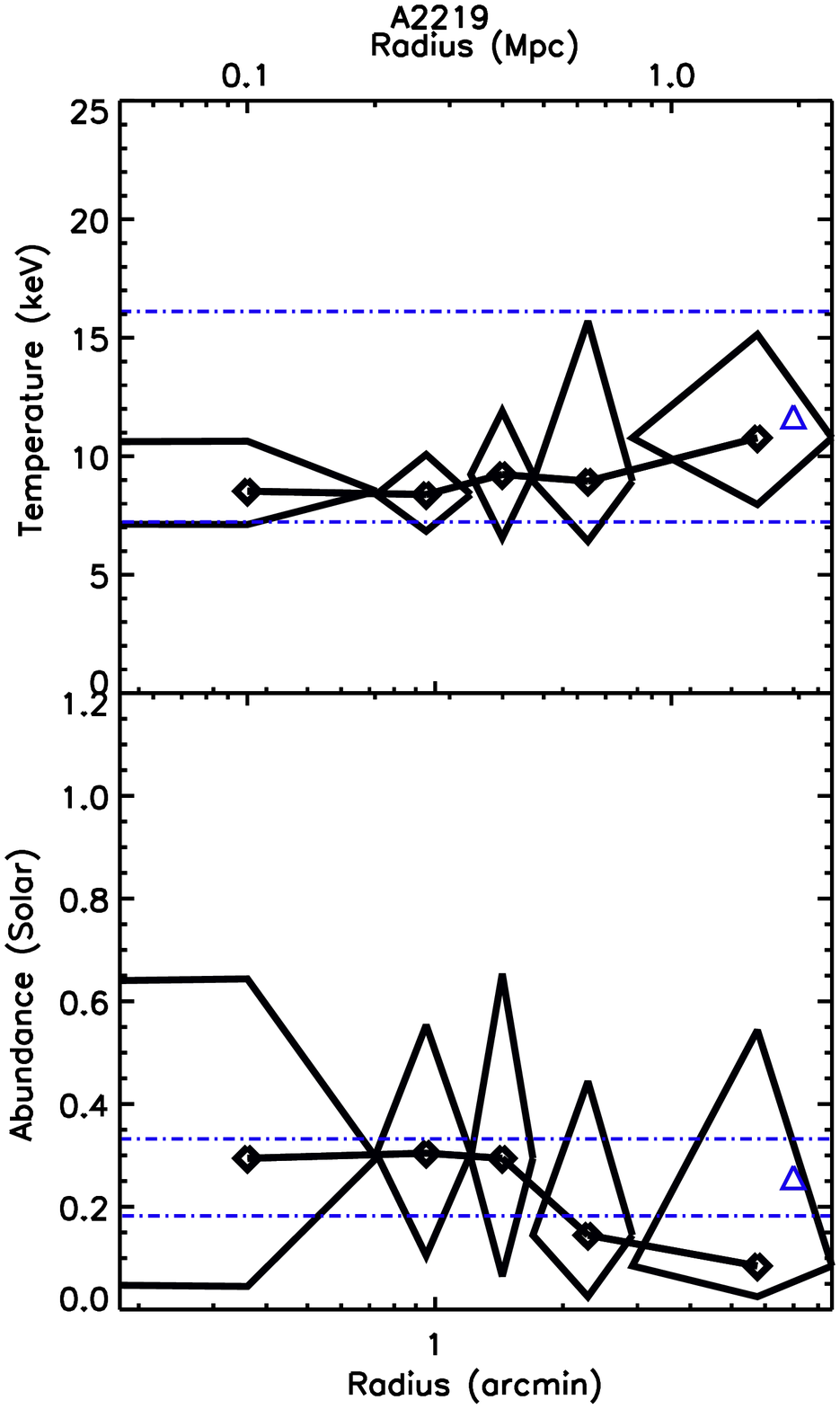,angle=0,width=\figwidth,height=\figheight}
    \psfig{figure=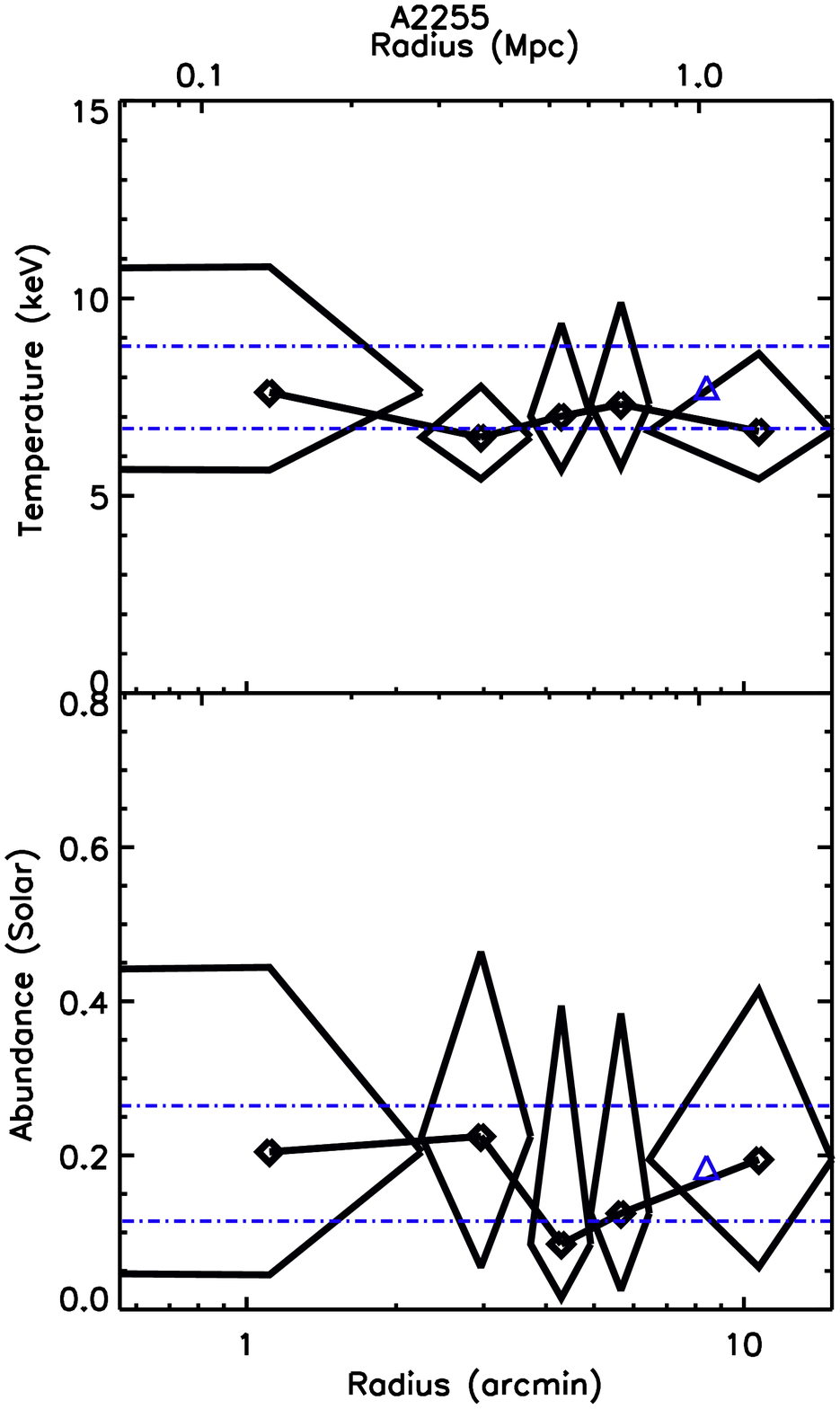,angle=0,width=\figwidth,height=\figheight}
    \psfig{figure=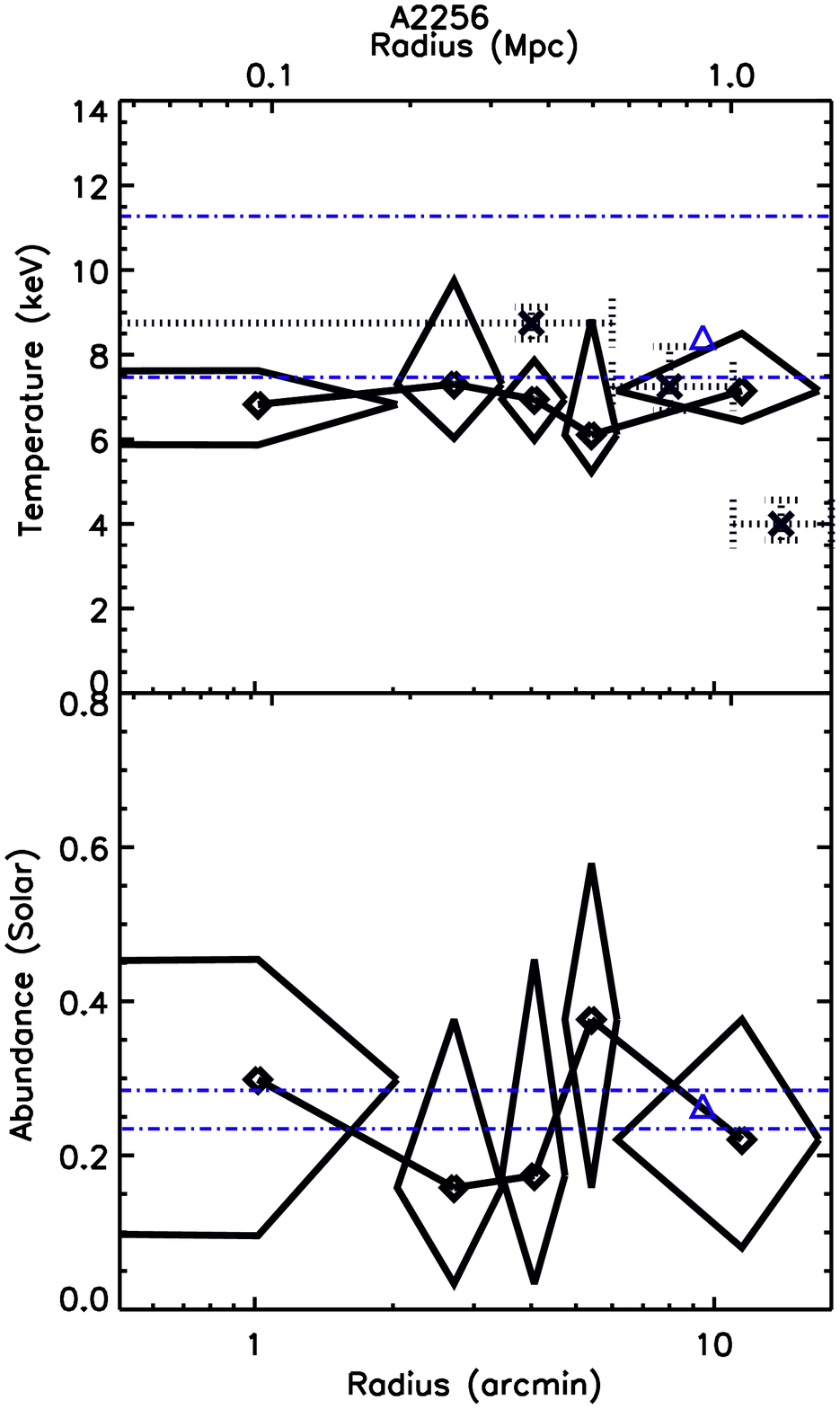,angle=0,width=\figwidth,height=\figheight}
  }
\parbox{\textwidth}{
    \psfig{figure=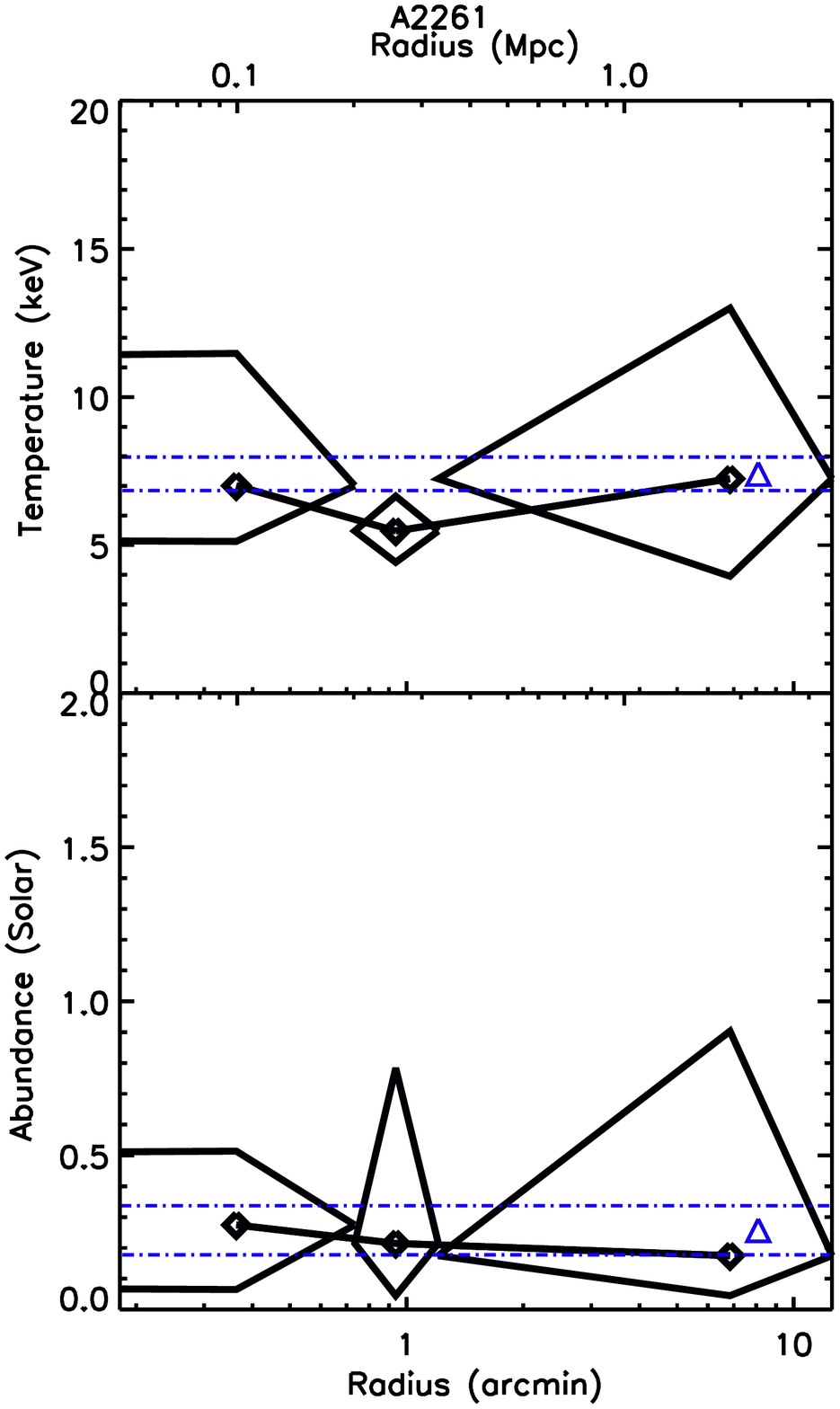,angle=0,width=\figwidth,height=\figheight}
    \psfig{figure=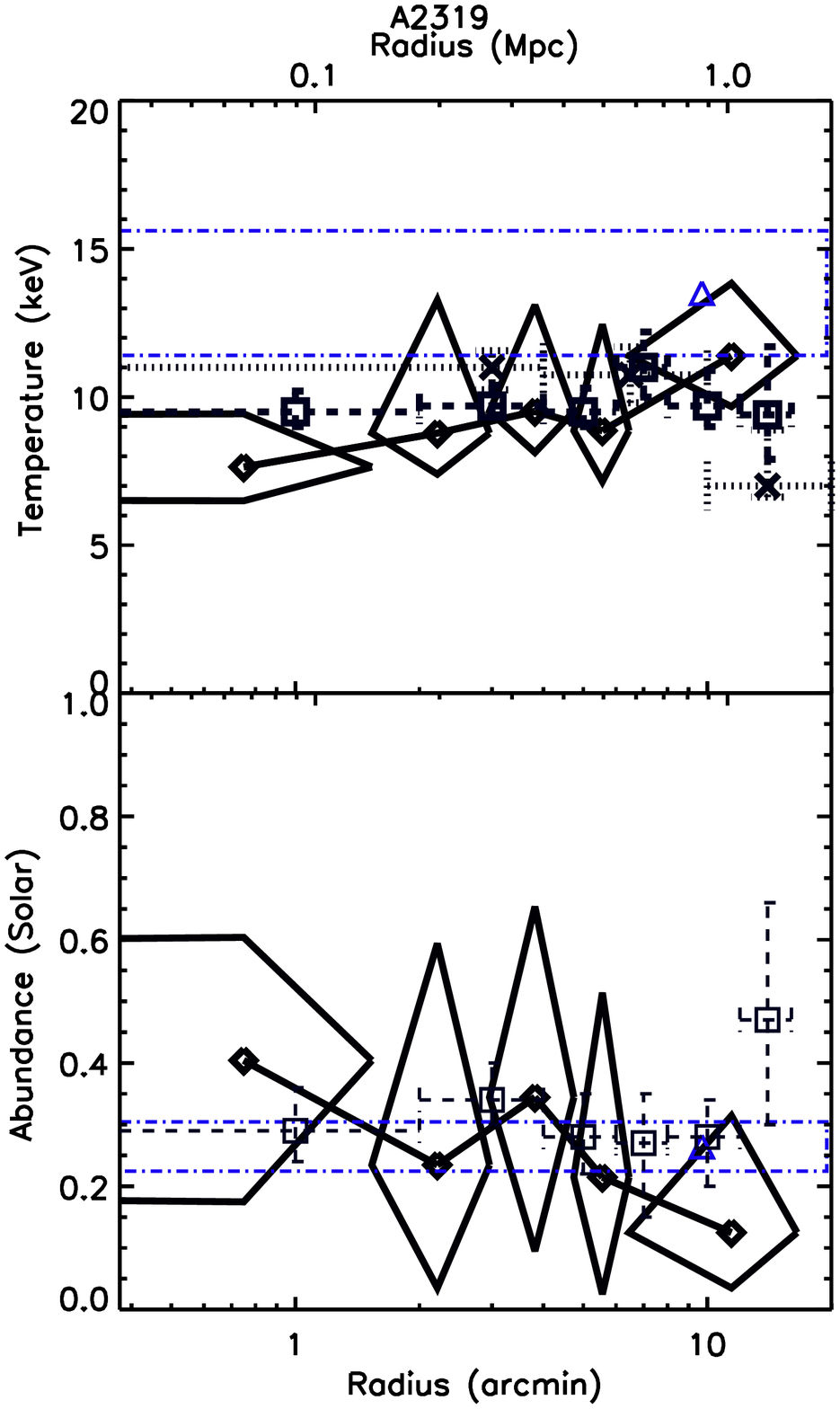,angle=0,width=\figwidth,height=\figheight}
    \psfig{figure=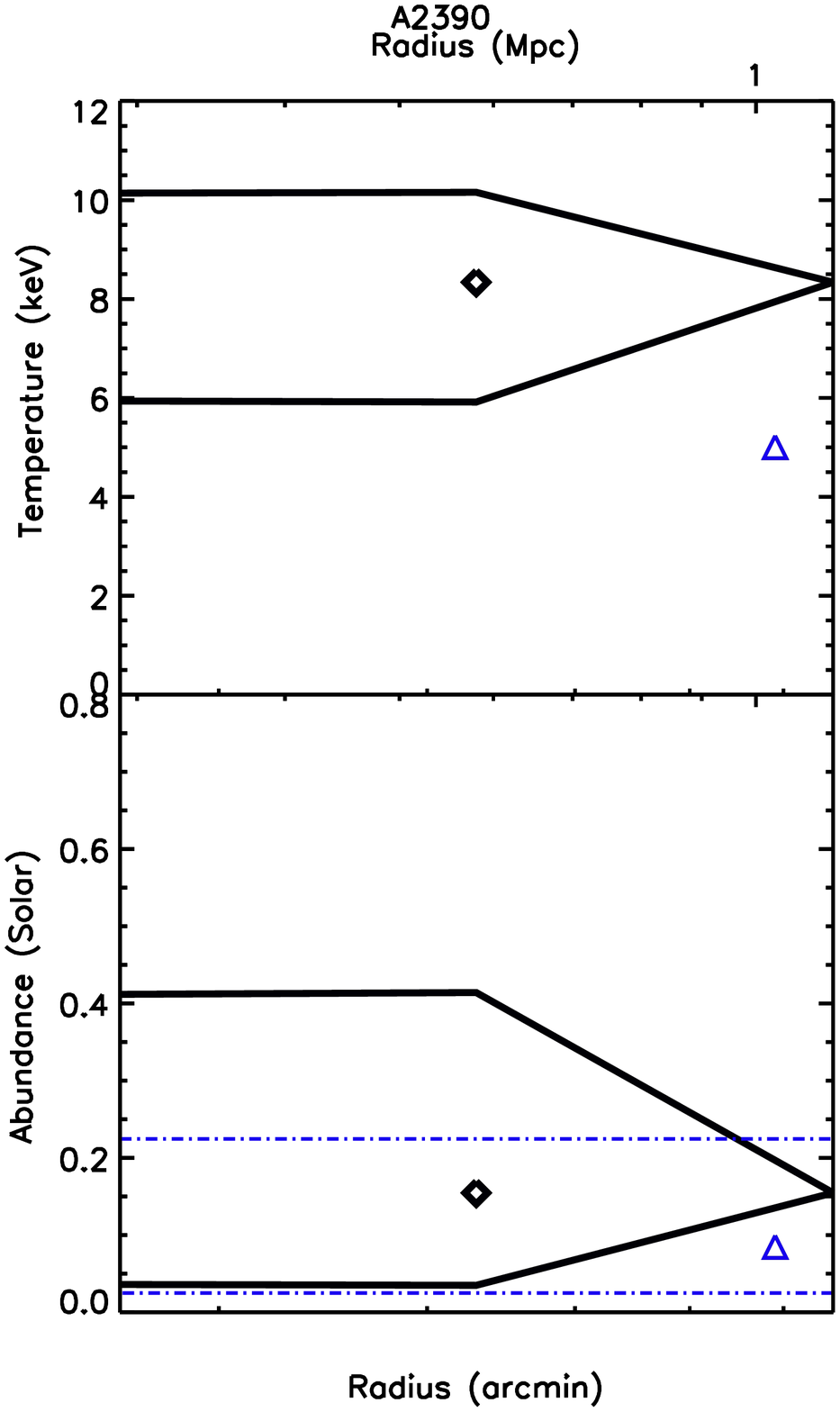,angle=0,width=\figwidth,height=\figheight}
    \psfig{figure=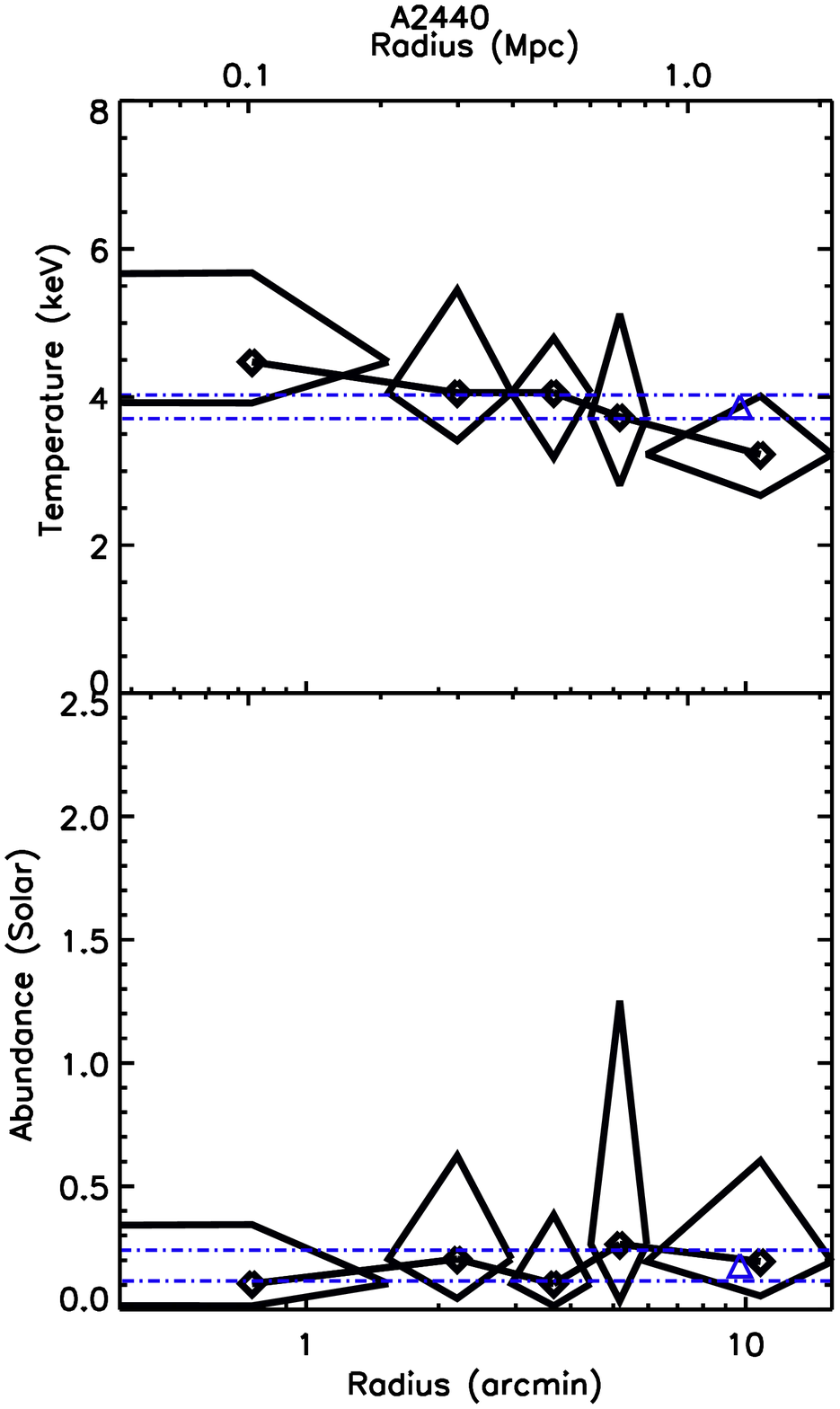,angle=0,width=\figwidth,height=\figheight}
  }
\end{figure*}
\clearpage
\begin{figure*}
\parbox{\textwidth}{
    \psfig{figure=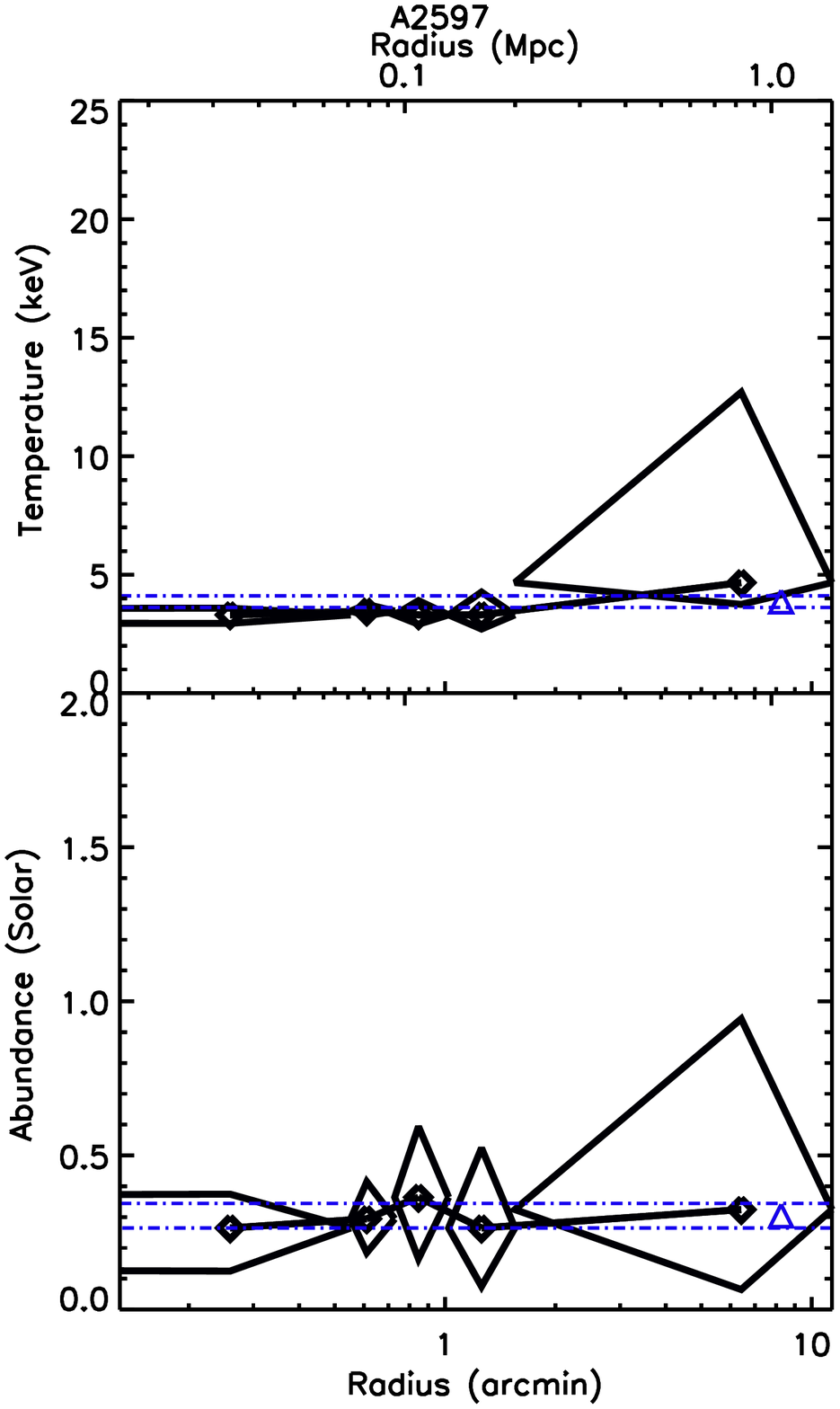,angle=0,width=\figwidth,height=\figheight}
    \psfig{figure=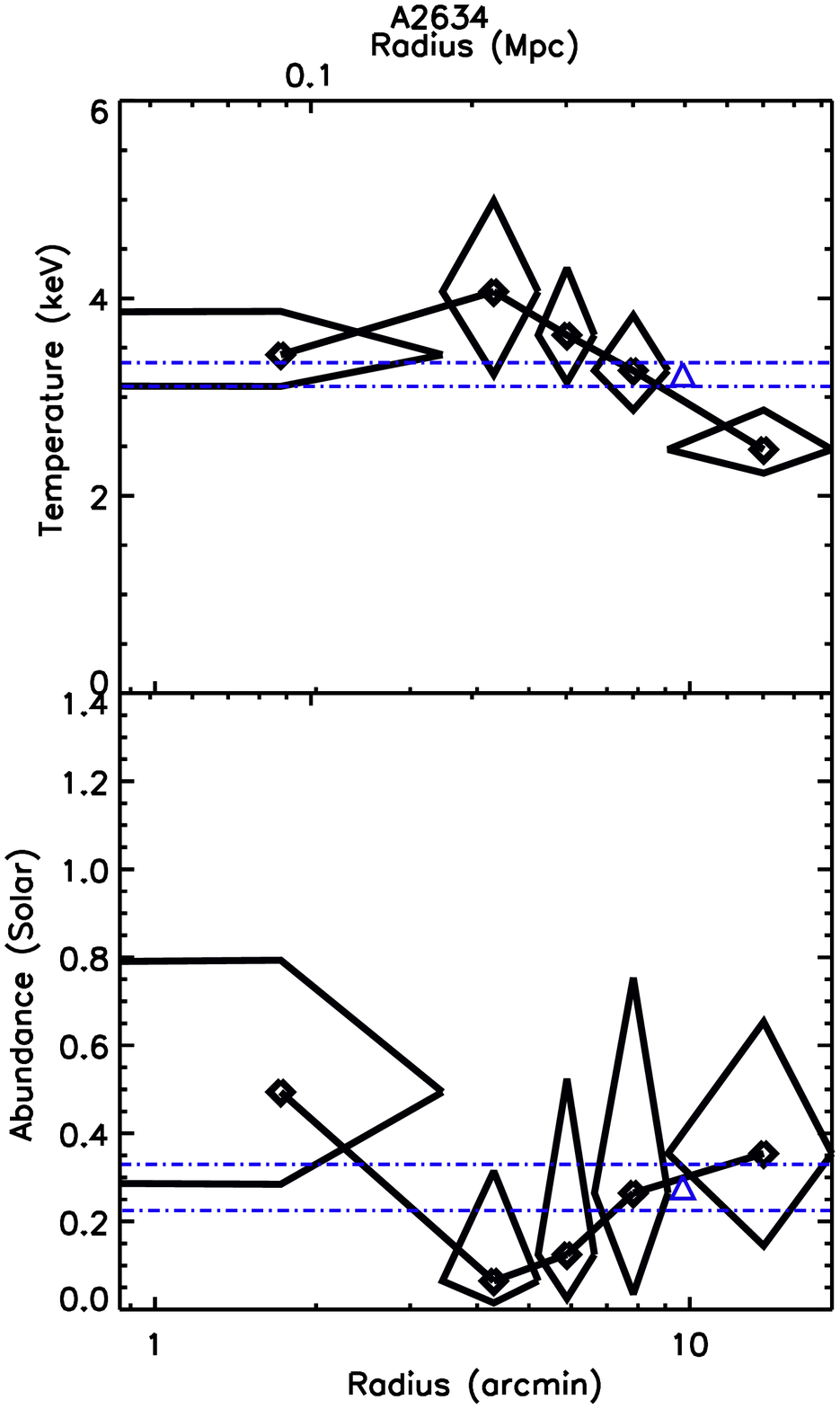,angle=0,width=\figwidth,height=\figheight}
    \psfig{figure=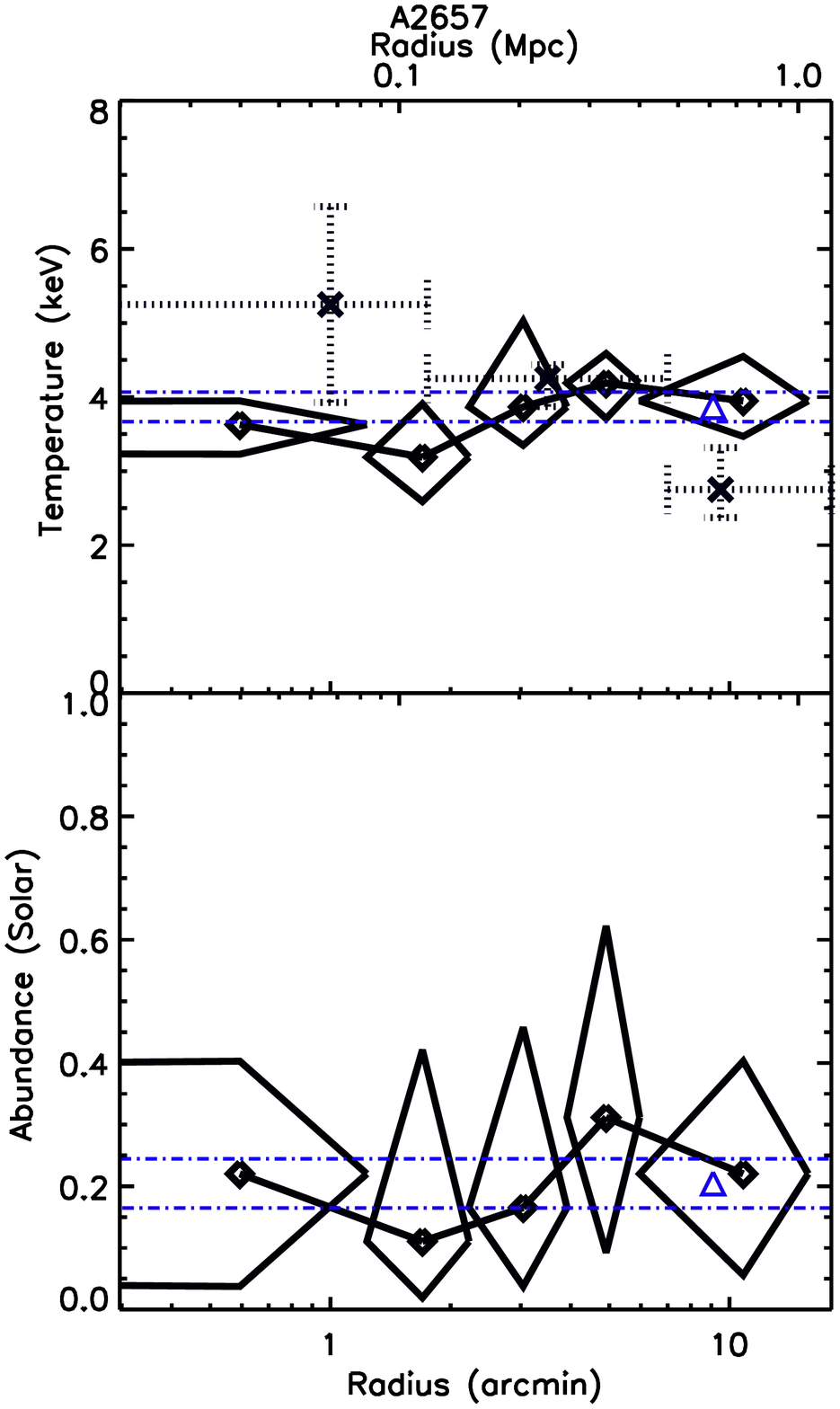,angle=0,width=\figwidth,height=\figheight}
    \psfig{figure=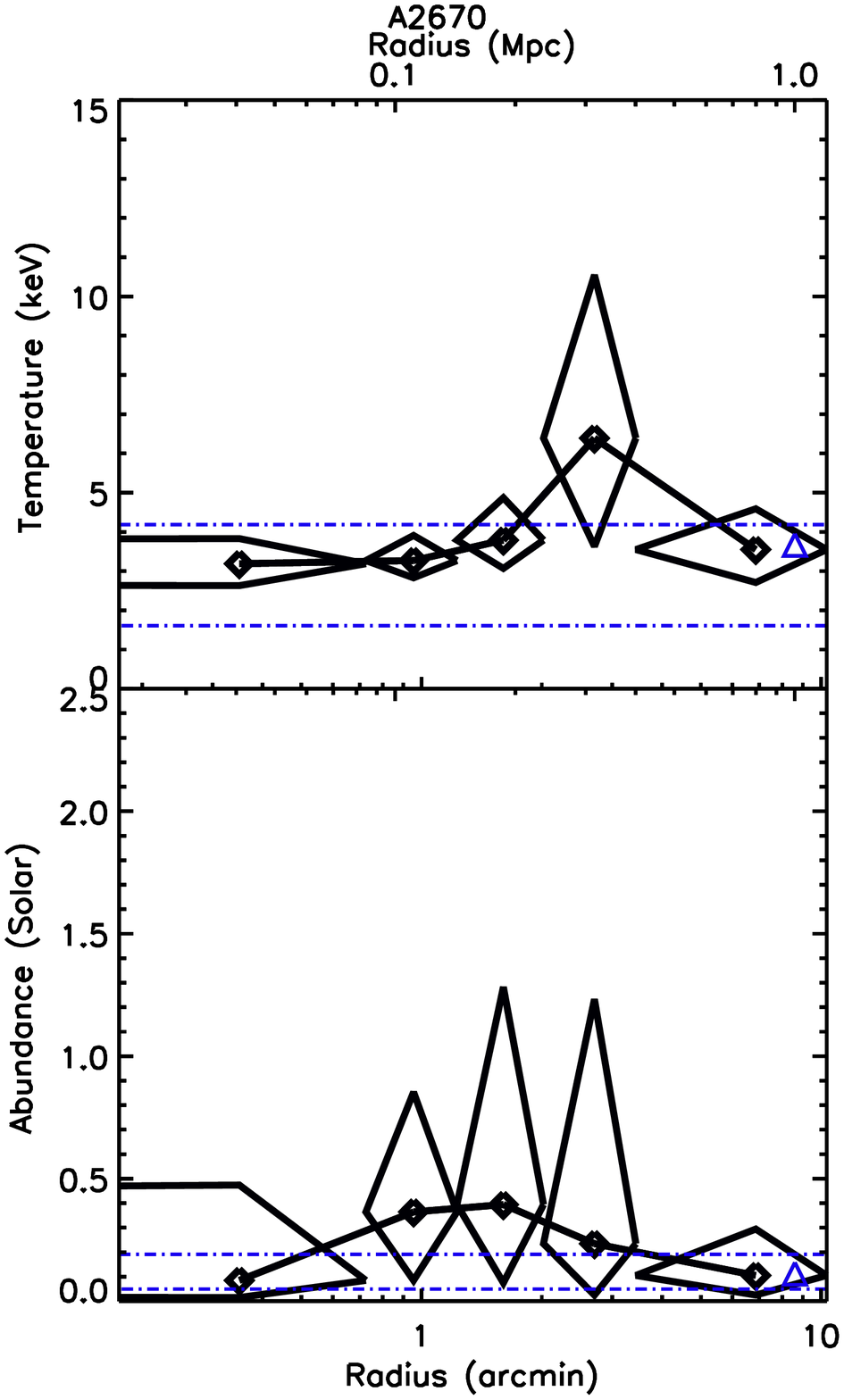,angle=0,width=\figwidth,height=\figheight}
  }
\parbox{\textwidth}{
    \psfig{figure=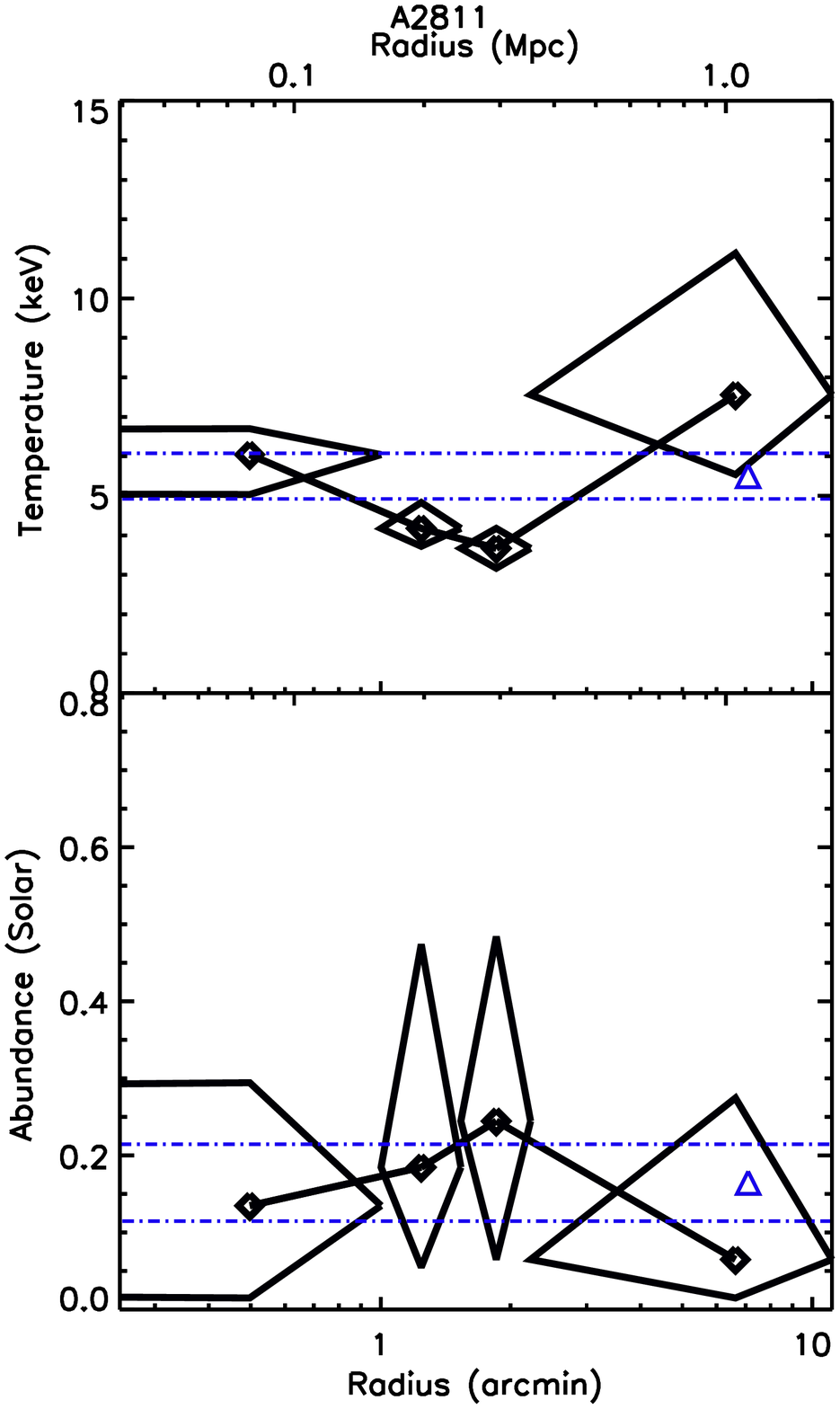,angle=0,width=\figwidth,height=\figheight}
    \psfig{figure=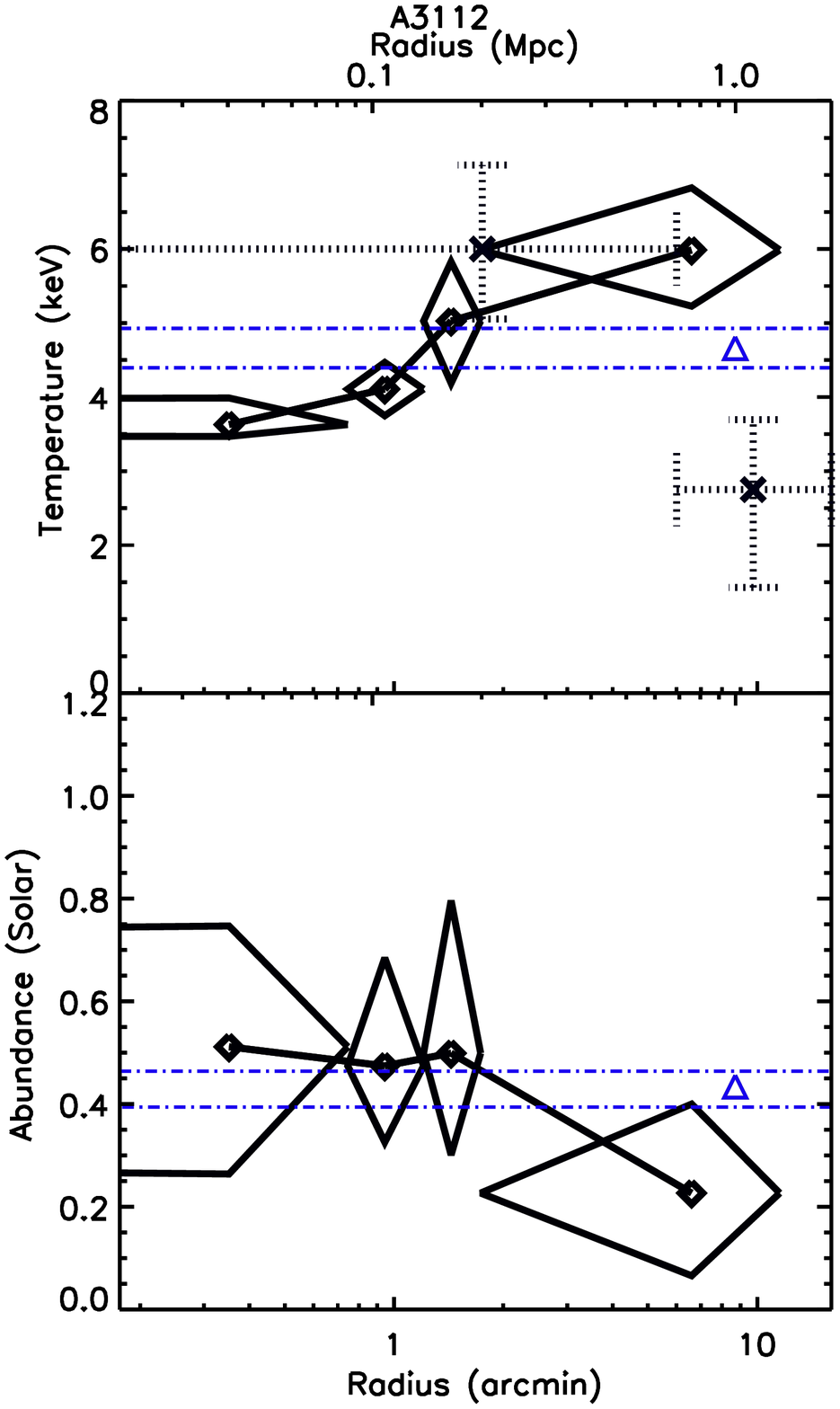,angle=0,width=\figwidth,height=\figheight}
    \psfig{figure=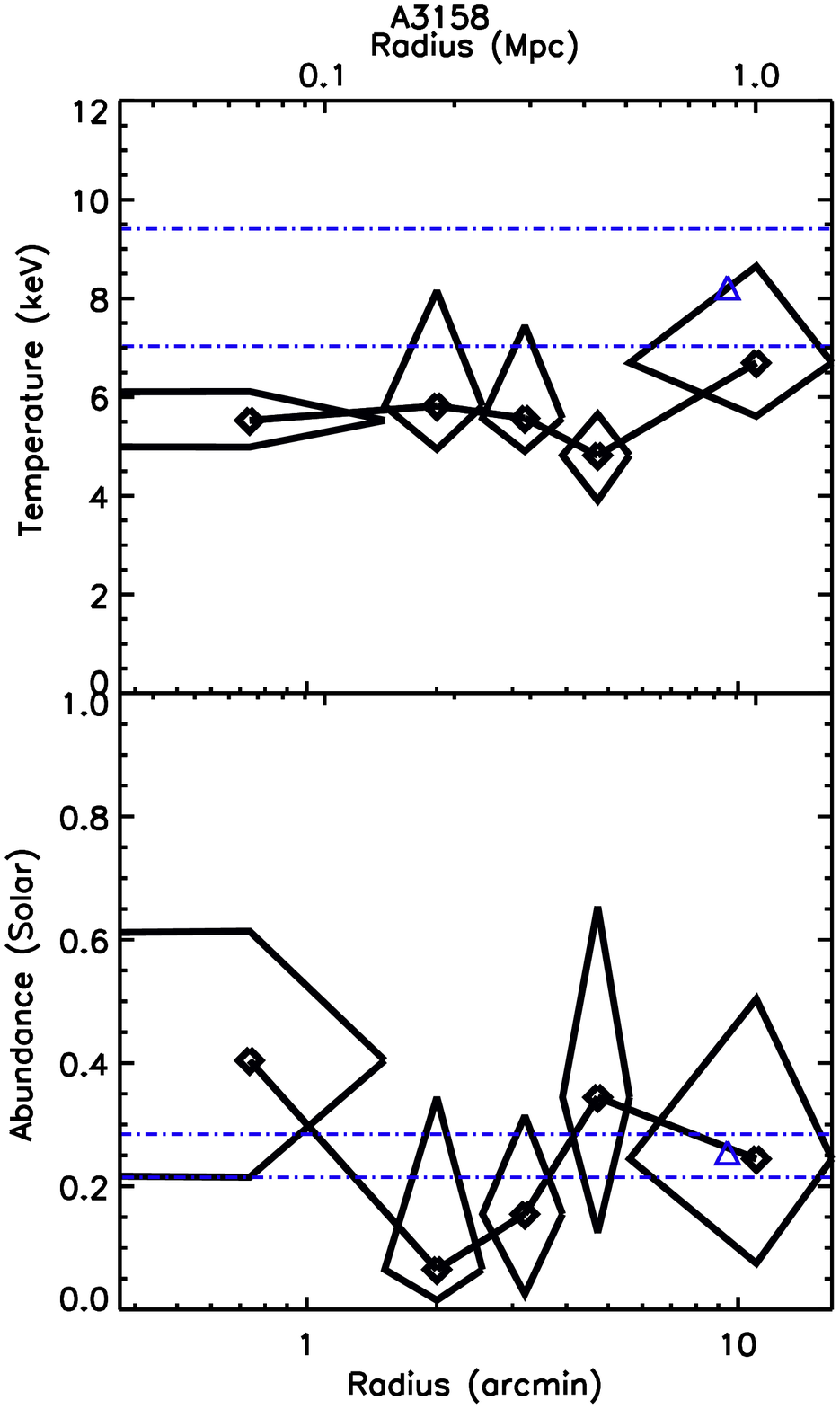,angle=0,width=\figwidth,height=\figheight}
    \psfig{figure=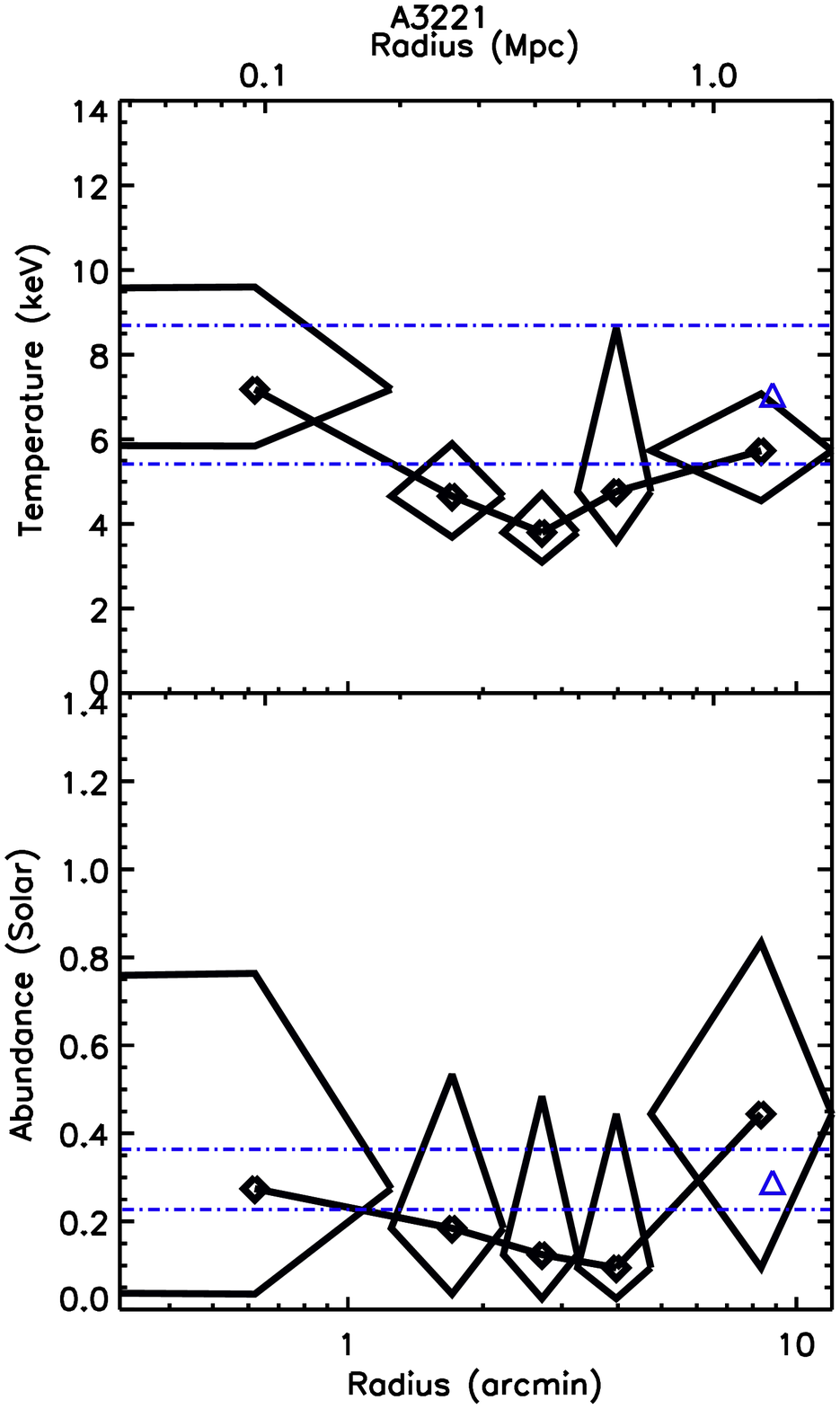,angle=0,width=\figwidth,height=\figheight}
  }
\parbox{\textwidth}{
    \psfig{figure=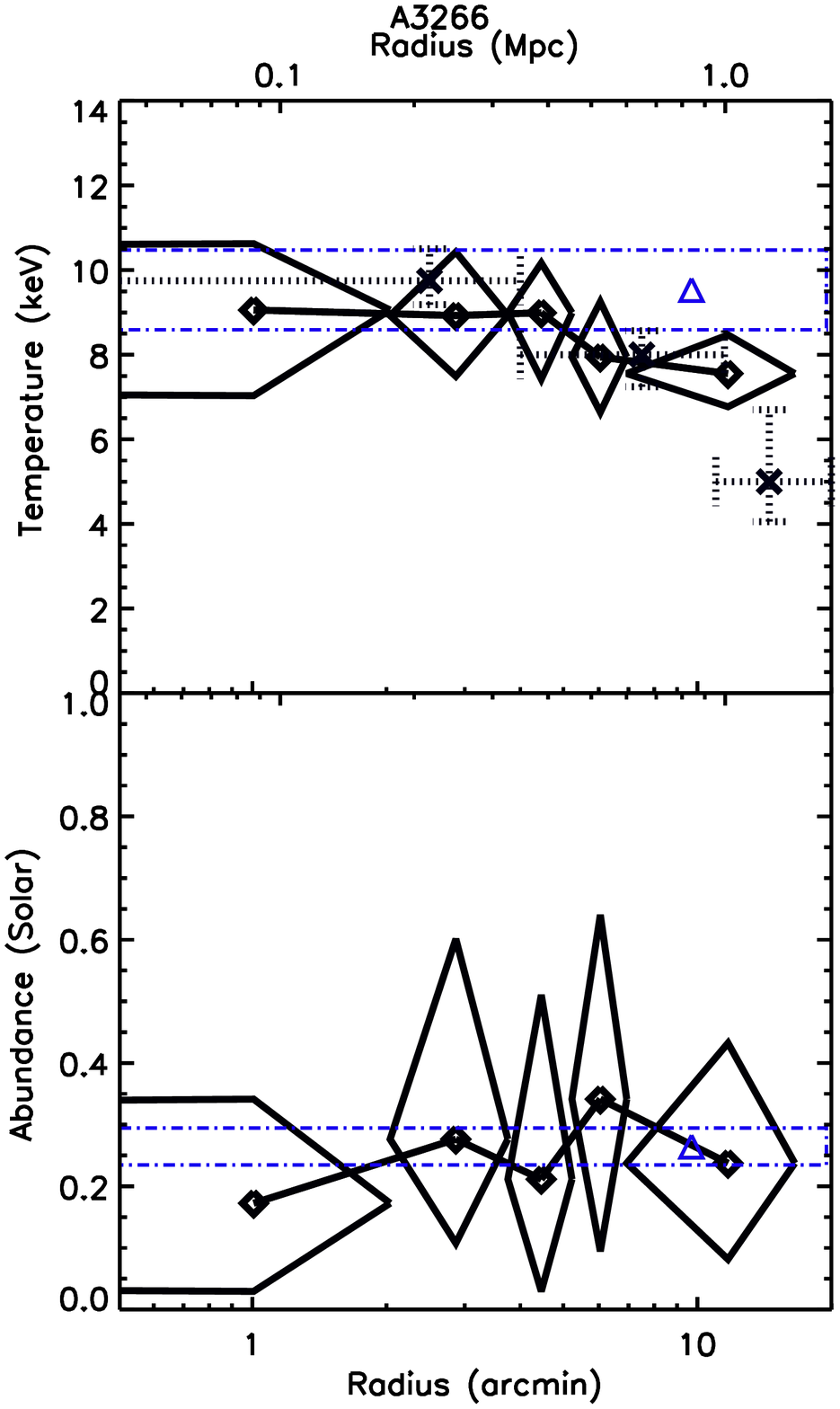,angle=0,width=\figwidth,height=\figheight}
    \psfig{figure=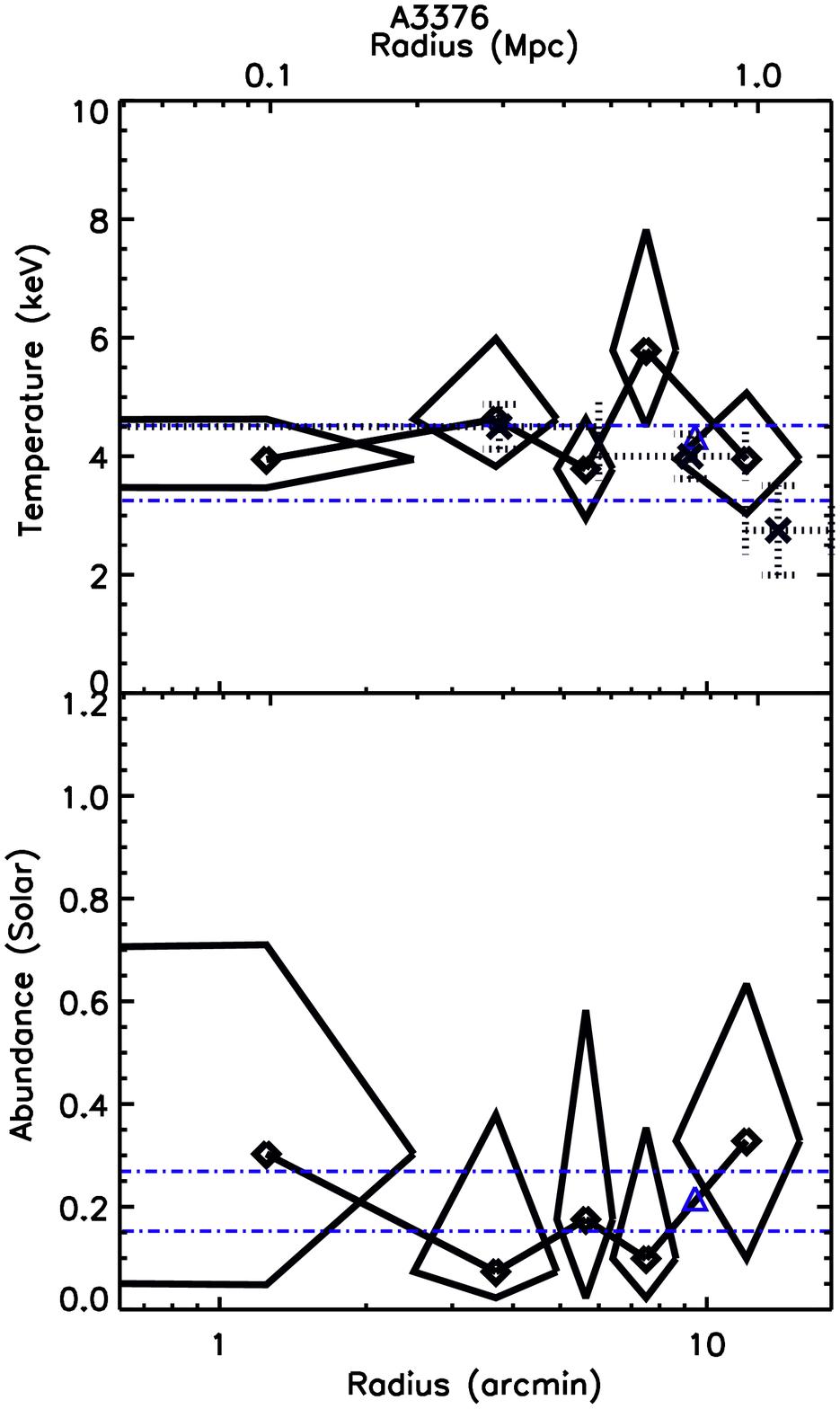,angle=0,width=\figwidth,height=\figheight}
    \psfig{figure=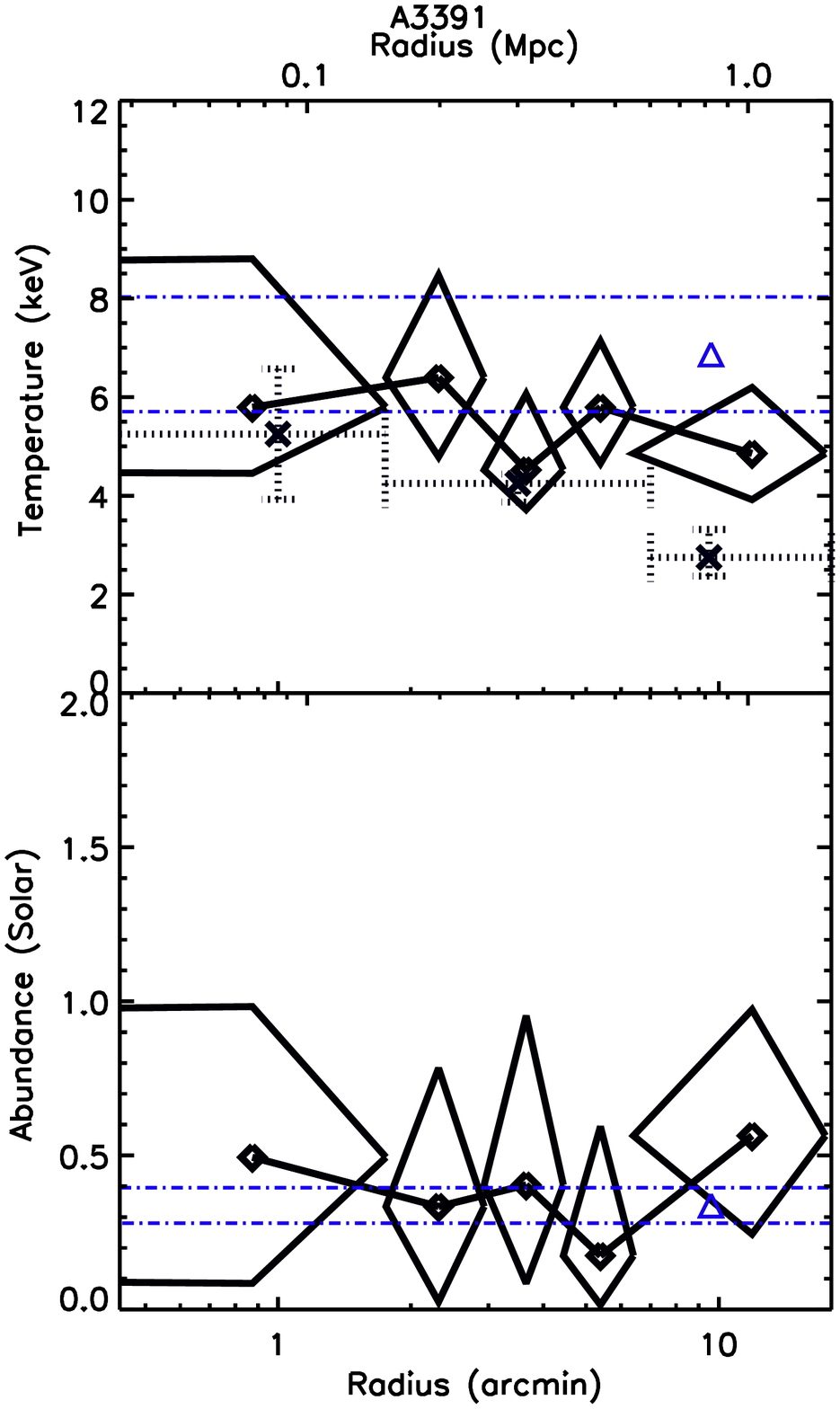,angle=0,width=\figwidth,height=\figheight}
    \psfig{figure=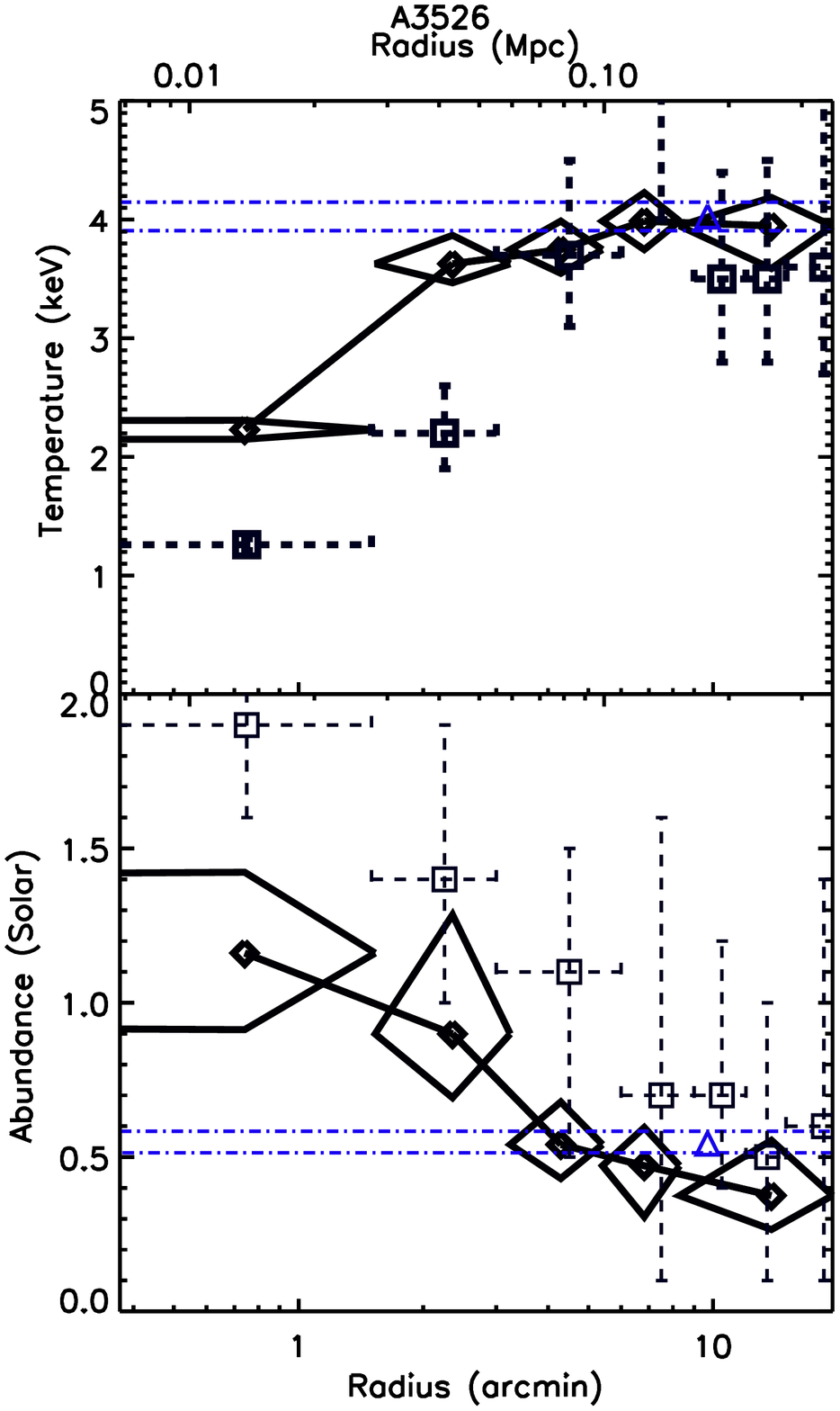,angle=0,width=\figwidth,height=\figheight}
  }
\end{figure*}
\clearpage
\begin{figure*}
\parbox{\textwidth}{
    \psfig{figure=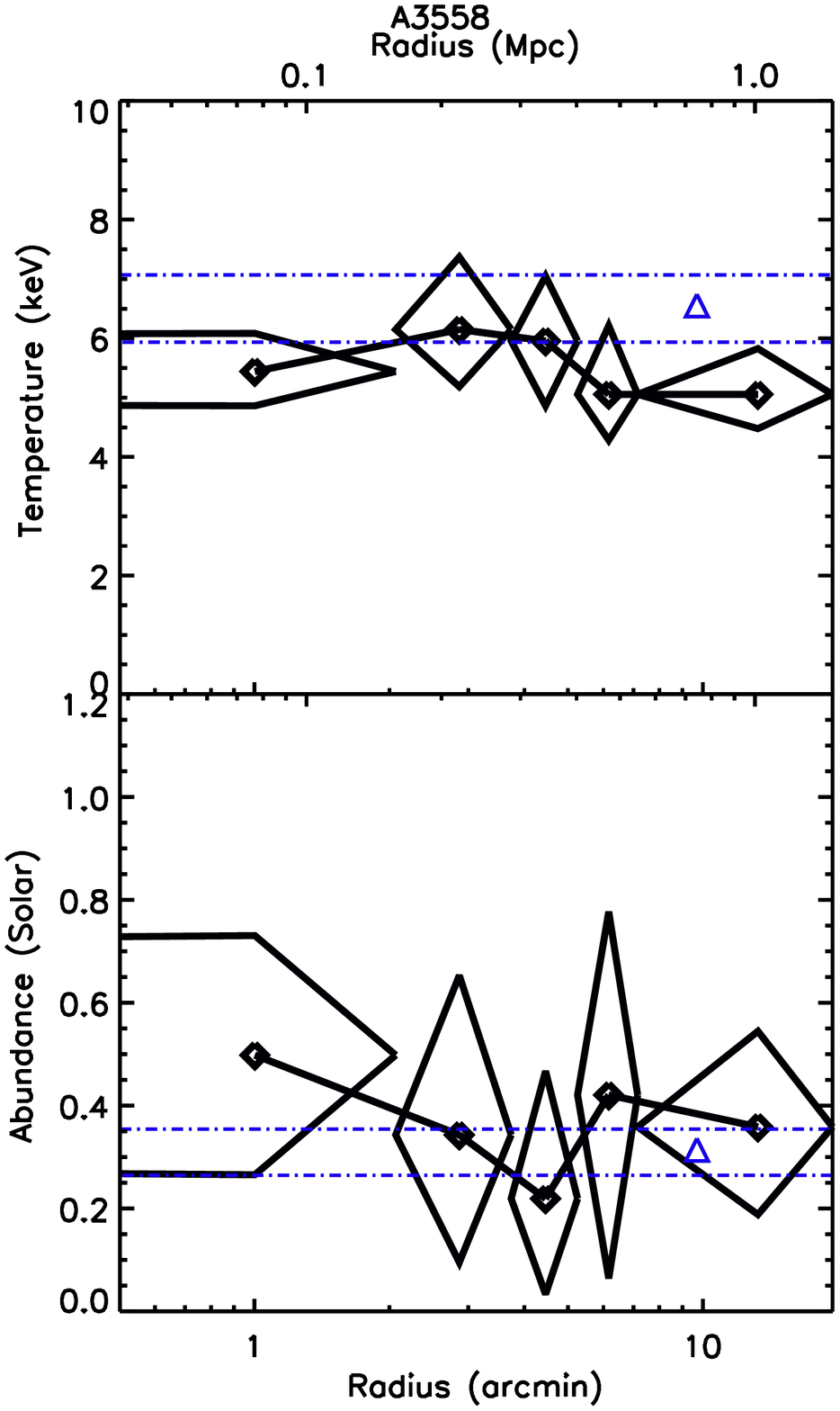,angle=0,width=\figwidth,height=\figheight}
    \psfig{figure=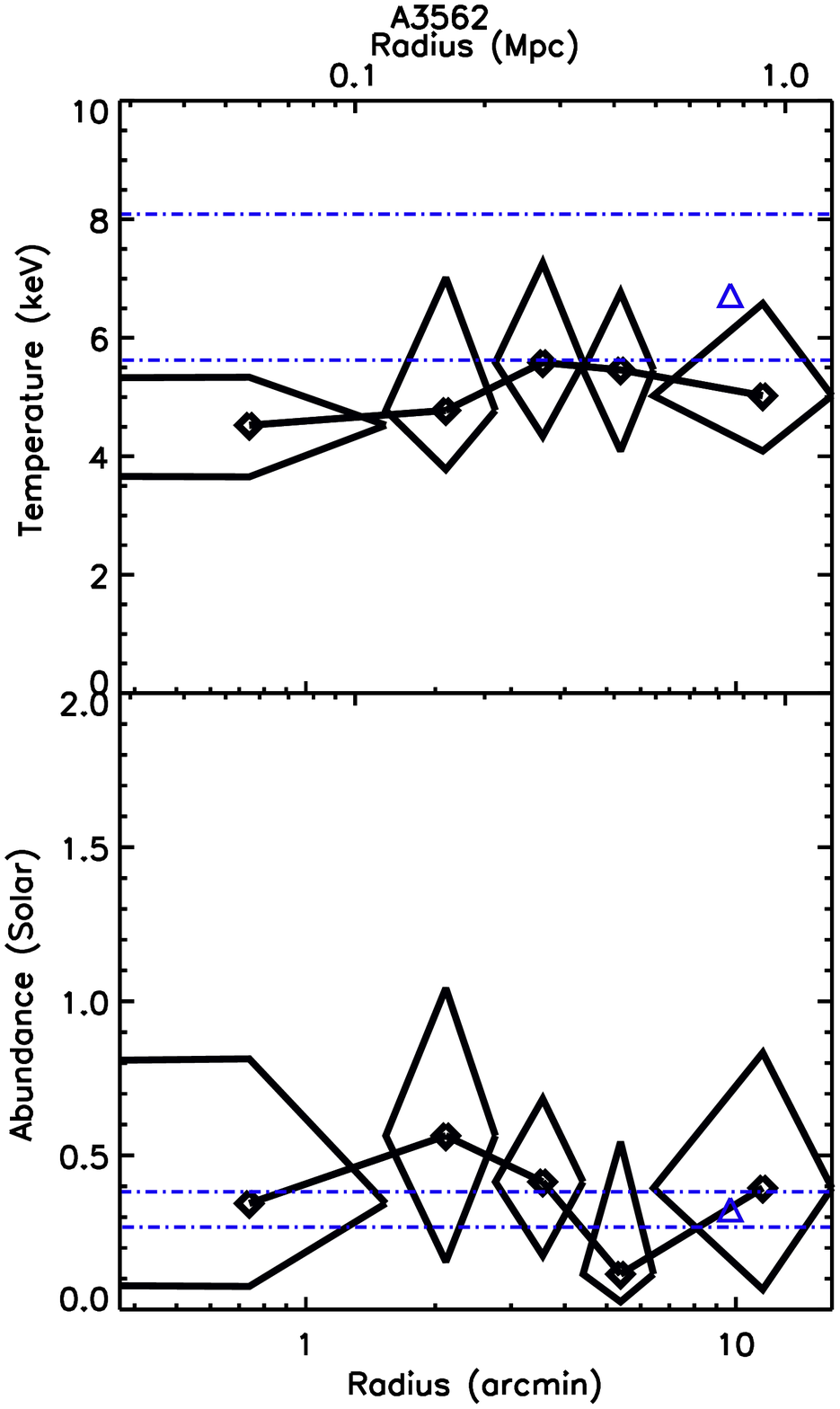,angle=0,width=\figwidth,height=\figheight}
    \psfig{figure=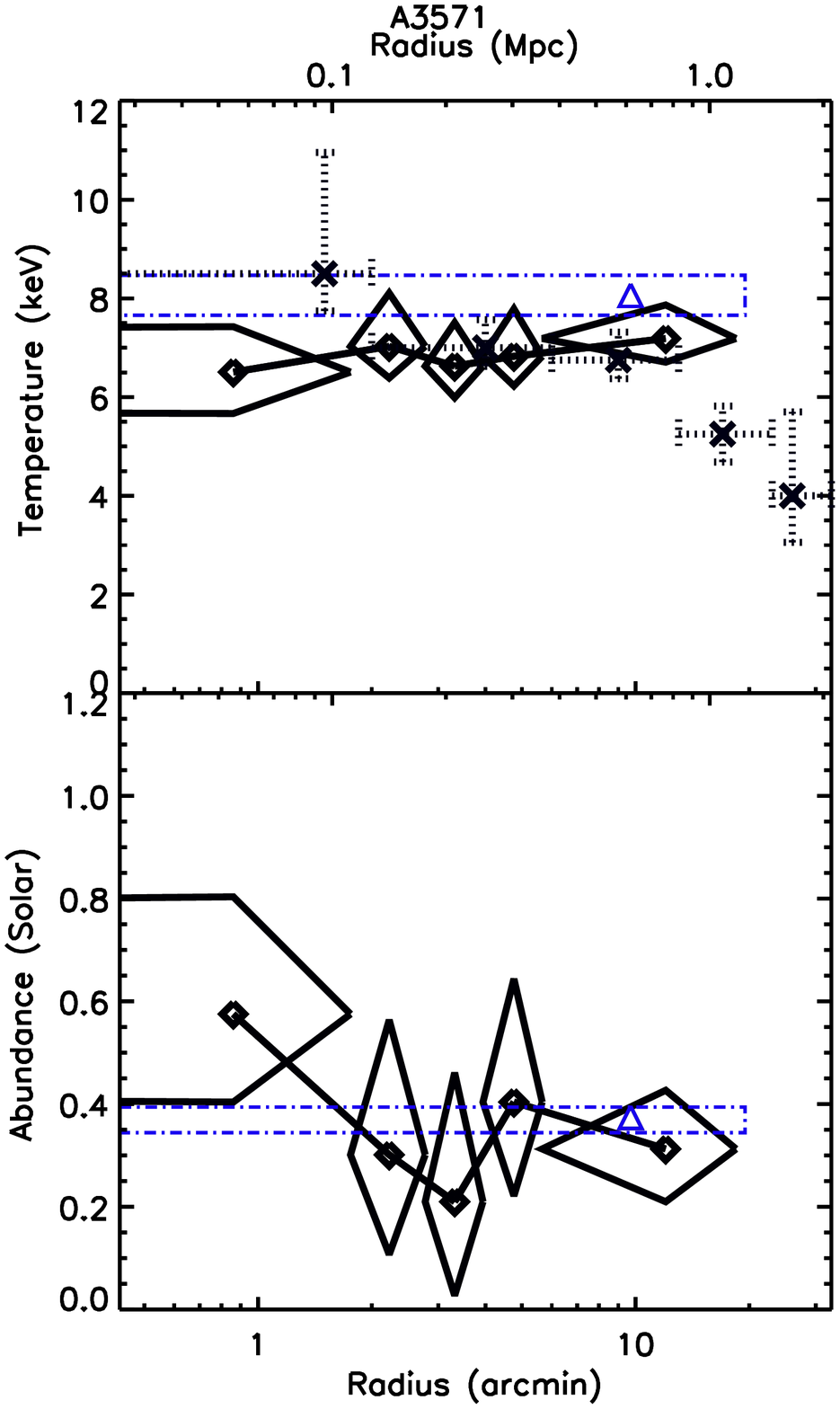,angle=0,width=\figwidth,height=\figheight}
    \psfig{figure=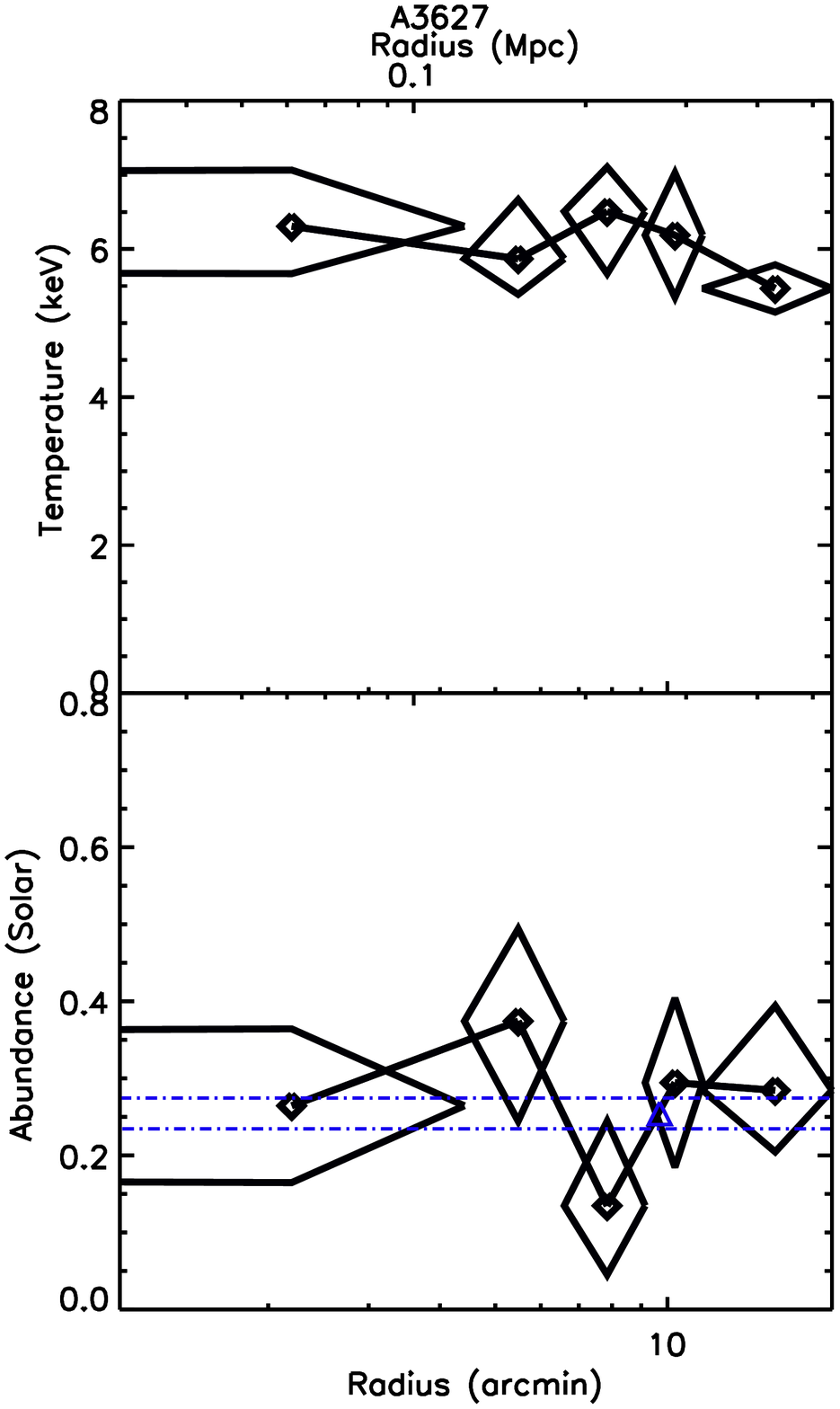,angle=0,width=\figwidth,height=\figheight}
  }
\parbox{\textwidth}{
    \psfig{figure=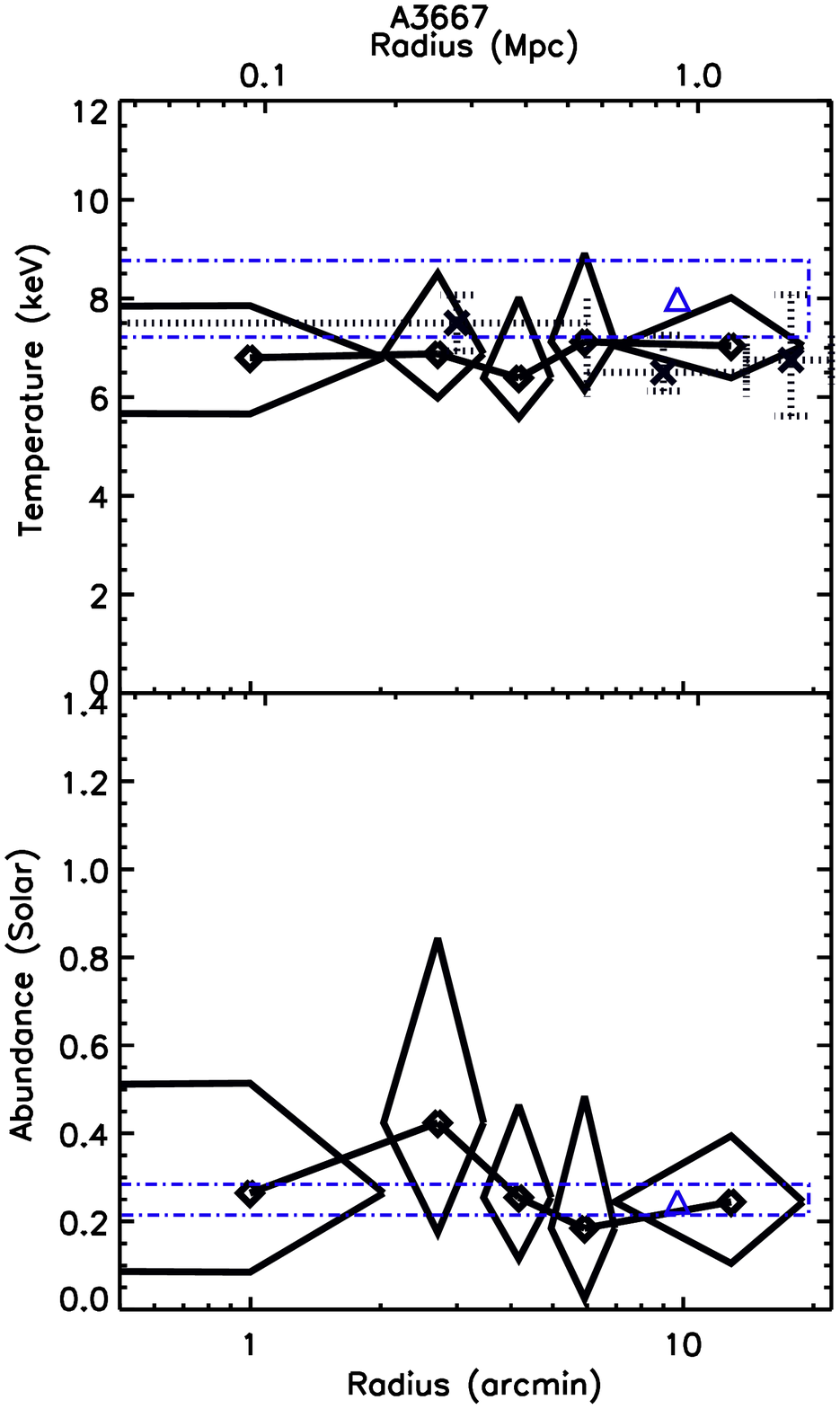,angle=0,width=\figwidth,height=\figheight}
    \psfig{figure=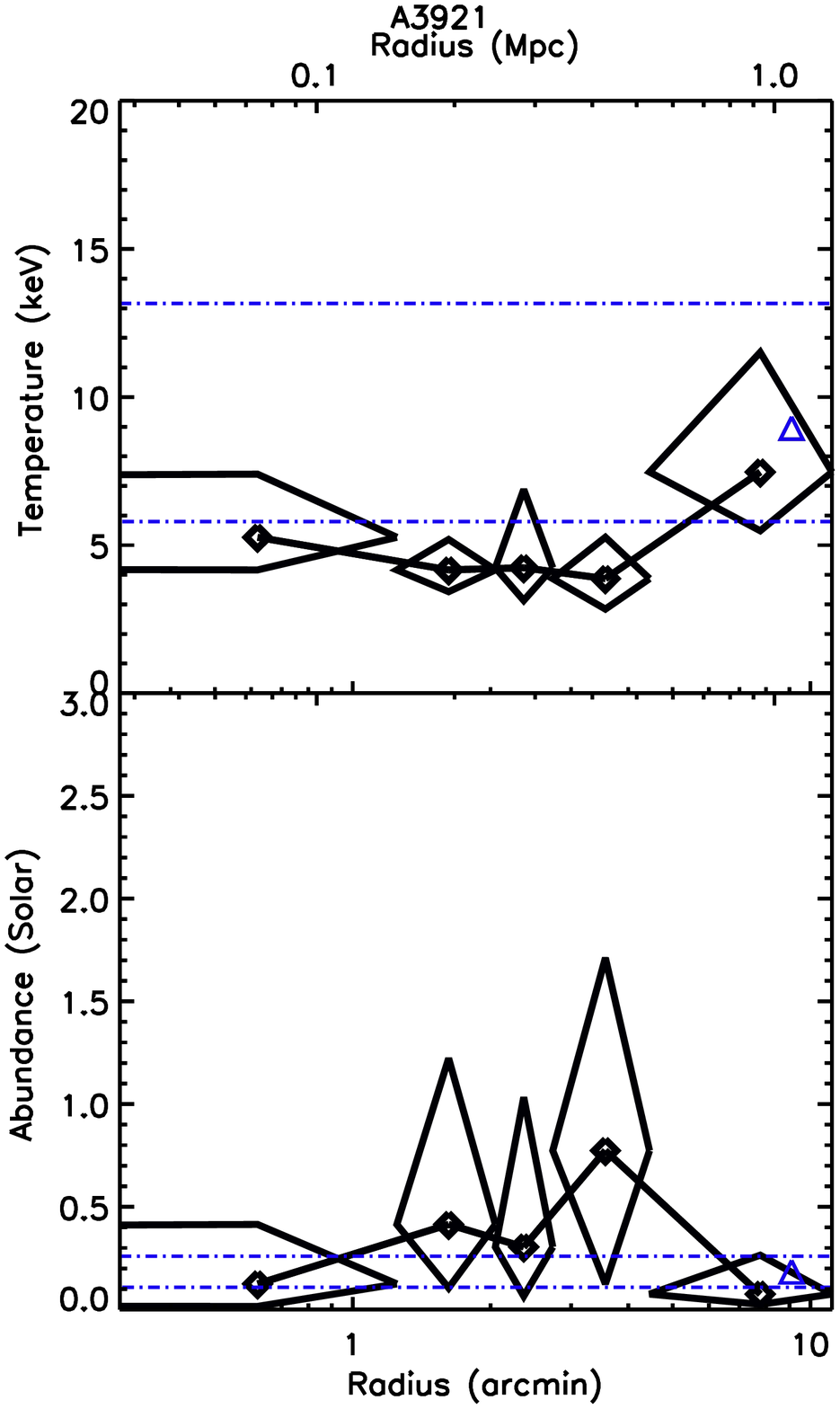,angle=0,width=\figwidth,height=\figheight}
    \psfig{figure=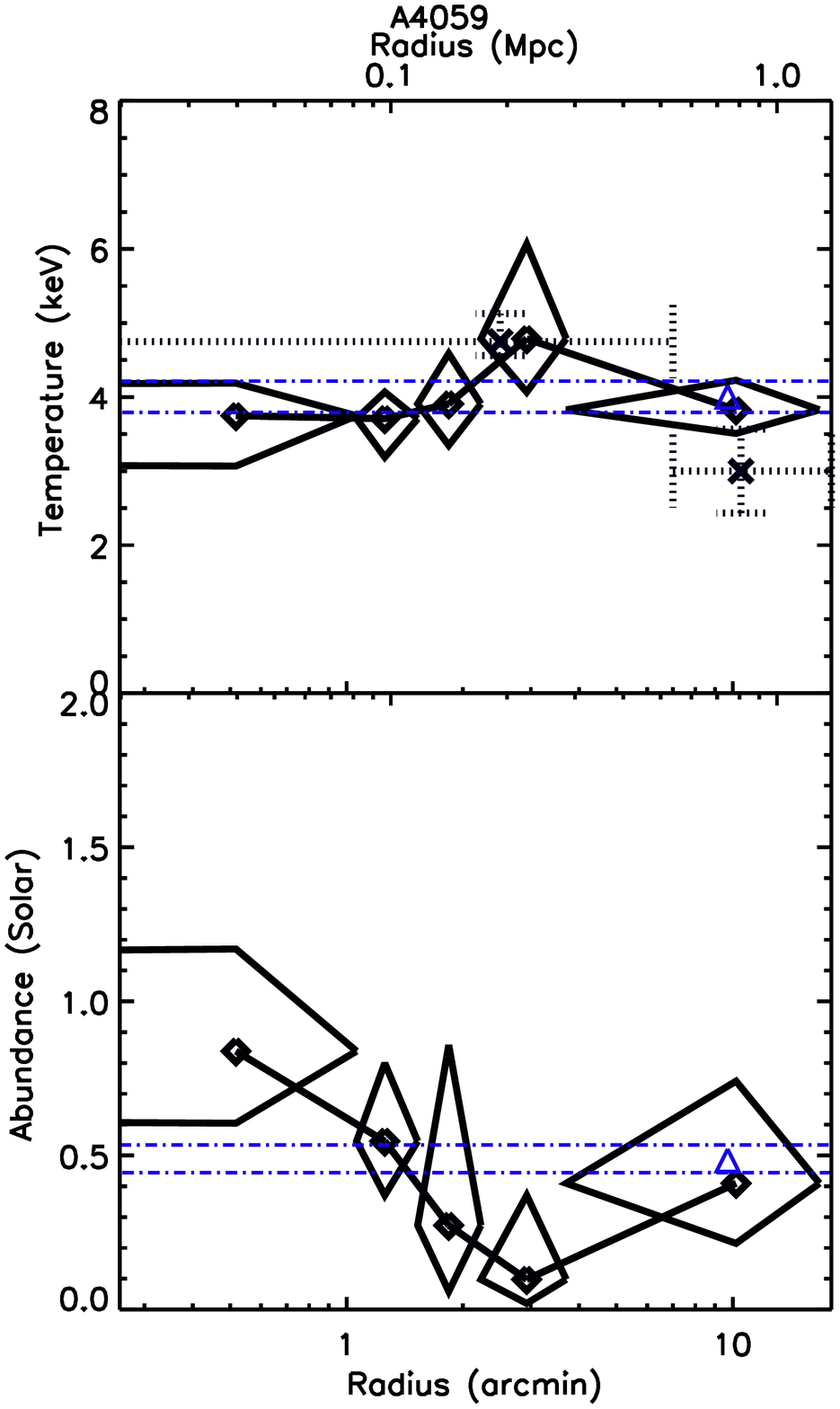,angle=0,width=\figwidth,height=\figheight}
    \psfig{figure=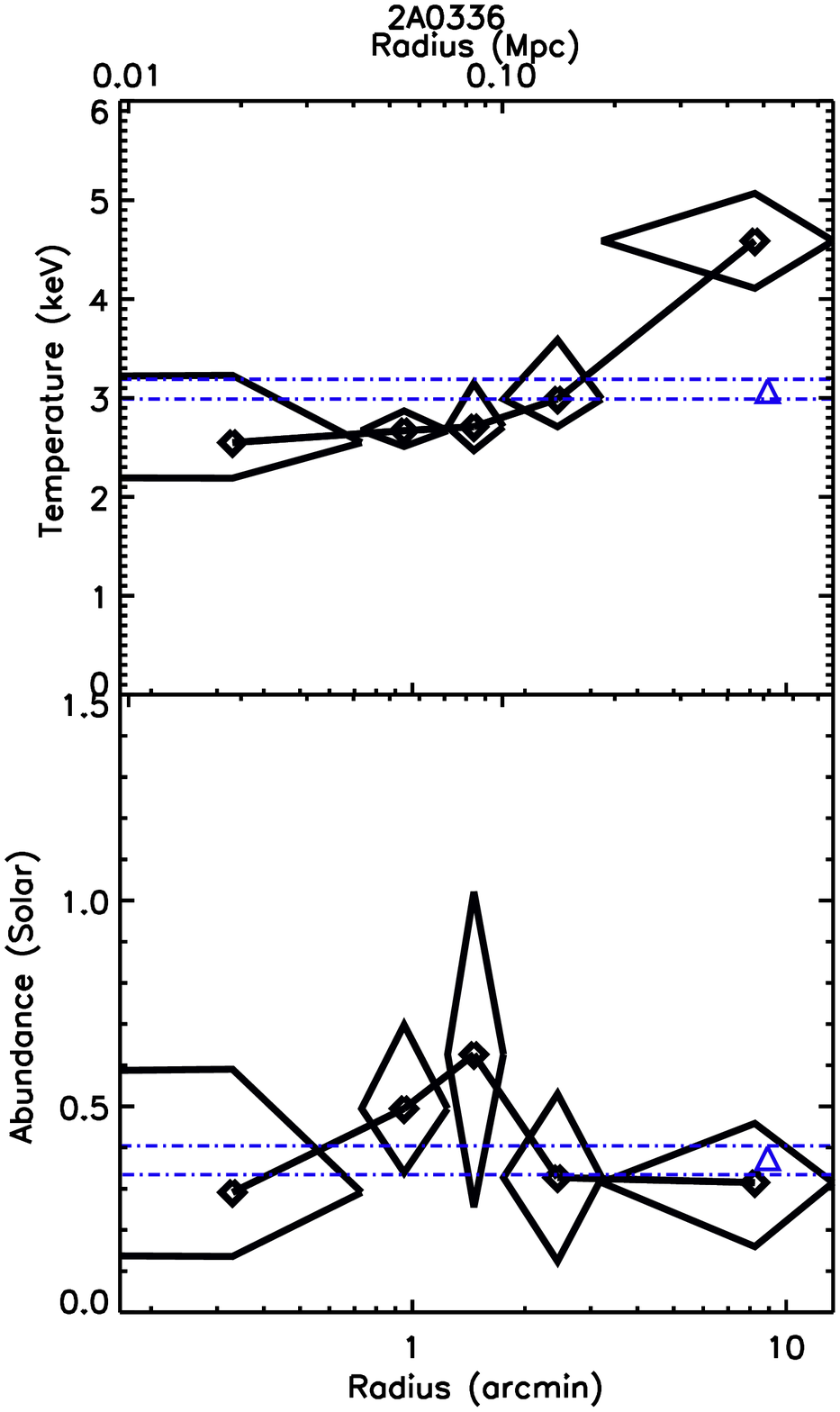,angle=0,width=\figwidth,height=\figheight}
  }
\parbox{\textwidth}{
    \psfig{figure=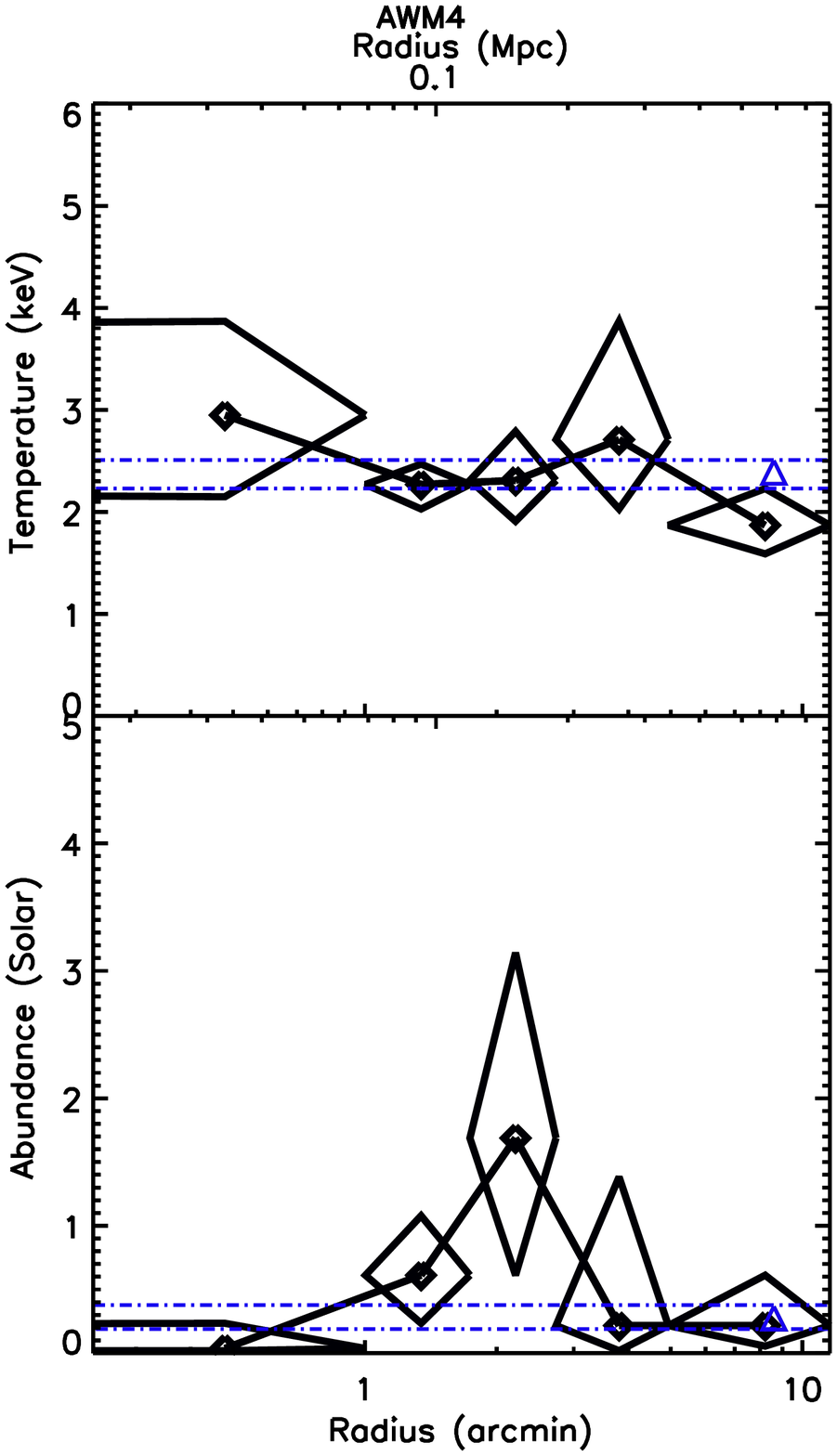,angle=0,width=\figwidth,height=\figheight}
    \psfig{figure=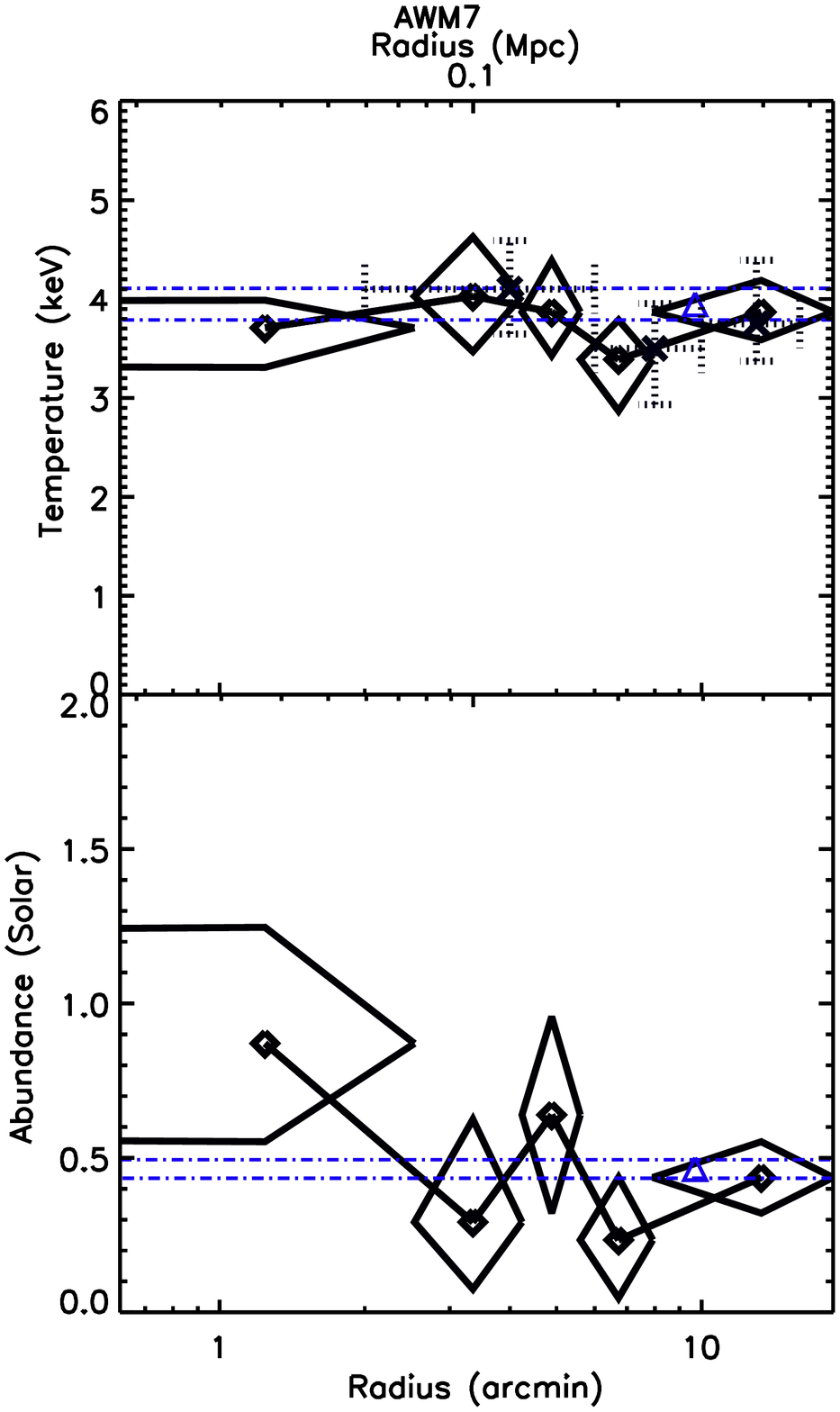,angle=0,width=\figwidth,height=\figheight}
    \psfig{figure=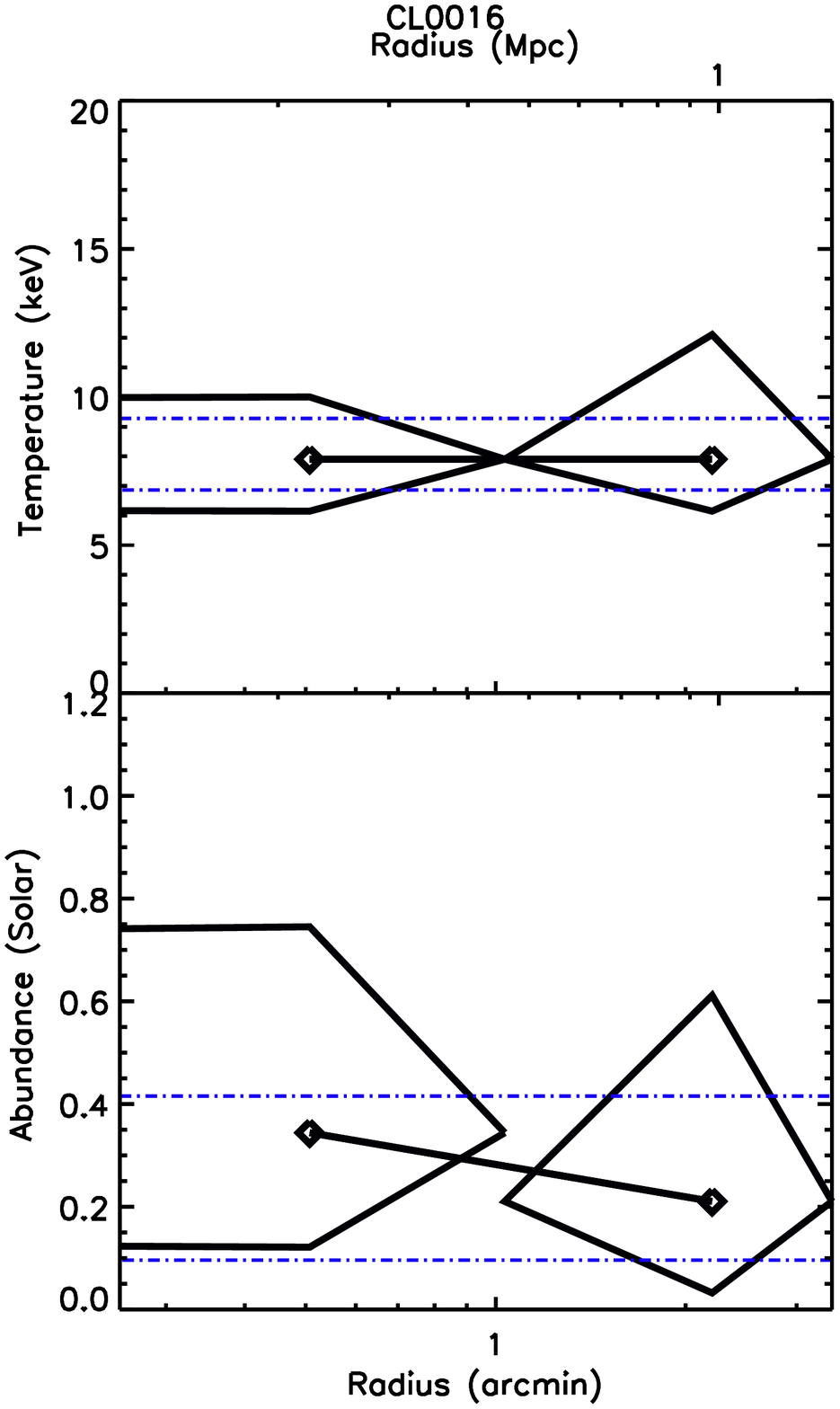,angle=0,width=\figwidth,height=\figheight}
    \psfig{figure=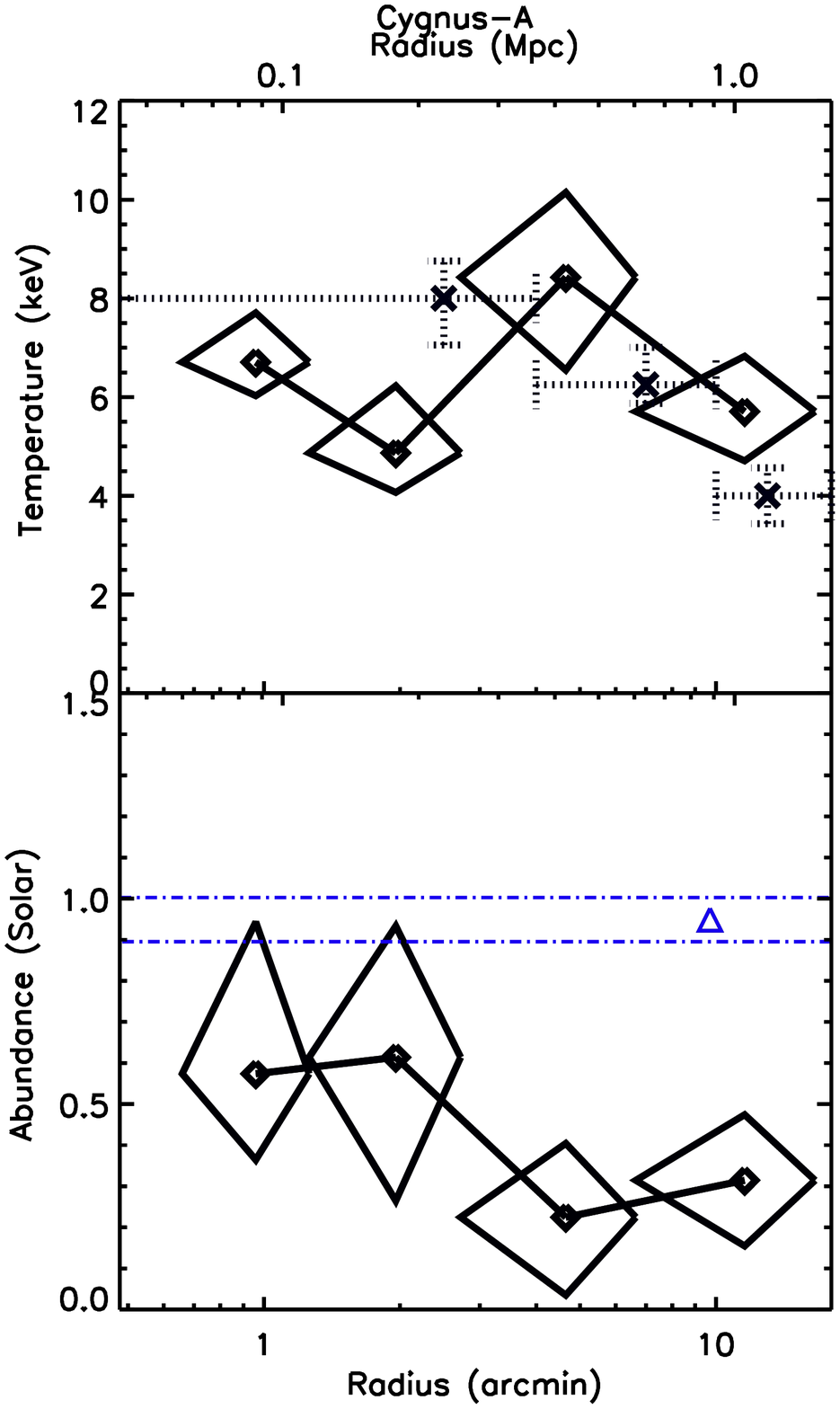,angle=0,width=\figwidth,height=\figheight}
  }
\end{figure*}
\clearpage
\begin{figure*}
\parbox{\textwidth}{
    \psfig{figure=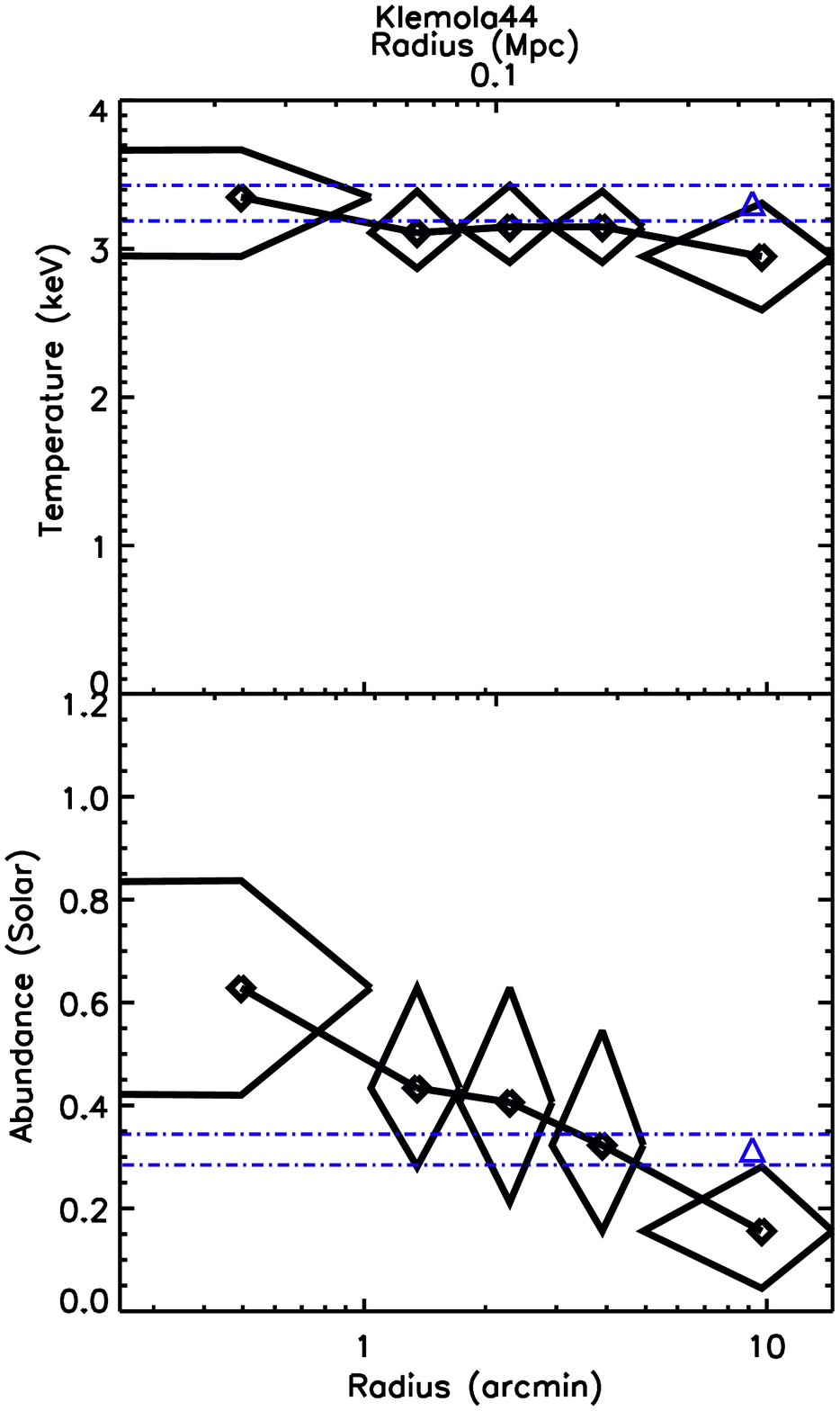,angle=0,width=\figwidth,height=\figheight}
    \psfig{figure=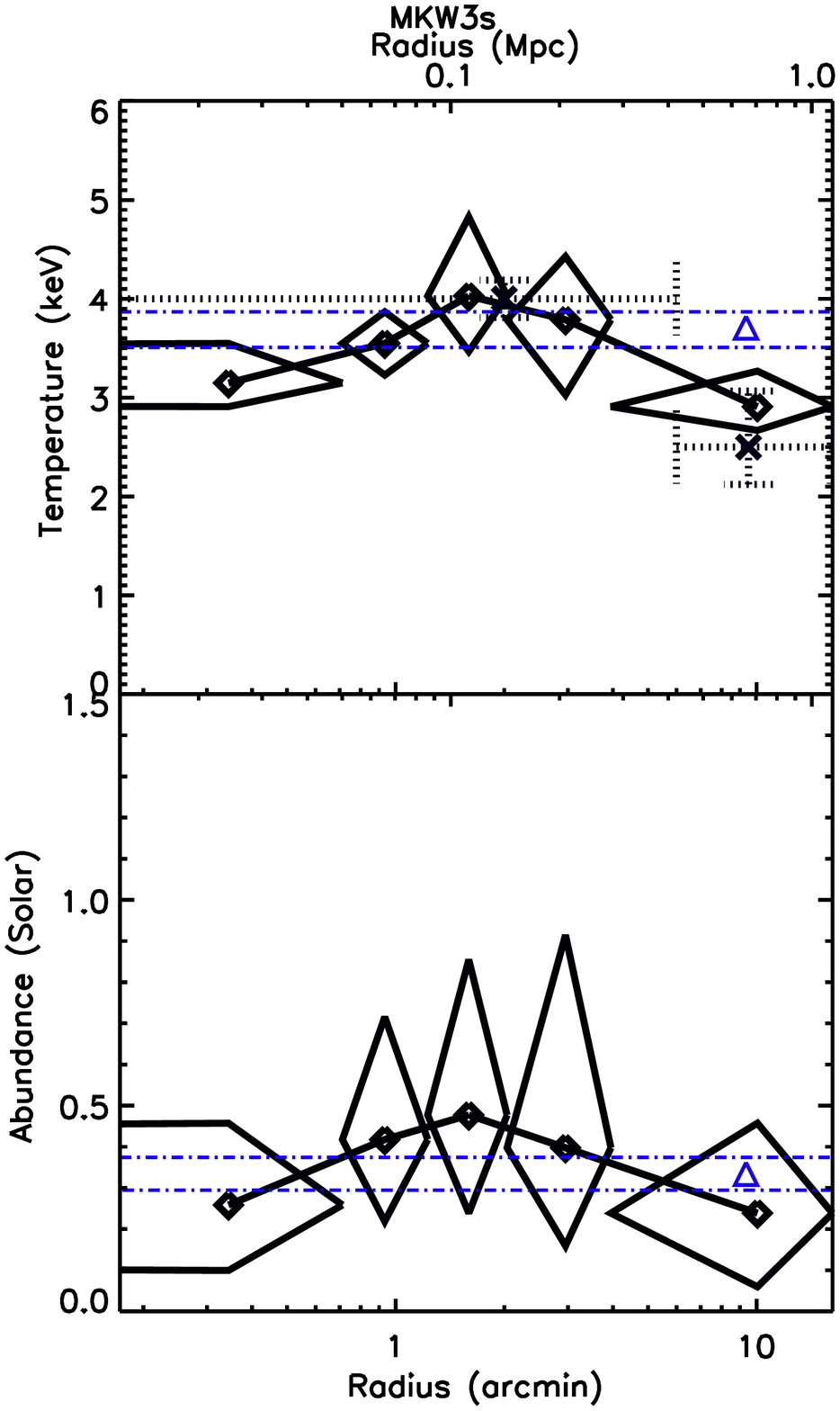,angle=0,width=\figwidth,height=\figheight}
    \psfig{figure=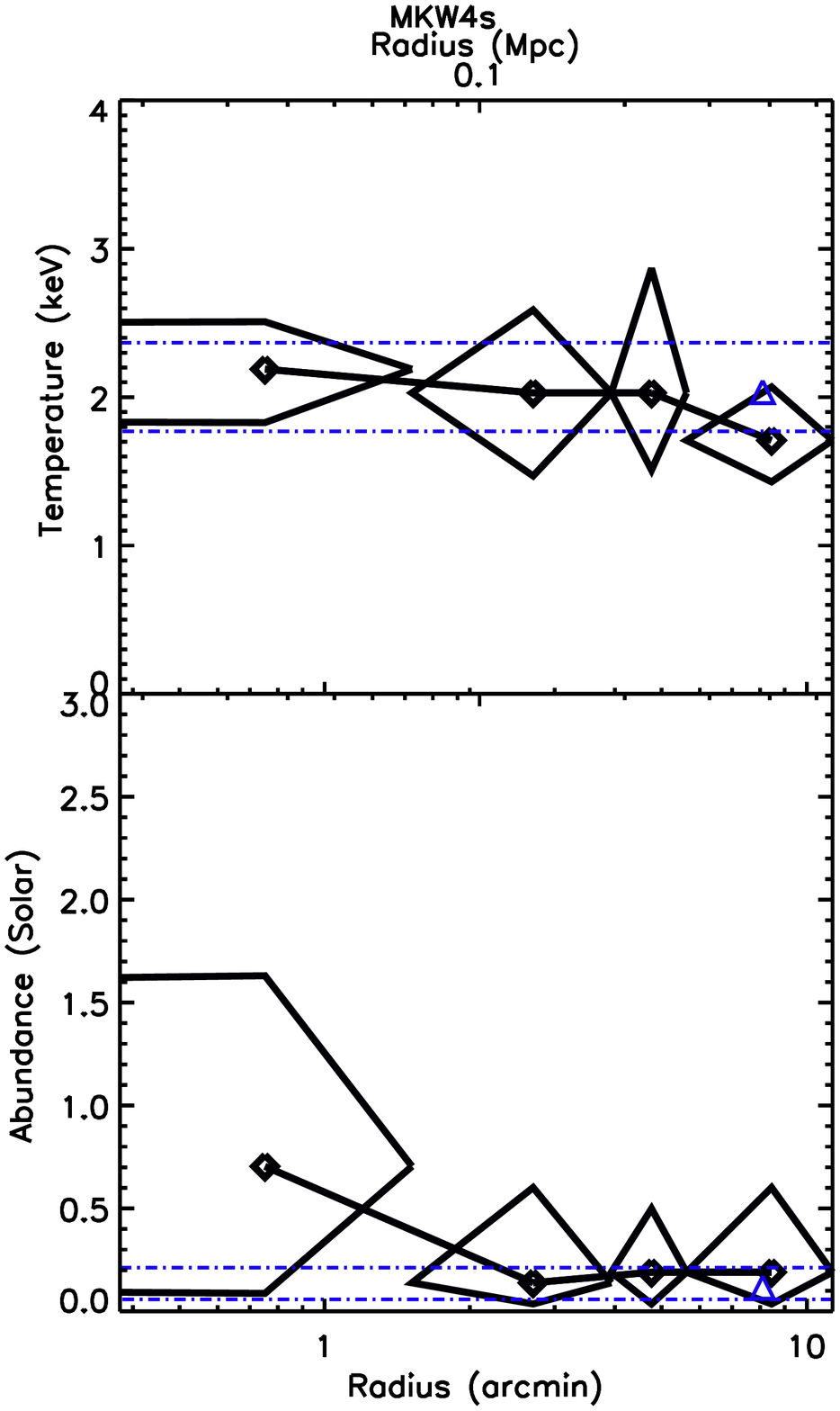,angle=0,width=\figwidth,height=\figheight}
    \psfig{figure=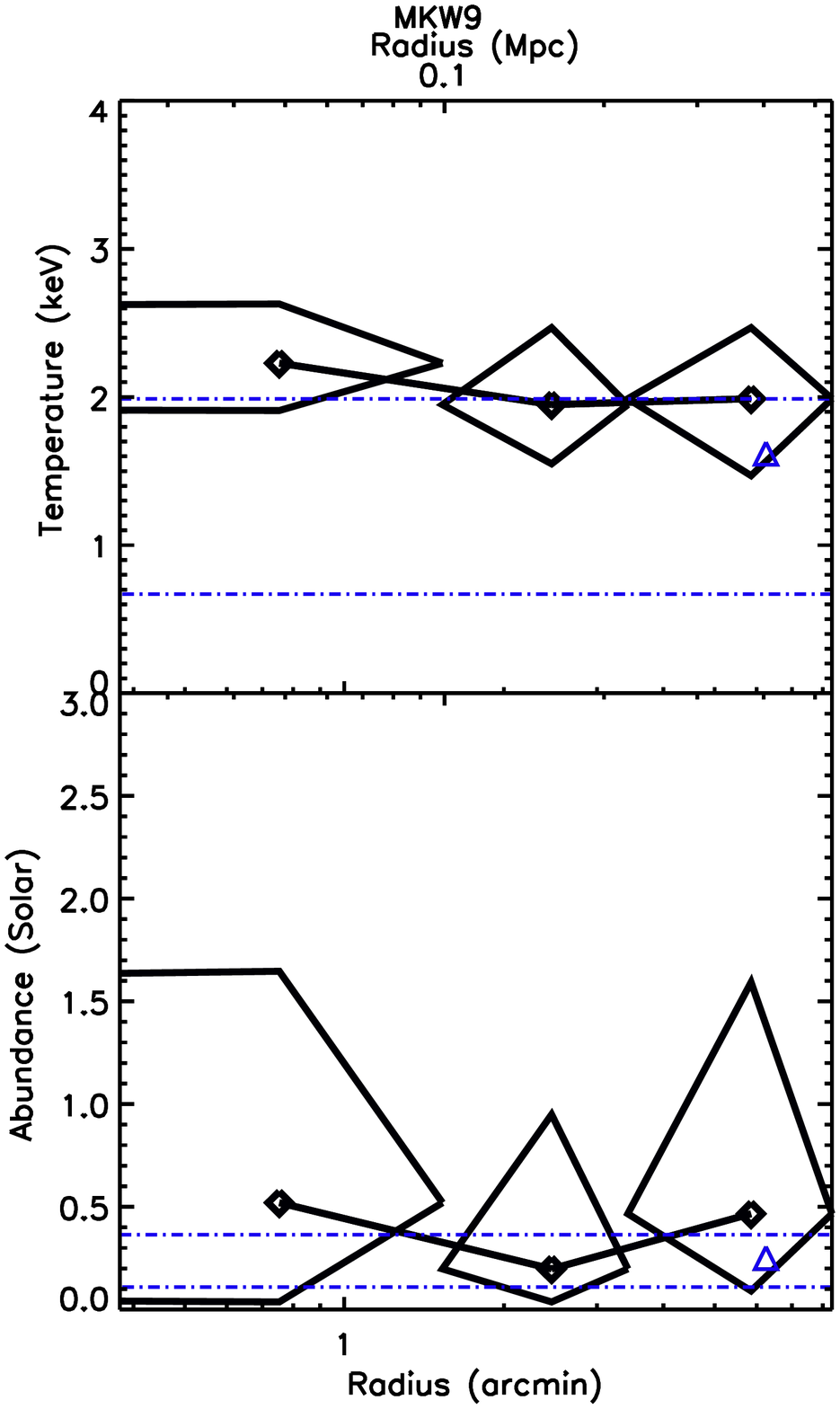,angle=0,width=\figwidth,height=\figheight}
  }
\parbox{\textwidth}{
    \psfig{figure=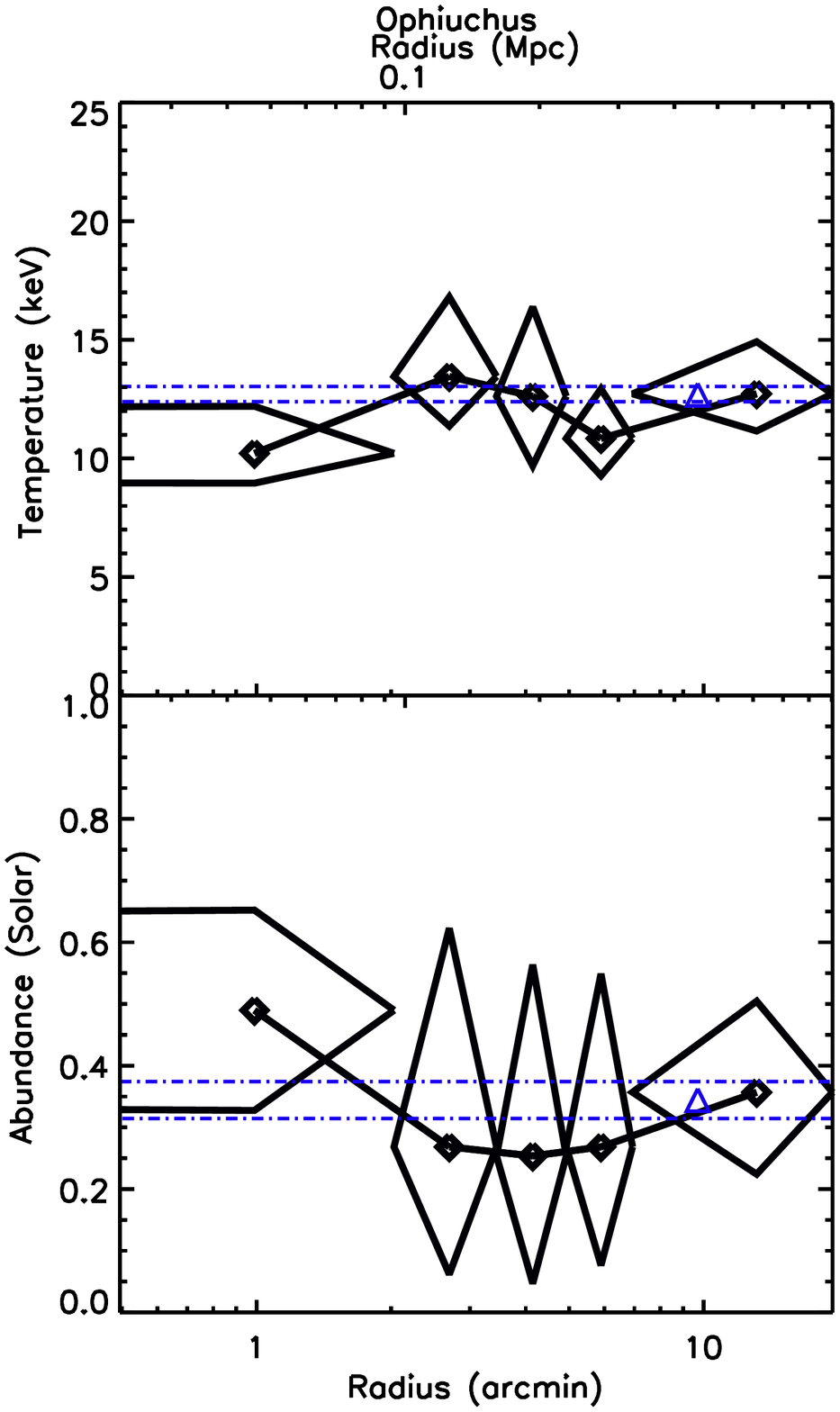,angle=0,width=\figwidth,height=\figheight}
    \psfig{figure=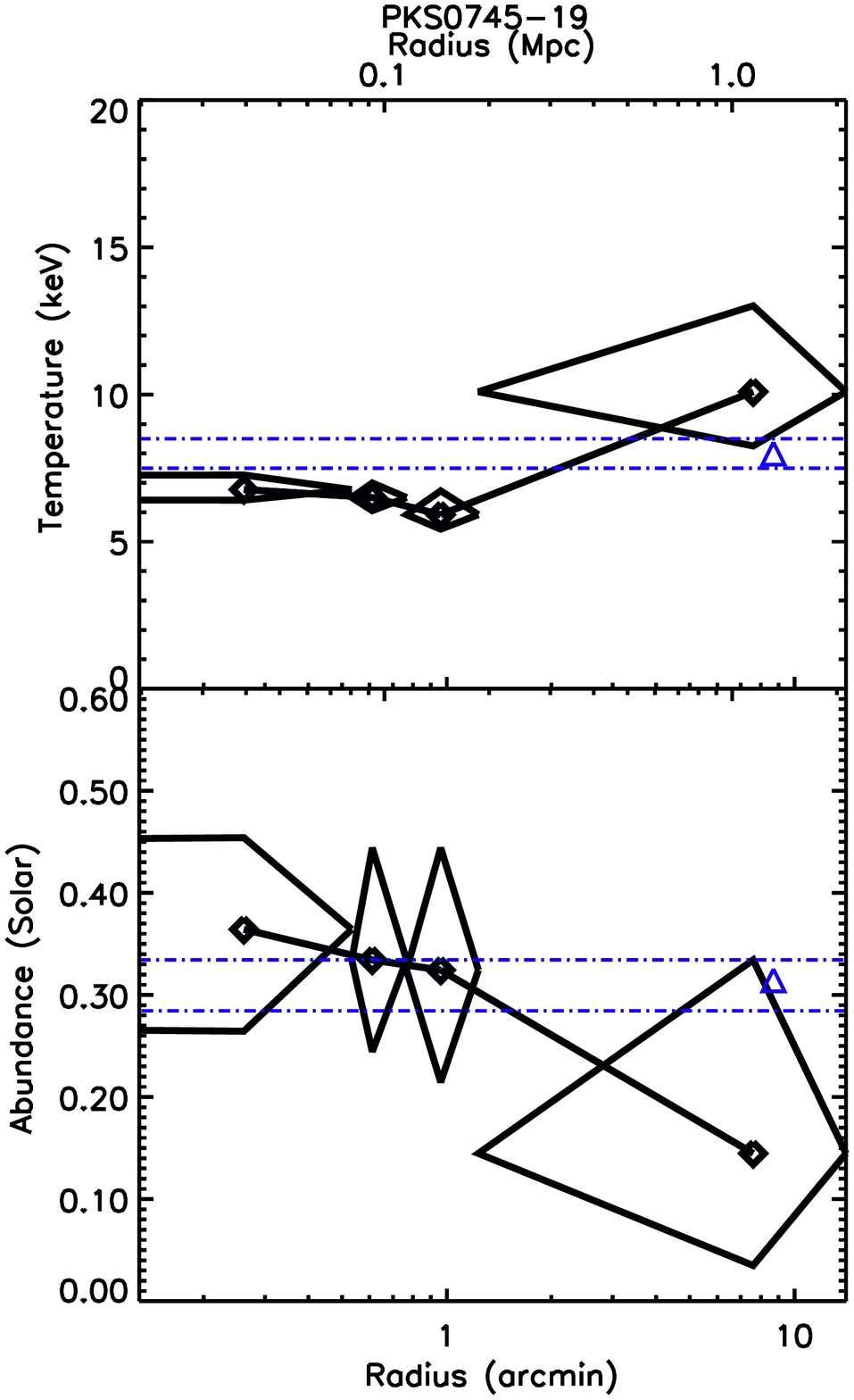,angle=0,width=\figwidth,height=\figheight}
    \psfig{figure=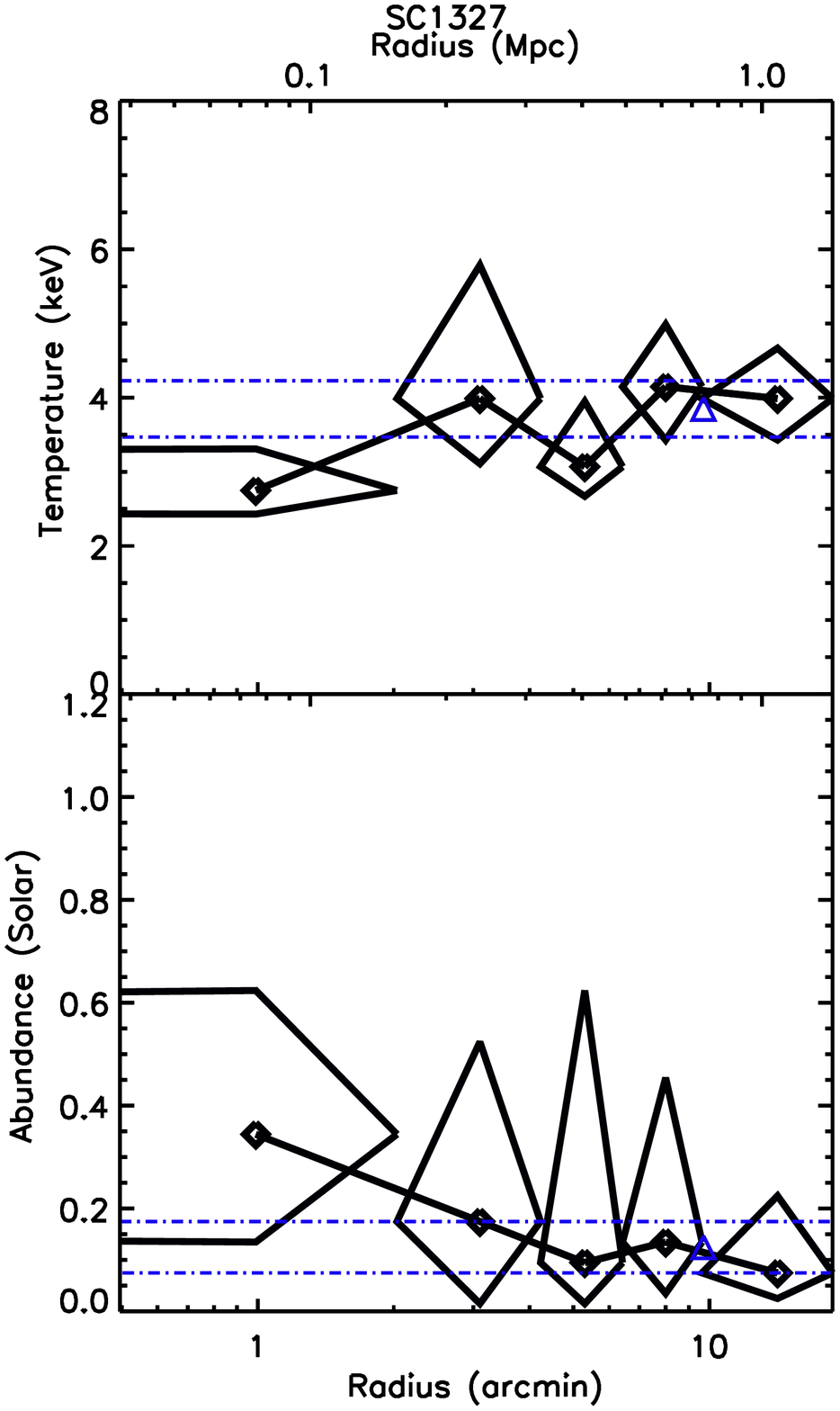,angle=0,width=\figwidth,height=\figheight}
    \psfig{figure=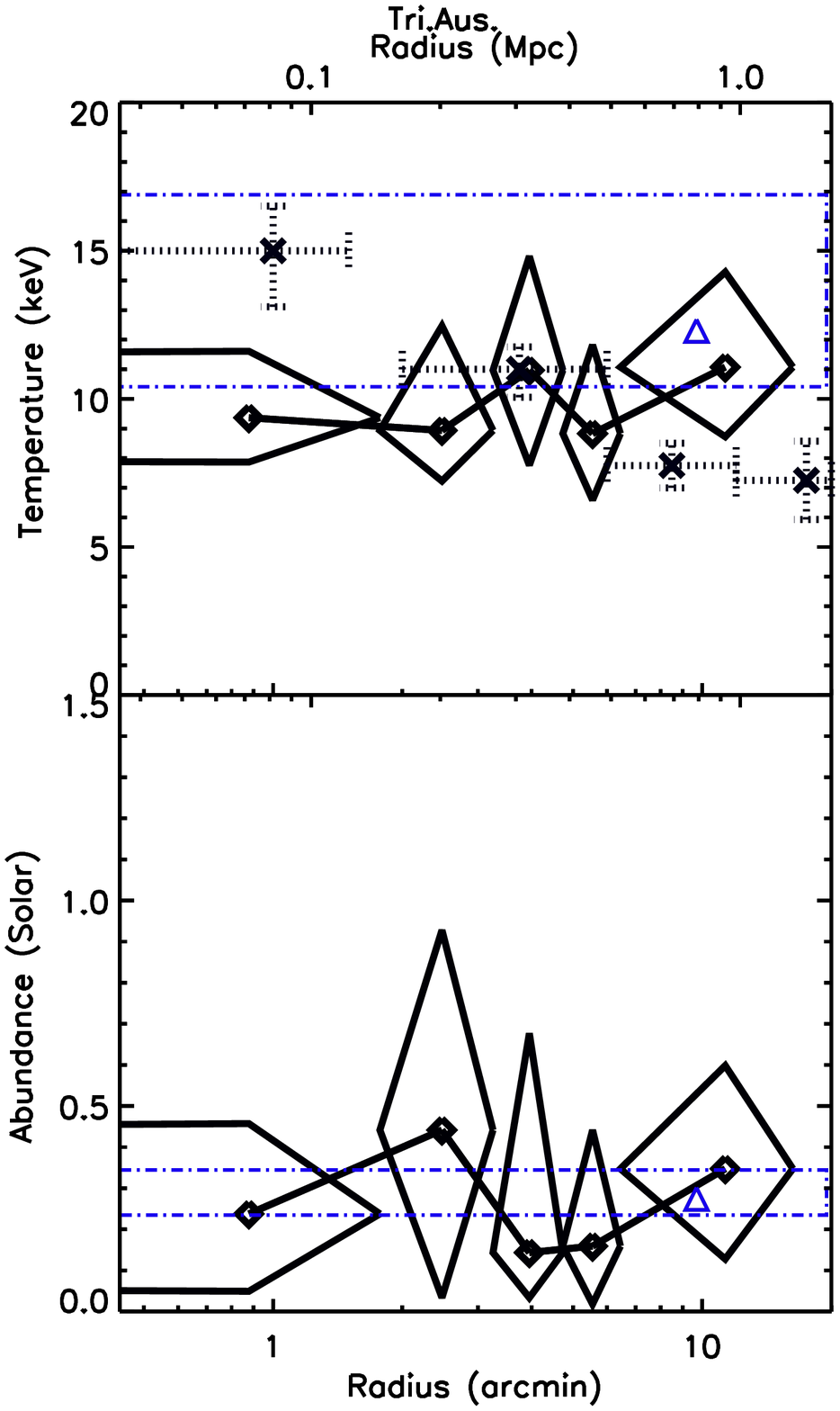,angle=0,width=\figwidth,height=\figheight}
  }
\parbox{\textwidth}{
    \psfig{figure=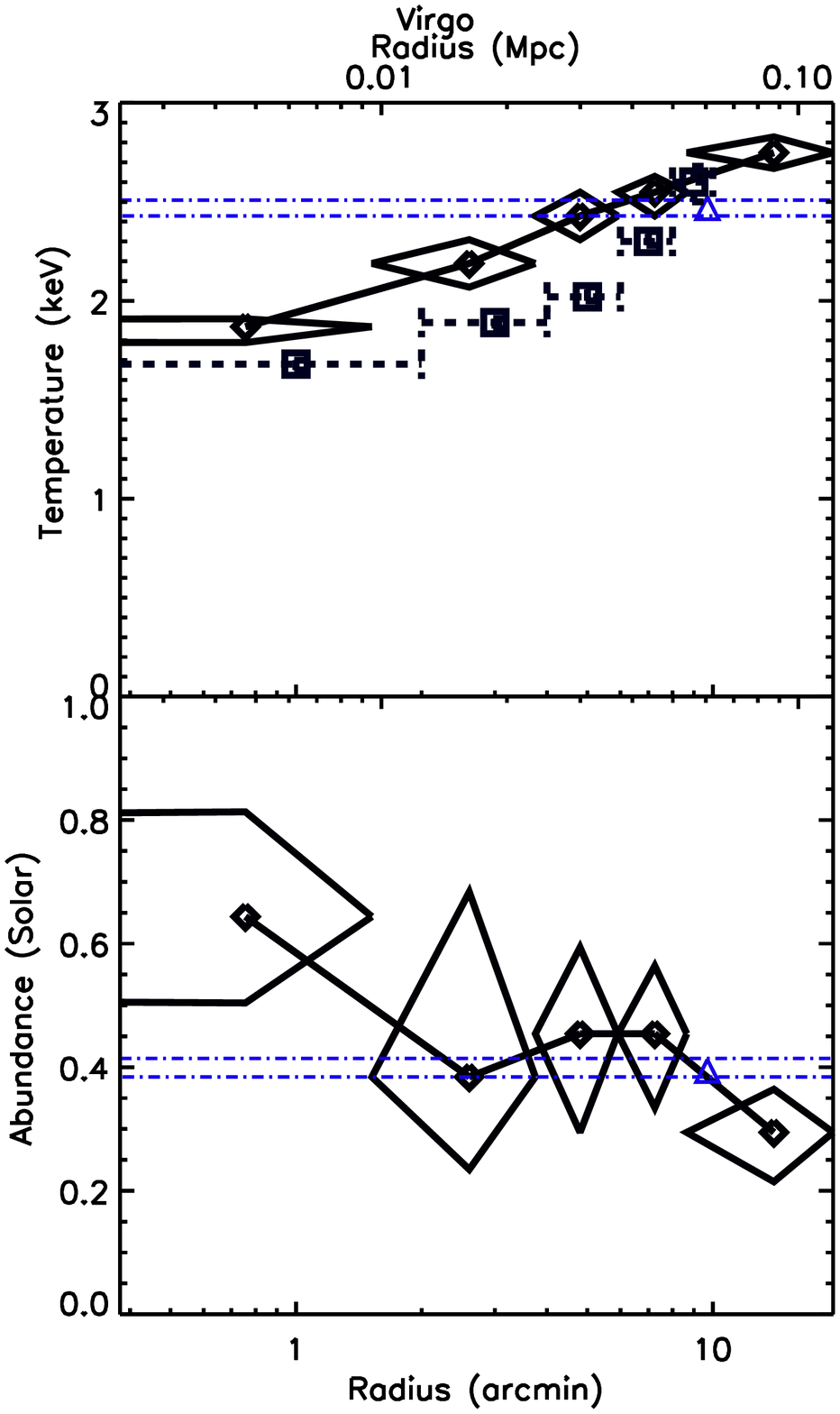,angle=0,width=\figwidth,height=\figheight}
    \psfig{figure=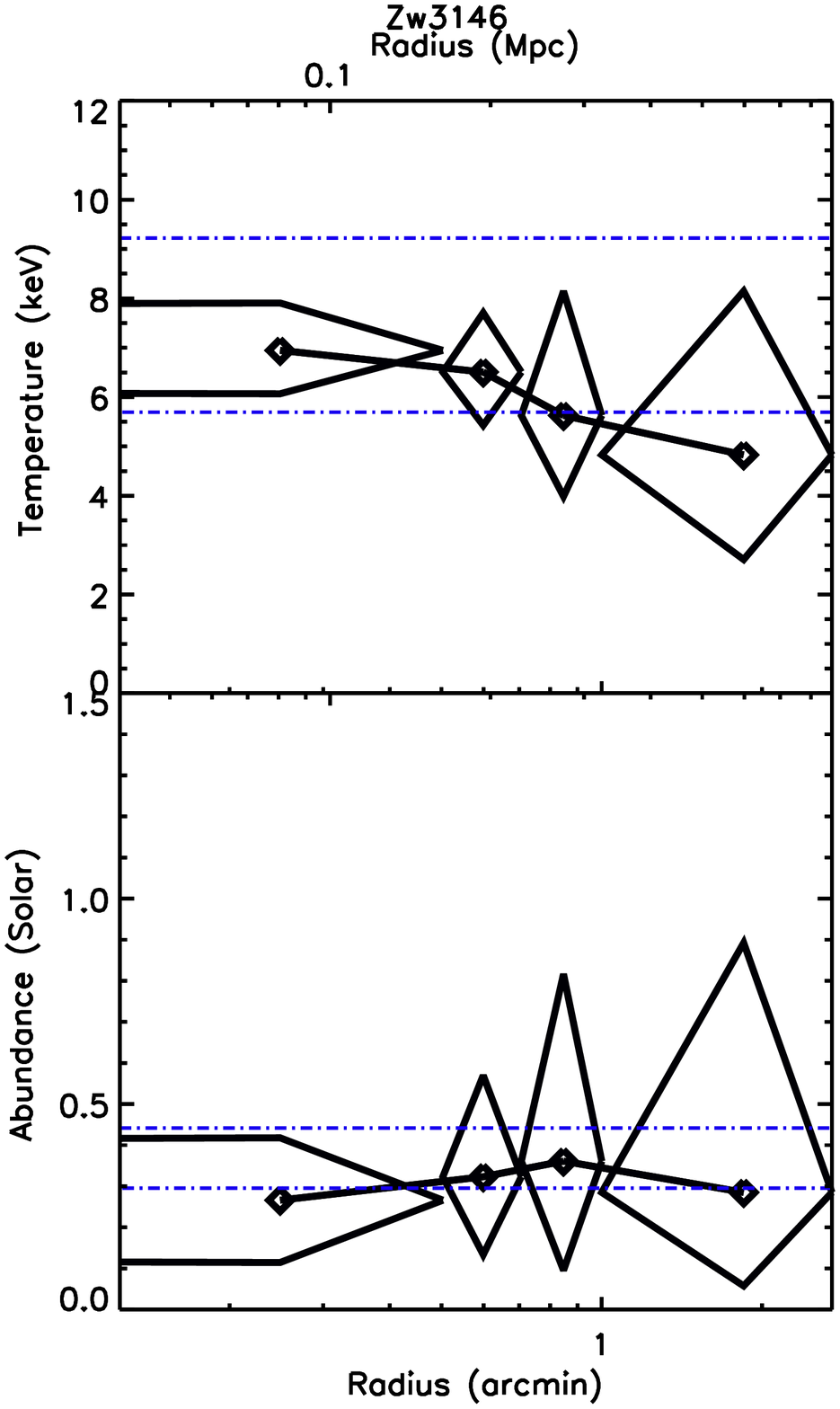,angle=0,width=\figwidth,height=\figheight}
  }
\end{figure*}
}

\def\figmm{
\begin{figure}
	\parbox{0.49\textwidth}{
		\psfig{figure=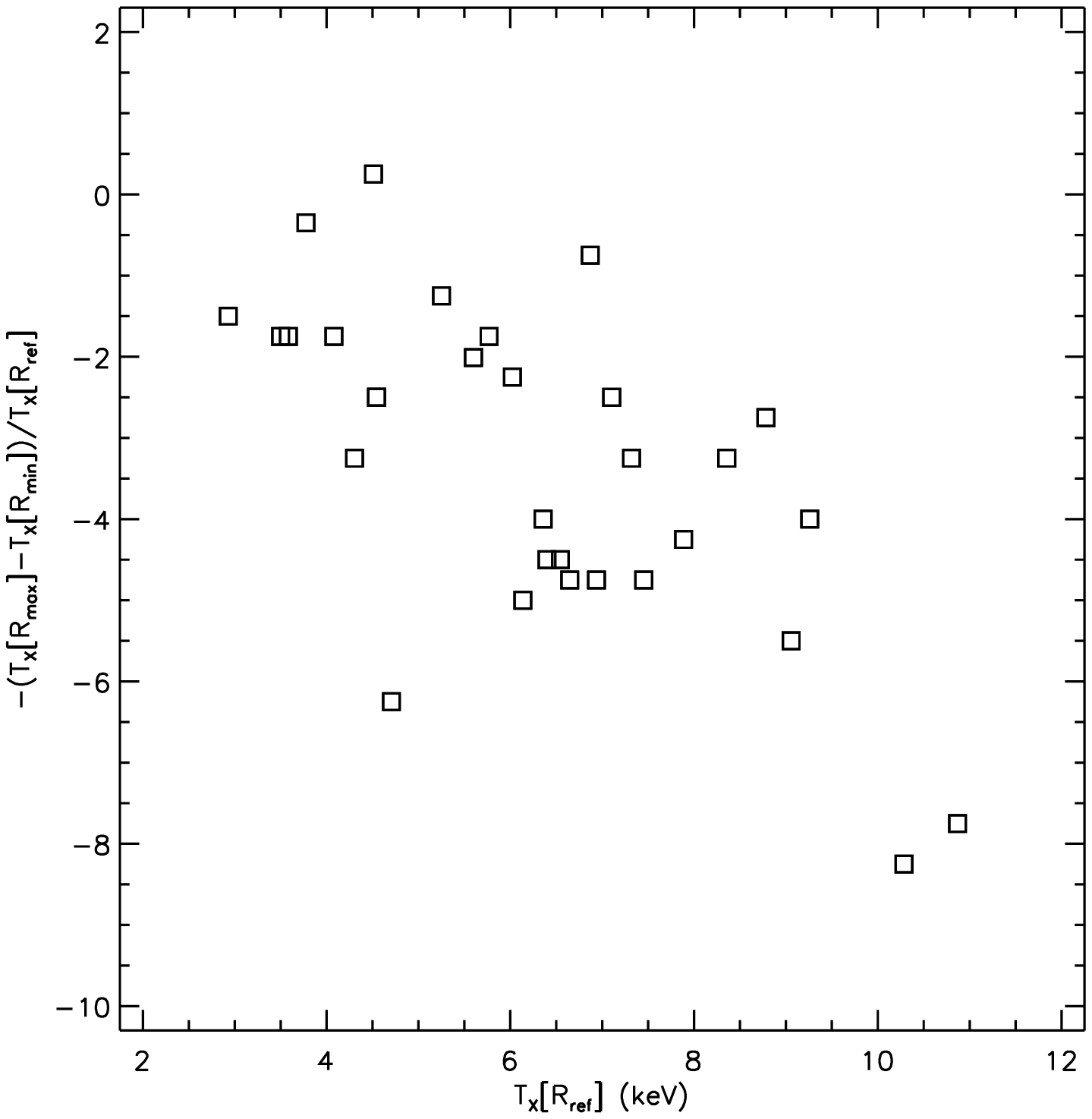,angle=0,width=0.49\textwidth,height=0.49\textwidth}
		\figsmallfont \normalsize } \parbox{0.49\textwidth}{
		\hfill\parbox{0.45\textwidth}{
		\captionfont\caption{\label{figure:figmm} This figure
		presents data from Markevitch \etal\ to investigate
		the possible systematic bias discussed in
		Section~\ref{section:isothermality}. The ordinate data
		shows the normalised temperature decline (\ie\ the
		difference in temperature between the inner and
		outermost radii, normalised by the inner temperature)
		plotted against the weighted-average temperature. The
		Pearson rank correlation test indicates that there is
		a marginal correlation in the data (1-$\sigma$)
		deviation from the null hypothesis; the deviation is
		3-$\sigma$ when the inner temperature is used as the
		abscissa).  From this plot we see that, in the
		Markevitch \etal\ sample, there is a tendency for
		hotter clusters to have larger (normalised) declines
		than the cooler clusters.  }}} \normalsize
\end{figure} }

\def\<{\thinspace}

\def\ROSAT{{\em ROSAT}}

\def\ASCA{{\em ASCA}}

\def\Mpc{{\rm\thinspace Mpc}}
\def\pMpc{\Mpc^{-1}}

\def\Msunpyr{\hbox{$\Msun\yr^{-1}\,$}}
\def\Msun{\hbox{$\rm\thinspace M_{\odot}$}}

\def\ctsps{\hbox{$\cts\s^{-1}\,$}}
\def\cts{\rm\thinspace cts}

\def\cm{{\rm\thinspace cm}}

\def\eg{{\it e.g.\ }}

\def\etal{{et al.}}

\def\qO#1{\thinspace q_{\rm 0}=#1}

\def\HO#1{\thinspace H_{\rm 0}=#1\thinspace \kmps\pMpc}
\def\ie{{\it i.e.}}

\def\keV{{\rm\thinspace keV}}

\def\kmps{\hbox{$\km\s^{-1}\,$}}
\def\km{{\rm\thinspace km}}

\def\pcmsq{\hbox{$\cm^{-2}\,$}}

\def\psqcm{\hbox{$\cm^{-2}\,$}}

\def\s{{\rm\thinspace s}}

\def\yr{{\rm\thinspace yr}}

\mathchardef\twiddle="2218

\def\arcmin{{\rm\thinspace arcmin}}

\def\Tx{\hbox{$T_{\rm X}\,$}}

\def\ne{\hbox{$n_{\rm e}\,$}}
\def\np{\hbox{$n_{\rm p}\,$}}

\def\Mdot{\hbox{$\dot M\,$}}

\def\nH{\hbox{$N_{\rm H}\,$}}

\def\ctsps{\hbox{$\cts\s^{-1}\,$}}

\def\singlespace {\smallskipamount=3pt plus1pt minus1pt
                  \medskipamount=6pt plus2pt minus2pt
                  \bigskipamount=12pt plus4pt minus4pt
                  \normalbaselineskip=12pt plus0pt minus0pt
                  \normallineskip=1pt
                  \normallineskiplimit=0pt
                  \jot=3pt
                  {\def\smallskip {\vskip\smallskipamount}}
                  {\def\medskip   {\vskip\medskipamount}}
                  {\def\bigskip   {\vskip\bigskipamount}}
                  {\setbox\strutbox=\hbox{\vrule 
                    height8.5pt depth3.5pt width 0pt}}
                  \parskip 6.0pt
                  \normalbaselines}
\def\middlespace {\smallskipamount=4.5pt plus1.5pt minus1.5pt
                  \medskipamount=9pt plus3pt minus3pt
                  \bigskipamount=18pt plus6pt minus6pt
                  \normalbaselineskip=18pt plus0pt minus0pt
                  \normallineskip=1pt
                  \normallineskiplimit=0pt
                  \jot=4.5pt
                  {\def\smallskip {\vskip\smallskipamount}}
                  {\def\medskip   {\vskip\medskipamount}}
                  {\def\bigskip   {\vskip\bigskipamount}}
                  {\setbox\strutbox=\hbox{\vrule 
                    height12.75pt depth5.25pt width 0pt}}
                  \parskip 9.0pt
                  \normalbaselines}
\def\doublespace {\smallskipamount=6pt plus2pt minus2pt
                  \medskipamount=12pt plus4pt minus4pt
                  \bigskipamount=24pt plus8pt minus8pt
                  \normalbaselineskip=24pt plus0pt minus0pt
                  \normallineskip=2pt
                  \normallineskiplimit=0pt
                  \jot=6pt
                  {\def\smallskip {\vskip\smallskipamount}}
                  {\def\medskip   {\vskip\medskipamount}}
                  {\def\bigskip   {\vskip\bigskipamount}}
                  {\setbox\strutbox=\hbox{\vrule 
                    height17.0pt depth7.0pt width 0pt}}
                  \parskip 12.0pt
                  \normalbaselines}

\def\defaultspace{\singlespace}
\defaultspace
\def\figsmallfont{\normalsize}
\def\captionfont{\normalsize}

\def\nsample{106} \title[ICM properties of \nsample\ galaxy clusters]
{Deconvolution of ASCA X-ray data: II. Radial
temperature and metallicity profiles for \nsample\ galaxy clusters}

\author[D.A.~White]{ 
\parbox[]{6.5in} {
	\large
	D.A.~White \\
	\footnotesize 
	Institute of Astronomy, Madingley Road, Cambridge CB3~OHA. 
		(E-mail: daw@ast.cam.ac.uk)\\
	}
}

\date{Received ***; in original form ***}

\defaultspace
\def\Tx{\hbox{$T_{\rm X}$}}

\def\Solar{\hbox{$\odot$}}

\def\MFSV{{\em {MFSV}}}
\def\SID{{\em {SID}}}

\def\nfrac{$\frac{1}{6}$th}

\begin{document}

\maketitle 


\begin{abstract}   

In Paper-I we presented a methodology to recover the spatial
variations of properties of the intracluster gas from
\ASCA\ X-ray satellite observations of galaxy clusters.  We
verified the correctness of this procedure by applying it to simulated
cluster datasets which we had subjected to the various contaminants
common in \ASCA\ data. In this paper we present the results which we
obtain when we apply this method to real galaxy cluster
observations. We determine broad-band temperature and cooling-flow
mass-deposition rates for the \nsample\ clusters in our sample, and
obtain temperature, abundance and emissivity profiles (\ie\ at least
two annular bins) for 98 of these clusters. We find that 90 percent of
these temperature profiles are consistent with isothermality at the
3-$\sigma$ confidence level. This conflicts with the prevalence of
steeply-declining cluster temperature profiles found by
\citeN{Markevitch:ASCA_Tx_similarity} from a sample of 30 clusters. In
Paper-III (in preparation) we utilise our temperature and emissivity
profiles to determine radial hydrostatic-mass properties for a
subsample of the clusters presented in this paper.

\end{abstract} 



\section{Introduction}

In Paper-I \citeN{White:deconv_i} we described our spectral-imaging
deconvolution (\SID) procedure which was developed to recover spatial
variations in the intracluster gas properties from \ASCA\ satellite
observations. Such a procedure is required because the X-ray mirrors
of this satellite have a point-spread function (PSF) which varies
considerably with position and, in particular, energy. This corrupts
the observed spectral properties of any extended source, \eg\ a galaxy
cluster, to the extent that any uncorrected spatially-resolved
spectral analysis will yield erroneous
results. \citeN{Takahashi:ASCA_PSF} showed that without correction an
isothermal cluster may appear to have a temperature profile which
increases with radius.

Our motivation was to create a procedure which would correct for this
effect and allow us obtain temperature, and thereby mass, profiles for
a large number of clusters. We also wished to address the claim made
by \citeN{Markevitch:ASCA_Tx_similarity} (hereafter \MFSV) for the
ubiquitous nature of temperature declines. With a sample of 30 objects
\MFSV\ parameterised the average temperature decline in clusters with
a polytropic index of $\gamma=1.24^{+0.20}_{-0.12}$, \ie\ $\Gamma\sim
4/3$. This result, if correct, is of great importance as it has been
generally been assumed that clusters are isothermal.

Steeply declining temperature profiles not only complicate the
interpretation of the temperature function, but also exacerbates the
discrepancy between the average baryon fraction in clusters and the
mean value of the Universe expected from primordial nucleosynthesis
calculations.  For a Flat Universe the mean baryon fraction is
calculated to be $0.06h_{50}^{-2}$\citeN{Walker:BBNS}, whereas the
typical baryon content of galaxy clusters determined from X-ray data
(\eg\ \citeNP{White:baryon}; \citeNP{Ettori:ROSAT_baryon}) is
$0.1-0.2h_{50}^{-3/2}$. This disparity, which was first highlighted by
\citeN{White:baryon_catastrophe}, can be resolved by moving to an Open
Universe model (ignoring a contribution from $\Omega_\Lambda$), \ie\
with values of $\Omega_0\approx0.2-0.3$. However, declining
temperature profiles imply less gravitational mass than equivalent
isothermal profiles, which leads to an increase in the generic cluster
baryon-fraction estimate, and a further decrease in the implied value
of $\Omega_0$.

In addition, \MFSV\ found that the average gradient is close to, and
consistent with (at 2-$\sigma$), the convective instability boundary
at $\Gamma\ge5/3$. While such instabilities could result from merger
activity, the ubiquitous nature of temperature declines implies that
most of the clusters in their sample are disturbed. However, \MFSV\
also found that approximately 60 percent of their sample exhibited
evidence for cooling flow activity. As cooling flows are found in
relaxed clusters [\citeN{Buote:morphology_ii}; \eg\ see Fig.~7 where
they plot the mass-flow rate against the ratio of the quadrapole to
monopole moments of the \ROSAT\ images of clusters, and
\citeN{Buote:Omega_substructure} where they quantify the relationship
between the ratio of these moments and the dynamical state of a
cluster], these two observations are contradictory.

Further doubts have been raised by \citeN{Irwin:ROSAT_Tx_profiles}
following their comparison of \MFSV\ results with the work from other
authors (\ie\ \citeNP{Ikebe:ASCA_Hydra-A}: A780;
\citeNP{Fujita:ASCA_A399andA401}: A399 and A401;
\citeNP{Ezawa:ASCA_AWM7}: AWM7; \citeNP{Ohashi:ASCA_four_clusters}:
3A0336+098, MKW3s, A1795 and PKS2354$-$35).  Most of these other
studies find isothermal temperature profiles, even for those where
\MFSV\ find a temperature
decline. \shortciteANP{Irwin:ROSAT_Tx_profiles} also presented their
own investigation of \ROSAT\ colour profiles of many of the clusters
in the \MFSV\ sample, and they found that these were generally
consisent with isothermal temperature profiles (excluding the core
regions of cooling flows), even in clusters common to the
\MFSV\ analysis. 

Recently, temperature profiles determinations have been derived from
Beppo-SAX data (A2319 -- \citeNP{Molendi:SAX_A2319}; Virgo --
\citeNP{DAcri:SAX_Virgo}; \citeNP{Molendi:SAX_A426}). 
Although the data from this satellite also require correction for a
broad (half-power radius of $1-2\arcmin$, depending on the detector)
spatial point-spread function (PSF), the PSF does not vary strongly
with energy (thus even without correction the SAX data could, in
principle, be anlysed to correctly determine whether a cluster is
isothermal or not). Of the SAX results listed above, the Virgo and
Perseus cluster data only cover the cooling flow regions, while A2319
is not a cooling flow and appears isothermal in the SAX data. However,
the authors are reluctant to claim any disparity with the Markevitch
\etal\ results on the basis of this one cluster observation.

Given the above results and the issues discussed, it is clear that the
\MFSV\ \ASCA\ results need to be checked by independent means. Before
we discuss our analysis we shall describe the method used by \MFSV,
and compare the relative advantages and disadvantages.

The procedure used by \MFSV\ requires an initial assumption for the
cluster spectrum and the spatial emissivity profile. By convolving
these with the energy- and position-dependent PSF they produce a
spectral-image model for the cluster which they compare with the
observational data. After modifying the source spectrum at different
positions they attempt converge to consistency between the model and
data. The disadvantages of this method are that they need to assume a
source spectum and that they require a constraint on the spatial
distribution of counts, \ie\ the emissivity profile. The latter they
obtain from \ROSAT\ data, however the energy range of X-rays detected
by \ROSAT\ is much softer ($0.2-2\keV$) than that used in the analysis
of \ASCA\ datasets (\MFSV\ use $\sim2-10\keV$ photons), and so the
question arises as to whether the emissivity profile of the cluster in
the \ROSAT\ energy band is suitable for use as a constraint in the
\ASCA\ data analysis. For clusters where the spectrum changes
significantly between these two energy bands, \ie\ in clusters which
have a strong cooling flow, excess absorption in the core region
\cite{White:SSS_abs}, or even a strong temperature decline, this
assumption is questionable.

In the following sections we present the results from our own analysis
of \ASCA\ GIS data, using the spectral-imaging deconvolution method
which we described in Paper-I (\citeANP{White:deconv_i}). This method
is essentially non-parametric as it does not require the spectrum of
the object to be specified, nor external (\ROSAT) data to describe the
emissivity profile. The main assumption of our method, namely that a
fixed spatially-invariant PSF may be used for the image deconvolution
has been tested for in Paper-I and shown to have a negligible impact
on the results. Our other tests in that paper also show that the
method can be used on observational data contaminated by events from
the cosmic and instrumental background. We are confident that our
procedure can reliably extract the intrinsic radial properties of the
ICM for a large number of clusters, and so we have applied it to GIS
data on \nsample\ galaxy clusters observed by \ASCA. Our methodolgy
yields temperature and abundance profiles (\ie\ at least 2 annular
bins) for 92 percent (98 clusters) of the sample.


\section{Sample}

In compiling this sample of \nsample\ clusters we have attempted to
obtain temperature profiles for as many clusters as possible. Although
this sample has not been selected according a flux limit we have
deliberately tried to include the 50 brightest clusters
\cite{Edge:EXOSAT_Lx_evolution} (only a couple are missing), and all
those clusters which were analysed by \MFSV.

All our data are pipeline-processed observations which have been
obtained from the HEASARC database\footnote{{\tt
http://heasarc.gsfc.nasa.gov/W3Browse/}} at the Goddard Space Flight
Centre. Some of the GIS3 data (\ie\ A478, A586, A2029, A2142, A2063,
and 2A0336) were affected by problems with the analogue-to-digital
converter and we have eliminated the GIS3 results (as results from
these data are generally discrepant with the GIS2 results at large
radii). Any contaminating sources which were evident in the original
GIS images were masked-out before the deconvolution analysis\footnote{
Cluster requiring source contamination removal were: A85, A370, A400,
A665, A697, A854, A1413, A1763, A1775, A1795, A1835, A1895, A1995,
A2034, A2063, A2029, A2104, A2142, A2219, A2440, A2634, A2811, A3391,
A3558, A3562, AWM4, Klemola~44, MKW4s MKW9, and SC1327 (Shapley).}. As
we analyse our spectra in circularly symmetric annuli we have
implicitly assumed spherical symmetry. However, some of the clusters
in our sample are clearly assymetric\footnote{Clusters with some
notable asymmetry in the GIS images: A539 (two core sources), A2151
(possible subtructure or contaminating source on one side of the
cluster at larger radii), A2256 (subcluster merger apparent through
elongated core), A2319 (offset of core with respect to the outer
regions), A2440 (possible multiple core sources), A3266 (slightly
offset core), A3376 (elongated), A3627 (elongated), Ophuichus (core
offset) and Cygnus-A (AGN point source and offset core with respect to
outer regions of cluster).}, even within the limited spatial
resolution of \ASCA. This should be bourne in mind when interpreting
the results from these clusters.

The range of background-subtracted fluxes in our sample is
approximately 1,000 to several 100,000 counts, and count rates of 0.04
to $15\ctsps$. The detailed numbers (from the non-deconvolved data
within $20\arcmin$ of the centre of the field, or the maximum radius
above that background) are given in Table~\ref{table:obsdat}, together
with details on the observation sequence number used and the exposure,
after cleaning the event list, of the observation.

\renewcommand{\baselinestretch}{1.5}
\begin{table}
\caption{\label{table:obsdat}Observational Dataset Statistics}
\begin{tabular}{lllcrrr} \hline \\
\multicolumn{1}{l}{\#} &
\multicolumn{1}{l}{Object} &
\multicolumn{1}{l}{Sequence} &
\multicolumn{1}{c}{Inst.} &
\multicolumn{1}{r}{Exposure} &
\multicolumn{1}{r}{Cluster} &
\multicolumn{1}{r}{Count} \\ 
\multicolumn{1}{l}{} &
\multicolumn{1}{l}{} &
\multicolumn{1}{l}{Number} &
\multicolumn{1}{c}{} &
\multicolumn{1}{r}{(k.sec.)} &
\multicolumn{1}{r}{Counts} &
\multicolumn{1}{r}{Rate} \\ \\ \hline
1 & A85 & 81024000  & G3 & 27.89 & 54,641 & 1.959 \\ 
  &   &   & G2 & 27.89 & 55,400 & 1.987 \\ 
  &   & 81024010  & G2 & 13.35 & 24,108 & 1.806 \\ 
  &   &   & G3 & 13.35 & 27,161 & 2.034 \\ 
2 & A119 & 83045000  & G2 & 35.42 & 26,367 & 0.744 \\ 
  &   &   & G3 & 35.41 & 26,902 & 0.760 \\ 
3 & A262 & 81031000  & G2 & 21.51 & 18,747 & 0.871 \\ 
  &   &   & G3 & 21.51 & 21,887 & 1.017 \\ 
4 & A370 & 80010000  & G3 & 35.60 & 1,924 & 0.054 \\ 
  &   &   & G2 & 35.61 & 1,698 & 0.048 \\ 
5 & A399 & 82008000  & G2 & 29.76 & 24,422 & 0.821 \\ 
  &   &   & G3 & 29.76 & 24,396 & 0.820 \\ 
6 & A400 & 83037000  & G2 & 51.43 & 16,897 & 0.329 \\ 
  &   &   & G3 & 51.40 & 19,288 & 0.375 \\ 
7 & A401 & 82010000  & G2 & 32.87 & 48,468 & 1.475 \\ 
  &   &   & G3 & 32.87 & 49,866 & 1.517 \\ 
8 & A426 & 80007000  & G2 & 16.81 & 245,322 & 14.598 \\ 
  &   &   & G3 & 16.80 & 281,366 & 16.744 \\ 
9 & A478 & 81015000  & G2 & 34.39 & 50,480 & 1.468 \\ 
10 & A496 & 80003000  & G2 & 40.05 & 66,827 & 1.669 \\ 
  &   &   & G3 & 39.99 & 75,115 & 1.879 \\ 
11 & A521 & 84071000  & G2 & 42.71 & 4,800 & 0.112 \\ 
  &   &   & G3 & 42.78 & 5,685 & 0.133 \\ 
12 & A539 & 83003000  & G2 & 32.68 & 16,250 & 0.497 \\ 
  &   &   & G3 & 32.68 & 18,326 & 0.561 \\ 
13 & A548 & 84034000  & G2 & 28.40 & 7,884 & 0.278 \\ 
  &   &   & G3 & 28.36 & 8,212 & 0.290 \\ 
14 & A576 & 84001000  & G2 & 48.56 & 25,222 & 0.519 \\ 
  &   &   & G3 & 48.53 & 29,476 & 0.607 \\ 
15 & A586 & 81009010  & G2 & 18.33 & 3,470 & 0.189 \\ 
16 & A611 & 84063000  & G2 & 56.98 & 4,340 & 0.076 \\ 
  &   &   & G3 & 56.98 & 4,493 & 0.079 \\ 
17 & A644 & 83022000  & G2 & 57.50 & 60,991 & 1.061 \\ 
  &   &   & G3 & 57.50 & 70,044 & 1.218 \\ 
18 & A665 & 80035000  & G3 & 40.03 & 9,835 & 0.246 \\ 
  &   &   & G2 & 40.05 & 10,473 & 0.261 \\ 
19 & A697 & 84031000  & G2 & 65.86 & 8,183 & 0.124 \\ 
  &   &   & G3 & 65.85 & 10,731 & 0.163 \\ 
20 & A754 & 82057000  & G2 & 21.60 & 40,426 & 1.872 \\ 
  &   &   & G3 & 21.77 & 46,503 & 2.136 \\ 
21 & A773 & 82001000  & G2 & 41.10 & 5,055 & 0.123 \\ 
  &   &   & G3 & 41.09 & 6,552 & 0.159 \\ 
22 & A780 & 80015000  & G3 & 18.64 & 20,003 & 1.073 \\ 
  &   &   & G2 & 18.64 & 20,646 & 1.107 \\ 
23 & A854 & 83006000  & G2 & 45.99 & 3,571 & 0.078 \\ 
  &   &   & G3 & 45.99 & 3,744 & 0.081 \\ 
24 & A963 & 80000000  & G2 & 30.31 & 4,553 & 0.150 \\ 
  &   &   & G3 & 30.31 & 5,537 & 0.183 \\ 
25 & A990 & 84070000  & G3 & 74.07 & 14,236 & 0.192 \\ 
  &   &   & G2 & 74.33 & 12,493 & 0.168 \\ 
26 & A1060 & 80004000  & G2 & 34.72 & 50,897 & 1.466 \\ 
  &   &   & G3 & 34.70 & 50,315 & 1.450 \\ 
\hline
\end{tabular}
\end{table}
\begin{table}
\parbox{0.5\textwidth}{\vspace{0.6\baselineskip}{\bf Table \ref{table:obsdat}.} Observational Dataset Statistics -- contd.}\vspace{0.6\baselineskip}
\begin{tabular}{lllcrrr} \hline \\
\multicolumn{1}{l}{\#} &
\multicolumn{1}{l}{Object} &
\multicolumn{1}{l}{Sequence} &
\multicolumn{1}{c}{Inst.} &
\multicolumn{1}{r}{Exposure} &
\multicolumn{1}{r}{Cluster} &
\multicolumn{1}{r}{Count} \\ 
\multicolumn{1}{l}{} &
\multicolumn{1}{l}{} &
\multicolumn{1}{l}{Number} &
\multicolumn{1}{c}{} &
\multicolumn{1}{r}{(k.sec.)} &
\multicolumn{1}{r}{Counts} &
\multicolumn{1}{r}{Rate} \\ \\ \hline
27 & A1068 & 84064000  & G3 & 35.63 & 9,234 & 0.259 \\ 
  &   &   & G2 & 35.64 & 7,651 & 0.215 \\ 
28 & A1204 & 82002000  & G2 & 32.37 & 3,502 & 0.108 \\ 
  &   &   & G3 & 32.36 & 4,696 & 0.145 \\ 
29 & A1246 & 83007000  & G2 & 42.19 & 4,469 & 0.106 \\ 
  &   &   & G3 & 41.67 & 5,185 & 0.124 \\ 
30 & A1367 & 81029000  & G2 & 10.62 & 8,863 & 0.834 \\ 
  &   &   & G3 & 10.62 & 10,066 & 0.947 \\ 
  &   & 81029010  & G3 & 8.61 & 8,250 & 0.959 \\ 
  &   &   & G2 & 8.61 & 7,527 & 0.875 \\ 
  &   & 81030000  & G2 & 7.94 & 7,184 & 0.905 \\ 
  &   &   & G3 & 7.94 & 7,693 & 0.969 \\ 
  &   & 81030010  & G2 & 10.75 & 9,848 & 0.916 \\ 
  &   &   & G3 & 10.75 & 9,857 & 0.917 \\ 
31 & A1413 & 81008000  & G3 & 37.51 & 15,388 & 0.410 \\ 
  &   &   & G2 & 37.51 & 16,237 & 0.433 \\ 
32 & A1553 & 84069000  & G2 & 34.62 & 9,220 & 0.266 \\ 
  &   &   & G3 & 34.63 & 10,822 & 0.313 \\ 
33 & A1576 & 83014000  & G2 & 39.65 & 2,504 & 0.063 \\ 
  &   &   & G3 & 39.65 & 3,141 & 0.079 \\ 
34 & A1650 & 84021000  & G2 & 55.08 & 31,696 & 0.575 \\ 
  &   &   & G3 & 55.13 & 37,382 & 0.678 \\ 
35 & A1651 & 82036000  & G2 & 35.59 & 20,878 & 0.587 \\ 
  &   &   & G3 & 35.58 & 25,782 & 0.725 \\ 
36 & A1656 & 80016000  & G2 & 7.75 & 44,653 & 5.760 \\ 
  &   &   & G3 & 7.75 & 45,159 & 5.825 \\ 
37 & A1682 & 84075000  & G2 & 39.32 & 3,946 & 0.100 \\ 
  &   &   & G3 & 39.34 & 4,806 & 0.122 \\ 
38 & A1689 & 80005000  & G2 & 38.15 & 20,542 & 0.538 \\ 
  &   &   & G3 & 38.15 & 21,708 & 0.569 \\ 
39 & A1704 & 81007000  & G2 & 19.60 & 1,626 & 0.083 \\ 
  &   &   & G3 & 19.60 & 2,058 & 0.105 \\ 
40 & A1736 & 83061000  & G3 & 17.38 & 8,980 & 0.517 \\ 
  &   &   & G2 & 17.39 & 7,058 & 0.406 \\ 
41 & A1763 & 83044000  & G2 & 39.44 & 6,374 & 0.162 \\ 
  &   &   & G3 & 39.38 & 7,278 & 0.185 \\ 
42 & A1772 & 81013000  & G3 & 42.55 & 2,723 & 0.064 \\ 
  &   &   & G2 & 42.56 & 2,400 & 0.056 \\ 
43 & A1774 & 83049000  & G2 & 32.92 & 5,020 & 0.152 \\ 
  &   &   & G3 & 32.91 & 5,586 & 0.170 \\ 
44 & A1775 & 85056000  & G2 & 32.78 & 9,963 & 0.304 \\ 
  &   &   & G3 & 32.77 & 9,446 & 0.288 \\ 
45 & A1795 & 80006000  & G3 & 34.77 & 59,068 & 1.699 \\ 
  &   &   & G2 & 34.77 & 58,694 & 1.688 \\ 
46 & A1835 & 82052000  & G3 & 17.84 & 6,839 & 0.383 \\ 
  &   &   & G2 & 17.84 & 5,753 & 0.322 \\ 
  &   & 82052010  & G3 & 15.68 & 6,070 & 0.387 \\ 
  &   &   & G2 & 15.69 & 5,486 & 0.350 \\ 
47 & A1851 & 82007000  & G3 & 36.21 & 1,464 & 0.040 \\ 
  &   &   & G2 & 36.15 & 1,564 & 0.043 \\ 
48 & A1895 & 83033000  & G2 & 24.34 & 1,333 & 0.055 \\ 
  &   &   & G3 & 24.34 & 1,762 & 0.072 \\ 
\hline
\end{tabular}
\end{table}
\begin{table}
\parbox{0.5\textwidth}{\vspace{0.6\baselineskip}{\bf Table \ref{table:obsdat}.} Observational Dataset Statistics -- contd.}\vspace{0.6\baselineskip}
\begin{tabular}{lllcrrr} \hline \\
\multicolumn{1}{l}{\#} &
\multicolumn{1}{l}{Object} &
\multicolumn{1}{l}{Sequence} &
\multicolumn{1}{c}{Inst.} &
\multicolumn{1}{r}{Exposure} &
\multicolumn{1}{r}{Cluster} &
\multicolumn{1}{r}{Count} \\ 
\multicolumn{1}{l}{} &
\multicolumn{1}{l}{} &
\multicolumn{1}{l}{Number} &
\multicolumn{1}{c}{} &
\multicolumn{1}{r}{(k.sec.)} &
\multicolumn{1}{r}{Counts} &
\multicolumn{1}{r}{Rate} \\ \\ \hline
49 & A1914 & 84032000  & G2 & 35.86 & 12,676 & 0.353 \\ 
  &   &   & G3 & 35.85 & 15,781 & 0.440 \\ 
50 & A1942 & 83000000  & G2 & 34.32 & 2,695 & 0.079 \\ 
  &   &   & G3 & 34.31 & 3,186 & 0.093 \\ 
51 & A1995 & 82005000  & G3 & 28.31 & 2,041 & 0.072 \\ 
  &   &   & G2 & 28.31 & 1,617 & 0.057 \\ 
52 & A2029 & 81023000  & G2 & 34.06 & 62,198 & 1.826 \\ 
53 & A2034 & 84022000  & G3 & 41.70 & 16,517 & 0.396 \\ 
  &   &   & G2 & 41.70 & 14,105 & 0.338 \\ 
54 & A2052 & 85061000  & G3 & 42.89 & 40,880 & 0.953 \\ 
  &   &   & G2 & 42.90 & 35,308 & 0.823 \\ 
55 & A2063 & 81002000  & G2 & 21.63 & 17,975 & 0.831 \\ 
56 & A2065 & 84054000  & G3 & 24.22 & 16,757 & 0.692 \\ 
  &   &   & G2 & 24.29 & 18,372 & 0.756 \\ 
  &   & 84054010  & G2 & 22.78 & 15,222 & 0.668 \\ 
  &   &   & G3 & 22.75 & 17,031 & 0.748 \\ 
57 & A2104 & 84072000  & G2 & 52.79 & 11,151 & 0.211 \\ 
  &   &   & G3 & 52.78 & 13,399 & 0.254 \\ 
58 & A2107 & 85060000  & G3 & 28.01 & 13,766 & 0.492 \\ 
  &   &   & G2 & 28.01 & 13,568 & 0.484 \\ 
59 & A2142 & 81004000  & G2 & 14.64 & 28,341 & 1.935 \\ 
60 & A2147 & 83074000  & G2 & 37.09 & 34,801 & 0.938 \\ 
  &   &   & G3 & 36.94 & 34,953 & 0.946 \\ 
61 & A2151 & 83030000  & G3 & 28.17 & 10,499 & 0.373 \\ 
  &   &   & G2 & 28.17 & 8,726 & 0.310 \\ 
62 & A2163 & 80024000  & G2 & 31.60 & 18,783 & 0.594 \\ 
  &   &   & G3 & 31.15 & 18,324 & 0.588 \\ 
63 & A2199 & 80023000  & G2 & 33.44 & 76,799 & 2.296 \\ 
  &   &   & G3 & 33.44 & 76,282 & 2.281 \\ 
64 & A2204 & 82045000  & G2 & 16.05 & 11,723 & 0.731 \\ 
  &   &   & G3 & 16.05 & 14,290 & 0.891 \\ 
  &   & 82045010  & G2 & 20.07 & 13,280 & 0.662 \\ 
  &   &   & G3 & 20.07 & 16,741 & 0.834 \\ 
65 & A2218 & 80001000  & G2 & 38.08 & 8,451 & 0.222 \\ 
  &   &   & G3 & 38.07 & 8,820 & 0.232 \\ 
66 & A2219 & 82037000  & G3 & 35.76 & 12,926 & 0.361 \\ 
  &   &   & G2 & 35.76 & 10,303 & 0.288 \\ 
67 & A2255 & 84012000  & G3 & 47.08 & 23,912 & 0.508 \\ 
  &   &   & G2 & 47.10 & 22,826 & 0.485 \\ 
  &   & 84012010  & G2 & 39.23 & 17,084 & 0.435 \\ 
  &   &   & G3 & 39.44 & 19,711 & 0.500 \\ 
68 & A2256 & 10004030  & G3 & 26.53 & 43,010 & 1.621 \\ 
  &   &   & G2 & 26.53 & 40,018 & 1.508 \\ 
  &   & 80002000  & G3 & 36.44 & 55,265 & 1.517 \\ 
  &   &   & G2 & 36.45 & 59,364 & 1.629 \\ 
69 & A2261 & 84062000  & G3 & 20.22 & 5,358 & 0.265 \\ 
  &   &   & G2 & 20.25 & 5,589 & 0.276 \\ 
70 & A2319 & 80041000  & G2 & 14.57 & 36,891 & 2.532 \\ 
  &   &   & G3 & 14.56 & 44,361 & 3.046 \\ 
  &   & 80041010  & G2 & 13.96 & 38,099 & 2.729 \\ 
  &   &   & G3 & 13.96 & 35,475 & 2.541 \\ 
\hline
\end{tabular}
\end{table}
\begin{table}
\parbox{0.5\textwidth}{\vspace{0.6\baselineskip}{\bf Table \ref{table:obsdat}.} Observational Dataset Statistics -- contd.}\vspace{0.6\baselineskip}
\begin{tabular}{lllcrrr} \hline \\
\multicolumn{1}{l}{\#} &
\multicolumn{1}{l}{Object} &
\multicolumn{1}{l}{Sequence} &
\multicolumn{1}{c}{Inst.} &
\multicolumn{1}{r}{Exposure} &
\multicolumn{1}{r}{Cluster} &
\multicolumn{1}{r}{Count} \\ 
\multicolumn{1}{l}{} &
\multicolumn{1}{l}{} &
\multicolumn{1}{l}{Number} &
\multicolumn{1}{c}{} &
\multicolumn{1}{r}{(k.sec.)} &
\multicolumn{1}{r}{Counts} &
\multicolumn{1}{r}{Rate} \\ \\ \hline
71 & A2390 & 82032000  & G3 & 8.81 & 3,898 & 0.443 \\ 
  &   &   & G2 & 8.81 & 3,210 & 0.364 \\ 
  &   & 82032010  & G3 & 7.80 & 2,998 & 0.384 \\ 
  &   &   & G2 & 7.80 & 2,478 & 0.318 \\ 
  &   & 82032020  & G3 & 5.21 & 1,956 & 0.376 \\ 
  &   &   & G2 & 5.21 & 1,665 & 0.320 \\ 
72 & A2440 & 81033000  & G3 & 40.00 & 10,324 & 0.258 \\ 
  &   &   & G2 & 40.01 & 9,859 & 0.246 \\ 
73 & A2597 & 83062000  & G2 & 42.18 & 22,577 & 0.535 \\ 
  &   &   & G3 & 42.16 & 22,831 & 0.541 \\ 
74 & A2634 & 83002000  & G2 & 41.23 & 16,626 & 0.403 \\ 
  &   &   & G3 & 41.23 & 17,192 & 0.417 \\ 
75 & A2657 & 84002000  & G2 & 48.49 & 24,886 & 0.513 \\ 
  &   &   & G3 & 48.43 & 29,192 & 0.603 \\ 
76 & A2670 & 82049000  & G2 & 29.16 & 6,527 & 0.224 \\ 
  &   &   & G3 & 29.15 & 7,264 & 0.249 \\ 
77 & A2811 & 84003000  & G2 & 49.84 & 14,200 & 0.285 \\ 
  &   &   & G3 & 49.82 & 15,262 & 0.306 \\ 
78 & A3112 & 81003000  & G3 & 36.62 & 29,570 & 0.808 \\ 
  &   &   & G2 & 36.62 & 26,438 & 0.722 \\ 
79 & A3158 & 84020000  & G2 & 35.82 & 27,696 & 0.773 \\ 
  &   &   & G3 & 35.83 & 33,619 & 0.938 \\ 
80 & A3221 & 83048000  & G2 & 25.01 & 9,615 & 0.384 \\ 
  &   &   & G3 & 25.02 & 11,067 & 0.442 \\ 
81 & A3266 & 83023000  & G2 & 33.51 & 48,161 & 1.437 \\ 
  &   &   & G3 & 33.50 & 50,999 & 1.522 \\ 
82 & A3376 & 84056000  & G3 & 22.42 & 10,403 & 0.464 \\ 
  &   &   & G2 & 22.40 & 11,271 & 0.503 \\ 
83 & A3391 & 72019000  & G2 & 21.18 & 11,591 & 0.547 \\ 
  &   &   & G3 & 21.18 & 12,932 & 0.610 \\ 
84 & A3526 & 80032000  & G2 & 19.36 & 61,758 & 3.190 \\ 
  &   &   & G3 & 19.36 & 61,773 & 3.191 \\ 
  &   & 83026000  & G2 & 72.31 & 209,227 & 2.893 \\ 
  &   &   & G3 & 72.31 & 241,614 & 3.341 \\ 
85 & A3558 & 82046000  & G3 & 15.53 & 26,330 & 1.695 \\ 
  &   &   & G2 & 15.54 & 26,917 & 1.732 \\ 
  &   & 83058000  & G2 & 29.22 & 36,039 & 1.233 \\ 
  &   &   & G3 & 29.22 & 40,019 & 1.370 \\ 
86 & A3562 & 84041000  & G2 & 20.04 & 12,291 & 0.613 \\ 
  &   &   & G3 & 20.03 & 13,202 & 0.659 \\ 
87 & A3571 & 82047000  & G2 & 24.33 & 74,343 & 3.056 \\ 
  &   &   & G3 & 24.33 & 75,252 & 3.093 \\ 
88 & A3627 & 84005000  & G2 & 41.47 & 83,424 & 2.011 \\ 
  &   &   & G3 & 41.44 & 93,158 & 2.248 \\ 
89 & A3667 & 83054000  & G2 & 19.68 & 28,130 & 1.429 \\ 
  &   &   & G3 & 19.68 & 31,778 & 1.615 \\ 
90 & A3921 & 83048010  & G3 & 19.51 & 8,714 & 0.447 \\ 
  &   &   & G2 & 19.49 & 7,730 & 0.397 \\ 
91 & A4059 & 82030000  & G3 & 35.77 & 28,134 & 0.786 \\ 
  &   &   & G2 & 35.77 & 27,977 & 0.782 \\ 
92 & 2A0336 & 82029000  & G2 & 20.07 & 40,550 & 2.021 \\ 
  &   &   & G3 & 20.06 & 42,397 & 2.113 \\ 
\hline
\end{tabular}
\end{table}
\begin{table}
\parbox{0.5\textwidth}{\vspace{0.6\baselineskip}{\bf Table \ref{table:obsdat}.} Observational Dataset Statistics -- contd.}\vspace{0.6\baselineskip}
\begin{tabular}{lllcrrr} \hline \\
\multicolumn{1}{l}{\#} &
\multicolumn{1}{l}{Object} &
\multicolumn{1}{l}{Sequence} &
\multicolumn{1}{c}{Inst.} &
\multicolumn{1}{r}{Exposure} &
\multicolumn{1}{r}{Cluster} &
\multicolumn{1}{r}{Count} \\ 
\multicolumn{1}{l}{} &
\multicolumn{1}{l}{} &
\multicolumn{1}{l}{Number} &
\multicolumn{1}{c}{} &
\multicolumn{1}{r}{(k.sec.)} &
\multicolumn{1}{r}{Counts} &
\multicolumn{1}{r}{Rate} \\ \\ \hline
  &   & 82040000  & G2 & 46.16 & 66,679 & 1.445 \\ 
93 & AWM4 & 83072000  & G2 & 51.42 & 10,188 & 0.198 \\ 
  &   &   & G3 & 51.12 & 12,491 & 0.244 \\ 
94 & AWM7 & 80036000  & G2 & 14.95 & 38,107 & 2.549 \\ 
  &   &   & G3 & 14.95 & 37,574 & 2.513 \\ 
95 & CL0016 & 80025000  & G3 & 32.09 & 1,827 & 0.057 \\ 
  &   &   & G2 & 32.29 & 1,816 & 0.056 \\ 
  &   & 84016000  & G3 & 58.36 & 3,189 & 0.055 \\ 
  &   &   & G2 & 58.91 & 3,155 & 0.054 \\ 
96 & Cygnus-A & 70003000  & G3 & 24.24 & 51,157 & 2.110 \\ 
  &   &   & G2 & 24.26 & 44,618 & 1.839 \\ 
  &   & 70003010  & G3 & 36.53 & 64,741 & 1.772 \\ 
  &   &   & G2 & 36.53 & 74,347 & 2.035 \\ 
97 & Klemola44 & 83004000  & G2 & 55.31 & 45,646 & 0.825 \\ 
  &   &   & G3 & 55.16 & 53,551 & 0.971 \\ 
98 & MKW3s & 80011000  & G3 & 29.67 & 24,396 & 0.822 \\ 
  &   &   & G2 & 29.69 & 24,279 & 0.818 \\ 
99 & MKW4s & 83008000  & G2 & 48.32 & 4,704 & 0.097 \\ 
  &   &   & G3 & 48.33 & 5,602 & 0.116 \\ 
100 & MKW9 & 83009000  & G3 & 46.64 & 3,988 & 0.086 \\ 
  &   &   & G2 & 46.66 & 3,753 & 0.080 \\ 
101 & Ophiuchus & 80027000  & G2 & 8.75 & 76,982 & 8.798 \\ 
  &   &   & G3 & 8.75 & 79,839 & 9.124 \\ 
102 & PKS0745-19 & 81016000  & G2 & 40.67 & 48,774 & 1.199 \\ 
  &   &   & G3 & 40.67 & 54,352 & 1.336 \\ 
103 & SC1327 & 83059000  & G2 & 30.03 & 14,735 & 0.491 \\ 
  &   &   & G3 & 29.87 & 15,792 & 0.529 \\ 
104 & Tri.Aus. & 83060000  & G2 & 11.85 & 37,459 & 3.162 \\ 
  &   &   & G3 & 11.85 & 38,543 & 3.253 \\ 
  &   & 83060010  & G2 & 6.95 & 15,877 & 2.283 \\ 
  &   &   & G3 & 6.95 & 18,521 & 2.666 \\ 
105 & Virgo & 60033000  & G2 & 16.87 & 134,917 & 7.995 \\ 
  &   &   & G3 & 16.87 & 137,748 & 8.163 \\ 
106 & Zw3146 & 80014000  & G3 & 33.66 & 8,233 & 0.245 \\ 
  &   &   & G2 & 33.65 & 7,740 & 0.230 \\ 
\hline
\end{tabular}
\end{table}


\section{Spectral-Imaging Analysis}\label{section:spectral}

We apply our spectral-imaging deconvolution process to the observed
GIS events between $1-9\keV$ in energy [a limit imposed by the PSF
images stored in the calibration databse (CALDB) at Goddard Space
Flight Centre (GSFC)] using exactly the same methodology applied to
the simualated datasets presented in Paper-I (\ie\ we run the \SID\
procdure on 10 randomisations of the observed events and then average
the spectral-fit results obtained from all these deconvolved
datasets). We also employ more conservative parameters (see details
below) than applied to the simulated data in Paper-I, to yield fewer
annular bins. This ensures that the results from the real data are
less affected by systematic effects, albeit with some loss of spatial
resolution.

As we are attempting to determine the spectral properties of regions
of low surface-brightness it is important that we account for the
contamination of the cluster data by instrumental and cosmic
X-rays. This is done using the blanksky observations, contained in the
CALDB at GSFC. The differences between the Galactic column density
applicable for the blanksky field and each cluster are unlikely to
cause significant problems because the lowest energy used in the
analysis is $1\keV$. (Some possible exceptions to this may occur in
very high Galactic column density clusters, where we sometimes see
complications in fitting cooling flow models, as discussed in
Section~\ref{section:spectral}). As we noted in Paper-I, we have
combined the blanksky observations, which are divided according to
COR\_MIN values, so that the resulting background datafile has the
same distribution of COR\_MIN values as that of the cluster
observation.

One important issue we had to resolve in the spectral analysis was a
way to determine an automatic and objective way of selecting the sizes
of the annular regions for the X-ray event extractions. In optimal
circumstances one would choose a bin size which yields a constant
signal-to-noise in each radial bin. However, the Maximum-Likelihood
image deconvolution introduces systematic variations in regions where
the surface-brightness changes relatively slowly and has less flux
(\ie\ in the outer regions of cooling flow clusters or non
cooling-flow clusters). Using fixed bin-widths of size sufficient to
smooth over these these systematic variations then results in a loss
of spatial resolution for more distant clusters. Consequently, we have
chosen a procedure which yields approximately the same number of bins
in each cluster, provided the cluster is sufficiently bright, by
requiring a fixed ratio of the total background-subtracted flux to be
present in each annulus. Our chosen ratio of \nfrac\ leads to a
maximum of approximately five or six radial bins. If this prescription
results in less than 2,000 background-subtracted counts in any
annulus, then the ratio is progressively reduced by a factor of 1.25
until this criterion is met (providing the background-subtracted
number of counts does not fall below a hard limit of 1,000
counts)\footnote{Note, in the simulations tests in Paper-I the nominal
fraction of the total number of background-subtracted counts per
annulus was set to 0.1, the soft-limit on the minimum number of counts
in an annulus was 1,000 and the hard-limit was 500 counts. In this
analysis these parameters are: 0.167, 2,000 and 1,000 counts,
respectively.}. As will be seen, this procedure gives reasonable
spatial resolution for distant clusters, and does not oversample
bright ones. The distribution of the number of annuli for the results
are as follows: 1 bin -- 8 clusters; 2 bins -- 5 clusters; 3 bins -- 8
clusters; 4 bins -- 17 clusters; 5 bins -- 67 clusters; 6 bins -- 1
cluster. Thus, 93 percent of the clusters in the sample have more than
a single annulus in their combined radial profiles, 80 percent have 4
or more, and 64 percent have more than 5 annuli.

Having obtained annular spectra for each cluster, we then fit what we
consider to be the simplest physically-plausible spectral model to
determine the radial variations in the properties of the intracluster
gas: a single thermal emission component (MEKAL: \citeNP{Mewe:MEKALa};
\citeNP{Mewe:MEKALb}), modified by foreground absorption
\cite{Morrison:wabs} according to the Galactic column density
\cite{Stark:nH}. This model is applied to data between
1 and $9\keV$\footnote{The GIS2 data which come from observations
where the GIS3 data were contaminated by the analogue-digital
conversion problem were fitted $2.5-9\keV$. This resulted in better
behaviour for the temperature profiles, especially in A2029. Note, the
GIS3 results for A773 and A854 were also neglected as these gave
implausibly high temperatures in the outer regions.}. In all the
deconvolved-data spectral fits we include 10 percent systematic errors
to account for some uncertainty in the accuracy of the \ASCA\
PSF. (This produces $\chi_\nu^2$ values which are slightly less than
unity compared to those fits without the 10 percent systematics --
which are generally around one.)
 
	\figegs

We also calculate an average broad-beam temperature for each cluster
using the non-deconvolved data, because we do not require any spatial
information. This has the advantage that we can cross-check our
deconvolved results against the `unprocessed' data (see the plots in
Appendix~\ref{section:resfig} which show these average temperatures
overplotted on the deconvolved profiles). As well as the single-phase
model we also fit a model which includes emission from a cooling
flow. This latter model introduces only one additional free parameter,
the mass deposition rate (\ie\ normalisation), because the other
parameters such as the temperature from which the gas cools and the
metallicity are tied to that of the single phase compononent. By
fitting the data from both GIS detectors simultaneously we obtain the
single-phase model parameter constraints, as indicated in
Table~\ref{table:obsresmk}, and cooling flow model parameters, as
listed in \ref{table:obsrescf}. (Note, the 10 percent systematics are
not included on the broad-band fits because these are not derived from
the deconvolved datasets.)

Although a cooling flow spectrum has been fitted to all the data, this
does not mean that it is an applicable model for all
datasets. Section~\label{section:restab} includes a description of the
values in Tables~\ref{table:obsresmk} and \ref{table:obsrescf} which
should be compared to determine whether the additional parameter in
the cooling flow model provides a statistically significant
improvement over the single-phase model. We find that in approximately
half our sample the decrease in $\chi^2_\nu$, due to the addition of
this extra cooling flow component, is significant at more than 90
percent (the significance at 68 percent confidence may be judged from
the errors on the mass-deposition rate in Table~\ref{table:obsrescf}).

In Section~\ref{section:results} we discuss particular examples of the
results from the deconvolution analysis and quantify the overall
isothermality of our cluster sample. Firstly, we discuss individual
graphical examples of the results for five clusters which show the
deconvolved radial temperature, metallicity, and model normalisation
(divided by the region area) profiles for each GIS detector. We also
overplot the broad-beam average results from the non-deconvolved data
which indicate the temperature obtained from the cooling flow spectral
model. Similar graphical presentations, excluding the emissivity
profiles, for the results for the whole sample of \nsample\ clusters
are given in Appendix~\ref{section:resfig}. Broad-beam temperatures
(from original data) are also overplotted for the cooling flow model
-- this has the advantage that it shows the `ambient' cluster
temperature, \ie\ the effect of the cooling flow on the cluster's
temperature is corrected for.

For the clusters where temperature profiles are published by
\citeANP{Markevitch:ASCA_Tx_similarity} (and references within) we
have also overplotted their data (after converting their 90 percent
uncertainties to our 1-$\sigma$ limits) on the temperature
profiles. Some results (including some abundance profiles) from other
satellites such as \ROSAT\ (A478 -- \citeNP{Allen:A478_PSPC}; A3526 --
\citeNP{Allen:Centaurus_PSPC}; A1795 -- \citeNP{Briel:ROSAT_A1795})
and Beppo-SAX (A426 -- \citeNP{Molendi:SAX_A426}; A2319 --
\citeNP{Molendi:SAX_A2319}; Virgo -- \citeNP{DAcri:SAX_Virgo}) are
also plotted.

	\figcmp

\subsubsection{Creating Averaged Radial Profiles From Many Datasets}
\label{section:errimg}

The radial profiles which we present in the various figures, are
created from the averaging of many datasets. These come from: (i) our
10 Monte-Carlo runs which we have employed to average the systematic
variations in single-runs of the \SID\ procedure; (ii) the
observations from the two GIS detectors, and (iii) occasionally
multiple observations of a cluster.

To perform this 2-d averaging, we create a grid which covers the
ranges of values in both dimensions. Then, for example, in an
temperature versus radius plot we take each temperature data point,
and its associated errors, and create a Gaussian distribution to
describe the function in the y-axis direction. Similarly, for the
radial data we create an x-axis function using a triangular function
to represent the emission-weighted radius (an alternative would be to
use a top-hat function but this weights all positions within the
radial extent of the bin equally). Then we combine these x- and
y-functions and construct a 2-d probability distribution (normalised
to unity over its entirety).  This is then placed at the appropriate
position in our 2-d grid. This process is repeated for all data points
to build up an image of the total 2-d probability distribution
function. We then simply sample this grid on any chosen scale to
obtain our averaged radial profile. We chose a sampling which yields
approximately the same binning as the best of the original datasets in
each cluster.

	\figiso


\section{Results}
\label{section:results}

\subsection{Discussion Of Individual Examples}

Fig.~\ref{figure:figegs} shows the radial profiles and broad-beam
averages for the Perseus (\ie\ A426 -- the brightest X-ray cluster)
and Coma Berenices (A1656) clusters, highlighting the difference
between classic examples of cooling and non cooling-flow clusters,
respectively.  The temperature decline due to the cooling flow in A426
is clearly detected, and there is some evidence for an abundance
gradient. In comparison, A1656 presents an isothermal temperature
profile with a flat abundance profile, and has a less sharply peaked
emissivity profile.

Fig.~\ref{figure:figcmp} presents a direct comparison of our results
with those from \ROSAT\ data by \citeN{Allen:Centaurus_PSPC} for the
Centaurus cluster (A3526). Although our \ASCA\ determination of the
temperature profile exhibits a slightly shallower decline in the core
of the Centaurus cluster, and slightly lower overall metallicity when
compared with the \ROSAT\ results, the general consistency between
these results is good.  [We also note that the \ROSAT\ results were
obtained fitting a \citeN{Raymond:plasma} plasma model, while we have
used the MEKAL model used in this analysis. Thus, the differences,
especially in the abundance profiles, will partially be due to the
plamsa codes and the abundance definitions, and also the lack of
excess absorption in our model.] As the \ROSAT\ data do not require
the spectral-imaging deconvolution, the fact the general trends are in
agreement indicates that our method can recover the intrinsic radial
properties of the intracluster gas.  Perhaps the most noticeable
feature is the strong abundance gradient, confirming the \ROSAT\
detection by \citeN{Allen:Centaurus_PSPC} and the early \ASCA\
analysis by \citeN{Fukazawa:ASCA_Centaurus}.

The results for Triangulum Australis and A2065 clusters illustrate
particular examples of the discrepancy between some of our results and
those of \MFSV. It is clear that our profiles appear to be isothermal
while the \MFSV\ results show their characteristc temperature decline.
This is typical for most of the clusters common to both analyses (see
Appendix~\ref{section:resfig}). There are notable examples where our
results show good agreement at all radii, but these tend to be those
clusters where \MFSV\ find relatively flat temperature profiles (\eg\
AWM7). Overall it appears that we find that our temperature profiles
are flatter (see two particular examples in Fig.~\ref{figure:figiso}
and others in Appendix~\ref{section:resfig}); at intermediate radii
the agreement is often good but our temperature determinations are
generally cooler in the core and hotter in the outermost regions. Only
in a handful of clusters (\eg\ A1553, A1689, A1774, A2034, A2218) do
we see significant systematic declines with increasing radius, and
many of these are clusters with poorer data which may suffer from
uncertainties in background subtraction. 

In our final example we plot our results against those of Markevitch
\etal, and those from the recent Beppo-SAX result on A2319 by
\citeANP{Molendi:SAX_A2319} (1999). Our temperature profile is
relatively flat (the slight increase with radius is probably not very
significant given the relatively large uncertainties), and it agrees
fairly well with the SAX data which have much smaller errors. Clearly
the SAX data shows that this cluster has an isothermal temperature
profile. For the Markevitch \etal\ data, the outer point in the
temperature profile is low compared to their inner values, and is
typical of their their temperature decline trend. As an aside, our
\SID\ adundance profile also agrees well with the SAX result.

The most significant apparent departures that we see from
isothermality are the declines in the single-phase temperature within
the core regions of some cooling flow clusters -- see for example the
Figs.~\ref{figure:figegs}(a) and \ref{figure:figcmp} for the Perseus
and Centaurus clusters respectively. Without any extra information we
would be unable to claim that the underlying ambient temperature for
these clusters is isothermal. However, when we fit the cooling flow
emission model to the broad-band spectra (\ie\ data which have not
been deconvolved) we find that the `representative' temperature of the
model increases to a consistent level with the spatially-resoved (\ie\
deconvolved) single-phase temperature found in the outer regions of
the cluster -- see Fig.~\ref{figure:figegs}. In other words, when we
correct for the cooling flow `contamination' we find that the mean
temperature is the same for the whole cluster as it is for the outer
regions.

A similar result is seen in most of the other clusters where a cooling
flow causes a significant drop in the average single-phase temperature
profile, as can be seen in a comparison for of the average
single-phase temperature and the cooling flow corrected temperature in
Tables~\ref{table:obsresmk} and \ref{table:obsrescf}, and the plots in
Appendix~\ref{section:resfig}. (Notable cooling flow exceptions are:
A478, PKS0745 and 2A0336, but in these cases the agreement can be
obtained by allowing excess absorption on the cooling flow component
of the spectrum in the core of the cluster -- which is not
unreasonable given that these clusters have high column densities
which can affect the spectrum above the $1\keV$ lower limit.)

	\figsax

\subsection{Quantifying The Ubiquity Of Cluster Isothermality}
\label{section:isothermality}

From our \SID\ analysis it is visually apparent that the clusters in
our sample are generally consistent with isothermality. Even in
cooling flow clusters where the single-phase temperature exhibits a
decline on the core temperature, a broad-spatial temperature
determination, corrected for the cooling flow effect, shows that the
`ambient' temperature in the core is the same as in the outer regions;
and therefore that these cooling flow clusters are also, essentially,
isothermal. 

We have attempted to quantify the significance of the isothermality in
our sample by fitting each cluster's (having more than a single
annular bin) average temperature profile with a simple power-law
function. The distribution of the best-fit slopes and the significance
of these deviations from a flat profile (zero slope) are shown in
Fig.~\ref{figure:figavg}(a). The cumulative distribution illustrates
that approximately 90 percent of the clusters in our sample are
consistent with having isothermal temperature profiles, within the
3-$\sigma$ confidence limit (and no account has been made here for the
obvious core temperature drop in some of the cooling flow
clusters). Admittedly, the errrors on the temperature profiles are
large, but the sample's slopes are distributed fairly symmetrically
around a value of zero (\ie\ a flat profile).


	\figavg

As we find that our sample is essentially consistent with a population
of isothermal clusters, we are left with the question as to the source
of the discrepancy between our method and that of \MFSV.  As we noted
in Paper-I, their method uses the \ROSAT\ emissivity profile over
$0.2-2\keV$ while their \ASCA\ analysis is over $2-10\keV$. If the
emissivity profile is significantly different between these two energy
bands then this procedure will yield incorrect results. If the \MFSV\
results are biased by such an effect the we would expect hotter
clusters to be more severely affected. We can test to see whether such
systematic effect exists in the \MFSV\ results by plotting the slope
of their temperature profiles against the average
temperature. Fig.~\ref{figure:figmm} shows that their hotter clusters
do have, on average, steeper temperature declines. Although this weak
trend could be physical in nature, it is also possible that it is due
to the assumption used in the \MFSV\ method.

\subsection{Abundance Gradients}
\label{section:abundgrad}

From our two example cooling flow clusters, especially the Centaurus
cluster (A3526), we see that their abundance profiles decline
systematically with radius. (Note, metallicity determinations are less
affected by the scattering problem as the abundances are measured with
respect to the local continuum.) Some other clusters which show a
decline are: A85$^*$, A478$^*$, A496$^*$, A426$^*$, A576, A586,
A990$^*$, A1413, A2029, A2034$^*$, A2052, A2199$^*$, A3526$^*$, A0459,
Cygnus-A$^*$, Klemola 44$^*$, PKS-0745$^*$, and Virgo$^*$. Those
clusters with the superscripted asterix are those where we find that
the cluster has a significant cooling flow, on the basis of the
mass-deposition rate values in Table~\ref{table:obsrescf}. Clearly
most of these are cooling flow clusters.

This list of objects was constructed from a visual inspection of the
plots in Appendix~\ref{section:resfig}. For a more subjective
investigation of the prevalence of abundance gradients in the overall
sample, we have fitted all the metalliticy profiles with a power-law
in the same manner as we fitted the temperature profiles in
Section~\ref{section:isothermality}. Fig.~\ref{figure:figavg}(b) shows
average abundance-profile slope, for those clusters with two or more
annuli, is approximately $-0.2$. The uncertainties on these fits is
such that 90 percent of this sample are consistent with a flat profile
at only $1.5\sigma$, and so this deviation is not statistically
significant. This does not mean that there is not a decline in a
sub-class of clusters, such as cooling flows, as this question is not
addressed by our statisical test. In fact, our impression, from our
quick visual inspection of the radial profiles for cooling flows,
would support the hypothesis that abundance gradients are more
prevalent in cooling flow clusters. However, this clearly requires
further study.


\section{Conclusions}

We have determined the projected radial temperature, metallicity and
emissivity profiles for \nsample\ clusters of galaxies, which have
been obtained using our spectral-imaging deconvolution procedure. We
present plots of the temperature and abundance profiles for all these
clusters (93 percent of the sample have two or more annular
bins). 

We developed this method was developed to correct for the energy and
position dependent point-spread function of the \ASCA\ setellite,
which would otherwise confuse any spatially-resolved spectral analysis
of extended sources.  Our primary goal is to use this method to
extract cluster temperature profiles from \ASCA\ data in order that we
place constraints on the mass properties of a large number of clusters
(Paper-III, in preparation).

In this paper we concentrate on investigating whether we find
the same ubiquitous trend for declining temperature profiles which
\citeN{Markevitch:ASCA_Tx_similarity} find from their analysis of 30
galaxy clusters. If their result is correct, it is of fundamental
importance because temperature profiles have implications for cluster
properties and cosmology. For example, cluster baryon fraction
estimates will increase due to the smaller total mass implied by
declining temperature profiles as compared to isothermality. This
exacerbates the discrepancy between primordial nucleosythesis
constraints and clusters baryon fractions (\eg\
\citeNP{White:baryon_catastrophe};
\citeNP{White:baryon}), and implies smaller values for Cosmological
Density parameter. 

However, we personally have reservations about the ubiquity and
steepness of these temperature declines. The average polytropic index
in their sample is $\gamma=1.24^{+0.20}_{-0.12}$ (\ie\ $\Gamma\sim
4/3$), which is statistcally consistent (at $\sim2$-$\sigma$) with the
convective instability boundary at $\Gamma=5/3$. Given that this near
convective instability could be due to merger events, this would be
inconsistent with the proportion of cooling flows which they find in
their sample (60 percent), because the cooling flow process would be
destroyed in a strong merger event (\ie\ see
\shortciteANP{Buote:morphology_ii} and
\shortciteANP{Buote:Omega_substructure}). 

Further doubts have been raised by
\citeN{Irwin:ROSAT_Tx_profiles}. They have compared the results from
\shortciteANP{Markevitch:ASCA_Tx_similarity} with other analyses (\ie\
\citeNP{Ikebe:ASCA_Hydra-A};
\citeNP{Fujita:ASCA_A399andA401}; \citeNP{Ezawa:ASCA_AWM7};
\citeNP{Ohashi:ASCA_four_clusters}) -- all of which seem to show
isothermal profiles where \shortciteANP{Markevitch:ASCA_Tx_similarity}
find a temperature decline. \shortciteANP{Irwin:ROSAT_Tx_profiles}
also present their own analysis of \ROSAT\ X-ray `colour' profiles
from which they conclude that the clusters are consistent with
isothermality, excluding core regions which may be contaminated by the
effects of cooling flows.

With our \nsample\ clusters, which includes all those in the
\shortciteANP{Markevitch:ASCA_Tx_similarity} sample, we find that
our spectral-imaging deconvolution procedure generally yields
isothermal temperature profiles. In only a handful of clusters do we
find any obvious systematic decline (although the addition of a
cooling flow component improves the spectral fits in approximately 50
percent of the sample). The significance of our result is such that
when we fit power-law functions to each of our cluster temperature
profiles, approximately 90 percent are consistent with isothermality
at the 3-$\sigma$ limit.  For those clusters common to both samples we
find that although there is often reasonable agreement at intermediate
radii, our core temperatures are cooler and our outer temperatures are
hotter. In strong cooling flow clusters, where we detect a drop in the
single-phase temperature of the gas due to the effect of cooling, we
also find that correcting for this effect indicates that the `ambient'
core temperature is consistent with the outer regions of the
cluster. This indicates that even in cooling flows, the underlying
temperature profile is consistent with being isothermal.

	\figmm

Our procedure also allows some constraints to be placed on the
abundance profiles, although the systematic uncertainties are much
larger than in the temperature profiles. There are examples of a clear
metallicity enhancements within the core regions of some clusters,
such as that for the Centaurus cluster -- cofirmining the previous
detections by \citeN{Allen:Centaurus_PSPC}; and
\citeN{Fukazawa:ASCA_Centaurus}]. While this effect is visually most
noticeable in cooling flow clusters, the decline (a power-law slope of
$-0.2$) is not statistically significant (because of the
uncertainties) in the sample as a whole.

In Paper-III \citeN{White:deconv_iii} we present the constraints on
gravtiational masses, baryon fractions and mass-profile model
parameters, which we have been able to obtain for a subsample of the
clusters in this paper.


\section{Acknowledgements}

D.A. White acknowledges support from the P.P.A.R.C., and would like to
thank D.A.~Buote for discussions while developing the spectral-imaging
procedure. D.A. White thanks H.E. Ebeling and S.W. Allen for useful
discussion and R.M. Johnstone for assistance.

This research has made use of: (i) data obtained through the High
Energy Astrophysics Science Archive Research Center Online Service,
provided by the NASA/Goddard Space Flight Center; (ii) the NASA/IPAC
Extragalactic Database (NED) which is operated by the Jet Propulsion
Laboratory, California Institute of Technology, under contract with
the National Aeronautics and Space Administration.


\bibliography{/data/soft3/astrobibv2.1/bibtex/mnrasmnemonic,/data/daw/text/macros/biblio}
\bibliographystyle{/data/soft3/astrobibv2.1/bibtex/mnrasv2}


\appendix

\section{Tabular Results}\label{section:restab}

Tables~\ref{table:obsresmk} and \ref{table:obsrescf} present the
results of broad-band spectral analysis on the (non-deconvolved)
original \ASCA\ data. The GIS2 and GIS3 data in each observation have
been fitted simultaneously over the $1-9\keV$ energy
band. Table~\ref{table:obsresmk} shows the best-fit spectral
parameters resulting from fitting a single thermal-emission component
(XSPEC's MEKAL model), while Table~\ref{table:obsrescf} gives the
results for a cooling flow model (see Section~\ref{section:spectral}
for details). The MEKAL model has three free parameters: temperature,
abundance and normalisation. The cooling flow model only has one more
additional parameter: the mass-deposition rate (\ie\ normalisation).

All parameters uncertainties are quoted as 1-$\sigma$ standard
deviations.  For clusters with more than one observation we have
indicated the weighted-average of the parameters determined from these
multiple observation constraints.

Note, the central regions of Cygnus-A appear to be contaminated by a
strong AGN. Although this has been removed in the temperature profile
plots in Appendix~\ref{section:resfig}, no correction has been applied
to these broad-band fits and so these fits may be affected by the AGN
contribution -- note the exceptionally large temperatures in the
cooling flow model.

The first seven columns headers in each table indicate: `Object' --
cluster name; `Seq. Num.' -- \ASCA\ observation sequence number;
`Inst.' -- instrument used in the fit; `$A_{\rm out}$ $(')$' -- outer
radius of fitted region in arcminutes; $z$ -- cluster redshift;
$R_{\rm out}$ $(\Mpc)$ -- outer metric radius of fitted region (for
$\HO{50}$, $\qO{0.5}$ cosmology); $\nH/10^{20}$ $(\psqcm)$ -- column
density normalised to $10^{20}\psqcm$. For the single thermal
component fits the additional column headers are: $\Tx$ $(\keV)$ --
best fit temperature; $Z$ $(\odot)$ -- metallicity as a fraction of
Solar; $Norm.$ $(10^{-14}/(4\pi D_L^2 \int \ne \np dV))$;
$\chi^2_{\nu}$ -- goodness-of-fit (for 3 free parameters in the
single-phase model and 4 in the cooling flow model); and $N_{\rm PI}$
-- the number of PI channels.

For the cooling flow data we present $\dot{M}$ $(\Msunpyr)$ -- the
normalisation of the additional cooling flow spectral component.  We
have used the F-test to calculate the $\chi^2_{\nu}$ value which would
be required for a 90 percent significant improvements in the the
spectral fit obtained by including the cooling flow component. This
value is given in the final column of Table~\ref{table:obsresmk}, and
should be compared with the value in the penultimate column of
Table~\ref{table:obsrescf}. If the latter value falls below the former
then the cooling flow component provides an improvement in the
spectral fit at more than 90 percent significance.

Table~\ref{table:obsresmk} starts on page~\pageref{table:obsresmk} ;
Table~\ref{table:obsrescf} on page page~\pageref{table:obsrescf}.

\renewcommand{\baselinestretch}{1.5}
\begin{table*}
\caption{\label{table:obsresmk}Broad-Band Single-Phase Plasma Fit Results}
\scriptsize
\begin{center}
\begin{tabular}{llccccccccccc} \hline \\
\multicolumn{1}{l}{Object} &
\multicolumn{1}{l}{Seq.} &
\multicolumn{1}{l}{Inst.} &
\multicolumn{1}{c}{$A_{\rm out}$} &
\multicolumn{1}{c}{$z$} &
\multicolumn{1}{c}{$R_{\rm out}$} &
\multicolumn{1}{c}{$\nH/10^{20}$} &
\multicolumn{1}{c}{$\Tx$} &
\multicolumn{1}{c}{$Z$} &
\multicolumn{1}{c}{$Norm.$} &
\multicolumn{1}{c}{$\chi^2_{\nu}$} &
\multicolumn{1}{c}{$N_{\rm PI}$} &
\multicolumn{1}{c}{$\chi^2_{\nu,{\rm reqd.}}$} \\ 
\multicolumn{1}{l}{} &
\multicolumn{1}{l}{Num.} &
\multicolumn{1}{l}{} &
\multicolumn{1}{c}{$(')$} &
\multicolumn{1}{c}{} &
\multicolumn{1}{c}{$(\Mpc)$} &
\multicolumn{1}{c}{$(\pcmsq)$} &
\multicolumn{1}{c}{$(\keV)$} &
\multicolumn{1}{c}{$(\Solar)$} &
\multicolumn{1}{c}{$(\frac{10^{-14}}{4\pi D_{\rm L}^2} \int n_{\rm e} n_{\rm H} dV)$} &
\multicolumn{1}{c}{} &
\multicolumn{1}{c}{} &
\multicolumn{1}{c}{for CF mod.} \\ \hline \\
A85 & 81024000 & G2G3 & $ 19.4$ & $ 0.0521$ & $ 1.61$ & $ 3.01$
 & $ 5.94^{+ 0.10}_{ -0.07}$
 & $ 0.38^{+ 0.02}_{ -0.02}$
 & $ 0.101^{+ 0.000616}_{-0.000595}$
 & $ 1.0765$
 & $1068$
 & $ 1.0765$ \\
    & 81024010 & G2G3 & $ 19.4$ &   & $ 1.61$ &  
 & $ 5.89^{+ 0.14}_{ -0.11}$
 & $ 0.38^{+ 0.03}_{ -0.03}$
 & $ 0.0992^{+ 0.000878}_{-0.000873}$
 & $ 0.9710$
 & $ 836$
 & $ 0.9709$ \\
   & Combined & & & &  & & $ 5.92\pm  0.11$ & $ 0.38\pm  0.03$ & $ 0.100\pm  0.000740$& \\
A119 & 83045000 & G2G3 & $ 19.4$ & $ 0.0443$ & $ 1.39$ & $ 3.45$
 & $ 5.62^{+ 0.12}_{ -0.12}$
 & $ 0.28^{+ 0.04}_{ -0.03}$
 & $ 0.0484^{+ 0.000470}_{-0.000466}$
 & $ 1.1023$
 & $ 907$
 & $ 1.1022$ \\
\hline
\end{tabular}
\end{center}
\end{table*}
\begin{table*}
\parbox{0.5\textwidth}{\vspace{0.6\baselineskip}{\bf Table \ref{table:obsresmk}.} Broad-Band Single-Phase Plasma Fit Results -- contd.}\vspace{0.6\baselineskip}
\scriptsize
\begin{center}
\begin{tabular}{llccccccccccc} \hline \\
\multicolumn{1}{l}{Object} &
\multicolumn{1}{l}{Seq.} &
\multicolumn{1}{l}{Inst.} &
\multicolumn{1}{c}{$A_{\rm out}$} &
\multicolumn{1}{c}{$z$} &
\multicolumn{1}{c}{$R_{\rm out}$} &
\multicolumn{1}{c}{$\nH/10^{20}$} &
\multicolumn{1}{c}{$\Tx$} &
\multicolumn{1}{c}{$Z$} &
\multicolumn{1}{c}{$Norm.$} &
\multicolumn{1}{c}{$\chi^2_{\nu}$} &
\multicolumn{1}{c}{$N_{\rm PI}$} &
\multicolumn{1}{c}{$\chi^2_{\nu,{\rm reqd.}}$} \\ 
\multicolumn{1}{l}{} &
\multicolumn{1}{l}{Num.} &
\multicolumn{1}{l}{} &
\multicolumn{1}{c}{$(')$} &
\multicolumn{1}{c}{} &
\multicolumn{1}{c}{$(\Mpc)$} &
\multicolumn{1}{c}{$(\pcmsq)$} &
\multicolumn{1}{c}{$(\keV)$} &
\multicolumn{1}{c}{$(\Solar)$} &
\multicolumn{1}{c}{$(\frac{10^{-14}}{4\pi D_{\rm L}^2} \int n_{\rm e} n_{\rm H} dV)$} &
\multicolumn{1}{c}{} &
\multicolumn{1}{c}{} &
\multicolumn{1}{c}{for CF mod.} \\ \hline \\
A1851 & 82007000 & G2G3 & $ 6.9$ & $ 0.2143$ & $ 1.83$ & $ 1.75$
 & $ 4.26^{+ 0.52}_{ -0.46}$
 & $ 0.08^{+ 0.18}_{ -0.08}$
 & $ 0.00326^{+ 0.000209}_{-0.000198}$
 & $ 0.9885$
 & $ 319$
 & $ 0.9880$ \\
A1895 & 83033000 & G2G3 & $ 8.1$ & $ 0.2250$ & $ 2.23$ & $ 1.97$
 & $ 5.17^{+ 0.57}_{ -0.52}$
 & $ 0.13^{+ 0.14}_{ -0.13}$
 & $ 0.00493^{+ 0.000212}_{-0.000211}$
 & $ 0.8868$
 & $ 175$
 & $ 0.8857$ \\
A1914 & 84032000 & G2G3 & $ 15.2$ & $ 0.1712$ & $ 3.44$ & $ 0.95$
 & $ 10.53^{+ 0.51}_{ -0.50}$
 & $ 0.25^{+ 0.06}_{ -0.06}$
 & $ 0.0280^{+ 0.000342}_{-0.000342}$
 & $ 0.9259$
 & $ 742$
 & $ 0.9258$ \\
A1942 & 83000000 & G2G3 & $ 16.9$ & $ 0.2240$ & $ 4.63$ & $ 2.58$
 & $ 5.73^{+ 0.73}_{ -0.60}$
 & $ 0.28^{+ 0.17}_{ -0.15}$
 & $ 0.00722^{+ 0.000316}_{-0.000306}$
 & $ 0.9548$
 & $ 459$
 & $ 0.9546$ \\
\hline
\end{tabular}
\end{center}
\end{table*}
\begin{table*}
\parbox{0.5\textwidth}{\vspace{0.6\baselineskip}{\bf Table \ref{table:obsresmk}.} Broad-Band Single-Phase Plasma Fit Results -- contd.}\vspace{0.6\baselineskip}
\scriptsize
\begin{center}
\begin{tabular}{llccccccccccc} \hline \\
\multicolumn{1}{l}{Object} &
\multicolumn{1}{l}{Seq.} &
\multicolumn{1}{l}{Inst.} &
\multicolumn{1}{c}{$A_{\rm out}$} &
\multicolumn{1}{c}{$z$} &
\multicolumn{1}{c}{$R_{\rm out}$} &
\multicolumn{1}{c}{$\nH/10^{20}$} &
\multicolumn{1}{c}{$\Tx$} &
\multicolumn{1}{c}{$Z$} &
\multicolumn{1}{c}{$Norm.$} &
\multicolumn{1}{c}{$\chi^2_{\nu}$} &
\multicolumn{1}{c}{$N_{\rm PI}$} &
\multicolumn{1}{c}{$\chi^2_{\nu,{\rm reqd.}}$} \\ 
\multicolumn{1}{l}{} &
\multicolumn{1}{l}{Num.} &
\multicolumn{1}{l}{} &
\multicolumn{1}{c}{$(')$} &
\multicolumn{1}{c}{} &
\multicolumn{1}{c}{$(\Mpc)$} &
\multicolumn{1}{c}{$(\pcmsq)$} &
\multicolumn{1}{c}{$(\keV)$} &
\multicolumn{1}{c}{$(\Solar)$} &
\multicolumn{1}{c}{$(\frac{10^{-14}}{4\pi D_{\rm L}^2} \int n_{\rm e} n_{\rm H} dV)$} &
\multicolumn{1}{c}{} &
\multicolumn{1}{c}{} &
\multicolumn{1}{c}{for CF mod.} \\ \hline \\
A3667 & 83054000 & G2G3 & $ 19.4$ & $ 0.0585$ & $ 1.79$ & $ 4.00$
 & $ 7.13^{+ 0.14}_{ -0.14}$
 & $ 0.25^{+ 0.03}_{ -0.03}$
 & $ 0.0943^{+ 0.000730}_{-0.000728}$
 & $ 0.9051$
 & $ 903$
 & $ 0.9050$ \\
A3921 & 83048010 & G2G3 & $ 18.2$ & $ 0.0784$ & $ 2.18$ & $ 2.94$
 & $ 5.73^{+ 0.24}_{ -0.23}$
 & $ 0.20^{+ 0.06}_{ -0.06}$
 & $ 0.0251^{+ 0.000435}_{-0.000429}$
 & $ 0.9901$
 & $ 548$
 & $ 0.9899$ \\
A4059 & 82030000 & G2G3 & $ 19.4$ & $ 0.0478$ & $ 1.49$ & $ 1.10$
 & $ 3.90^{+ 0.07}_{ -0.07}$
 & $ 0.50^{+ 0.04}_{ -0.04}$
 & $ 0.0441^{+ 0.000545}_{-0.000543}$
 & $ 0.9779$
 & $ 866$
 & $ 0.9778$ \\
2A0336 & 82029000 & G2G3 & $ 19.4$ & $ 0.0350$ & $ 1.12$ & $ 17.20$
 & $ 3.08^{+ 0.03}_{ -0.03}$
 & $ 0.39^{+ 0.03}_{ -0.03}$
 & $ 0.124^{+ 0.00130}_{ -0.00131}$
 & $ 1.0012$
 & $ 885$
 & $ 1.0011$ \\
\hline
\end{tabular}
\end{center}
\end{table*}

\cleardoublepage
\renewcommand{\baselinestretch}{1.5}
\begin{table*}
\caption{\label{table:obsrescf}Broad-Band Cooling Flow Spectral Fit Results}
\scriptsize
\begin{center}
\begin{tabular}{llccccccccccc} \hline \\
\multicolumn{1}{l}{Object} &
\multicolumn{1}{l}{Seq.} &
\multicolumn{1}{l}{Inst.} &
\multicolumn{1}{c}{$A_{\rm out}$} &
\multicolumn{1}{c}{$z$} &
\multicolumn{1}{c}{$R_{\rm out}$} &
\multicolumn{1}{c}{$\nH/10^{20}$} &
\multicolumn{1}{c}{$\Tx$} &
\multicolumn{1}{c}{$Z$} &
\multicolumn{1}{c}{$Norm.$} &
\multicolumn{1}{c}{$\Mdot$} &
\multicolumn{1}{c}{$\chi^2_{\nu}$} &
\multicolumn{1}{c}{$N_{\rm PI}$} \\ 
\multicolumn{1}{l}{} &
\multicolumn{1}{l}{Num.} &
\multicolumn{1}{l}{} &
\multicolumn{1}{c}{$(')$} &
\multicolumn{1}{c}{} &
\multicolumn{1}{c}{$(\Mpc)$} &
\multicolumn{1}{c}{$(\pcmsq)$} &
\multicolumn{1}{c}{$(\keV)$} &
\multicolumn{1}{c}{$(\Solar)$} &
\multicolumn{1}{c}{$(\frac{10^{-14}}{4\pi D_{\rm L}^2} \int n_{\rm e} n_{\rm H} dV)$} &
\multicolumn{1}{c}{$(\Msunpyr)$} &
\multicolumn{1}{c}{} &
\multicolumn{1}{c}{} \\ \hline \\ 
A85 & 81024000 & G2G3 & $ 19.4$ & $ 0.0521$ & $ 1.61$ & $ 3.01$
 & $ 6.47^{+ 0.33}_{ -0.29}$
 & $ 0.39^{+ 0.02}_{ -0.02}$
 & $ 0.0910^{+ 0.00514}_{ -0.00557}$
 & $ 147^{+ 72}_{ -72}$
 & $ 1.0737$
 & $1068$ \\
    & 81024010 & G2G3 & $ 19.4$ &   & $ 1.61$ &  
 & $ 7.35^{+ 0.82}_{ -0.57}$
 & $ 0.38^{+ 0.04}_{ -0.04}$
 & $ 0.0732^{+ 0.00923}_{ -0.0102}$
 & $ 341^{+ 93}_{ -104}$
 & $ 0.9598$
 & $ 836$ \\
   & Combined & & & &  & & $ 6.74\pm  0.50$ & $ 0.38\pm  0.03$ & $ 0.0846\pm  0.00753$& $ 229\pm  85$& \\
A119 & 83045000 & G2G3 & $ 19.4$ & $ 0.0443$ & $ 1.39$ & $ 3.45$
 & $ 6.05^{+ 0.55}_{ -0.43}$
 & $ 0.27^{+ 0.03}_{ -0.04}$
 & $ 0.0436^{+ 0.00473}_{ -0.00445}$
 & $ 52^{+ 42}_{ -51}$
 & $ 1.1024$
 & $ 907$ \\
\hline
\end{tabular}
\end{center}
\end{table*}
\begin{table*}
\parbox{0.5\textwidth}{\vspace{0.6\baselineskip}{\bf Table \ref{table:obsrescf}.} Broad-Band Cooling Flow Spectral Fit Results -- contd.}\vspace{0.6\baselineskip}
\scriptsize
\begin{center}
\begin{tabular}{llccccccccccc} \hline \\
\multicolumn{1}{l}{Object} &
\multicolumn{1}{l}{Seq.} &
\multicolumn{1}{l}{Inst.} &
\multicolumn{1}{c}{$A_{\rm out}$} &
\multicolumn{1}{c}{$z$} &
\multicolumn{1}{c}{$R_{\rm out}$} &
\multicolumn{1}{c}{$\nH/10^{20}$} &
\multicolumn{1}{c}{$\Tx$} &
\multicolumn{1}{c}{$Z$} &
\multicolumn{1}{c}{$Norm.$} &
\multicolumn{1}{c}{$\Mdot$} &
\multicolumn{1}{c}{$\chi^2_{\nu}$} &
\multicolumn{1}{c}{$N_{\rm PI}$} \\ 
\multicolumn{1}{l}{} &
\multicolumn{1}{l}{Num.} &
\multicolumn{1}{l}{} &
\multicolumn{1}{c}{$(')$} &
\multicolumn{1}{c}{} &
\multicolumn{1}{c}{$(\Mpc)$} &
\multicolumn{1}{c}{$(\pcmsq)$} &
\multicolumn{1}{c}{$(\keV)$} &
\multicolumn{1}{c}{$(\Solar)$} &
\multicolumn{1}{c}{$(\frac{10^{-14}}{4\pi D_{\rm L}^2} \int n_{\rm e} n_{\rm H} dV)$} &
\multicolumn{1}{c}{$(\Msunpyr)$} &
\multicolumn{1}{c}{} &
\multicolumn{1}{c}{} \\ \hline \\ 
A1851 & 82007000 & G2G3 & $ 6.9$ & $ 0.2143$ & $ 1.83$ & $ 1.75$
 & $ 4.23^{+ 0.60}_{ -0.42}$
 & $ 0.08^{+ 0.18}_{ -0.08}$
 & $ 0.00326^{+ 0.000187}_{-0.000599}$
 & $ 0^{+ 150}_{ 0}$
 & $ 0.9916$
 & $ 319$ \\
A1895 & 83033000 & G2G3 & $ 8.1$ & $ 0.2250$ & $ 2.23$ & $ 1.97$
 & $ 5.18^{+ 1.70}_{ -0.52}$
 & $ 0.13^{+ 0.14}_{ -0.13}$
 & $ 0.00492^{+ 0.000735}_{ -0.00193}$
 & $ 0^{+ 411}_{ 0}$
 & $ 0.8920$
 & $ 175$ \\
A1914 & 84032000 & G2G3 & $ 15.2$ & $ 0.1712$ & $ 3.44$ & $ 0.95$
 & $ 11.70^{+ 3.57}_{ -1.58}$
 & $ 0.26^{+ 0.07}_{ -0.06}$
 & $ 0.0252^{+ 0.00312}_{ -0.00586}$
 & $ 256^{+ 390}_{ -256}$
 & $ 0.9266$
 & $ 742$ \\
A1942 & 83000000 & G2G3 & $ 16.9$ & $ 0.2240$ & $ 4.63$ & $ 2.58$
 & $ 15.60^{+ 3.44}_{ -4.65}$
 & $ 0.30^{+ 0.20}_{ -0.19}$
 & $( 3.44^{+ 6.94}_{ -3.44})\times 10^{-08}$
 & $ 817^{+ 118}_{ -741}$
 & $ 0.9543$
 & $ 459$ \\
\hline
\end{tabular}
\end{center}
\end{table*}
\begin{table*}
\parbox{0.5\textwidth}{\vspace{0.6\baselineskip}{\bf Table \ref{table:obsrescf}.} Broad-Band Cooling Flow Spectral Fit Results -- contd.}\vspace{0.6\baselineskip}
\scriptsize
\begin{center}
\begin{tabular}{llccccccccccc} \hline \\
\multicolumn{1}{l}{Object} &
\multicolumn{1}{l}{Seq.} &
\multicolumn{1}{l}{Inst.} &
\multicolumn{1}{c}{$A_{\rm out}$} &
\multicolumn{1}{c}{$z$} &
\multicolumn{1}{c}{$R_{\rm out}$} &
\multicolumn{1}{c}{$\nH/10^{20}$} &
\multicolumn{1}{c}{$\Tx$} &
\multicolumn{1}{c}{$Z$} &
\multicolumn{1}{c}{$Norm.$} &
\multicolumn{1}{c}{$\Mdot$} &
\multicolumn{1}{c}{$\chi^2_{\nu}$} &
\multicolumn{1}{c}{$N_{\rm PI}$} \\ 
\multicolumn{1}{l}{} &
\multicolumn{1}{l}{Num.} &
\multicolumn{1}{l}{} &
\multicolumn{1}{c}{$(')$} &
\multicolumn{1}{c}{} &
\multicolumn{1}{c}{$(\Mpc)$} &
\multicolumn{1}{c}{$(\pcmsq)$} &
\multicolumn{1}{c}{$(\keV)$} &
\multicolumn{1}{c}{$(\Solar)$} &
\multicolumn{1}{c}{$(\frac{10^{-14}}{4\pi D_{\rm L}^2} \int n_{\rm e} n_{\rm H} dV)$} &
\multicolumn{1}{c}{$(\Msunpyr)$} &
\multicolumn{1}{c}{} &
\multicolumn{1}{c}{} \\ \hline \\ 
A3667 & 83054000 & G2G3 & $ 19.4$ & $ 0.0585$ & $ 1.79$ & $ 4.00$
 & $ 8.11^{+ 0.82}_{ -0.73}$
 & $ 0.25^{+ 0.03}_{ -0.03}$
 & $ 0.0810^{+ 0.00867}_{ -0.00880}$
 & $ 196^{+ 106}_{ -122}$
 & $ 0.9038$
 & $ 903$ \\
A3921 & 83048010 & G2G3 & $ 18.2$ & $ 0.0784$ & $ 2.18$ & $ 2.94$
 & $ 9.96^{+ 7.83}_{ -2.74}$
 & $ 0.19^{+ 0.08}_{ -0.07}$
 & $ 0.0101^{+ 0.00771}_{ -0.0101}$
 & $ 319^{+ 55}_{ -115}$
 & $ 0.9824$
 & $ 548$ \\
A4059 & 82030000 & G2G3 & $ 19.4$ & $ 0.0478$ & $ 1.49$ & $ 1.10$
 & $ 4.05^{+ 0.23}_{ -0.19}$
 & $ 0.49^{+ 0.04}_{ -0.04}$
 & $ 0.0416^{+ 0.00282}_{ -0.00360}$
 & $ 51^{+ 65}_{ -51}$
 & $ 0.9783$
 & $ 866$ \\
2A0336 & 82029000 & G2G3 & $ 19.4$ & $ 0.0350$ & $ 1.12$ & $ 17.20$
 & $ 3.15^{+ 0.10}_{ -0.08}$
 & $ 0.38^{+ 0.03}_{ -0.03}$
 & $ 0.118^{+ 0.00597}_{ -0.00674}$
 & $ 75^{+ 89}_{ -75}$
 & $ 1.0015$
 & $ 885$ \\
\hline
\end{tabular}
\end{center}
\end{table*}

\cleardoublepage

\section{Temperature And Abundance Profile Results}\label{section:resfig}

In the following figures we plot the averaged radial temperature and
metallicity profiles (created using the procedure detailed in
Section~\ref{section:errimg}) for the \nsample\ clusters in our
sample.  The y-axis data have 1-$\sigma$ error bars, and the x-axis
error-bars represent the extrema of the radii for each annular
bin. Note, the bottom of each panel shows the radius in $\arcmin$
while the top shows the radius is $\Mpc$ assuming $\HO{50}$,
$\qO{0.5}$.

The solid (heavier) diamond symbols with with diamond error bars are
our averaged \SID\ results. The triangle symbol with a dot-dashed
line for the limits which cover the whole radial range of the data are
the broad-beam fits to cooling flow model.

Results from \citeN{Markevitch:ASCA_Tx_similarity} (and the papers
referenced therein which use the same method) [A85, A119, A399, A401,
A478, A665, A754, A780, A1650, A1651, A1795, A2029, A2065, A2142,
A2163, A2256, A2319, A2657, A3112, A3266, A3376, A3391, A3395, A3571,
A3667, A4059, AWM7, Cygnus-A, MKW3s, A3558, Triangulum Australis] are
shown by cross symbols with conventional error-bar limits as a dotted
line.  Additional data from \ROSAT\ [A478, A1795, A3526] and Beppo-SAX
[A426, A2319, Virgo] are shown by square symbols with a dashed line.
(References for the \ROSAT\ and Beppo-SAX data are given at the end of
Section~\ref{section:spectral}.)

  \resfig


\end{document}